%% file: main.tex
\documentclass[12pt]{nmsuth01}
\usepackage{latexsym}
\usepackage{amsmath,amsthm}
\usepackage{mathtools}
\usepackage{amssymb}
\usepackage{braket}
\usepackage[all]{xy}
\usepackage{graphicx}
\usepackage{caption}
\usepackage{subcaption}
\usepackage[colorlinks=true,linktocpage=true,linkcolor=blue,citecolor=blue]{hyperref}

\usepackage[margin=1in]{geometry}

\usepackage{multirow}

\theoremstyle{plain}

\theoremstyle{definition}
\theoremstyle{remark}

\newtheorem*{convention*}{Convention}
\newlength{\doublespace}
\usepackage{float}

\usepackage[usenames,dvipsnames]{color}

\usepackage[usenames,dvipsnames]{color}

\usepackage[usenames,dvipsnames]{color}

\extrafloats{100}

\usepackage{gensymb}

\usepackage{feynman} 

\begin{document}
\setlength{\baselineskip}{\doublespace}
\pagenumbering{roman}
\pagestyle{empty}

%
\include{title}
\pagestyle{plain}
%
\include{approval}
%
\include{dedica}
%
\include{ackno}
%
\include{vita}
\include{abstract}

\tableofcontents
\newpage



\clearpage
\phantomsection
\addcontentsline{toc}{section}{LIST OF FIGURES}
\listoffigures
\newpage

\pagenumbering{arabic}
%

%
\input{chp1_Intro}

\newpage
\input{chp2_Cherenkov}

\newpage
\input{chp3_Jetdrift}
\newpage
\input{chp4_Conclusion}
\newpage
%

\section*{Acknowledgments}

This material is based upon work supported by the U.S. Department of Energy, Office of Science, Office of Nuclear Physics under Award Number DE-SC0024560.  This work utilized resources from the New Mexico State University High Performance Computing Group, which is directly supported by the National Science Foundation (OAC-2019000), the Student Technology Advisory Committee, and New Mexico State University and benefits from inclusion in various grants (DoD ARO-W911NF1810454; NSF EPSCoR OIA-1757207; Partnership for the Advancement of Cancer Research, supported in part by NCI grants U54 CA132383 (NMSU)).  This research also utilized services provided by the OSG Consortium \cite{osg07, osg09, https://doi.org/10.21231/906p-4d78, https://doi.org/10.21231/0kvz-ve57}, which is supported by the National Science Foundation awards \#2030508 and \#1836650.
The open-source APE software suite and associated resources are publicly available on \href{https://github.com/Jopacabra/ape}{GitHub}.

\newpage
\addcontentsline{toc}{section}{REFERENCES}
\bibliographystyle{custom_style}
\bibliography{references, Refs} 
\newpage 



\clearpage
\appendix

\addcontentsline{toc}{section}{APPENDIX} 

\input{chp5_Appendix}
\input{chp6_glossary} 
\end{document}

%% file: title.tex
\renewcommand{\rmdefault}{2}
~\\~\\
\vspace{-48pt}
\begin{center}

MEDIUM CHARACTERIZATION WITH HARD PROBES: \\ FROM CHERENKOV LIGHT IN QED TO JET DRIFT IN QCD \\

\vspace{0.075in}
BY\\
\vspace{-10pt}
HASAN REJOANUR RAHMAN, B.S., M.S.
\end{center}

\vspace{.65in}
\begin{center}
A dissertation submitted to the Graduate School\\
\vspace{-10pt}
in partial fulfillment of the requirements\\
\vspace{-10pt}
for the degree \\
\vspace{-10pt}
Doctor of Philosophy
\end{center}

\vspace{.65in}
\begin{center}
Major: Physics\\
\vspace{-10pt}
Concentration: Nuclear Theory \\
\vspace{-10pt}
\end{center}

\vspace{.65in}
\begin{center}
NEW MEXICO STATE UNIVERSITY\\
\vspace{-10pt}
LAS CRUCES, NEW MEXICO\\
\vspace{-10pt}
(April 2026)
\end{center}

%% file: approval.tex
\noindent

\setlength{\baselineskip}{\doublespace}
\vspace{0.1in}
\begin{flushleft}

\small
Hasan Rejoanur Rahman

\vspace{4pt}
\hrule
\vspace{4pt}

\large
\textit{Candidate}

\vspace{0.3in}

\small
\textit{Physics}

\vspace{4pt}
\hrule
\vspace{4pt}

\large
\textit{Major}

\vspace{0.4in}

\small
This Dissertation is approved on behalf of the faculty of New Mexico State University, and it is acceptable in quality and form for publication:
\begin{center} 
\end{center}


\large
\textit{Approved by the thesis Committee:}

\vspace{0.2in}

\small
\textit{Matthew D. Sievert}
\vspace{4pt}
\hrule
\vspace{4pt}
\large
\textit{Chairperson}

\vspace{0.2in}

\small
\textit{Vassili Papavassiliou}
\vspace{4pt}
\hrule
\vspace{4pt}
\large
\textit{Committee Member}

\vspace{0.2in}

\small
\textit{Moire Prescott}
\vspace{4pt}
\hrule
\vspace{4pt}
\large
\textit{Committee Member}

\vspace{0.2in}

\small
\textit{Ludi Miao}
\vspace{4pt}
\hrule
\vspace{4pt}
\large
\textit{Committee Member}

\vspace{0.2in}


\end{flushleft}






%% file: dedica.tex
\begin{center}
DEDICATION
\end{center}

I dedicate this work to my late father Mohammad Khalilur Rahman Miah, mother Hosne Ara Bhuiya, and brother Hasan Shahriar Rahman. 


%% file: ackno.tex
\begin{center}
ACKNOWLEDGMENTS
\end{center}



I want to thank my advisor, Dr.~Matthew Sievert, and co-advisor, Dr.~Vassili Papavassiliou, for their encouragement, motivation, and patience throughout this project. Dr.~Sievert provided extensive guidance during all phases of the research discussed in this work and greatly facilitated the timely completion of this dissertation. I would also like to thank all of our collaborators, including Dr.~Ivan Vitev from Los Alamos National Laboratory. I wish to remember and thank all of my teachers from elementary school through the graduate level, especially Dr.~Khorshed Ahmad Kabir, who was my greatest motivation for pursuing graduate studies and a career in high-energy nuclear physics. I also thank Dr.~Arshad Momen, Dr.~Kamrul Hassan, Dr.~Jose Ba\~nuelos, Dr.~Marian Manciu, Dr.~Heinz Nakotte, Dr.~Michael Engelhardt, Dr.~Moire Prescott, and many others for their inspirational roles along my academic journey. A special thanks goes to my colleague and friend, Dr.~Joseph Bahder, for helping me above and beyond with our simulation package (APE) and for insightful discussions on heavy-ion collisions and jet physics. Moreover, my friend Manuel Ca\~nas deserves special thanks for providing initial help and tutorials on programming in Python, which were instrumental for the computational portion of this work. 


I want to express deepest gratitude to my family: my late father Mohammad Khalilur Rahman Miah, mother Hosne Ara Bhuiya, brother Hasan Shahriar Rahman, sister-in-law Sara Ahmed, wife Mehjabeen Sultana; my in-laws Mohammad Mahbubul Alam and Syeda Sharmin Sultana; and my nephew Shayer, nieces Shayna and Shanaya – for their endless support, love, and understanding. I also thank all of my uncles and aunts, Shukur Mahmud, Abul Kalam, Jahanara Bhuiya, Arfan Bhuiya, Amena Begum, Abul Kashem, Bedana Begum, Rokeya Begum, Rowshon Ara Bhuiya, Arman Bhuiya, Ajharul Islam, Rehena Bhuiya, and Arif Bhuiya, for their guidance and support. I am also grateful to all of my cousins, including Dalour Beg, Hashem Beg, Badrun Nahar, Jerin Jahan, Salahuddin Sumon, and Afsana Nijhu; and to my friends, Mahmudul Hassan Bhuiyan, Fahad Hasan, Tahsin Hassan Prithu, Naeem Hasan, Mir Mehedi Faruk, Fatema Farjana, Nowmi Tabassum, Shouvik Sarker, Ariful Haque, Tahmina Akter, Tahneen Jahan Neelam, Robbie Billings, Weeam Albaltan, Wajdi Al Smadi, Bryson Stemock, Humairat Rahman, Mahtab Uddin Munna, Md.~Minuddin, Rony, Sajid, Tomal, Hector, Sandra, Abhi, Javed, Jake, Jimmy, Raisul, Koushik, Dipon, Bony, Sarhan, and many others, for their encouragement and support. Lastly, I would like to mention my cats Negro, Luna, and Luna's kittens, Starfish and Calypso, for their love, patience and wonderful company. 


%% file: vita.tex
\begin{center}
            VITA
\end{center}
\begin{flushleft}
\begin{tabular}{ll}
June 6, 1987 &  Born at Tangail, Bangladesh
\\
& \\
2005-2010        &  B.S., University of Dhaka, Dhaka
\\
& \\
2016-2018        &  M.S., \& Graduate Teaching Assistant, Dept. of Physics
\\
                 &  The University of Texas at El Paso, El Paso, Texas.  
\\

& \\
2018-2021        &  Ph.D. \& Graduate Teaching Assistant, Department of Physics, 
\\               &  New Mexico State University, Las Cruces, New Mexico.

\& \\
2021-2026        &  Ph.D. Candidate \& Graduate Research Assistant, Department of Physics, 
\\               &  New Mexico State University, Las Cruces, New Mexico. 
\\
\end{tabular}
\end{flushleft}
\vspace{0.1in}
\begin{center}
PROFESSIONAL  AND HONORARY SOCIETIES
\end{center}
\begin{flushleft}

\begin{itemize}
\item   American Physical Society (APS)
\item   Physics Graduate Students Organization (PGSO), New Mexico State University
\item   Bangladesh Astronomical Society
\end{itemize}

\end{flushleft}
\vspace{0.1in}
\begin{center}PUBLICATIONS AND PRESENTATIONS
\end{center}

\begin{itemize}

\item   Bahder, J., Rahman, H., Sievert, M. D., \& Vitev, I. (2026). Signatures of jet drift in quark-gluon plasma hard-probe observables. Physical Review Research, 8(1), L012016.

\item   Vitev, I. M., Rahman, H., Bahder, J., \& Sievert, M. (2025, May). Jet Drift in Heavy-Ion Collision: Acoplanarity and v2. In EPJ Web of Conferences (Vol. 339, No. LA-UR-25-24325). Los Alamos National Laboratory (LANL), Los Alamos, NM (United States).

\item   Bahder, J., Rahman, H., Sievert, M., \& Vitev, I. (2025). Jet Drift in Heavy Ion Collisions. In EPJ Web of Conferences (Vol. 339, p. 02009). EDP Sciences.

\item   Antiporda, L., Bahder, J., Rahman, H., \& Sievert, M. D. (2022). Jet drift and collective flow in heavy-ion collisions. Physical Review D, 105(5), 054025.


\item   Sievert, M., Rahman, H., Bahder, J., \& Antiporda, L. (2021). Asymmetric Energy Loss of Jets Due to Collective Flow in Heavy-Ion Collisions. In APS Division of Nuclear Physics Meeting Abstracts (Vol. 2021, pp. EH-002).

\item   Rahman, H., \& Banuelos, J. (2019). Solvent and Concentration Effects Governing the Hierarchical Organization of Asphaltenes: A Small-Angle X-Ray Scattering Study. Bulletin of the American Physical Society, 64.



\end{itemize}

\hspace{\parindent}


\begin{center}
FIELD OF STUDY
\end{center}
\begin{flushleft}
Major Field: Theoretical Nuclear \& High-Energy Physics
\newline

\end{flushleft}

%% file: abstract.tex
\begin{center}
ABSTRACT
\end{center}
\vspace{0.3in}
\begin{center}
MEDIUM CHARACTERIZATION WITH HARD PROBES: \\ FROM CHERENKOV LIGHT IN QED TO JET DRIFT IN QCD \\

\vspace{0.075in}
BY\\
\vspace{-10pt}
HASAN REJOANUR RAHMAN, B.S., M.S.
\end{center}

\vspace{0.3in}
\begin{center}
Doctor of Philosophy

New Mexico State University

Las Cruces, New Mexico, 2026

Dr.~Matthew D.~Sievert, Advisor

\end{center}
\vspace{0.3in}
\hspace{\parindent}

This dissertation presents a unified framework for medium characterization with hard probes spanning from Cherenkov light in quantum electrodynamics (QED) to jet drift in quantum chromodynamics (QCD). 
We first develop a dispersive fit to the refractive index $n(\lambda)$ of liquid argon (LAr) by incorporating anomalous  dispersion at the 106.6 nm resonance for the first time. We show that the angular distribution of Cherenkov radiation is highly sensitive to the peak of the refractive index and contributes a significant excess over isotropic scintillation in certain angular bins.  This work is important for precision Particle Identification (PID) for experiments like DUNE and CCM.
Transitioning to high-energy nuclear collisions, we utilize ``jet drift'' -- the flow-induced deflection of partons -- as a tomographic probe of the Quark-Gluon Plasma (QGP). Using the Anisotropic Parton Evolution (APE) Monte Carlo simulation across various collision systems (PbPb, AuAu, and UU), we disentangle how the jet modification depends on medium size, temperature, and geometry. We show that jet drift exhibits distinct systematics in observables like the elliptic flow ($v_2$) and dihadron acoplanarity ($\Delta\phi$), which helps disentangle it from conventional energy loss. 
Together, these studies demonstrate how the angular and kinematic signatures of hard probes revolutionize our ability to resolve the fundamental properties of matter.

%% file: chp1_Intro.tex
\section{INTRODUCTION} \label{intro}
\hspace{\parindent}


%
\subsection{Hard Probes of a Medium}
%

``Hard probes'' is a term that generally refers to high-energy or high-momentum  particles that propagate through a medium.  Underlying the concept of a hard probe is the physics principle of separation of energy scales, which makes it possible to distinguish between the hard probe and the medium itself.  Hard probes are not a part of the medium itself, but their interactions with the medium can be used to extract information about it. The ways in which the properties of a hard probe are modified by the medium can, in principle, tell us about the medium it propagates through.  Some of the common ways that hard probes can interact with a medium are through collisional energy loss, collisional transverse momentum diffusion, and stimulated radiation.

The origin of the method of using a known high-energy probe to characterize an unknown medium may be traced back at least as far as Rutherford's gold foil experiment, in which he bombarded a gold foil with $\alpha$ and $\beta$ radiation to learn about the atomic structure of gold \cite{Rutherford1911}.  Today, hard probes are used to characterize exotic media like the quark-gluon plasma (QGP) produced in ultra-relativistic collisions of heavy ions like gold or lead.  Characterizing the QGP using hard probes has been a cornerstone of high-energy nuclear physics for over two decades.  The interactions of hard probes with a medium are also used as an experimental basis for particle identification (PID) in high-energy collisions, such as through the Cherenkov radiation which is stimulated as they pass through the detector.

In this dissertation, we will present two studies of hard probes and their potential for medium characterization.  The first one is the classic example of Cherenkov light, a hallmark of quantum electrodynamics (QED).  Cherenkov light is a specific type of medium-induced photon radiation that only occurs when a particle moves faster than the speed of light in a medium.  Because it is very sensitive to the speed of light in the medium, the Cherenkov radiation produced by hard probes carries useful information about the medium's optical properties.  These include the refractive index $n(\lambda)$ and the absorption coefficient $\alpha(\lambda)$, both of which are in general dispersive (frequency dependent).  In Chapter.~\ref{cherenkov}, we will show how Cherenkov yield and angular distributions in liquid argon can be used to characterize the refractive index $n(\lambda)$ and consider the implications for particle identification (PID).

The second study of medium characterization using hard probes that we will explore is the quenching and transverse momentum drift of hard probes in the QGP produced in heavy-ion collisions.  Jet quenching is a phenomenon in which high-energy quarks and gluons undergo energy loss in the QGP, largely due to medium-induced bremsstrahlung radiation of gluons in quantum chromodynamics (QCD) called the Landau-Pomeranchuk-Migdal effect \cite{Landau:1953um, Migdal:1956tc}.  Transverse momentum diffusion also occurs when jets propagate through the QGP, including both isotropic broadening and anisotropic ``jet drift'' in the direction of the local fluid velocity \cite{Sadofyev:2021ohn}. ``Jet tomography'' based on jet quenching has been a cornerstone of heavy-ion physics for over two decades. In Chapter.~\ref{jetdrift}, we will show how the novel physics of jet drift can be used to characterize the distinct effects of QGP temperature, path length, and geometry.

The unifying theme of this dissertation is that hard probes can be used to extract medium information across a range of conditions and energy scales: from QED interactions at the MeV scale to QCD interactions at tens of GeV.  Despite these differences, the fundamental physics of hard probes is largely universal: energy loss, transverse momentum diffusion, and stimulated radiation.  The theory of hard probes in a medium is fairly simple, yet also so powerful that it can be applied to a wide range of physics questions and energy scales.  This universality makes it possible to recognize the fundamental similarities between superficially different problems and identify new opportunities for innovation, such as the ones described in this dissertation.


%
\subsection{Characterizing LAr with Cherenkov Radiation}
%

%
\subsubsection{Neutrinos: A Window to the Mysteries of the Universe} 
%

Neutrinos are fundamental particles that are abundantly produced in the core of young stars, supernovae, and from cosmic rays that bombard every inch of the Earth’s surface at nearly the speed of light. They can also be created artificially in nuclear reactors and particle accelerators. Despite being abundant in the universe, neutrinos can penetrate through large amounts of matter without interacting, which makes them difficult to detect. Nevertheless, neutrinos hold clues to some of the biggest questions in physics: the origin of matter, unification of forces, black hole formation, and more (see, e.g., Ref.~\cite{Valle:2013}). The majority of the neutrinos detected on Earth originate from nuclear reactions taking place in our sun \cite{Bahcall_2005}.

Wolfgang Pauli first postulated the neutrinos in 1930 to explain how beta decay could conserve energy, momentum, and angular momentum \cite{1978PhT....31i..23B}. He considered neutrinos to be emitted from the nucleus together with the electron in the beta decay process. Later, in Fermi's theory of beta decay, a neutron could decay to a proton, electron, and the smaller neutral particle now called an electron antineutrino: $n^0 \longrightarrow p^+ + e^- + \bar{\nu_e}$ \cite{1934ZPhy...88..161F}.

Neutrinos are fermions, a spin $\frac{1}{2}$ elementary particle described by the Standard Model of Particle Physics, denoted by $\nu$. Unlike their charged leptonic counterparts (electron, muon, and tau), neutrinos do not have any electric charge or color charge, which restricts them to interacting via the weak interaction and gravity only. For this reason, the elusive neutrinos are a major focus of research in nuclear and particle physics.  For each neutrino, there also exists a corresponding antiparticle, called an antineutrino, which also has spin of $\frac{1}{2}$ and no electric charge. Antineutrinos are distinguished from neutrinos by having opposite-signed lepton number and weak isospin, and right-handed instead of left-handed chirality. To conserve total lepton number in a nuclear beta decay, electron neutrinos only appear together with positrons (anti-electrons), whereas electron antineutrinos only appear with electrons \cite{FClose:2012, RJayawardhana:2015}.

Since neutrinos provide independent proof of physics beyond the Standard Model (BSM), they are essential to our comprehension of the cosmos. Neutrinos are numerous, constituting the second most prevalent elementary particle in the cosmos (after photons). They played an important role in galaxy formation and setting up the large-scale structures in the early universe despite their tiny masses \cite{LESGOURGUES2006307}. As neutrinos decoupled from other matter just seconds after the Big Bang, the Cosmic Neutrino Background (CNB) carries information from a much earlier epoch of the universe compared to the well-known Cosmic Microwave Background (CMB) counterpart (which formed 380,000 years later) \cite{Lesgourgues:2006nd, Weinberg:2008zzc}. Neutrinos may also carry answers to questions of leptogenesis related to the observed asymmetry between matter and antimatter \cite{FUKUGITA198645}. Many fundamental questions still remain about neutrinos, including their absolute masses (despite the $\sim$ eV upper limits set by experiments like KATRIN \cite{KATRIN:2024cdt}). Whether they are Majorana or Dirac fermions and whether beyond the Standard Model ``sterile neutrinos'' exist, neutrinos persist as a focal point for the most significant enduring questions in modern physics.



%
\vspace{0.25cm}
\noindent\textit{Interactions of Neutrinos with Matter} 
%

The weak interaction is the main method of neutrino detection, consisting of both neutral-current interactions (mediated by the $Z$ boson) and charged-current interactions (mediated by the $W^\pm$ bosons).  

In a neutral-current interaction, the neutrino enters and then leaves the detector after having transferred some of its energy and momentum to a `target' particle.  If the interaction takes place in a scintillation detector, then the recoil of a scattered charged particle can be detected through the subsequent emission of scintillation light.  Moreover, if the target charged particle is sufficiently lightweight (e.g.~an electron), it might be accelerated to a relativistic speed and consequently emit Cherenkov radiation as well. All three neutrino flavors (electronic, muonic, and tauonic) can participate, regardless of the neutrino energy.  Neutral-current interactions occur equally for all neutrino flavors, while charged-current interactions produce different charged leptons depending on the flavor of the neutrino. \cite{Weinberg:1995mt}

In a charged current interaction, a high-energy neutrino transforms into its partner lepton (electron, muon, or tauon). However, if the neutrino does not have sufficient energy to create its heavier partner's mass, the charged current interaction is effectively unavailable to it. Neutrinos from the Sun and from nuclear reactors have enough energy to create electrons. Most accelerator-created neutrino beams can also create muons, and a very few can create tauons. A detector which can distinguish among these leptons can reveal the flavor of the neutrino incident to a charged current interaction; because the interaction involves the exchange of a W boson, the `target' particle also changes. An example is Neutrino-initiated inverse beta decay $\nu_e + n \rightarrow e^- + p^+$ \cite{Weinberg:1995mt}.

Neutrino interactions with matter are categorized into several key processes: Coherent Elastic Neutrino-Nucleus Scattering (CE$\nu$NS), Quasi-Elastic (QE) scattering, Inverse Beta Decay (IBD), and Resonance Production (RES). In liquid argon, the neutral-current CE$\nu$NS process occurs at low energies ($E_{\nu} \sim 50$ MeV) and acts as a ``standard candle'' measurement \cite{PhysRevD.9.1389, coherentcollaboration2016coherent}. At higher energies, inelastic interactions such as IBD become prominent; these charged-current processes trigger nucleon conversion and particle ejection.  Specifically, QE scattering (including IBD) is a vital source of correlated, low-energy protons in experimental detections \cite{Strumia_2003, Sobczyk_2011}. Finally, Resonance Production (RES), a charged current process, dominates at higher energy scales, typically between 0.5 GeV and 3.0 GeV \cite{athar2022neutrinosinteractionsmatter}.

%
\vspace{0.25cm}
\noindent\textit{Experimental Detection of Neutrinos} 
%

Numerous neutrino experiments are operating worldwide to detect different kinds of neutrinos with different energy ranges. The detectors can be classified by the material used to detect the neutrino interactions.  From this point of view, most of them are either scintillation-based detectors (CCM, JUNO, MiNER$\nu$A, NO$\nu$A, $SNO^+$ etc) or Cherenkov-based detectors (IceCube, MiniBooNE, Hyper-Kamiokande, T2K, SNO etc) but there are also detectors based on Time Projection Chambers (TPCs) (DUNE, ICARUS, MicroBooNE, etc), semiconductors, radiochemicals, and nuclear emulsions \cite{Katori:2016yel, Formaggio_2012}.  The various scintillation-based detectors use different kinds of scintillating materials, including argon, xenon, and mineral oil, in different states (solid, liquid, or gaseous). 
%
\begin{figure}[t]
\begin{centering}
\includegraphics[width=0.6\textwidth]{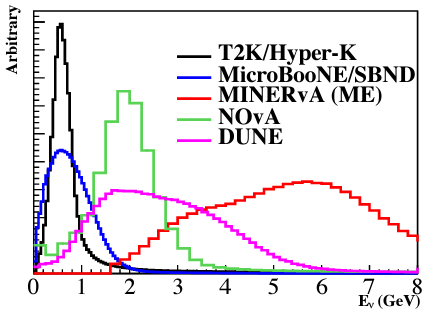}
\caption{Muon neutrino and muon anti-neutrino flux predictions from current and future accelerator based neutrino experiments (Hyper-Kamiokande, MicroBooNE, NOvA, DUNE, MINERvA with medium energy NuMI beamline) \cite{Katori:2016yel, Formaggio_2012}
\label{f:neutrinoexpeflux}
}
\end{centering}
\end{figure}
For example, scintillation-based detectors are often used to detect the nuclear recoil of argon atoms in a CE$\nu$NS process (see Ref.~\cite{PhysRevD.9.1389} for details), while Cherenkov detectors are intended to measure the lighter charged particles (like muons) produced directly in charged-current interactions like IBD. The energy distribution of neutrino fluxes of different ongoing and future neutrino experiment programs is shown in Fig.~\ref{f:neutrinoexpeflux}.  

Two examples of important modern neutrino experiments are the Coherent CAPTAIN-Mills (CCM) and the Deep Underground Neutrino Experiment (DUNE) (see MS thesis \cite{rahman2024} for details). The Coherent CAPTAIN (Cryogenic Apparatus for Precision Tests of Argon Interactions with Neutrinos)-Mills (CCM) experiment consists of an 800 MeV proton beam, a tungsten target, and a 10-ton liquid argon scintillation detector located at the Los Alamos Neutron Science Center at Los Alamos National Laboratory \cite{snowmass, Dunton:2022dez, shoemaker2021sailing}. CCM is located downstream from a neutrino beam produced by pion decay at rest and searches for signatures of BSM physics like sterile neutrinos, light dark matter, and axion-like particles.  The international DUNE experiment, hosted by the U.S. Department of Energy’s Fermi National Accelerator Laboratory (Fermilab) in Illinois, is composed of a high-intensity neutrino source, a near detector, and a far detector.  The neutrino source is generated from Fermilab's main injector, while the near detector is positioned just downstream of the neutrino source and the far detector is situated 1.5 km away at the Sanford Underground Research Facility (SURF) in South Dakota. DUNE's science program will address CP violation (the matter/antimatter imbalance), and its far detector will be a modular liquid argon time-projection chamber used to reconstruct the neutrino interactions \cite{LBNE:2013dhi}.

%
\subsubsection{Cherenkov Radiation}
%

When a charged particle passes through a dielectric medium at a speed greater than the phase velocity of light in that medium, Cherenkov radiation is emitted transverse to the direction of that incoming particle \cite{PhysRev.52.378}.  Cherenkov photons are emitted primarily at UV and near-UV wavelengths, contributing to their famous blue color. A classic example is the characteristic blue glow of an underwater nuclear reactor. Unlike other mechanisms of radiation, like scintillation, Cherenkov radiation has a distinctive angular pattern.  Cherenkov radiation is famously emitted along a Cherenkov cone, with the cone angle being dictated by the charged particle velocity and the refractive index of the medium.  This collimated property of the Cherenkov radiation helps to distinguish it from the isotropic scintillation light, as well as to distinguish different charged particles from each other.  For this reason, Cherenkov detectors are commonly used for particle identification. 

\begin{figure}[t!]
\centering
\begin{subfigure}{.35\textwidth}
\centering
\includegraphics[width=1\textwidth]{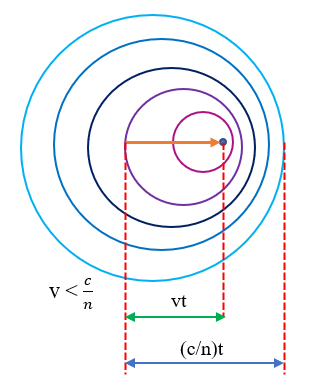}
\caption{$v < c/n$
\label{f:coherentwaveform2}
}
\end{subfigure}
\centering
\begin{subfigure}{.64\textwidth}
\centering
\includegraphics[width=1\textwidth]{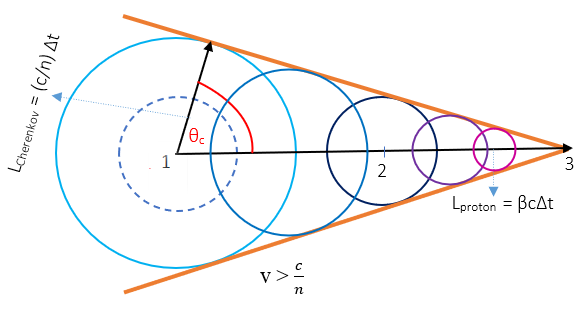}
\caption{$v > c/n$
\label{f:coherentwaveform1}
}
\end{subfigure}
\caption{Spherical wavelets of fields of a particle travelling less than (left fig) and greater than (right fig) the velocity of light in the medium. For $v > c/n$, an electromagnetic "shock" wave appears known as Cherenkov radiation, moving in the direction given by the Cherenkov angle. Deriving the Cherenkov condition from Coherent Electromagnetic wave construction by expanding the spherical wavefront at three instances of time denoted by 1,2, and 3 (right fig).
\label{f:Cherenkovsphericalwavelets}
}
\end{figure}

Cherenkov radiation is an instantaneous process \cite{Jackson:1998nia} that occurs at any wavelength $\lambda$ for which the speed $v = \beta c$ of the proton exceeds the speed of light $c_{LAr} (\lambda) = \frac{1}{n(\lambda)} c$ in medium (see Fig.~\ref{f:Cherenkovsphericalwavelets}).  Here $n(\lambda)$ is the index of refraction of liquid argon at wavelength $\lambda$ and $c$ is the speed of light in vacuum.  This Cherenkov condition $v > c_{LAr}$ may be expressed in terms of the index of refraction $n(\lambda)$ as
\begin{align}   \label{e:CherenCond1}
    \frac{v}{c_{LAr} (\lambda)}  
    = \frac{v}{c} \, \frac{c}{c_{LAr} (\lambda)} = \beta \, n(\lambda)
    > 1 \: .
\end{align}
This shows that the speed of the particle must be larger than the phase velocity of the electromagnetic fields at frequency $\omega$ (or wavelength $\lambda$) in order to have an emission of Cherenkov radiation of that frequency.

\begin{figure}[t!]
\begin{centering}
\includegraphics[width=0.42\textwidth]{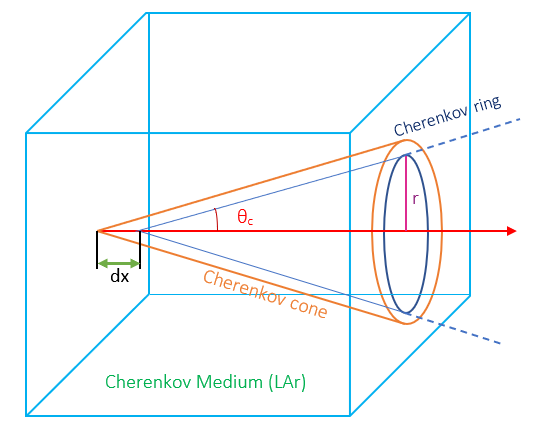}
\hspace{1cm}
\includegraphics[width=0.49\textwidth]{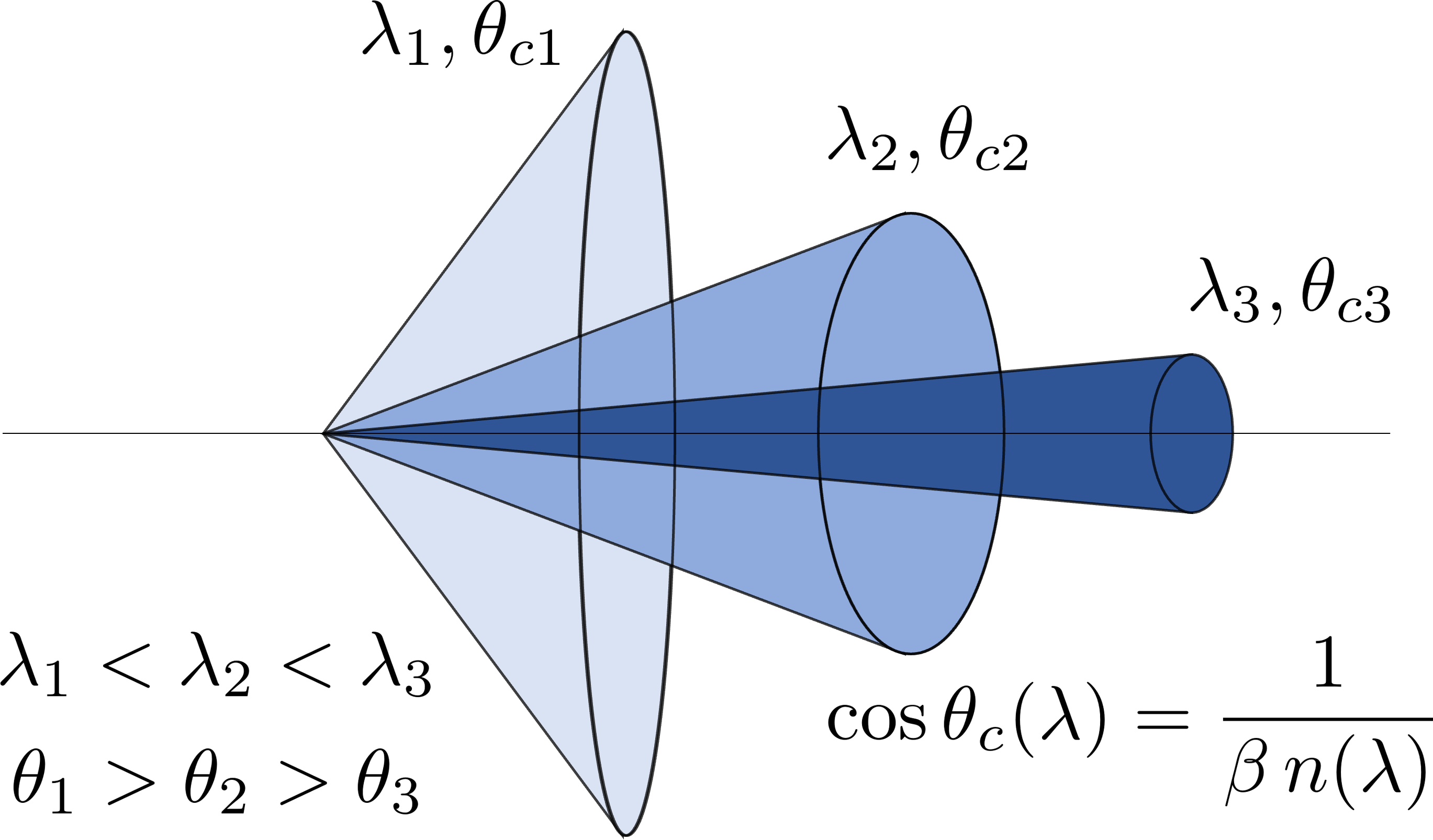}
\caption{Characteristic cone-shaped Cherenkov radiation of a charged particle exceeding the speed of light in a medium.  Left:  A single Cherenkov cone at fixed $\theta_c$; this occurs when the index of refraction $n(\lambda) \approx \mathrm{const}$(i.e., far away from a resonance). Right:  A superposition of many Cherenkov cones emitted at different angles $\theta_c$ corresponding to different wavelengths $\lambda$ with different indices of refraction $n(\lambda)$, shown here for the case of normal dispersion $dn/d\lambda < 0$.
\label{f:cherenkovcone}
}
\end{centering}
\end{figure}

Cherenkov radiation is produced by the coherent superposition of expanding spherical waves if the speed of the particle is greater than the speed of light in the medium, as illustrated in Fig~\ref{f:Cherenkovsphericalwavelets}.  Such a coherent waveform can only be constructed while the charged particle travels through a medium, where the speed of light is less than its vacuum value $c$.  In Fig.~\ref{f:coherentwaveform1}, an example of such a waveform is depicted produced by expanding the spherical wavefront at three instances of time denoted by 1, 2, and 3. Constructive interference (net emission of Cherenkov photons) occurs only at a specific angle $\theta_c$ relative to the trajectory given by the Cherenkov condition,
\begin{align}   \label{e:CherenAngle1}
    \cos\theta_c = \frac{L_{Cherenkov}}{L_{Proton}} = \frac{(c/n)\Delta t}{\beta c \Delta t}= \frac{1}{\beta n(\lambda)}    \: ,
\end{align}
unlike scintillation photons, which are emitted isotropically. Note that the Cherenkov angle $\theta_c$ describes the direction of \textit{propagation} of the wavefront, which is normal to the wavefront itself. Thus $\theta_c$ is \textit{not} the opening angle of the \textit{wavefront cone} shown in Fig.~\ref{f:coherentwaveform1}, but rather the opening angle of the cone defining the photon \textit{momenta} seen in Fig.~\ref{f:cherenkovcone}.  The emission angle $\theta_c$ can be interpreted qualitatively in terms of a \textit{shock} wavefront akin to the familiar shock wave (sonic boom) produced by an aircraft in supersonic flight.

The standard picture of Cherenkov radiation as drawn in the left panel of Fig.~\ref{f:cherenkovcone} shows a single Cherenkov cone at a fixed angle given by \eqref{e:CherenAngle1}.  This is the case if the index of refraction $n$ is a constant, as occurs at wavelengths $\lambda$ far away from any resonances.  But near a resonance (and in general), the index of refraction is a continuous function $n(\lambda)$ of wavelength, so that there can be a continuous range of frequencies $\lambda$ which radiate.  For the same proton velocity $\beta$, the various frequencies will have different indices of refraction $n(\lambda)$, resulting in the emission of each wavelength at its own Cherenkov angle \eqref{e:CherenAngle1}.  Thus, for a proton with instantaneous velocity $\beta$ moving through LAr, there will not be a single Cherenkov cone as in Fig.~\ref{f:cherenkovcone}, but rather a continuous distribution of Cherenkov photons of different wavelengths $\lambda$, all emitted at different angles $\theta_c (\lambda)$ (right panel of Fig.~\ref{f:cherenkovcone}).

%
\subsubsection{Optical Properties of Liquid Argon}
%

Liquid argon scintillators are commonplace in the search for neutrinos in state-of-the-art experiments like CCM and DUNE.  While the main goal of using liquid argon is to take advantage of its extraordinary scintillation properties, Cherenkov radiation is also produced from the secondary particles traveling in that medium and can contribute significantly to the yields of low-energy protons \cite{rahman2024}.  

Liquid argon is known to have a resonance at a wavelength of $106.6$ nm, at which the refractive index $n(\lambda)$ becomes large.  Correspondingly, the speed of light in the medium $c_{LAr}$ becomes small near the resonance, leading to the emission of Cherenkov light.  The intensity and angular pattern of that Cherenkov radiation are sensitive to the details of the shape of the refractive index near the resonance, even for low-energy protons \cite{rahman2024}.

Despite the importance of liquid argon for the neutrino program, aspects of its optical properties near the resonance remain poorly constrained.  The refractive index of LAr was first experimentally measured by Sinnock and Smith \cite{Sinnock:1969zz} for wavelengths between 361.2 and 643.9 nm, by the spectroscopic method of minimum deviation. In this method, a prism-shaped specimen was confined in an optical cell contained in a cryostat and located at the axis of rotation of a spectrometer mounted outside the cryostat on a coaxial turntable. The refractive index $n$ was determined from the angle of minimum deviation $D$ and the angle of the prism $A$.  These data are far from the resonance at 106.6 nm and provide little constraining power on the behavior of $n(\lambda)$ there. 

More recently, Babicz et al. \cite{Babicz:2020den} measured the propagation velocity of scintillation light in liquid argon $v_g$ at the scintillation wavelength $\lambda \sim 128$ nm wavelength for the first time in a dedicated experimental setup at CERN. The obtained result for the velocity was then used to derive the value of the refractive index $n$ and the Rayleigh scattering length $L$ for LAr in the VUV region. For the scintillation wavelength, $\lambda = 128$ nm, they found $n = 1.357 \pm 0.001$.  This measurement provided substantial constraining power on the behavior of $n(\lambda)$ near the resonance at 106.6 nm, providing a unique opportunity to realistically calculate the Cherenkov radiation in LAr for the first time. The details of the intensity and angular distribution of the resulting Cherenkov radiation is highly sensitive to the detailed form of the refractive index near the resonance, and a precise measurement of Cherenkov radiation could potentially set new constraints on the optical properties of LAr.  This subject is analyzed in detail in Ch.~\ref{cherenkov}.

%
\subsection{Characterizing the QGP with Jets}
%

%
\subsubsection{Heavy Ion Collisions: Recreating the Big Bang} 
%

Heavy-ion collisions are the basis of ``Hot QCD,'' which refers to the study of nuclear physics at extreme temperatures and densities. Hot QCD seeks to understand the internal constituents of nuclear matter - quarks and gluons, collectively called ``partons'' (point-like constituents within a hadron as proposed by R.~Feynman \cite{Feynman:1969ej}) and their dynamics by creating a thermal medium of extremely high temperature. At high enough temperatures ($155 \, \mathrm{MeV} \sim 1.8 \times 10^{12}$ K) \cite{Bazavov2012}, protons and neutrons melt into a deconfined plasma of partons. The resulting QGP is a nuclear fireball which obeys relativistic hydrodynamics \cite{Luzum:2013yya, Heinz:2024jwu}. The strong interactions of the partons cause the QGP to be a ``nearly perfect liquid,'' with a record low value for the viscosity-to-entropy-density ratio $\tfrac{\eta}{s}$.  Because of this tiny viscosity, the QGP flows like a nearly ideal fluid and retains a strong correlation with its initial geometry. The QGP is studied with ultrarelativistic collisions of heavy ions at the Relativistic Heavy Ion Collider (RHIC) and at the Large Hadron Collider (LHC).  See Ref.~\cite{Connors:2017ptx} for a review.

%
\subsubsection{Jets and Medium Modification} 
%

Characterizing the properties of the QGP is a central goal of hot QCD, but the lifetime of the QGP is too short(tens of femtoseconds) to be able to measure its properties with external probes. Therefore, we have to use the particles produced along with the QGP to characterize it, including both the soft sector (low-energy particles that ``are'' the medium) and the hard sector (high-energy hard probes that propagate through the medium).  The theory of jet quenching describes the energy loss of energetic quarks and gluons in the QGP, primarily arising from the medium-modified pattern of radiation caused by the Landau-Pomeranchuk-Migdal (LPM) effect \cite{LandauPomeranchuk1953, Migdal1956} which accounts for the suppression of radiation due to coherence effects in multiple scattering.

The most important measure of jet quenching is the nuclear modification factor $R_{AA}$ given by
\begin{align} \label{e:RAA}
    R_{AA}(p_T) & = \frac{dN^{AA}}{dyd^2p_T} \bigg{/} \langle n_{coll} \rangle\frac{dN^{pp}}{dyd^2p_T} \, , 
\end{align}
where $\langle n_{coll} \rangle$ is the number of binary collisions.  $R_{AA}$ is the ratio of hard particles produced in nucleus-nucleus (AA) collisions compared to the expected number from independent proton-proton (pp) collisions, which measures the suppression in the number of hard particles produced due to the medium.  The nuclear modification factor is a robust quantity which can be reproduced by a wide variety of calculations based on perturbative QCD and beyond \cite{Noronha-Hostler:2016eow, Molnar:2013eqa, Zhao:2021vmu, Zhang:2013oca, Kopeliovich:2012sc, Andres:2019eus}; see \cite{CMS:2024krd} for a direct multi-model comparison.  Another important measure of jet interactions with the QGP is the anisotropic flow of hard particles.  The Fourier harmonics of the azimuthal distribution of hard particles define the anisotropic flow coefficients $v_n^{hard}$ of hard particles,
\begin{align}
    \frac{R_{AA}(p_T, \phi)}{R_{AA}(p_T)} & \equiv  
    1 + 2 \sum_{n = 1}^\infty v_n^{hard} \, \cos [n (\phi - \psi_n^{hard} (p_T))] \, , 
\end{align}
where $\psi_n^{hard}$ is the angle of the $n^{th}$ harmonic hard particle event plane \cite{Noronha-Hostler:2016eow}.  This measures the correlation of the hard particle distribution in momentum space to the geometry of the QGP.  A third measure of the interactions of hard particles with the medium is the acoplanarity of two particles such as hadrons or jets.  The acoplanarity 
\begin{align}
    \Delta \phi = \pi - \left|\mathrm{mod}\big(\phi_1 - \phi_2 \: , \: \pi\big) \right|
\end{align}
measures the deviation of two hard particles from being initially back to back in the transverse plane. If two hard particles are exactly back to back (as they would be initially for a $2 \rightarrow 2$ process), they would have $\Delta\phi = 0$, with $\Delta \phi$ increasing as the particles are deflected away from the back-to-back limit.

The program of using measurements of energy loss and $R_{AA}$ and $v_n^{hard}$ to infer information about the size, shape, temperature, and density of the plasma seen by the hard probes has formed the basis of ``jet tomography'' in the field \cite{Vitev:2002pf, JET:2013cls, Cao:2020wlm}.  Other, more differential jet observables include jet shapes \cite{CMS:2013lhm, CMS:2019btm}, groomed jets \cite{Larkoski:2014wba}, and jet substructure \cite{Chien:2016led}.

%
\subsubsection{Jet Drift and Jet Tomography} 
%

In Ref.~\cite{Sadofyev:2021ohn}, a new mechanism of medium modification was introduced which goes significantly beyond the traditional picture of energy loss and isotropic momentum broadening.  ``Jet drift'' refers to an anisotropic net deflection $\langle \vec{q}_{drift} \rangle$ of hard partons in the direction of the flow velocity $\vec{u}_\bot$ transverse to the parton momentum.  Jet drift is a sub-eikonal (power-suppressed) correction in the small parameter $\mu/E \ll 1 $, where the Debye mass $\mu$ is a medium scale.  Jet drift is therefore insignificant for very high parton energy $E$ but introduces new physical effects at lower energies.  Jet drift affects different Feynman diagrams differently and is distinct from a simple boost of the leading-power result.  

As derived in Ref.~\cite{Sadofyev:2021ohn}, the net anisotropic momentum acquired by the jet drift effect is
\begin{align}\label{e:q_drift_moment}
    \left\langle \vec{q}_{drift} \right\rangle & = \hat{e}_\perp \int d\ell \, \frac{3}{E} \, \frac{\mu^{2} }{\lambda} \, \ln\frac{E}{\mu} \: 
    \frac{u_\perp }{1-u_\parallel } \: ,
\end{align}
where $\lambda$ is the mean free path, $\mu$ is the Debye screening  mass, and $\vec{u}$ is the collective flow velocity, with components $\parallel$ and $\perp$ to the parton momentum vector, and $\hat{e}_\perp$ is the corresponding unit vector. 

In Ref.~\cite{Antiporda:2021hpk}, we showed that in simple medium geometries, the aggregation of the locally anisotropic jet drift over the path length generates additional elliptic flow $v_2$ of hard and semi-hard partons.  More recently, we have presented the first study \cite{Bahder:2024jpa} of event-by-event jet drift via APE (Anisotropic Partonic Evolution), a new open-source Monte Carlo parton trajectory simulator, that incorporates a host of flexible model choices to survey the variability of the effect. APE is an in-house hybrid module that takes initial hard scattering events from PYTHIA \cite{Bierlich:2022pfr}, embeds them into the open-source Duke QCD medium model \cite{bernhard:2016tnd}, performs our calculations of standard energy loss and jet drift, fragments the partons into final-state hadrons, and implements a hadronic afterburner \cite{Bass:1998ca, Petersen:2008kb} all in one place, to mimic different stages of the heavy-ion collision.

We demonstrated in the paper \cite{Bahder:2024jpa} that despite conservative assumptions, the imprint of flow on hard probes is robust and endures averaging over numerous events, exhibited by $\sqrt{s}=5.02$~TeV PbPb collisions at the Large Hadron Collider (LHC). The findings showed that jet drift leads to an enhancement of both the elliptic flow $v_2$ of hard particles and the mean acoplanarity for low- and intermediate-momentum particles ($p_T \leq 10 \mathrm{GeV}$).  Both effects reflect the jets being ``pushed'' by the velocity of the flowing medium.  These observables also showed distinct dependence on medium properties like centrality, reflecting the different impacts these properties have on the two observables.  The hints obtained in Ref.~\cite{Bahder:2024jpa} that the different observables (acoplanarity and $v_2$) measure the effect of jet drift and quenching in different ways suggest that these features can be disentangled by comparing collision systems with different properties.  PbPb collisions at the LHC are geometrically very similar to the size and shape of collision geometries produced in AuAu collisions at RHIC, but probe much higher temperatures.  Similarly, UU collisions at RHIC occur at the same energy and temperature as AuAu collisions (193 GeV vs 200 GeV), but have substantially different geometries (especially in central collisions) due to the ellipsoidal deformation of uranium.  In Ch.~\ref{jetdrift}, we perform a study of the effects of jet drift and jet quenching on the acoplanarity and $v_2$ across PbPb, AuAu, and UU collisions to disentangle the contributions of the multiple underlying variables to the observables.  Finally, we summarize our conclusions in Ch.~\ref{conclusion}.

%% file: chp2_Cherenkov.tex
\section{CHERENKOV RADIATION IN LIQUID ARGON} \label{cherenkov}
\hspace{\parindent}

%
\subsection{Introduction}
\label{sec:intro}
%
%
%
\subsubsection{Cherenkov PID}
\label{sec:pid}
%
%
\begin{figure}[h] 
\centering
\begin{subfigure}{.48\textwidth}
\includegraphics[width=1\textwidth]{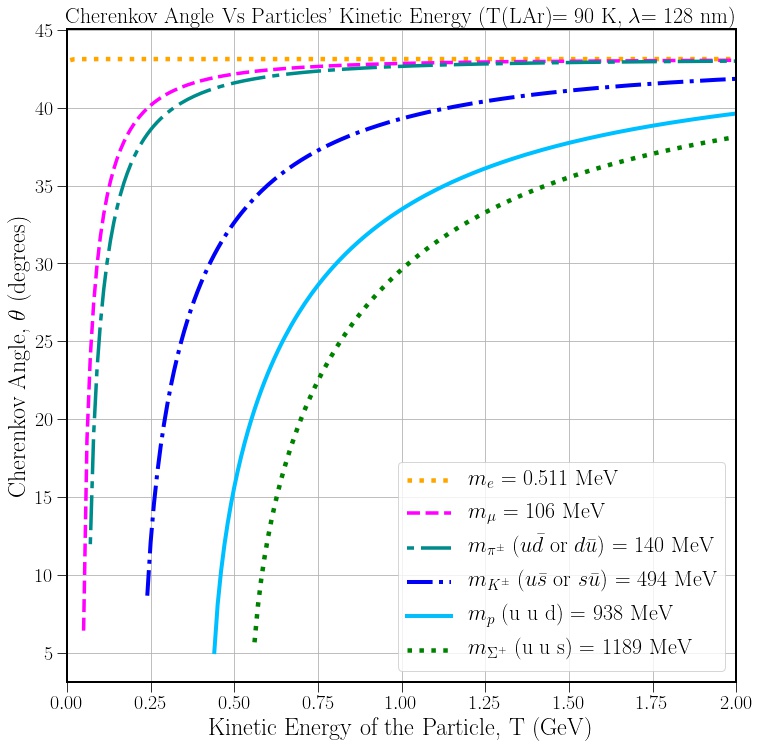}
\caption{
}
\label{f:thetavstforall}
\end{subfigure}
\begin{subfigure}{.47\textwidth}
\centering
\includegraphics[width=1\textwidth]        {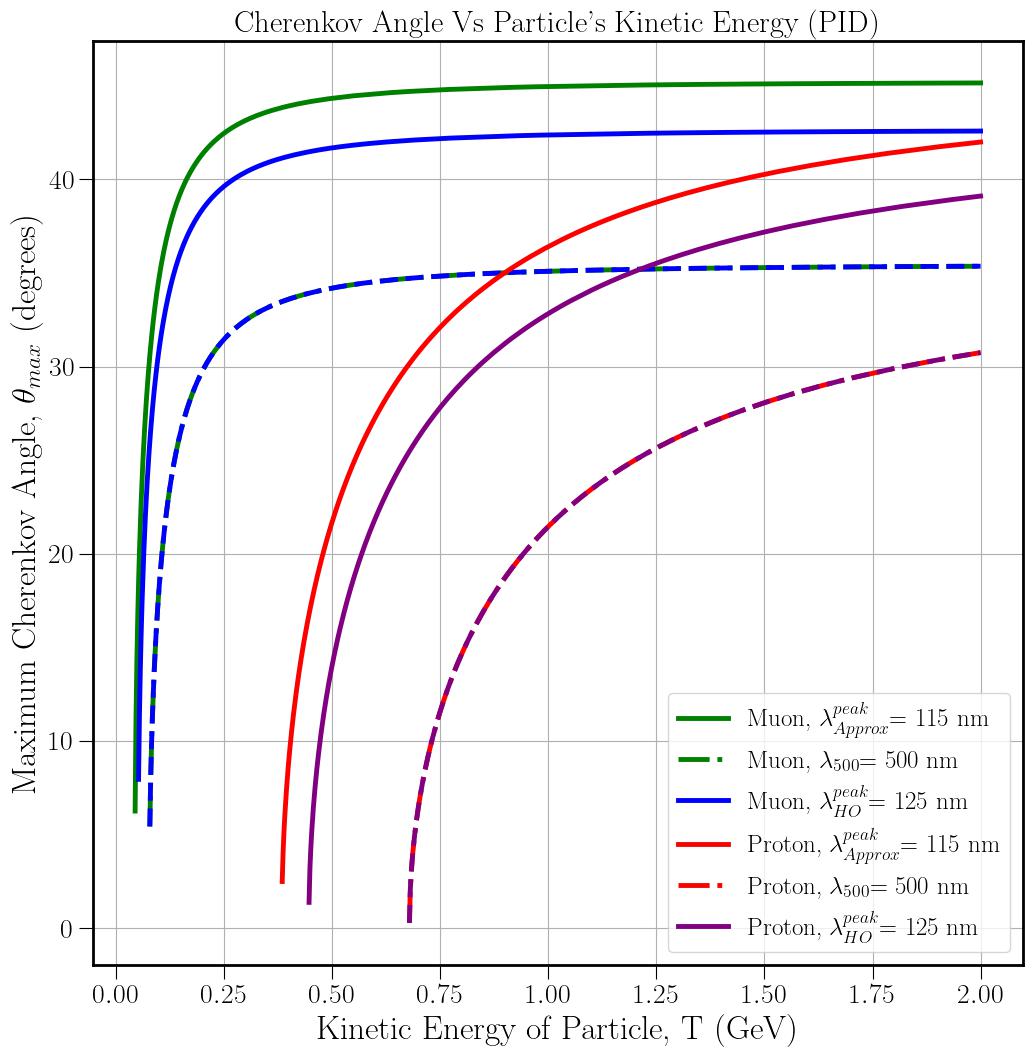}
\caption{
}
\label{f:pidhovsapprox}
\end{subfigure}
\caption{Method of particle identification which uses the Cherenkov angle $\theta_c$ to discriminate between particles.  Curves for different particles are shown in a nondispersive medium (a) and with dispersion (b).  
\label{f:pidtheory}
}
\end{figure}

Because of its strong directional properties and sensitivity to the velocity $\beta$, Cherenkov radiation is a powerful tool for particle identification (PID): the identification and discrimination between different particles based on their physical properties like mass, charge, and spin \cite{Lippmann_2012}.  The reason Cherenkov radiation is such a powerful discriminator between different particles is that the Cherenkov angle
\begin{align}   \label{e:CherenAngle2}
    \theta_C = \cos^{-1}(1 / \beta n)
\end{align}
depends on a particle's velocity $\beta$ at a given kinetic energy -- and therefore its mass.  As seen in Fig.~\ref{f:thetavstforall}, the shape of the curve relating the Cherenkov angle $\theta_c$ and the kinetic energy $T$ depends on the mass of the particle.  The Cherenkov angle \eqref{e:CherenAngle1} depends only on the velocity $\beta$ and the index of refraction $n$, so the mass enters only through the relation 
\begin{align}   \label{e:beta1}
\beta = \sqrt{1 - \frac{1}{(1 + \frac{T}{mc^2})^2}} \: .
\end{align}
between kinetic energy $T$ and velocity $\beta$. Since heavier particles have a lower velocity $\beta$ at the same kinetic energy, this corresponds to a Cherenkov angle $\theta_c$ closer to zero.  Thus, for light particles like electrons, the Cherenkov angle is nearly flat as a function of kinetic energy, while for heavier particles like protons, the angle changes significantly as a function of $T$.  These differences can then be used to distinguish between different species of charged particles through the separation of the curves in Fig.~\ref{f:thetavstforall}. This illustrates that Cherenkov radiation is a valuable method for PID, but it can only be applied to charged particles which emit Cherenkov photons. The Cherenkov radiation technique is most useful for identifying particles with high velocities, such as muons, pions, and electrons, and for distinguishing them from other particles that have similar energies but different masses and charges.

Fig.~\ref{f:thetavstforall} summarizes the familiar case of a nondispersive medium $n(\lambda) = const$, or at frequencies far from any resonance (left panel).  The ability to discriminate between particles comes from having Cherenkov radiation at one specific angle \eqref{e:CherenAngle1} with significant mass discrimination power (PID) through the $\beta$ (or $T$) dependence. Although far from a resonance, $n \approx const$, near a resonance there can be significant dispersive effects which cause $n(\lambda)$ to vary significantly.  This dispersion can be understood from fundamental physics principles in, e.g., the Lorentz oscillator model (HO model).

The immediate result shown in Fig.~\ref{f:pidhovsapprox} is that the Cherenkov radiation is smeared between the maximum and minimum values of $n(\lambda)$.  This results in the spread seen between the wavelengths emitting at the maximum angle (solid curves) and minimum angle (dashed curves).  There is also significant model dependence 
on the form of the refractive index $n(\lambda)$ (HO vs. Approx models), as described below.  So the consequence of using dispersive refractive index is that you can get a Cherenkov \textit{band} instead of a traditional single curve, which complicates the picture of PID.

\begin{figure}[t]
\centering
\begin{subfigure}{.48\textwidth}
\centering
\includegraphics[width=1\textwidth]{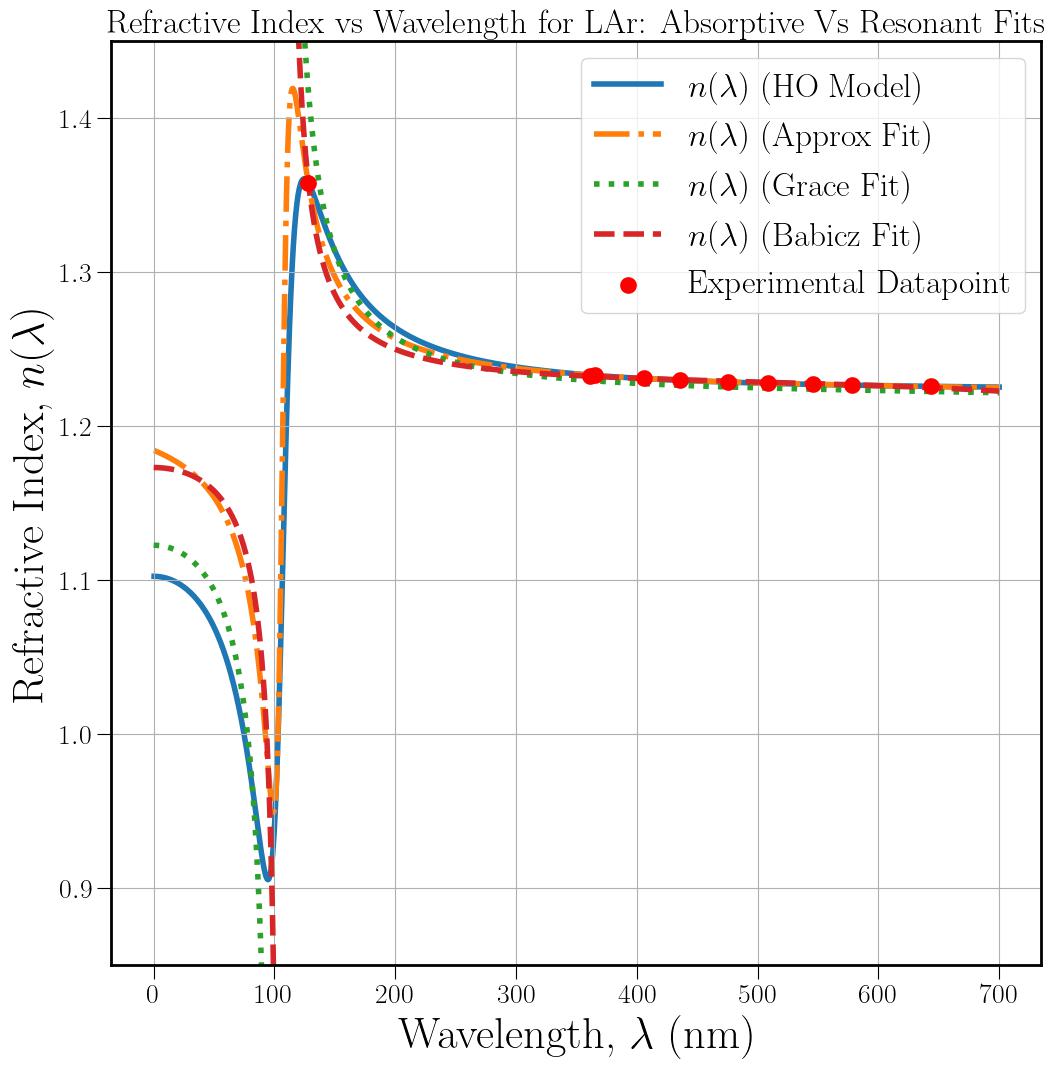}
\caption{
\label{f:nvslambdahofit}
}
\end{subfigure}
\begin{subfigure}{.49\textwidth}
\centering
\includegraphics[width=1\textwidth]
{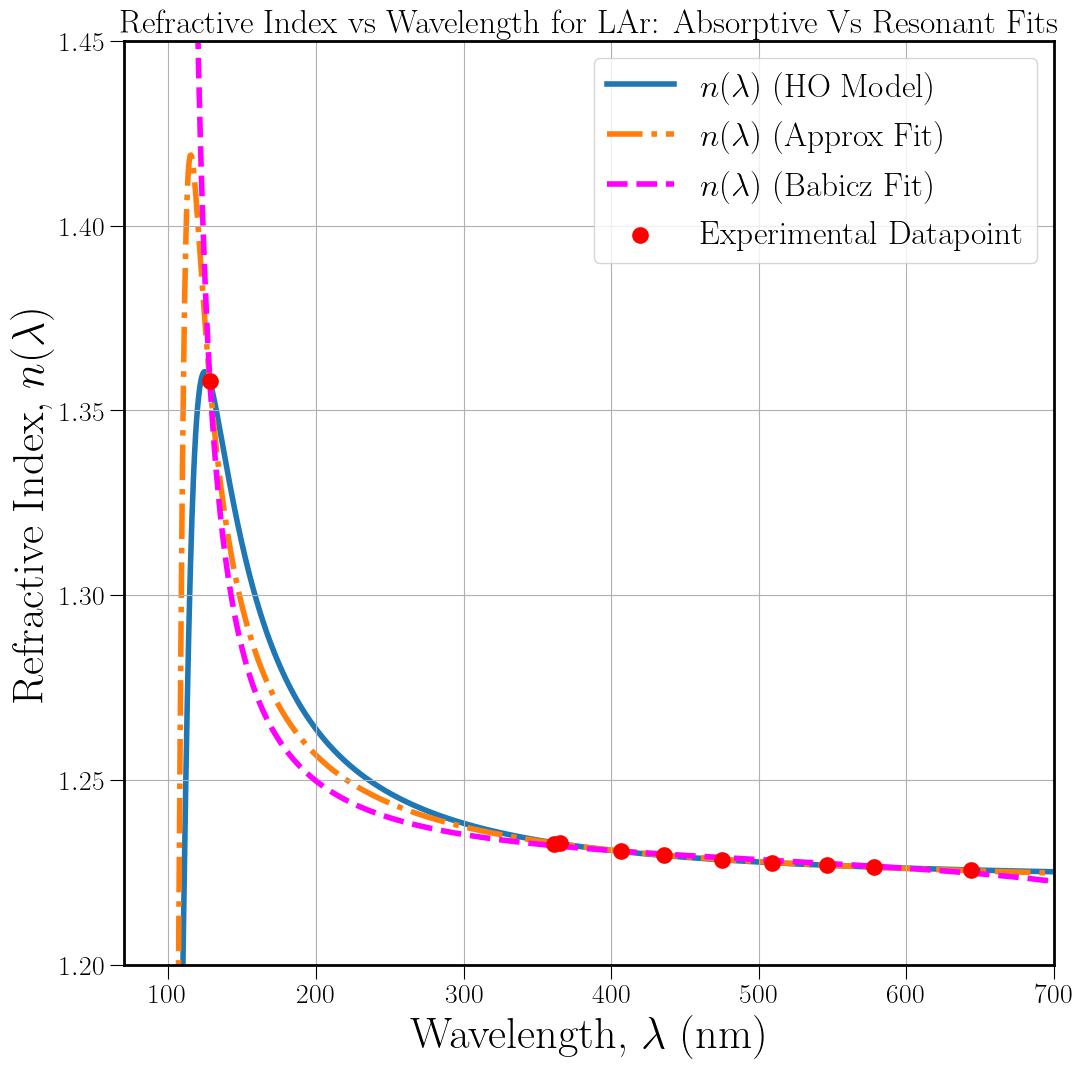}
\caption{
\label{f:nvslambdahofitvsBabicz}
}
\end{subfigure}
\caption{Data and fits to the refractive index $n(\lambda)$ of LAr.  A zoomed out plot (a) shows the behavior of $n(\lambda)$ as it crosses the resonance, while the region above the resonance (b) is of interest for the Cherenkov radiation.
\label{f:nvslambdahoandBabiczfit}
}
\end{figure}

%
\subsubsection{Fits to the LAr Refractive Index}
\label{sec:LAr_Fits}
%

Fig.~\ref{f:nvslambdahoandBabiczfit} shows the status of experimental data on the refractive index of liquid argon (LAr) and various fits to that data. Previously, the only phenomenological fit to the refractive index that included the new data point measured by Babicz et al. was a resonant fit for which $n(\lambda) \rightarrow \infty$ at the resonance:
\begin{align}   \label{e:nBabicz}
  n(\lambda) = \sqrt{1 + \frac{3x(\lambda)}{3-x(\lambda)}}  
\end{align}
with
\begin{align}   \label{e:Babiczxre}
    x(\lambda) = a_{0} + \frac {a_{UV}\lambda^{2}} {\lambda^{2} - \lambda_{UV}^{2}} + \frac {a_{IR}\lambda^{2}} {\lambda^{2} - \lambda_{IR}^{2}}   \: ,
\end{align}
where $a_{0} = 0.335 \pm 0.003$, $a_{UV} = 0.099 \pm 0.003$, $a_{IR} = 0.008 \pm 0.003$ are Sellmeier coefficients and $\lambda_{UV} = 106.6 \: \mathrm{nm}, \lambda_{IR} = 908.3 \: \mathrm{nm}$  \cite{Babicz:2020den}.  Such fits are unrealistic starting points for the calculation of the Cherenkov spectrum, as we will discuss. On the contrary, absorptive fits correct this theoretical deficiency by a proper treatment of anomalous dispersion near the resonance.  For a detailed treatment, see our previous work in Ref.~\cite{rahman2024}.

The first-principles Lorentz oscillator model (HO model) of the refractive index describes a dipole resonance with an oscillating electromagnetic field \cite{Jackson:1998nia, griffiths_2017}, leading to a refractive index of the form
\begin{align}   \label{e:nlambdauvhofinal}
     n_{HO} = a_{0(HO)} + a_{UV(HO)} \left(\frac{\lambda_{UV}^{-2} - \lambda^{-2}}{(\lambda_{UV}^{-2} - \lambda^{-2})^2 + \gamma_{UV}^2 \lambda^{-2}}\right) \: , 
\end{align}
with the corresponding absorption coefficient given by 
\begin{align}   \label{e:alphalambdauv}
    \alpha \cong \frac{Nq^2}{m \epsilon_0 c} \left( \frac{f_{UV} \gamma_{UV} \lambda^2 \lambda_{UV}^4}{(2 \pi c)^2 (\lambda^2 - \lambda_{UV}^2)^2 + \gamma_{UV}^2 \lambda^2 \lambda_{UV}^4} \right). 
\end{align}
The best-fit parameters for LAr were $a_{0(HO)} = 1.10232$, $a_{UV(HO)} = 0.00001058$, and $\gamma_{UV} = 0.002524 \: \mathrm{nm}^{-1}$ \cite{rahman2024}. The physics of the damped-driven harmonic oscillator leads to a refractive index has a finite peak $n_{peak}$ instead of diverging to infinity, and a region of anomalous dispersion where $\frac{dn}{d\lambda} > 0$.

Although the simple HO model follows from first principles, any similar form in which $n(\lambda)$ rises to a finite maximum before turning around will similarly capture the qualitative physics of absorption near a resonance which is responsible for the transition from ``normal dispersion'' $\frac{dn}{d\lambda} < 0$ to ``anomalous dispersion'' $\frac{dn}{d\lambda} > 0$.  For this reason, it is also useful to consider an alternative parameterization (Approx model) in the form: 
\begin{align}   \label{e:nmodifiedlambda}
    n(\lambda)  &=  a_0 + a_{approx} \:
    \frac{(\lambda_{UV}^{-1} - \lambda^{-1})}{(\lambda_{UV}^{-1} - \lambda^{-1})^2 + \Gamma^2}
\end{align}
with corresponding absorption coefficient
\begin{align}   \label{e:alphamodified2}
    \alpha(\lambda) &=
    a_{approx}
    \left(\frac{4\pi}{\lambda} 
    n(\lambda) \right)
    \:
    \frac{\Gamma}{(\lambda^{-1}_{UV} - \lambda^{-1} )^2 + \Gamma^{2}}
\end{align}
and best-fit parameters $a_{0} = 1.18416$, $a_{approx} = 0.000325985$, and $\Gamma = 0.000693652 \mathrm{nm}^{-1}$.  A more detailed summary on different fits of $n(\lambda)$ of LAr can be found in the MS thesis \cite{rahman2024}. This $n(\lambda)$ fit has been mentioned and implemented in the 2025 Coherent CAPTAIN-Mills experimental paper \cite{aguilararevalo2025measurementliquidargonscintillation} for the measurement of LAr scintillation pulse shape. 

While various absorptive fits are possible, allowing for a certain amount of model flexibility, the general shape of the resonance is known, and the peak of the resonance must be at least $n_{peak} \geq 1.36$ in order to agree with the measured data point at $128$ nm. Hence, a fit like the HO model yields the most conservative possible value for $n_{peak}$, while in other reasonable models like the Approx model, the value of $n_{peak}$ can be even larger.

The physics of a finite $n_{peak}$ and a region of anomalous dispersion substantially changes the shape of the resulting distribution of Cherenkov radiation from the nondispersive expectation of Fig.~\ref{f:thetavstforall}. As we will show, even the small differences between the HO and Approx models leads to significantly different predictions for the resulting distributions of Cherenkov radiation.  Precise measurements of Cherenkov radiation in LAr can therefore help set constraints on the form of the refractive index $n(\lambda)$.

%
\subsubsection{Instantaneous Distribution of Cherenkov Radiation}
\label{sec:distribution}
%

As seen from Eq.~\eqref{e:CherenCond1}, the refractive index $n(\lambda)$ of LAr is particularly important to determine the wavelength range over which Cherenkov emission will occur.  This further determines the angular distribution of the Cherenkov photons through the condition \eqref{e:CherenAngle1}.  

The number distribution $d^2 N / dx d\lambda$ of Cherenkov photons per unit distance $dx$ and per unit wavelength $d\lambda$ is given by the Frank-Tamm formula \cite{Workman:2022ynf},
\begin{align}   \label{e:FrankTamm1}
\frac{dN}{dx \, d\lambda \, d\cos\theta} = \frac{2 \pi \alpha z^2}{\lambda^2} \left( 1 - \frac{1}{\beta^2 n^2(\lambda)} \right) \: \delta\left(\cos\theta - \tfrac{1}{\beta(x) \, n(\lambda)}\right)     \: .
\end{align}
Here $\alpha = 1/137$ is the fine structure constant, $z$ is the charge of the proton in units of $e$, and the delta function enforces the Cherenkov condition \eqref{e:CherenAngle1} for a dispersive index of refraction $n(\lambda)$.  In this work we consider particles with $|z| = 1$.  The Frank-Tamm formula was first derived in 1937 and published to explain the radiation measured in 1934 by Cherenkov.

By integrating Eq.~\eqref{e:FrankTamm1} over the wavelengths $\lambda_{min} \leq \lambda \leq \lambda_{max}$ satisfying the Cherenkov condition \eqref{e:CherenCond1}, we obtain the \textbf{instantaneous angular distribution} (IAD) 
\begin{align}   \label{e:IAD1}
    \frac{dN}{d\Omega \, dx} &= \int\limits_{\lambda_{min}}^{\lambda_{max}} d\lambda \, \frac{dN}{dx \, d\lambda \, d\cos\theta} \: ,
\end{align}
and integration over the solid angle $d\Omega$ gives the instantaneous yield
\begin{align}   \label{e:IYield1}
 \frac{dN}{dx} = \int d\Omega \, \frac{dN}{d\Omega \, dx} = \int\limits_{\lambda_{min}}^{\lambda_{max}} d\lambda \, \frac{dN}{dx \, d\lambda}   \: .
\end{align}    

As detailed in Ref.~\cite{rahman2024}, after performing the integral over wavelengths $\lambda$, the resulting IAD is given by:
\begin{align} \label{e:IAD}   
\frac{dN}{d\Omega \, dx} &=
\frac{2 \alpha_{EM}}{\beta^2} \frac{1}{\lambda_\theta^2} \left(\frac{\sin^2 \theta}{\cos^3 \theta}\right)
\frac{1}{\left| \frac{dn^2}{d\lambda} \right|_{\lambda= \lambda_{\theta}}}
\: .
\end{align}
Here, the IAD $\frac{dN}{d\Omega \, dx}$ stands for the number $dN$ of Cherenkov photons emitted instantaneously, per unit path length $dx$, in a differential solid angle $d\Omega$, and $\lambda_\theta$ is the solution to the wavelength-dependent Cherenkov condition. One should also note that $\lambda_\theta$  will also appear from the derivative of $n^2$ in the denominator. The specific form of the IAD for a given $n(\lambda)$ depends on the fit function. Also recall that the solid angle is defined as,
\begin{align}
    d\Omega = \sin\theta \: d\theta d\phi = d\cos\theta d\phi
\end{align}
so that for an azimuthally symmetric distribution,
\begin{align}   
    \int d\phi \frac{dN}{d\Omega dx} = \frac{dN}{d\cos\theta dx} = 2\pi \frac{dN}{d\Omega dx} \: .
\end{align}
This explains the origin of the factor of $2\pi$ in Eq.~\eqref{e:IAD}.

\begin{figure}[t] 
\centering
\begin{subfigure}{.47\textwidth}
\centering
\includegraphics[width=1\textwidth]{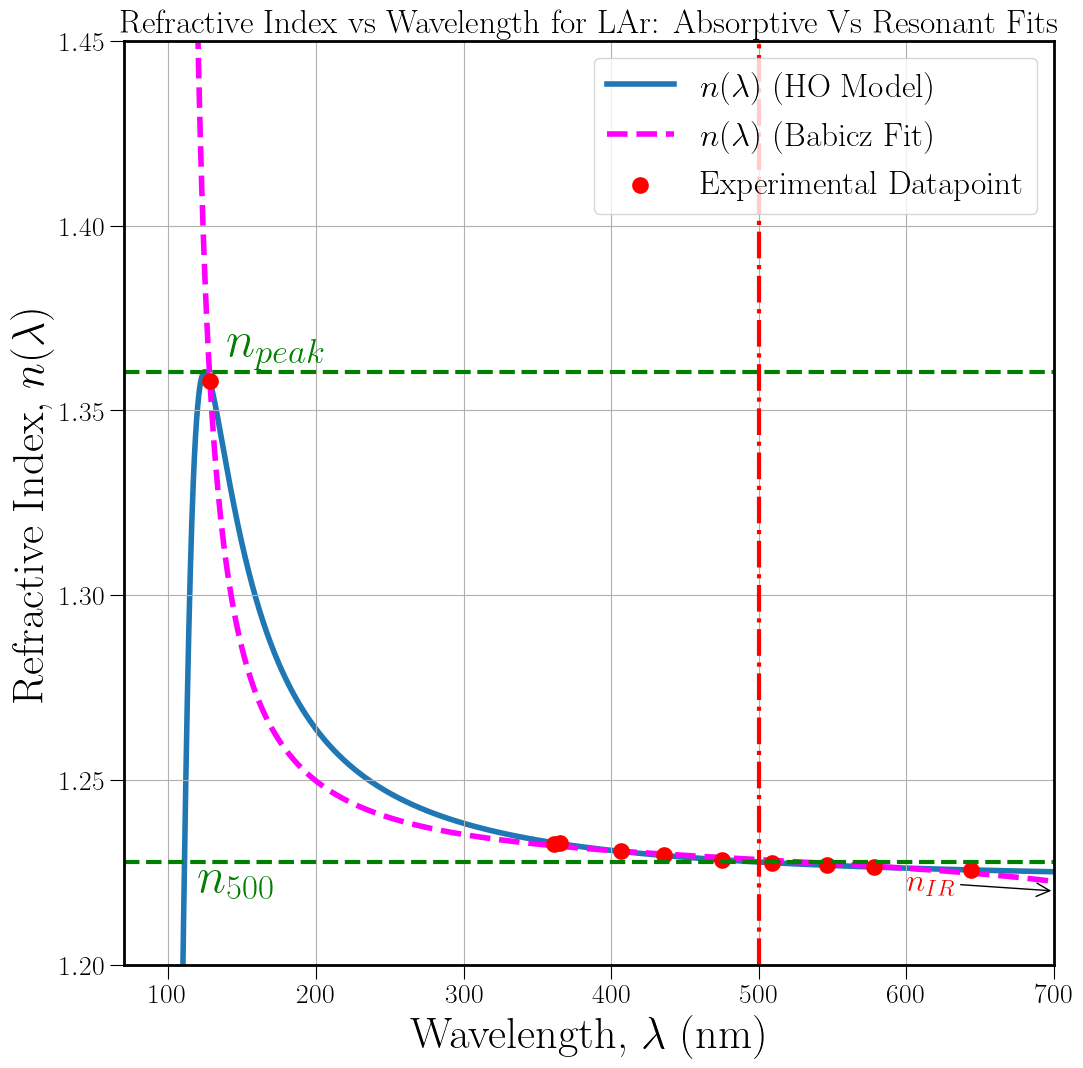}
\caption{
\label{f:Refractive_Index_Annotated}
}
\end{subfigure}
\begin{subfigure}{.48\textwidth}
\includegraphics[width=1\textwidth]{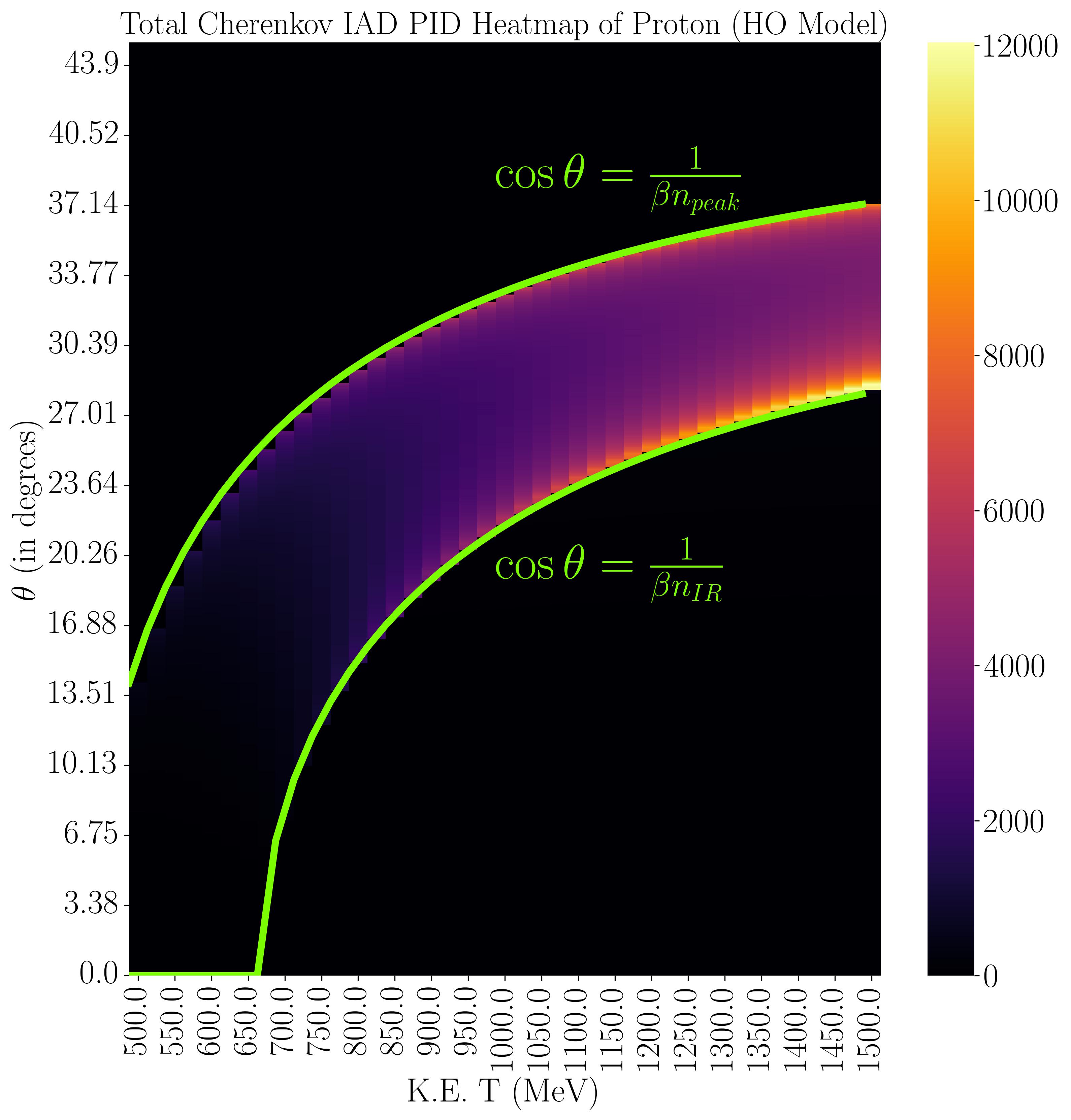}
\caption{
\label{f:IADPID_Annotated}
}
\end{subfigure}
\caption{Relationship between the key features $n_{peak}, n_{IR}$ of the refractive index (a) to the boundaries of the Cherenkov IAD (b). 
\label{f:TheoristPlots}
}
\end{figure}

The medium properties like the refractive index $n(\lambda)$ also play a very crucial role in realistically predicting the Cherenkov yield and angular distributions.   


When we combine these IADs for various charged particles, for different kinetic energies $T$, we can construct the detailed bands of Cherenkov radiation outlined in Fig.~\ref{f:pidtheory} and examine their implications for PID.

There are two general features of the IAD, as clearly seen in Fig.~\ref{f:TheoristPlots}:  (1) A maximum angle determined by $\cos\theta = 1 / \beta n_{peak}$ which is sensitive to the peak of the refractive index (UV feature), and (2) a minimum angle determined by $\cos\theta = 1 / \beta n_{IR}$ (IR feature). Whether $n_{IR}$ should be taken to be the position of the true asymptote or some earlier cutoff like $n_{500}$ depends upon detector conditions.  
[We chose $\lambda = 500$ for the IR cutoff to compare with the finite detector acceptance for neutrino experiments to match experimental data. Also, we have checked that the photon contribution for higher wavelengths ($\lambda > 500 nm$) is trivial.  

At large wavelengths, the dispersion of the refractive index is becoming small, with $n(\lambda) \approx const$, resulting in many photons of different wavelength $\lambda$ all coming out at approximately the same angle $\theta$.  This ``IR pileup'' phenomenon results is a sharp bright inner edge to the IAD. The IR pileup effect on the inner edge is universal: all models which differ in the UV will still converge to the same constant value $n_{IR}$ far from the resonance (see Fig.~\ref{f:nvslambdahoandBabiczfit}).  Even the resonant models like Babicz will correctly capture this feature of the AD, which is determined by the physics of ``normal'' dispersion.
  
On the other hand, the threshold effect on the outer edge is very model dependent: it is a direct reflection of the value of $n_{peak}$ which can change from model to model.  As a result, a precise measurement of this threshold edge can discriminate between models with different predictions for $n_{peak}$.  In the extreme case, the resonant models have no Cherenkov threshold whatsoever: for any $\beta > 0$, the Cherenkov spectrum extends out to $90^\circ$.  Hence the finite threshold is unique to absorptive fits, and the intensity of Cherenkov radiation near the outer edge is governed by the anomalous component of the dispersive refractive index.

%
\subsubsection{Energy Loss and Bethe-Bloch Formula}
\label{sec:bbformula}
%

A charged particle traveling through a medium loses energy through inelastic collisions with atomic electrons in the medium. The energy loss $- \frac{dE}{dx}$ per unit length is given by the Bethe-Bloch equation \cite{Workman:2022ynf} as a function of the proton velocity $\beta = v / c$ in units of the speed of light $c$:
\begin{align}   \label{e:BetheBloch1}
-\frac{dE}{dx} = K\frac{\rho Z}{A} \frac{z^2}{\beta^2} \left[ \ln \left(\frac{2m_{e}c^2\gamma^2\beta^2}{I} \right) - \beta^2  \right]   \: .
\end{align}
Here, $z$ is the charge of the particle in units of the fundamental charge $e$; $\rho \approx 1.38 \, \mathrm{g/cm}^3$ is the mass density of LAr per unit volume at $T = 89 \, \mathrm{K}$ \cite{BNL:2023,Sinnock:1969zz}, $Z=18$ is its atomic number, and $A=39.948$ is its atomic mass. For LAr, the mean ionization potential $I$ (the mean energy needed to ionize an electron out of the atom) is $23.6 \, \mathrm{eV}$ \cite{Workman:2022ynf}.  This means that, when a charged particle goes through liquid argon (LAr), it takes about 23.6 eV energy to ionize an electron on average. For minimally ionizing particles, the corresponding mean energy loss is about 2.1 MeV/cm, leading to about 100 nm average separation between two adjacent ions.  Finally, the constant $K$ is given by
\begin{align}   \label{e:constk}
K = \frac{4\pi \alpha_{EM}^2 (\hbar c)^2 N_{A} (10^3 \, \mathrm{kg}^{-1})}{m_{e} c^2} = 30.7 \: \mathrm{keV \, m^2/kg} = 0.0307 \: \mathrm{MeV \, m^2 / kg }     \: ,
\end{align}
where $\alpha_{EM} = 1/137$ is the fine structure constant, $\hbar$ is the reduced Planck constant, $N_A$ is Avogadro's number, and $m_e$ is the mass of the electron \cite{Workman:2022ynf}. 

\begin{figure}[ht]
\centering
\begin{subfigure}{.47\textwidth}
\centering
\includegraphics[width=1\textwidth]{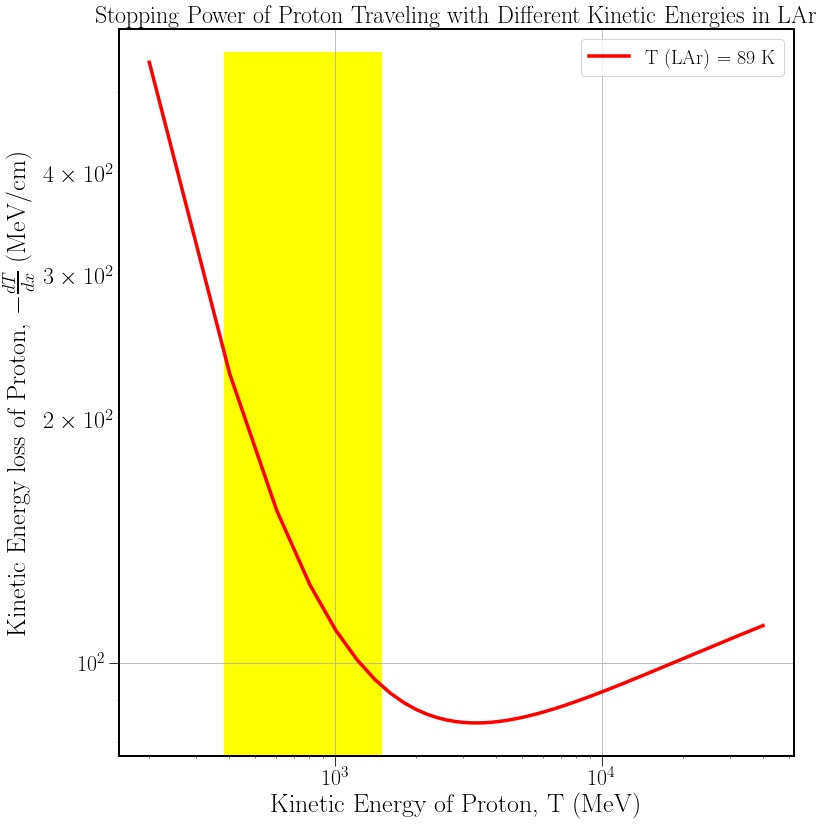}
\caption{Bethe-Bloch Formula
\label{f:dTbydxvsKElog}
}
\end{subfigure}
\begin{subfigure}{.48\textwidth}
\centering
\includegraphics[width=1\textwidth]{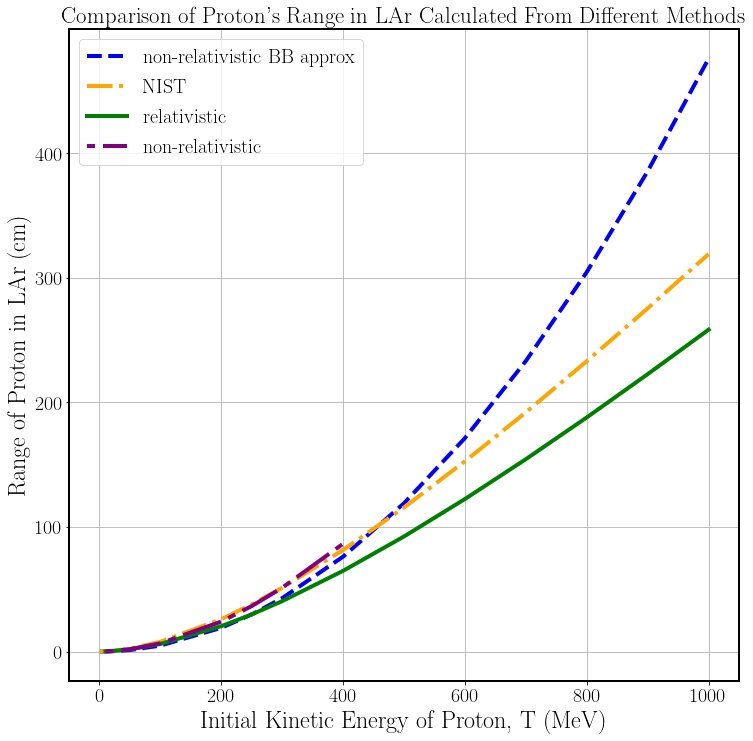}
\caption{Range of Proton
\label{f:rangeproton}
}
\end{subfigure}

\caption{(a) Stopping power $-\frac{dT}{dx}$ for a proton in LAr as a function of its kinetic energy $T$.  The range of interest for this work (383 - 1500 MeV) is highlighted in yellow.  (b)  Range of a proton in LAr with a given initial kinetic energy estimated from different methods (right).  
\label{f:dTbydxvsKE}
}
\end{figure}

After fixing the various parameters, the Bethe-Bloch equation \eqref{e:BetheBloch1} describes the energy loss (or ``stopping power'') of protons as a function of their instantaneous velocity $\beta = v/c$.  The Bethe-Bloch stopping power for LAr using the constants described above is shown in Fig.~\ref{f:dTbydxvsKE} as a function of the proton energy. This non-uniform stopping power leads to a non-uniform rate of loss of the proton kinetic energy.  In particular, we note that in the low-energy range of interest to us (383 - 1500 MeV, shaded in yellow in the left panel of Fig.~\ref{f:dTbydxvsKE}), the stopping power is a \textit{decreasing} function of the kinetic energy.
As one goes higher in $T$, the stopping power reaches a minimum before turning over to become an increasing function at high $T$.

The decrease in the kinetic energy of the proton corresponds to a decrease in $\beta$, such that the Bethe-Bloch equation constitutes an implicit differential equation for the velocity $\beta(x)$ as a function of distance $x$ propagated in LAr.  By solving the Bethe-Bloch equation with fully relativistic kinematics, we determine the velocity $\beta(x) = v(x) / c$ of the proton in units of the speed of light $c$ in vacuum.

One important note here is that the BB equation used for the energy loss only considers electromagnetic stopping power. The BB formula given by Eq.~\ref{e:BetheBloch1} depends on the charge squared ($z^2$) of the particle in the units of electron's charge, which is not distinguishable between a positively charged particle and a negatively charged one. Moreover, the nature of the energy loss is rather a continuous one -- treats protons and muons on the same footing, effectively just coming from their mass difference. It does not include any nuclear stopping effects, which gives the range calculated from it larger than some of the other treatment of energy loss \cite{Beacom:2003zu} in the literature. Including ``nuclear stopping'' with random, discrete energy loss events / Monte Carlo effect will substantially shorten the range of the particle travelling in LAr or other medium. In the next section, we show how integrated distributions are computed by superimposing the IADs from the initial $\beta$ to the subsequent $\beta$'s until it drops to the threshold for Cherenkov emission, in other words, integrating the IAD over the full range of the probing particle in the medium. It will be interesting to compare the integrated Cherenkov distributions emerging from different energy loss models (discrete vs continuous or EM vs nuclear stopping power) in the future.

As seen in Fig.~\ref{f:rangeproton}, relativistic corrections to the  kinetic energy \eqref{e:beta1} lead to significant changes in the particle range $R$ even for low-energy protons \cite{rahman2024}. Other data sources, i.e., NIST, shown in yellow dashed line, fall in between the relativistic (green solid) and non-relativistic treatment (purple dot-dashed) of range.

%
\subsubsection{Integrated Distribution of Cherenkov Radiation}
\label{sec:adpid}
%

\begin{figure}[h] 
\centering
\includegraphics[width=0.6\textwidth]{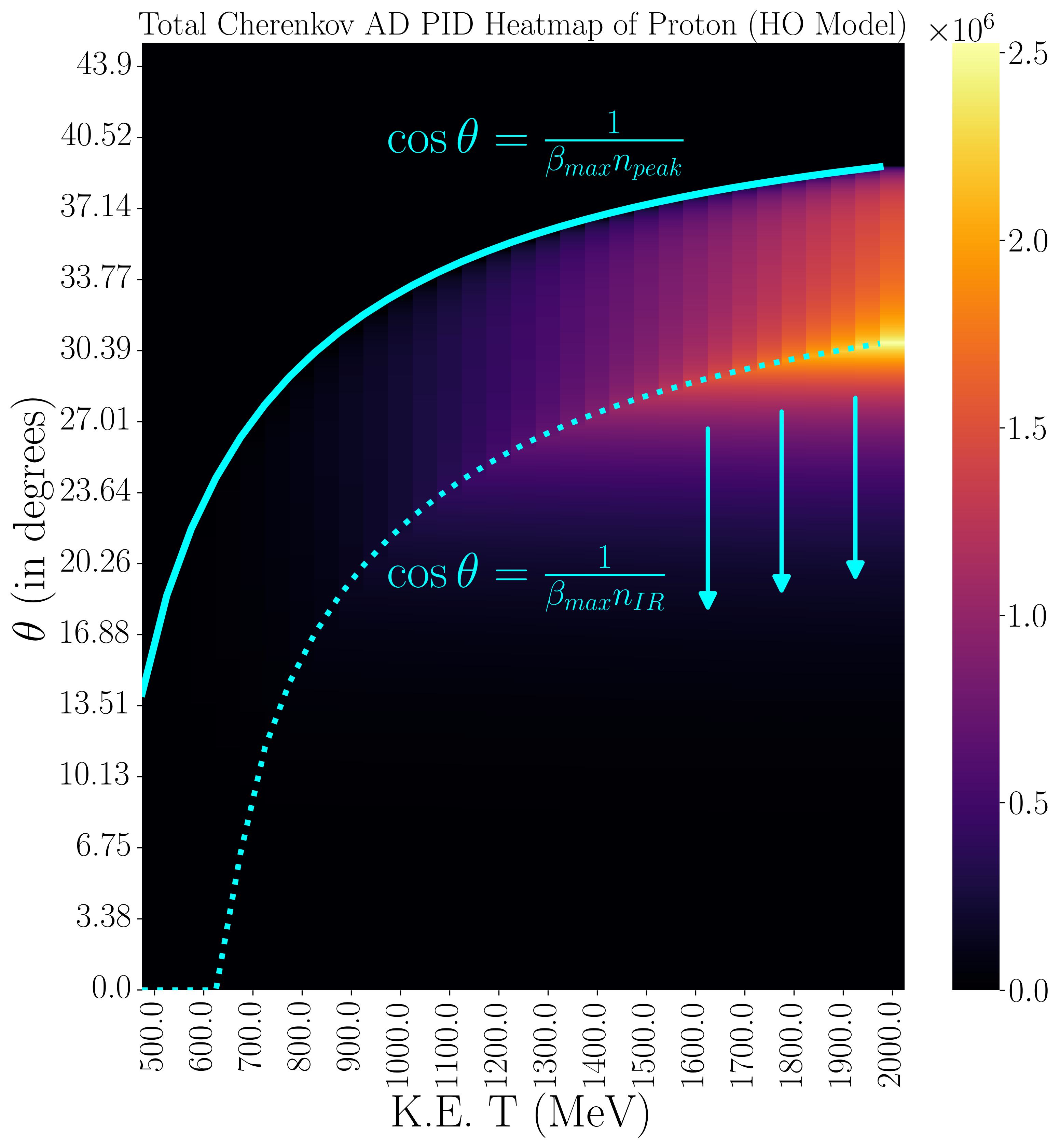}
\caption{General features of the integrated angular distribution.
\label{f:ADPID_Cartoon}
}
\end{figure}

Integrating over the trajectory $dx$ amounts to integrating over the decreasing velocity $\beta(x)$ of the particle over time. After numerically solving the Bethe-Bloch equation for $\beta(x)$, we the integrate from the initial position $x = 0$ to the range $x = R$ at which the particle comes to rest.  Integrating the IAD $\frac{dN}{d\Omega \, dx}$ gives the \textbf{angular distribution} (AD)
\begin{align}
    \frac{dN}{d\Omega} = \int\limits_0^R dx \, \frac{dN}{d\Omega \, dx} \: ,
\end{align}
and fully integrating the Cherenkov distribution gives the total yield
\begin{align}
    N = \int\limits_0^R dx \, \frac{dN}{dx} = \int d\Omega \frac{dN}{d\Omega} \: .
\end{align}

The total Cherenkov yield should be compared against the baseline of the typical scintillation yield.  As discussed in Ch.~\ref{intro}, liquid argon (LAr) is a prominent choice of detector material for neutrino physics, especially for its scintillation properties.  See Ref.~\cite{Segreto_2021} for a review.  Argon is an exceptionally good scintillator, with a scintillation yield given roughly by 
\begin{align} \label{e:scint1}
    N_{scint} &\approx T \times \left(\frac{40,000 \, \gamma}{\mathrm{MeV}}\right) \times \Big( 27.5\% \, \mathrm{prompt} \Big) 
    \notag \\ &=
    \left(\frac{T}{\mathrm{MeV}}\right) \times 11,000   \: ,
\end{align}
where charged particles in liquid argon release approximately $40,000$ photons per MeV of energy lost.  The scintillation photons are emitted from one of two channels: a ``fast component'' carrying 27.5\% of the total scintillation light with a time constant of 8 ns, and a ``slow component'' with a much longer time constant of 1.6 $\mu$s.  The fast component of the scintillation light is used as the primary signal for charged particle detection in the scintillator.  For Cherenkov light to be detectable above this baseline, the Cherenkov yield has to exceed the typical size $\sqrt{N_{scint}}$ of statistical fluctuations in the baseline \cite{rahman2024}.

The form of the angular distribution can be understood by superimposing the IAD seen in Fig.~\ref{f:IADPID_Annotated} as the particle decreases in kinetic energy.  Both the inner and outer edges of the Cherenkov spectrum move inward to smaller angles as the energy falls, as seen in Fig.~\ref{f:ADPID_Cartoon}. The result is that the outer edge of the AD is still a pure reflection of the initial, maximum velocity $\beta_{max}$ and is still sensitive to $n_{peak}$. The bright line reflecting IR pileup at the initial velocity $\cos\theta = 1 / \beta_{max} n_{IR}$ is also still visible. Smearing from the falling energy broadens the inner part of the AD but is much smaller than the huge signal coming from IR pileup at $\beta_{max}$. The main takeaway from the AD is that it smears -- but does not fully erase -- the structures created in the IAD at the initial velocity $\beta_{max}$. Just like the IAD, the outer edge of the AD is still a very sensitive model discriminator. 

The AD integrated over the trajectory may be experimentally more advantageous to measure than the IAD because of the better statistics coming from the higher yields.  The timing information necessary to extract the IAD may also not always be available, depending on the experimental setup. Nevertheless, both AD and IAD carry complementary information about the medium properties.

%
\subsection{Results}
\label{sec:results}
%


\subsubsection{Instantaneous and Total Cherenkov Yield}

\begin{figure}[t]
    \centering
        \begin{subfigure}{.48\textwidth}
        \includegraphics[width=1\textwidth]{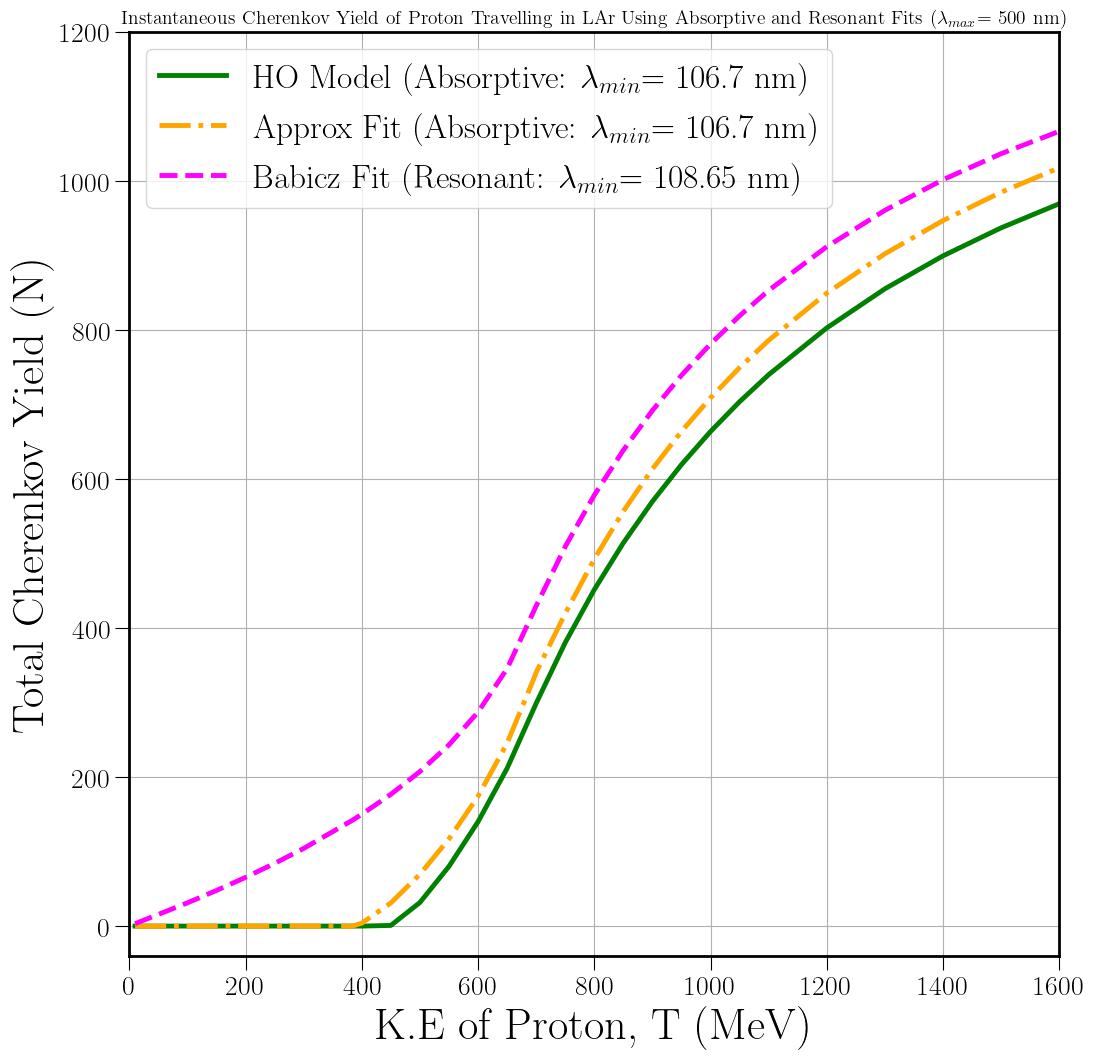}
        \caption{
        \label{f:abshoinstyieldallfits}
        }
        \end{subfigure} 
    \centering
        \begin{subfigure}{.48\textwidth}
        \centering
        \includegraphics[width=1\linewidth]{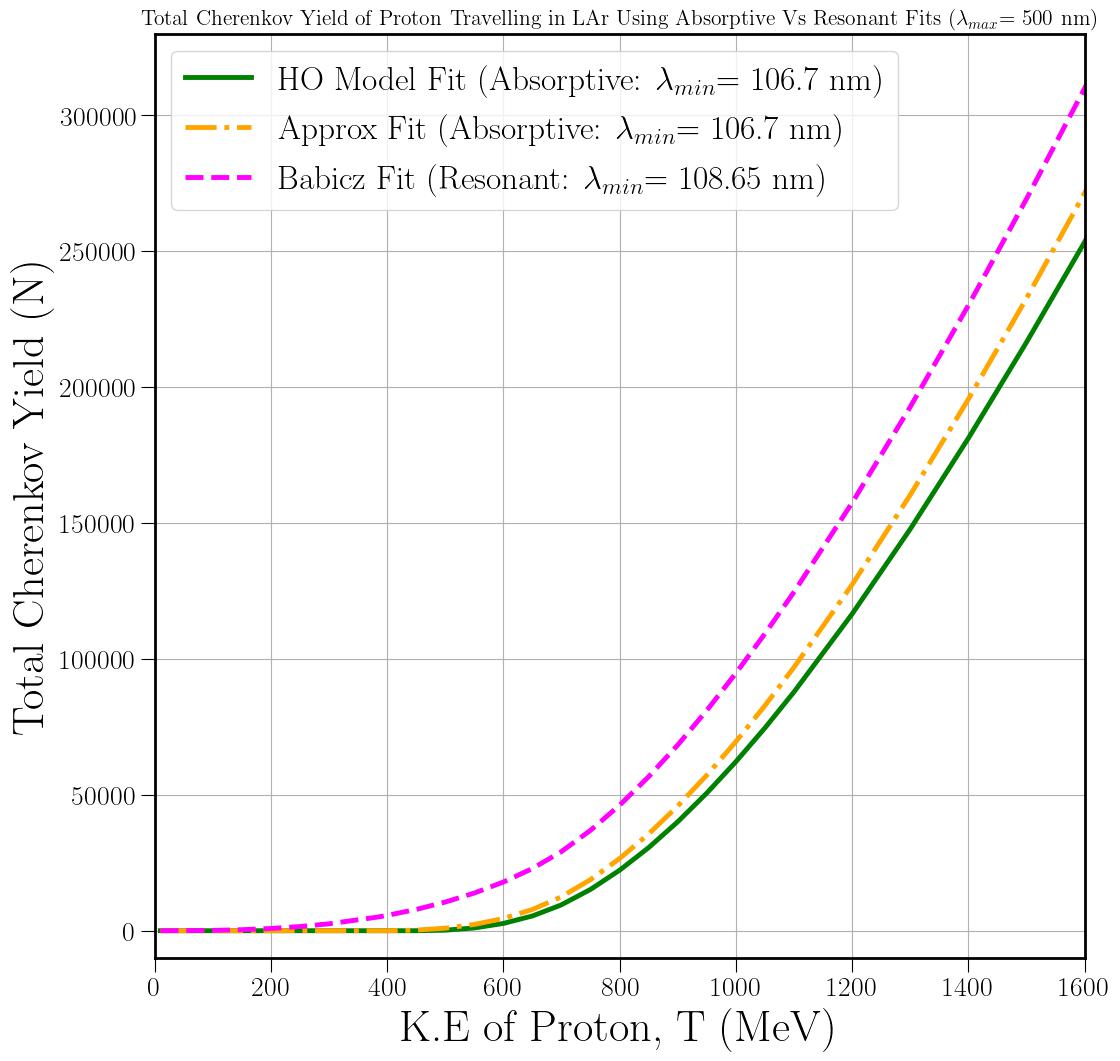}
        \caption{
        \label{f:abshointegyieldallfits}
        }
        \end{subfigure}
    \caption{Instantaneous (a) and integrated (b) Cherenkov yields for protons in LAr for different models of $n(\lambda)$.
    \label{f:Cherenkovnoabsvsnonabs}
    }
\end{figure}

In our previous work Ref.~\cite{rahman2024}, we performed detailed comparisons of the instantaneous and integrated yields of low-energy protons in LAr, for two resonant fits (Babicz \cite{Babicz:2020den}, Grace \cite{Grace:2015yta}) and two absorptive fits (HO, Approx). This study primarily focuses on the model discrimination power of the angular distributions, but there is still a nontrivial difference between models at the level of the yields.

In Fig.~\ref{f:Cherenkovnoabsvsnonabs}, we see that both instantaneous (left panel) and integrated yields (right panel) show that the absorptive fits have a finite threshold energy for Cherenkov radiation which depends upon $n_{peak}$ \cite{rahman2024}.  In contrast, the resonant fit by Babicz has no threshold, emitting Cherenkov photons even down to $T \rightarrow 0$.  This shows how qualitatively unrealistic the resonant fits would be if used to predict the Cherenkov radiation.  There is also a non-negligible difference in yields between the two absorptive models, so that even without angular resolution, the Cherenkov yield carries some model discrimination power between two realistic models of the LAr refractive index. Having calculated the yields is also a useful numerical benchmark for the later angular distributions, to test for self-consistency.


In my MS thesis \cite{rahman2024}, we compared integrated Cherenkov yield computed from the harmonic oscillator fit \eqref{e:nlambdauvhofinal} with the standard error $\sqrt{N_{scint}}$ in the scintillation background. Despite the reduction in yield compared to the resonant fits (i.e. Babicz), the absorptive fit \eqref{e:nlambdauvhofinal} still predicted that the total Cherenkov yield will exceed the detection threshold over the background for sufficiently high-energy protons.  Whereas for the resonant fits, this detection threshold was crossed very quickly, at $200 - 300$ MeV, which is unphysical,
for the absorptive fit, the crossing of the detection threshold was found to be delayed to $595$ MeV.


%
\subsubsection{Instantaneous Angular Distributions and PID}
\label{sec:Instantaneous}
%

\begin{figure}[t] 
\centering
\begin{subfigure}{.47\textwidth}
\centering
\includegraphics[width=1\textwidth]{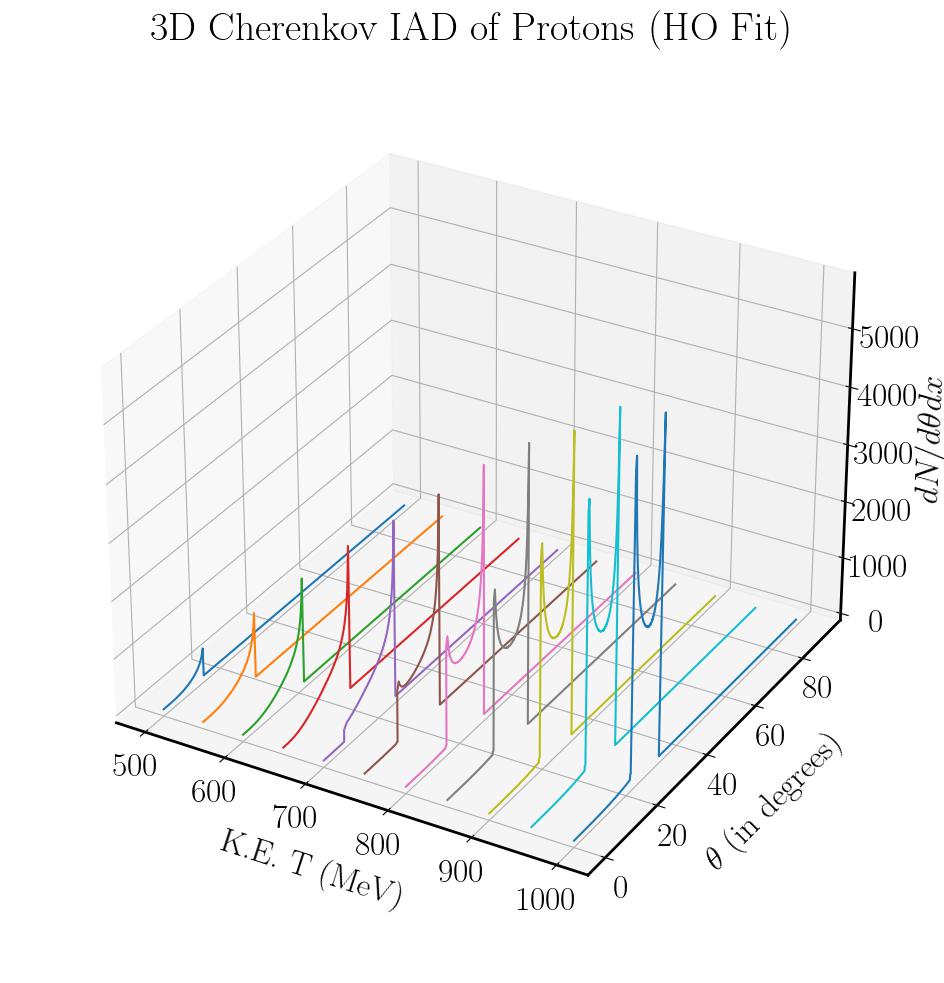}
\caption{
\label{f:proton3diadho}
}
\end{subfigure}       
\begin{subfigure}{.47\textwidth}
\centering
\includegraphics[width=1\textwidth]{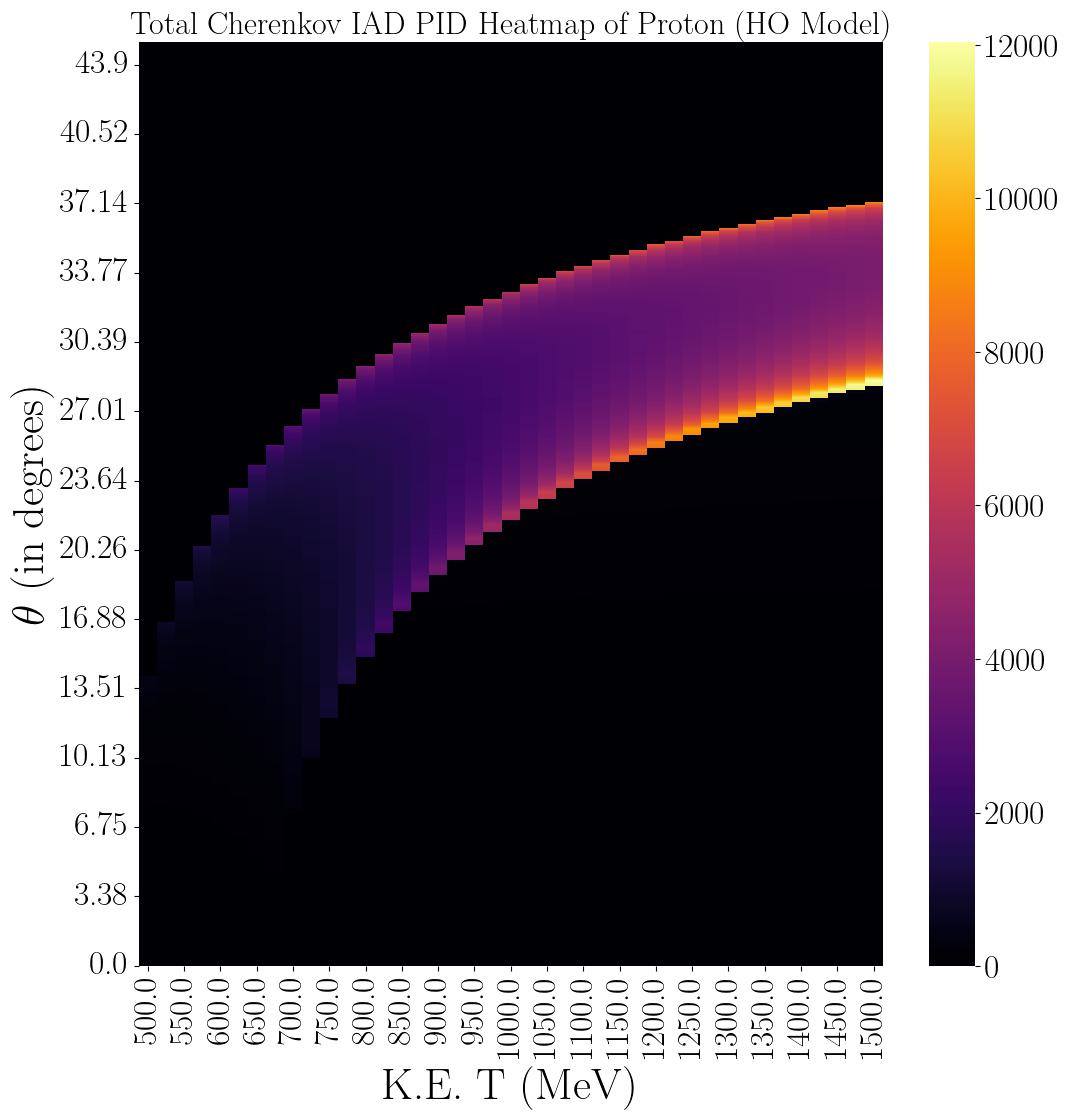}
\caption{
\label{f:protonpidiadheatmapho}
}
\end{subfigure}
\caption{Instantaneous AD of protons in LAr using the HO model. Angular distributions for select energies are shown (a), as well as the two-dimensional intensity map (b).
\label{f:protonpidiadho}
}
\end{figure}

\noindent

Fig.~\ref{f:protonpidiadho} shows the IAD of protons in LAr for various energies in the HO model of $n(\lambda)$, showing the qualitative change in profile with energy. The intensities shown here are orders of magnitude more statistically significant over the scintillation background, than the yield itself, as discussed in Ref.~\cite{rahman2024}.
    
Due to the absorptive form of the refractive index, the realistic Cherenkov IAD has a finite range $\theta_{min} \leq \theta \leq \theta_{max}$.  In contrast, resonant fits with $n_{peak} \rightarrow \infty$ radiate out to high angles, at all energies.  At low energies near the Cherenkov threshold, the IAD has a single peak located at the outer edge $\cos\theta = 1 / \beta n_{peak}$.  With increasing energy, that outer bright edge moves outward to larger $\theta$ and increases in brightness. At higher energies $\gtrsim 700 \, \mathrm{MeV}$, the long-wavelength modes far from the resonance begin to emit Cherenkov radiation, resulting in a rapid pileup of intensity at the inner edge $\cos\theta = 1 / \beta n_{IR}$. With increasing energy, the inner bright edge moves outward and greatly increases in intensity.

The two competing features, outer bright edge and inner bright edge, reflect opposite features of the refractive index.  The outer bright edge is a direct reflection of $n_{peak}$ and is therefore sensitive to different models of the behavior near the resonance; the inner bright edge is a reflection of the IR physics far from the resonance and is universal across even resonant models. Precisely comparing the height and shape of the inner and outer bright peaks can convey detailed information about the form of the refractive index.

\begin{figure}[t] 
\centering
\begin{subfigure}{.47\textwidth}
\centering
\includegraphics[width=1\textwidth]{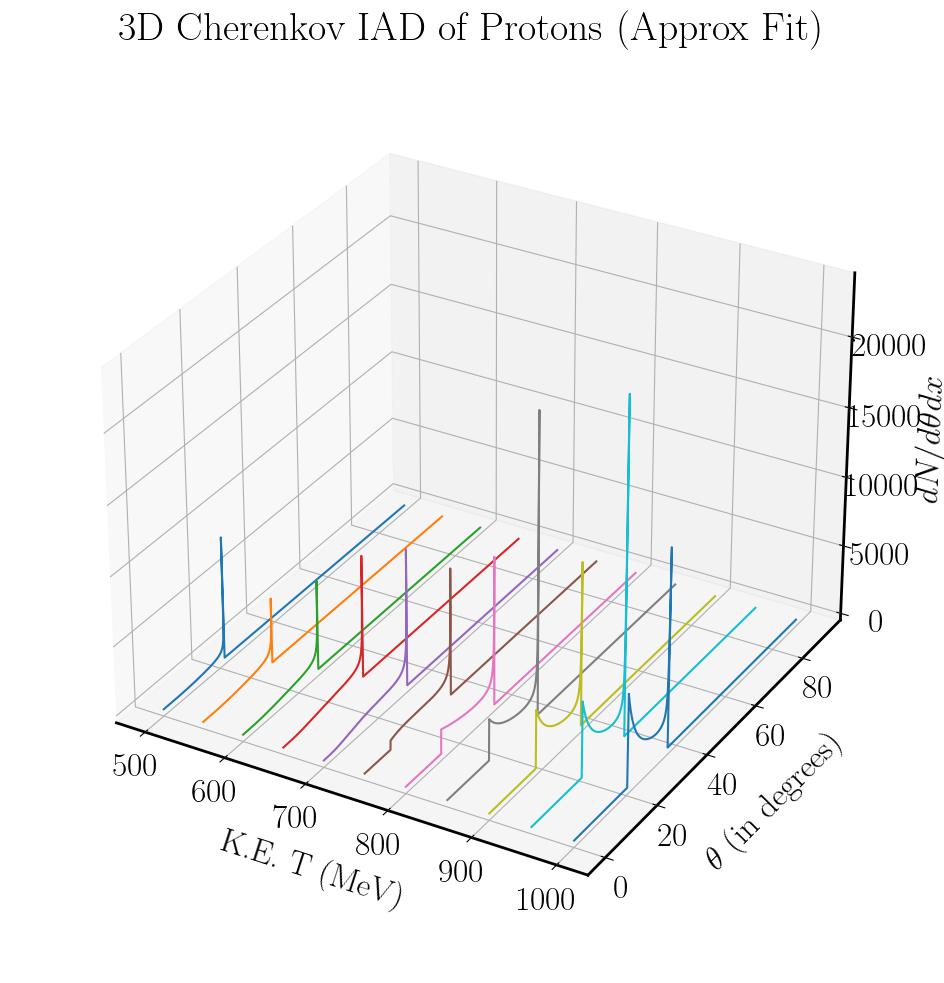}
\end{subfigure}       
\begin{subfigure}{.47\textwidth}
\centering
\includegraphics[width=1\textwidth]{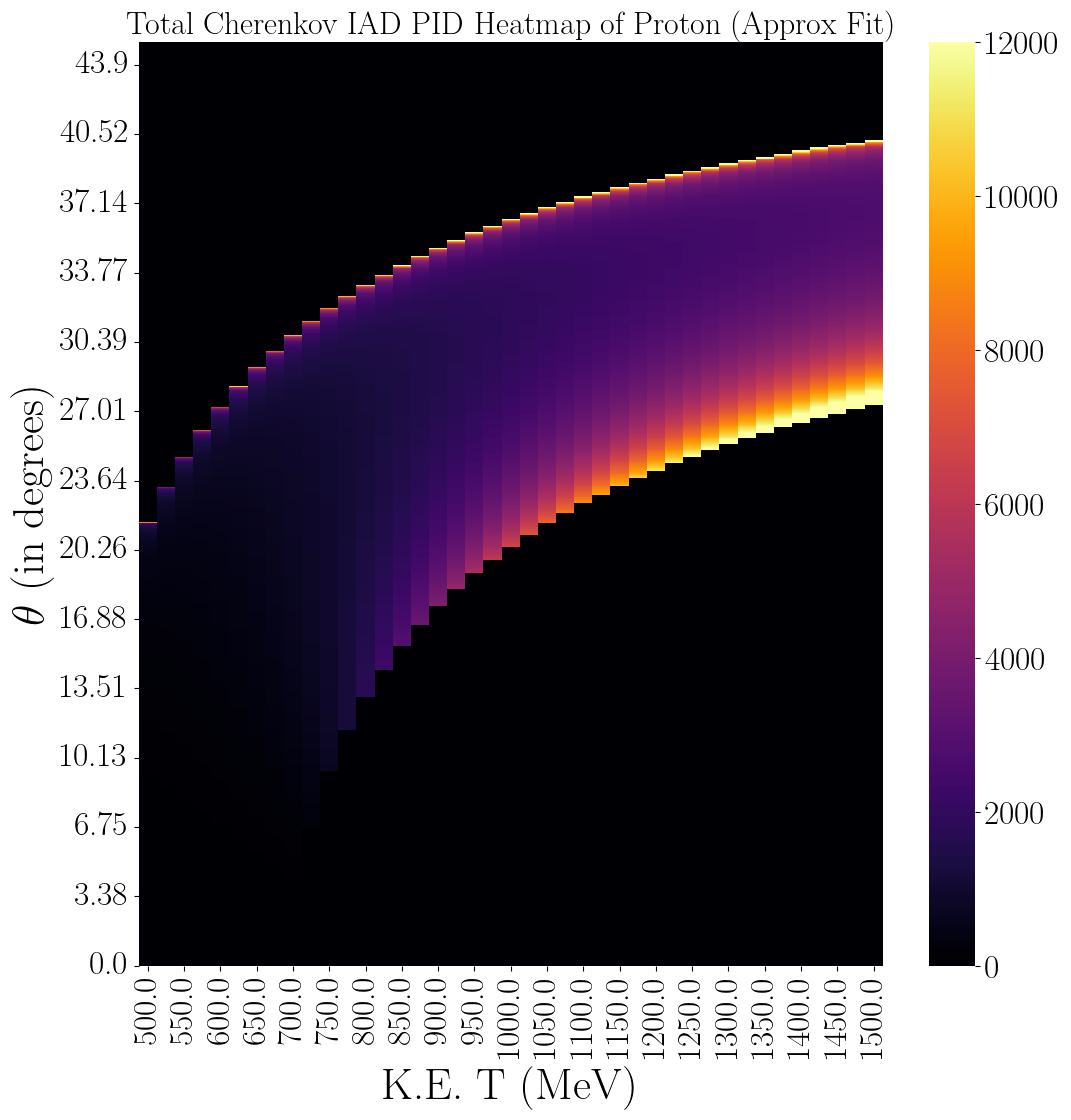}
\end{subfigure}
\caption{Instantaneous AD of protons in LAr using Approx fit (left) and associated PID (Cherenkov angle ($\theta_{c}$) versus proton's K.E.) heatmap (right). 
\label{f:protonpidiadapprox}
}
\end{figure}

\noindent

Just as in Fig.~\ref{f:protonpidiadho}, we present IAD for protons in the Approx model in Fig.~\ref{f:protonpidiadapprox}, showing the qualitative change in profile with energy.  
Although both the HO and Approx models are absorptive and have similar shapes (see Fig.~\ref{f:nvslambdahoandBabiczfit}), differences in model parameters, such as peak location and height, produce distinct angular distributions for the two fits. The outer bright edge in the Approx model occurs at larger angles than in HO, due to the increase in $n_{peak}$. 

On the other side, the inner bright edge in the Approx model occurs at about the same angles as in HO, due to the comparable values of $n_{IR}$. But the intensity on the inner edge takes longer to build up (needs higher energies).  This seems to indicate some amount of model discrimination power even in the IR. As a result, there are systematic differences in the height and shape of inner versus outer peak which reflect the IR and UV differences in the refractive index between HO and Approx models. Defining a quantitative measure of these differences can provide a powerful metric for model discrimination.

\begin{figure}[t] 
\begin{centering}
\includegraphics[width=0.50\textwidth]{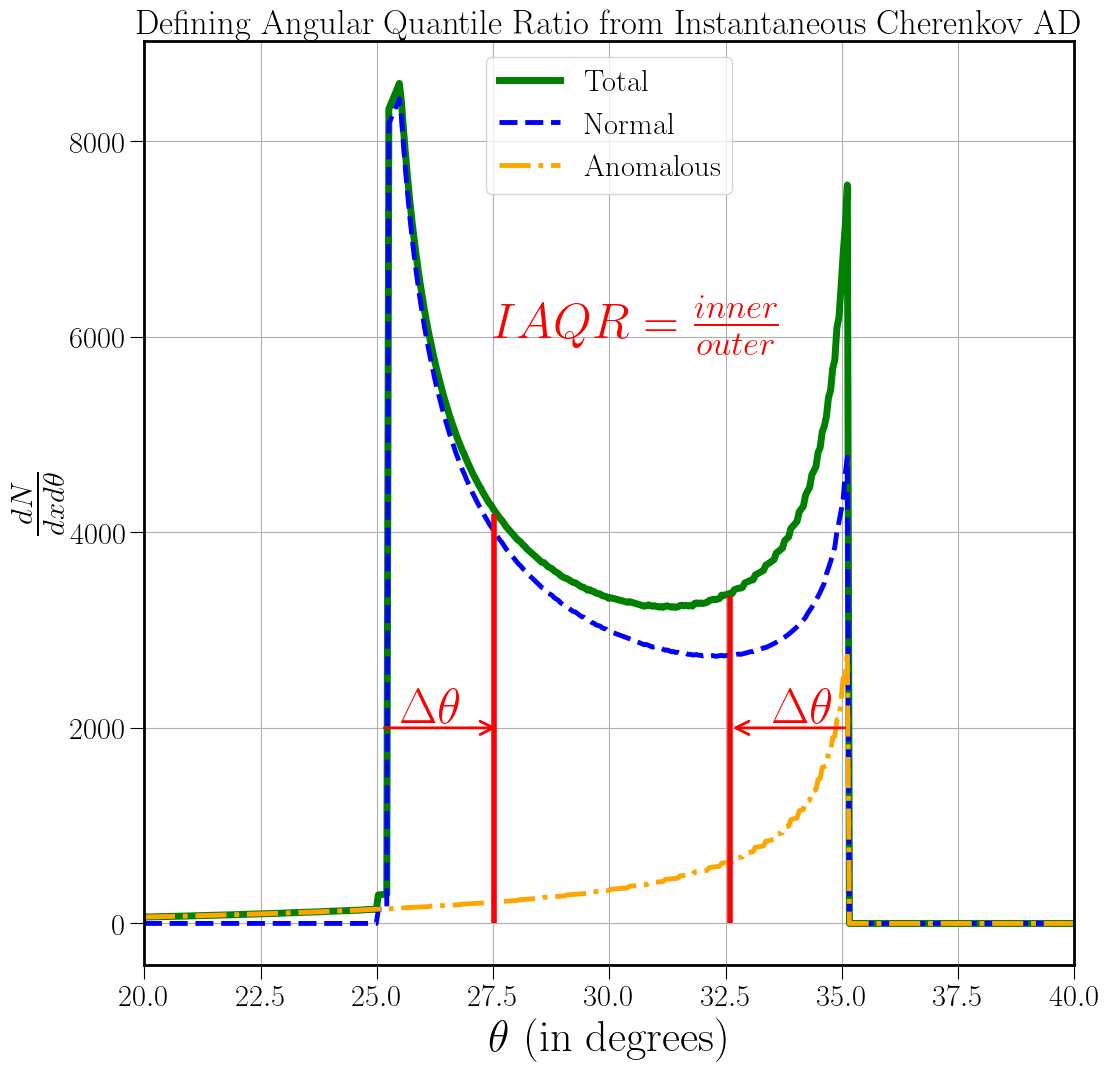}
\caption{Illustration of the angular quantile ratio (IAQR) for the Instantaneous AD.
\label{f:IAQR_Cartoon}
}
\end{centering}
\end{figure}

\noindent

Next, let us introduce a metric that we call the ``Instantaneous Angular Quantile Ratio (IAQR)'' as the ratio of the integrated intensities around the inner and outer edges of the distribution:
\begin{align}   \label{e:IAQR_defn}
    IAQR \equiv \frac{
    \int\limits_{\theta_{min}}^{\theta_{min} + \Delta\theta}{d\theta} \, \frac{dN}{dx \, d\theta}
        }{
    \int\limits_{\theta_{max} - \Delta\theta}^{\theta_{max}}{d\theta} \, \frac{dN}{dx \, d\theta}
        }
\end{align}
where $\theta_{max} (\theta_{min})$ are the maximum (minimum) angles at which nonzero Cherenkov light is observed, and $\Delta\theta$ is a predefined window, chosen to be a fraction (e.g. 10\%) of the total angular window.  This observable is defined to compare the relative intensities of the two peaks (inner / outer) so as to quantify the relative contributions of the IR / UV regions for a given refractive index, as illustrated in Fig.~\ref{f:IAQR_Cartoon}.  Note the sensitivity to the contribution of anomalous photons from the outer edge. The IAQR is an observable metric that can estimate the relative contribution of anomalous to normal Cherenkov yield. 

The power of the IAQR \eqref{e:IAQR_defn} is the flexibility of the angular cut $\Delta\theta$, which can be loosened or tightened as desired to maximize sensitivity. To illustrate this feature in Fig.~\ref{f:IAQR_Cartoon}, we can see that on the left side of the plot, the normal distribution is the dominant contribution to the total IAD, but the anomalous contribution also starts growing slowly with increasing Cherenkov angle $\theta$ and becomes quite significant near the $\theta_{max}$ region. Even though the normal part might still be higher in intensity, both normal and anomalous distributions contribute to the double peak structure seen for higher energetic particles (for both muons and protons that are discussed in this paper). We performed the analysis for three different choices of the angular cut $\Delta \theta$, which are 10\%, 5 \%, and 1 \% shown in red, blue, and green curves respectively. We will be interested in tuning $\Delta\theta$ to optimize the model discrimination power.

\begin{figure}[t] 
\begin{centering}
\includegraphics[width=0.5\textwidth]{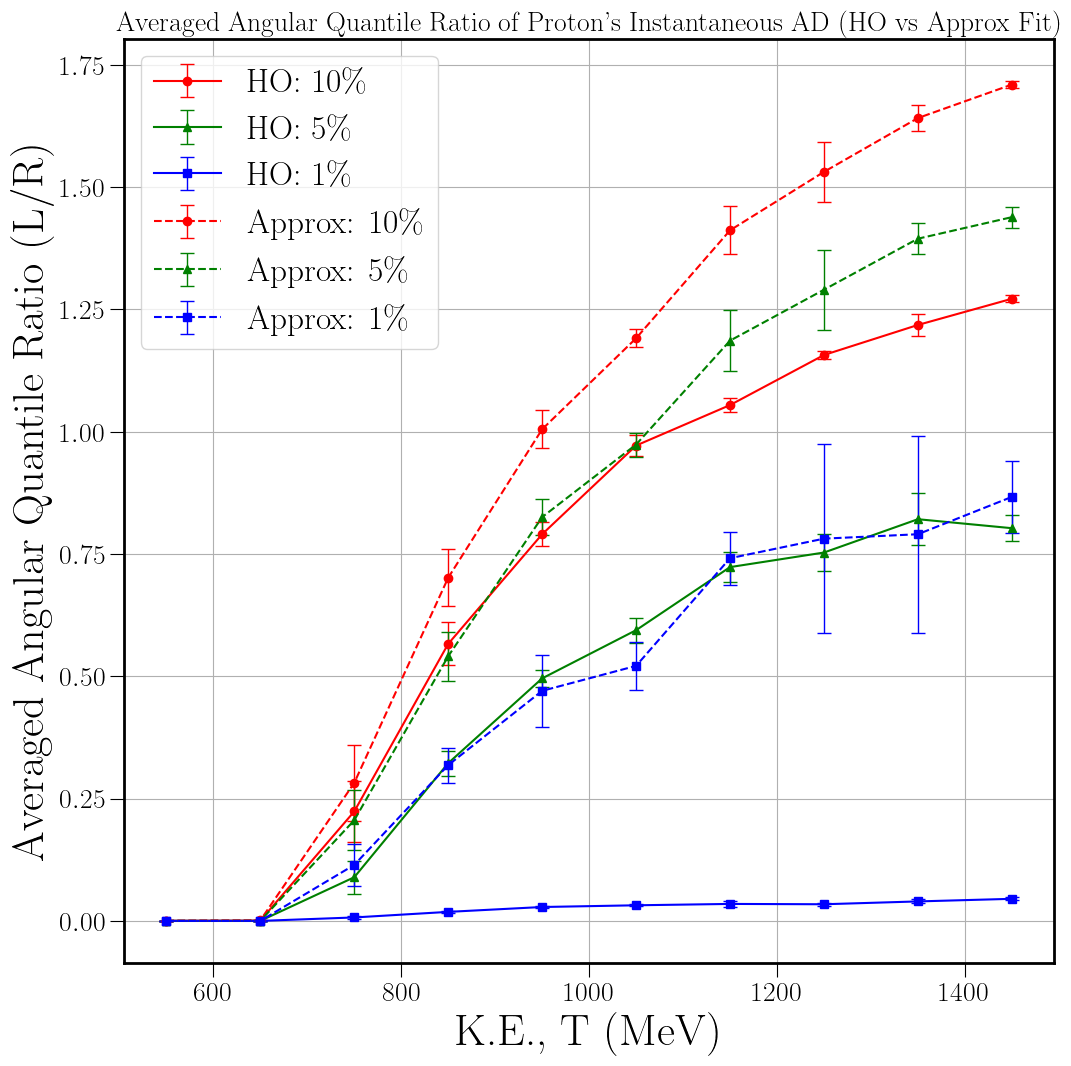}
\includegraphics[width=0.49\textwidth]{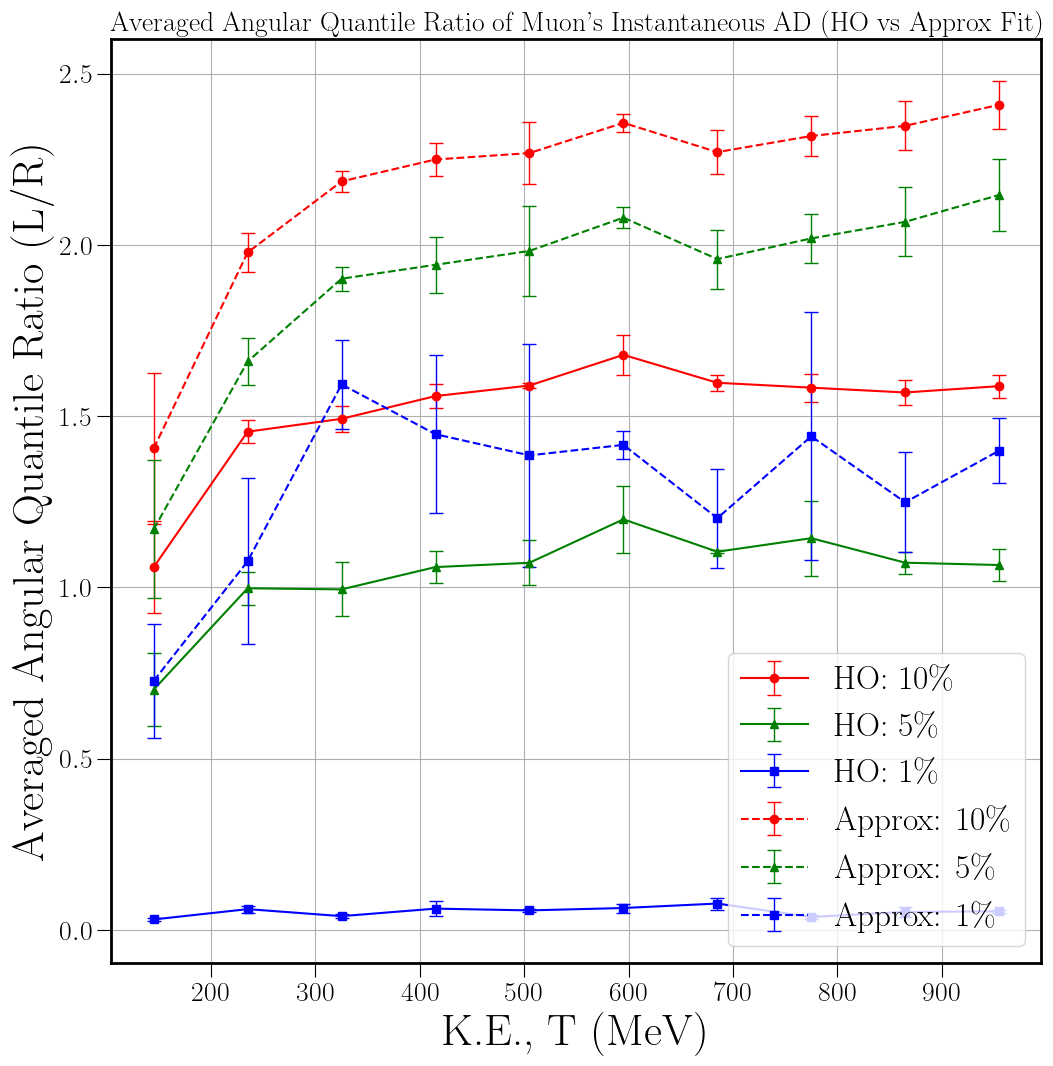}
\caption{Averaged Angular Quantile Ratio (IAQR) of Proton (left Plot) and Muon (Right Plot) from their respective instantaneous ADs.
\label{f:avgangquantratiop&muhovsapprox}
}
\end{centering}
\end{figure}

\begin{figure}[h] 
\begin{centering}
\includegraphics[width=0.495\textwidth]{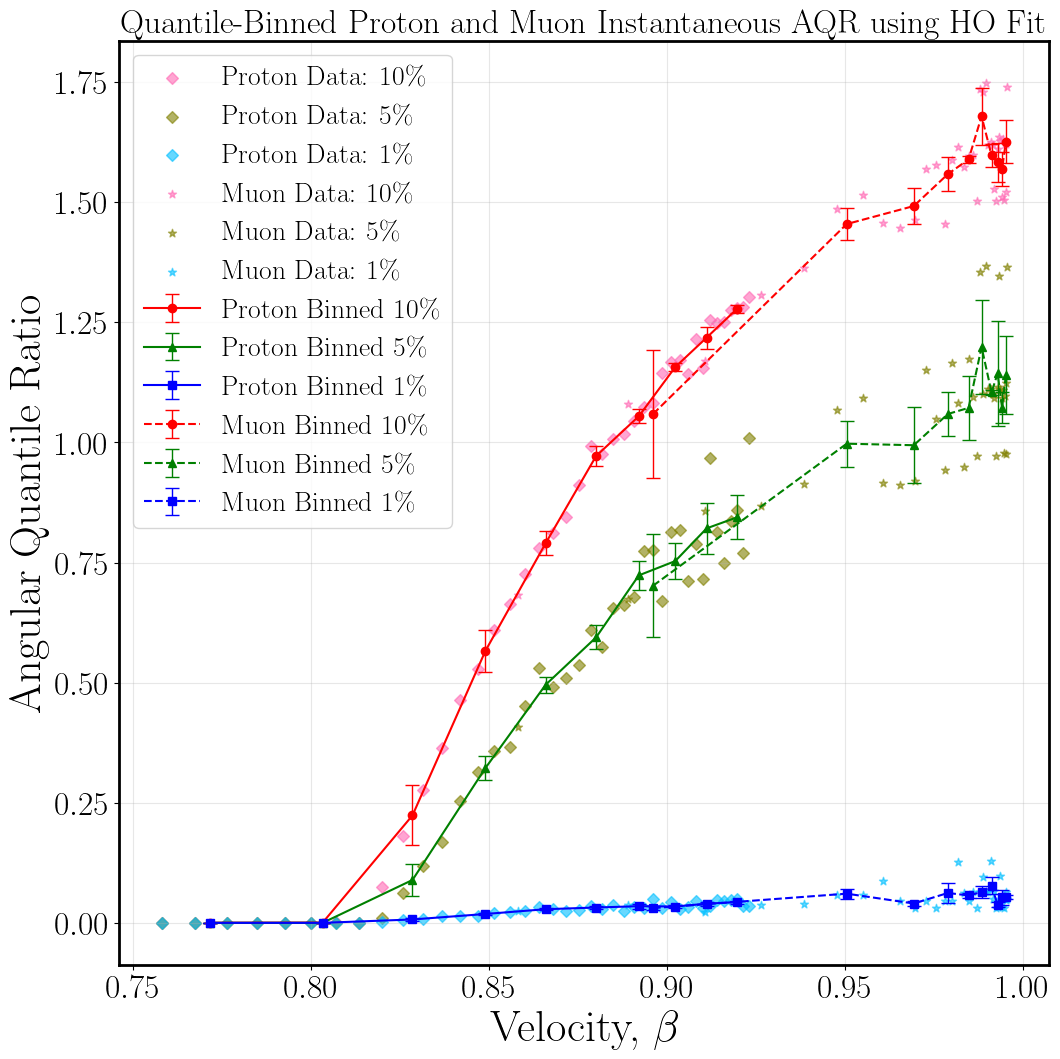}
\includegraphics[width=0.495\textwidth]{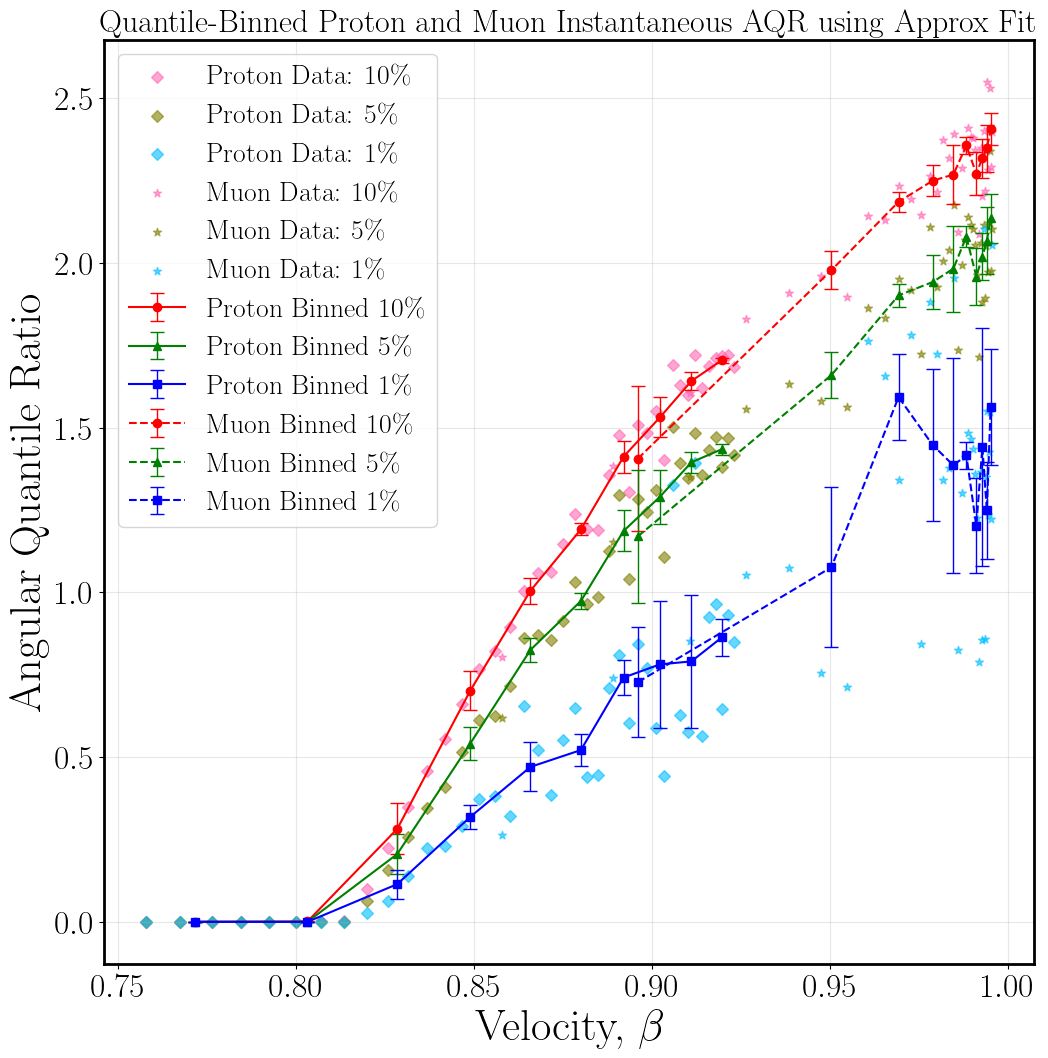}
\caption{Binned IAQR of proton and muon with jackknife error as a function of velocity $\beta$ plotted together with the respective raw IAQR data for HO fit (left panel) and Approx fit (right panel). 
\label{f:iAQRbetaplot}
}
\end{centering}
\end{figure}

In Fig.~\ref{f:avgangquantratiop&muhovsapprox}, we present a comparison of instantaneous IAQR for protons (left panel) and muons (right panel) for two different models of $n(\lambda)$ as a function of K.E., $T$. These plots were generated by binning the IAQR data in the respective bins of K.E., $T$ for three different choices of the angular cut $\Delta \theta$, which are 10\%, 5 \%, and 1 \% shown in red, blue, and green curves respectively. From both panel of this Fig, we see that the size of the AQR is significant to be measurable, with a signal strength that increases with energy but decreases with tighter cuts. Numerical uncertainty in the calculation is represented using jackknife resampling in bins of the kinetic energy. The AQR shows model discrimination ability (significant difference between IAQR for different models). The model discrimination power increases with energy and increases with tighter cuts.

Fig.~\ref{f:iAQRbetaplot}, IAQR for protons and muons have been merged onto the same plot for HO model (left panel) and Approx fit (right panel) of $n(\lambda)$. For both of these models, three choices of 10\%, 5 \%, and 1 \% angular cuts are shown in red, blue, and green curves, respectively. Proton (diamond shape) and muon (asterisks shape) raw IAQR data are shown on the plot along with the binned plots with jackknife error, where the solid lines represent protons and dashed lines connect the muon data. It is worth mentioning that we used quantile binning to generate this plot, rather than equally spaced (linear) binning, to ensure that each bin has a comparable data size, which is essential for specially the muon case to avoid running out of data in the lower $\beta$s. Fig.~\ref{f:iAQRbetaplot} shows that for both fits, different particles have the same instantaneous distribution in terms of $\beta$, but probe different energy ranges due to the mass for both fits. This means that muons probe the high-$\beta$ asymptotics of the IAQR, while the protons probe the threshold region. The IAQR successfully serves its intended purpose: it greatly enhances the model discrimination power for different choices of $n(\lambda)$, with different particles being sensitive to different regions or features. This is a reversal of the usual PID paradigm, where one uses knowledge of $n(\lambda)$ to distinguish between particle species; instead, we can use a range of different particle species to gain new knowledge about $n(\lambda)$.

\begin{figure}[t] 
\begin{centering}
\includegraphics[width=0.49\textwidth]{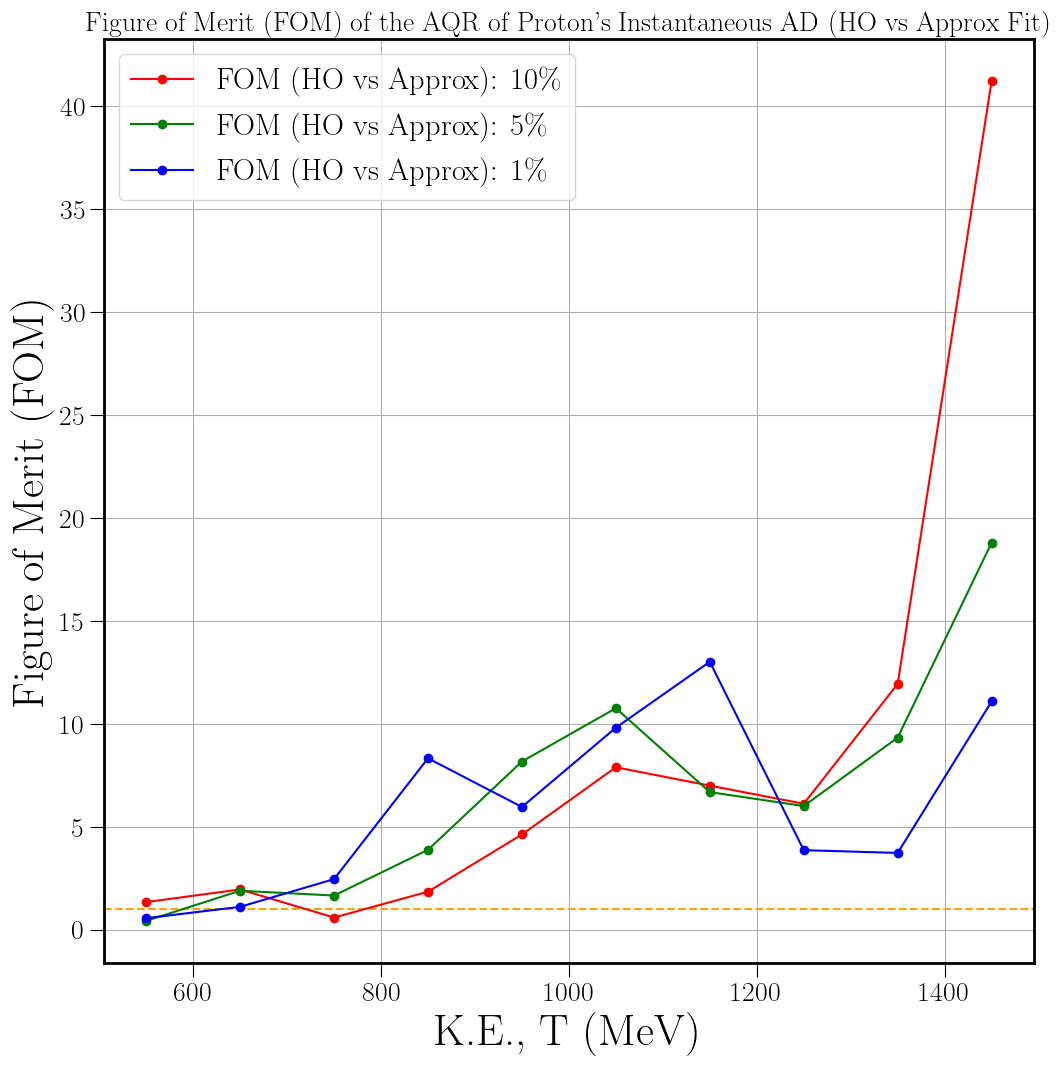}
\includegraphics[width=0.49\textwidth]{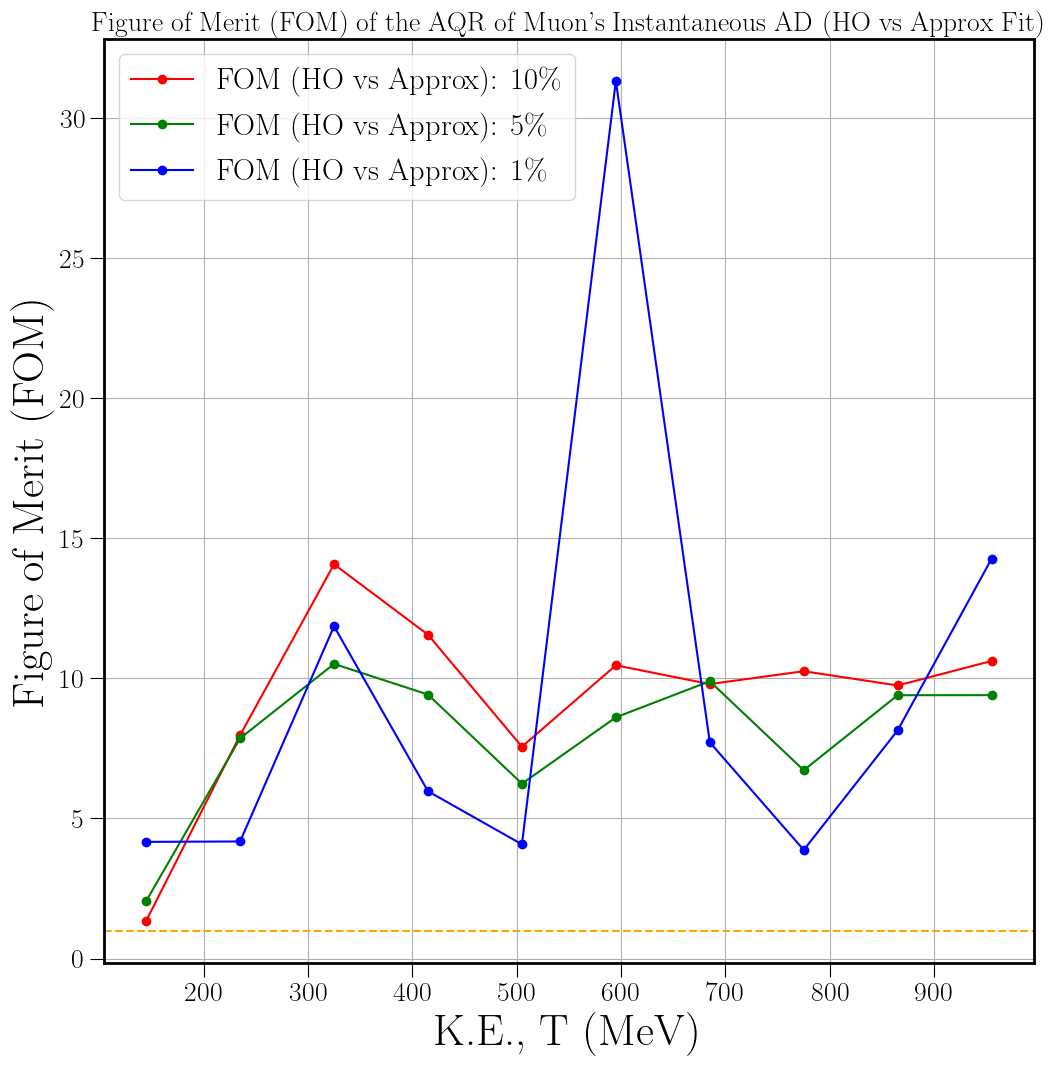}
\caption{Figure of Merit (FoM) of the IAQR for protons (left) and muons (right). The dashed orange horizontal line is set at $FoM=1$; points above this line indicate that the difference between the two models is greater than their combined uncertainty. 
\label{f:aqr_iAD_fom_proton}
}
\end{centering}
\end{figure}

Next, we define a Figure of Merit (FoM) describing the model discrimination power of the instantaneous AQR by the ratio of the model separation to the uncertainty:
\begin{align}
    \mathrm{FoM} = \frac{|(\mathrm{AQR})_1 - (\mathrm{AQR})_2|}{\sqrt{ (\Delta \mathrm{AQR})_1^2 + (\Delta \mathrm{AQR})_2^2}}
\end{align}

\noindent
In Fig.~\ref{f:aqr_iAD_fom_proton}, the FoM of the IAD is plotted which is a universal function of $\beta$ that increases with kinetic energy. Little dependence is seen on the tightness of the angular cut $\Delta\theta$ used (consistent with zero).  This suggests that the increased separation between AQR of the two models as the cut is tightened is being offset by the increased uncertainty. Although the uncertainty in this calculation is numerical, not statistical, both types of error increase with tighter cuts $\Delta\theta$ for the same reason: the shrinking phase space used to compute the observable.

The FoM suggests that higher energies have better model discrimination power and that the choice of $\Delta\theta$ can be made to optimize other considerations without negatively impacting the model discrimination power.

%
\subsubsection{Integrated Angular Distributions and PID}
\label{sec:Integrated}
%

\begin{figure}[t] 
\centering
\begin{subfigure}{.49\textwidth}
\centering
\includegraphics[width=1\textwidth]{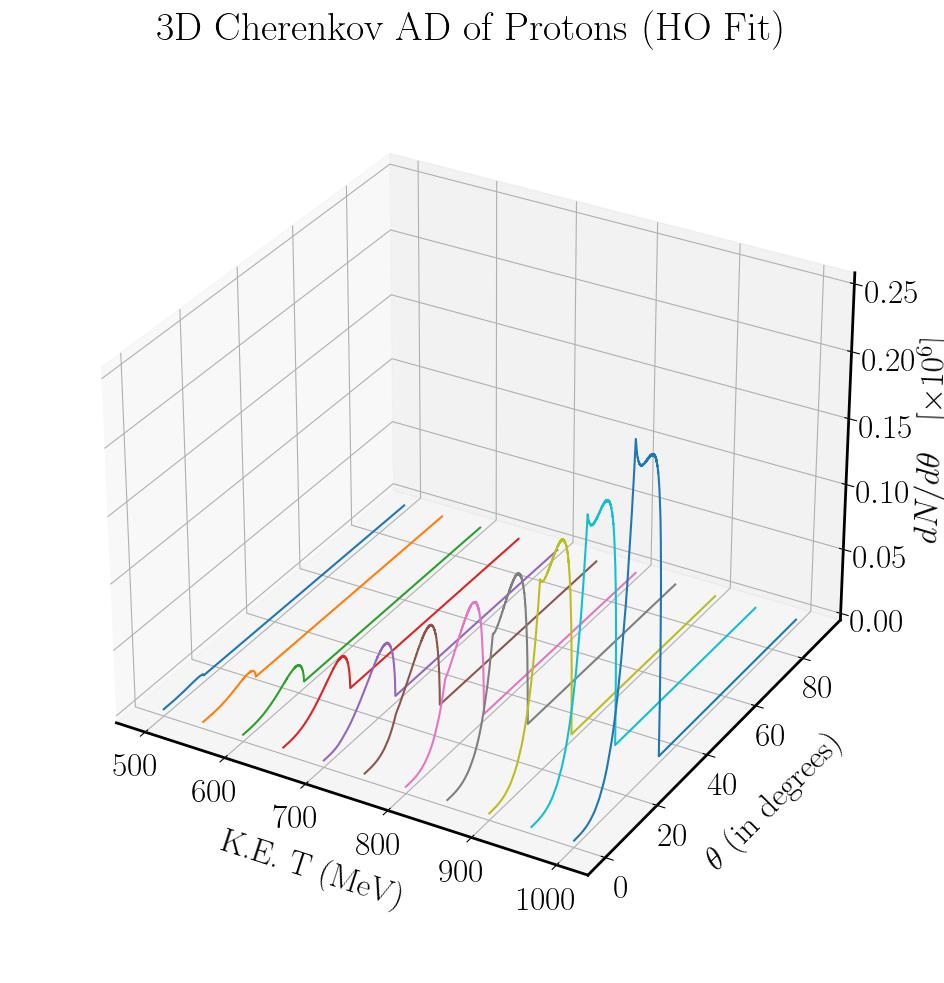}
\label{f:proton3dpidadho}
\end{subfigure}   
\begin{subfigure}{.49\textwidth}
\centering
\includegraphics[width=1\textwidth]{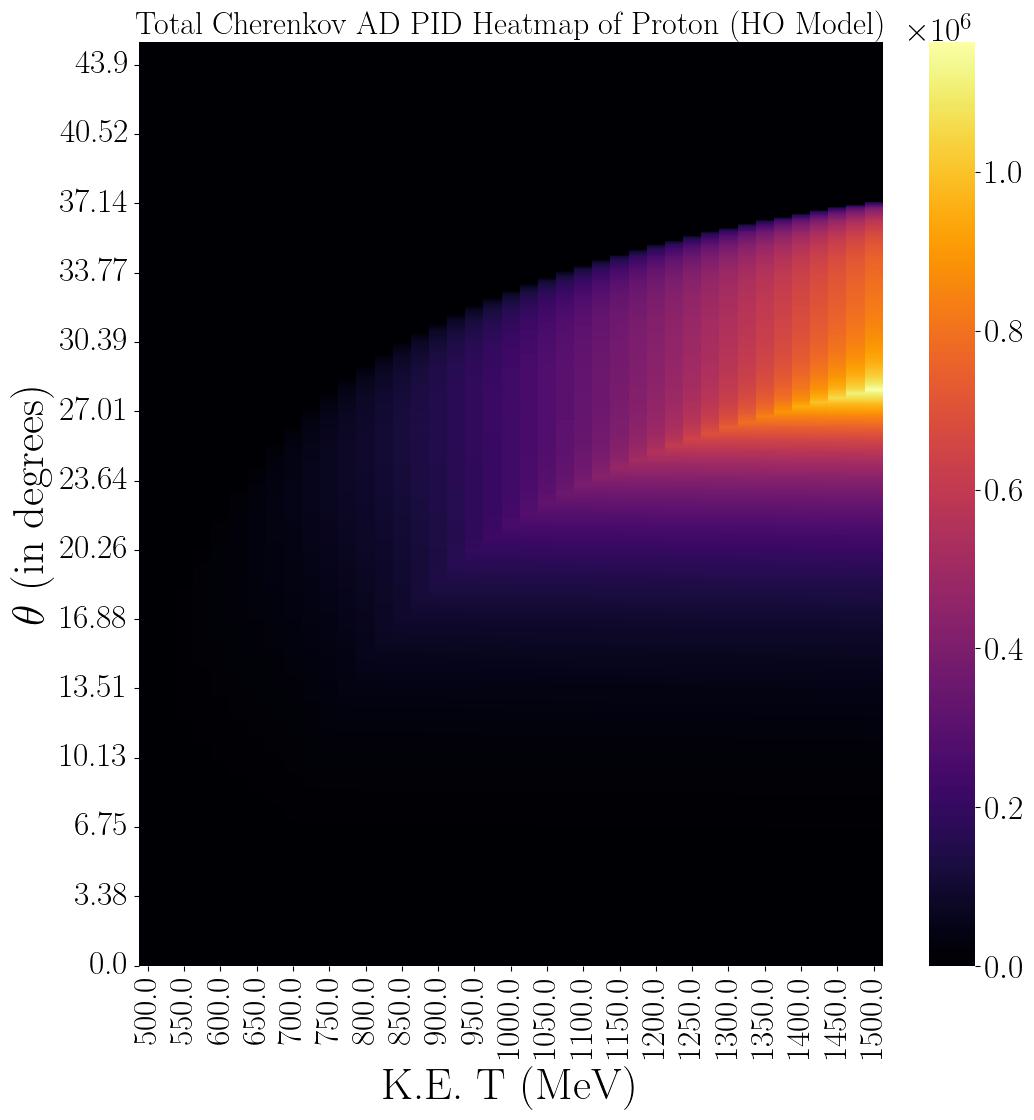}
\label{f:protonadheatmaptotal}
\end{subfigure}
\caption{Integrated AD of protons in LAr using HO model fit (left) and associated PID heatmap (right).
\label{f:protonadpidho}
}
\end{figure}

Now, we will extend our discussion of the AQR analysis at the level of integrated AD. It will be interesting to see if the AQR still be a useful metric for differentiating / distinguishing between the two models of $n(\lambda)$ as in Sec.~\ref{sec:Instantaneous}.   

To begin with, let's look at the integrated AD of proton generated from both the HO and approx model of $n(\lambda)$. 
In Fig.~\ref{f:protonadpidho}, the integrated angular distribution (AD) of a proton calculated from the HO model fit for various energies is shown on the left panel and the corresponding heatmap in the style of a PID plot is shown on the right panel. From the left panel plot, we can see that the leading outer edge is significantly smeared, but the outermost edge is still a clean signature of the initial velocity. The bright peak arising from IR pileup now no longer occurs on the innermost edge of the spectrum, because $\theta_{min}$ decreases as the particle loses energy. The bright IR peak still remains though, concentrated around $\cos\theta = 1 / \beta_{max} n_{IR}$, as with the instantaneous angular distribution.  

At higher energies, the outermost edge of the spectrum flattens into a ``shoulder,'' while the peak corresponding to IR pileup continues to grow. This suggests that a similar AQR comparison which highlights the peak and the shoulder regions can be even more sensitive to the form of the refractive index.

\begin{figure}[t] 
\centering
\begin{subfigure}{.49\textwidth}
\centering
\includegraphics[width=1\textwidth]{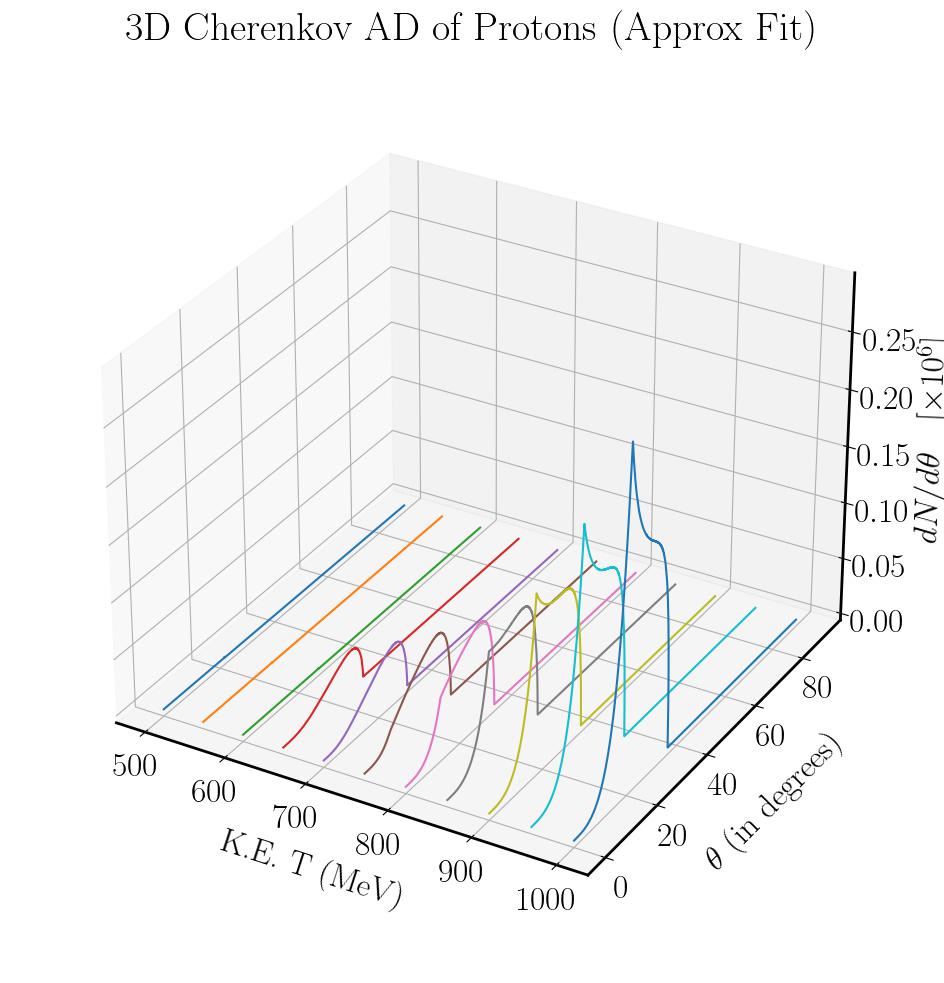}
\end{subfigure}   
\begin{subfigure}{.49\textwidth}
\centering
\includegraphics[width=1\textwidth]{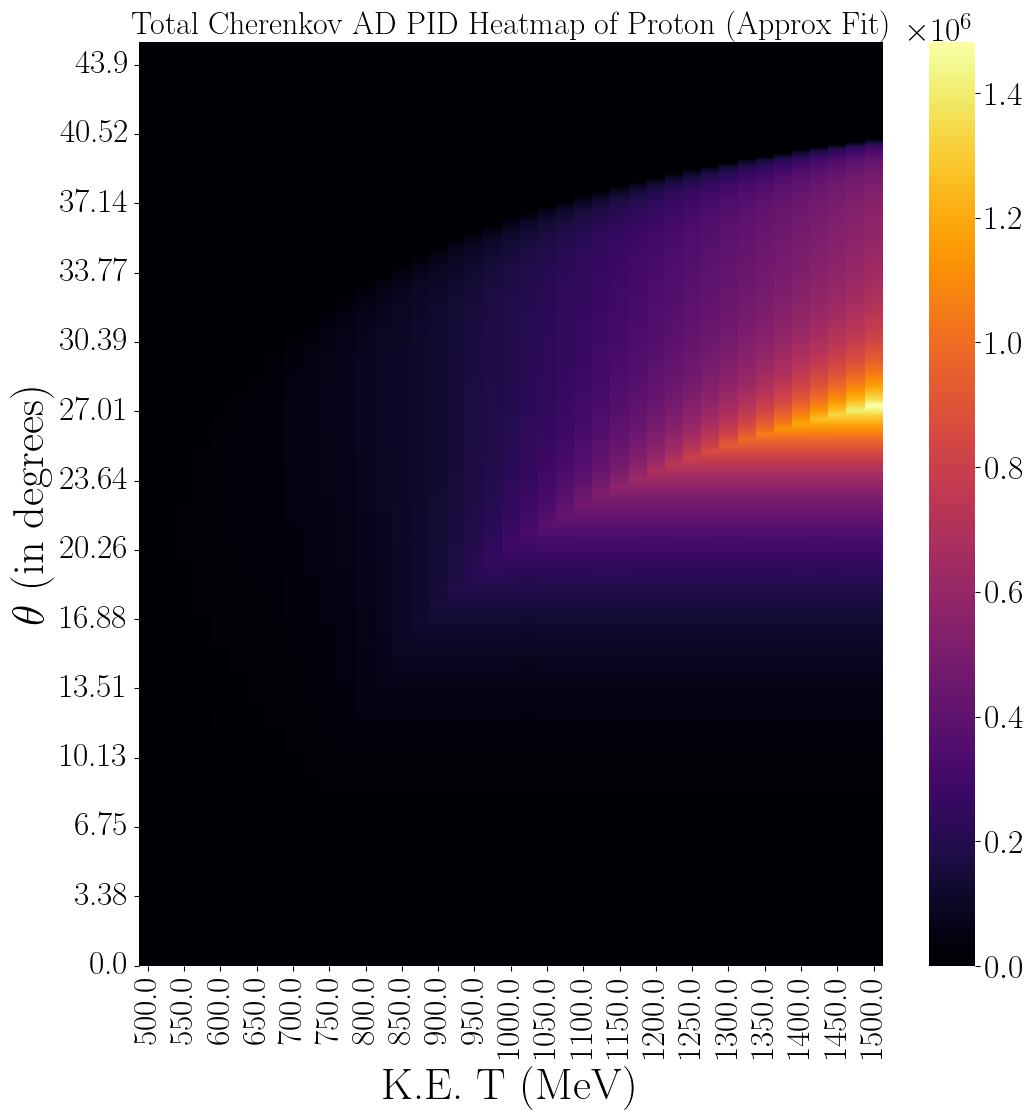}
\end{subfigure}
\caption{Integrated AD of protons in LAr using Approx fit (left) and associated PID heatmap (right).
\label{f:protonadpid_approx}
}
\end{figure}

In Fig.~\ref{f:protonadpid_approx}, we show the integrated AD computed using the Approx fit of $n(\lambda)$ (left panel) and its associated PID plot (right panel). Similar to the AD of the HO fit shown previously in Fig.~\ref{f:protonadpidho}, we see that the Approx fit produces a similar evolution of the AD with increasing $T$. Not only does the angular spread increase in the AD with growing $T$, but also a second peak emerges and grows originating from the IAD double peak structure for the corresponding $T$. The highest peak (and the corresponding brightest region shown in the PID plot in Fig.~\ref{f:protonadpid_approx}) originating from the IR pileup occurs at the inner region, which is then followed by a tail produced from the same smearing effect discussed in the HO case (visible in the PID plot as well- below the brightest region). On the higher angle emission side, we see a comparatively sharper fall (compared to Fig.~\ref{f:protonadpidho}) of the intensity from the brightest peak towards the outer UV-region peak before it falls to the outer boundary set by Cherenkov condition. The striking difference in the AD between these two models are the difference in the relative brightness around the brighter (IR) peak and the fainter (UV) shoulder of the AD. There is great potential to discriminate between models by zooming in on the two regions (IR peak versus UV shoulder). Experimental measurements or data with fine resolution in these regions can also help fine tune the model parameters of $n(\lambda)$.

\begin{figure}[t] 
\begin{centering}
\includegraphics[width=0.50\textwidth]{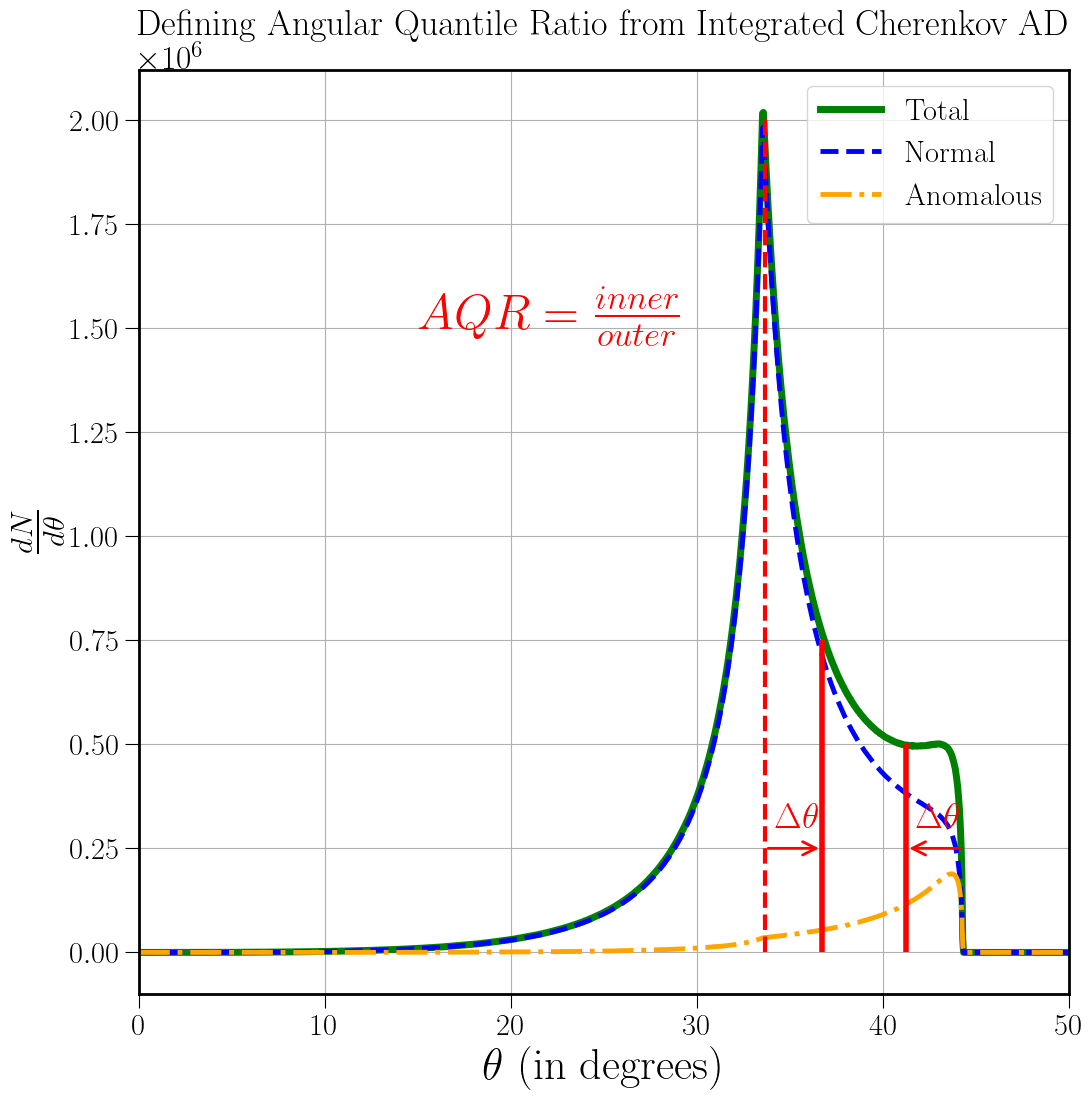}
\caption{Illustration of the angular quantile ratio (AQR) for the integrated AD.
\label{f:AQR_Cartoon_Integrated}
}
\end{centering}
\end{figure}

We define an analogous angular quantile ratio for the integrated distribution (AQR) in Fig.~\ref{f:AQR_Cartoon_Integrated} as the ratio of the integrated intensities around the inner bright peak and the outer edge of the distribution:
\begin{align}
    AQR \equiv \frac{
    \int\limits_{\theta_{IR}}^{\theta_{IR} + \Delta\theta}{d\theta} \, \frac{dN}{d\theta}
        }{
    \int\limits_{\theta_{max} - \Delta\theta}^{\theta_{max}}{d\theta} \, \frac{dN}{dx}
        }
\end{align}
where $\theta_{max} = \cos^{-1}(1/\beta_0 n_{UV})$ and $\theta_{IR} = \cos^{-1} (1 / \beta_0 n_{IR})$ are the maximum and minimum angles of the initial spectrum, which correspond to the outer edge and bright peak of the final distribution.  As before, the integration range $\Delta\theta$ is a predefined window, chosen to be a fraction (e.g. 10\%) of the total angular window $\theta_{max} - \theta_{IR}$.
This observable is defined to compare the relative intensities of the IR peak with the UV shoulder in the integrated distributions. Since the UV shoulder is significantly smeared, while the IR peak is not, this observable is differently sensitive to $n(\lambda)$ and possesses different systematics.

\begin{figure}[h] 
\begin{centering}
\includegraphics[width=0.49\textwidth]{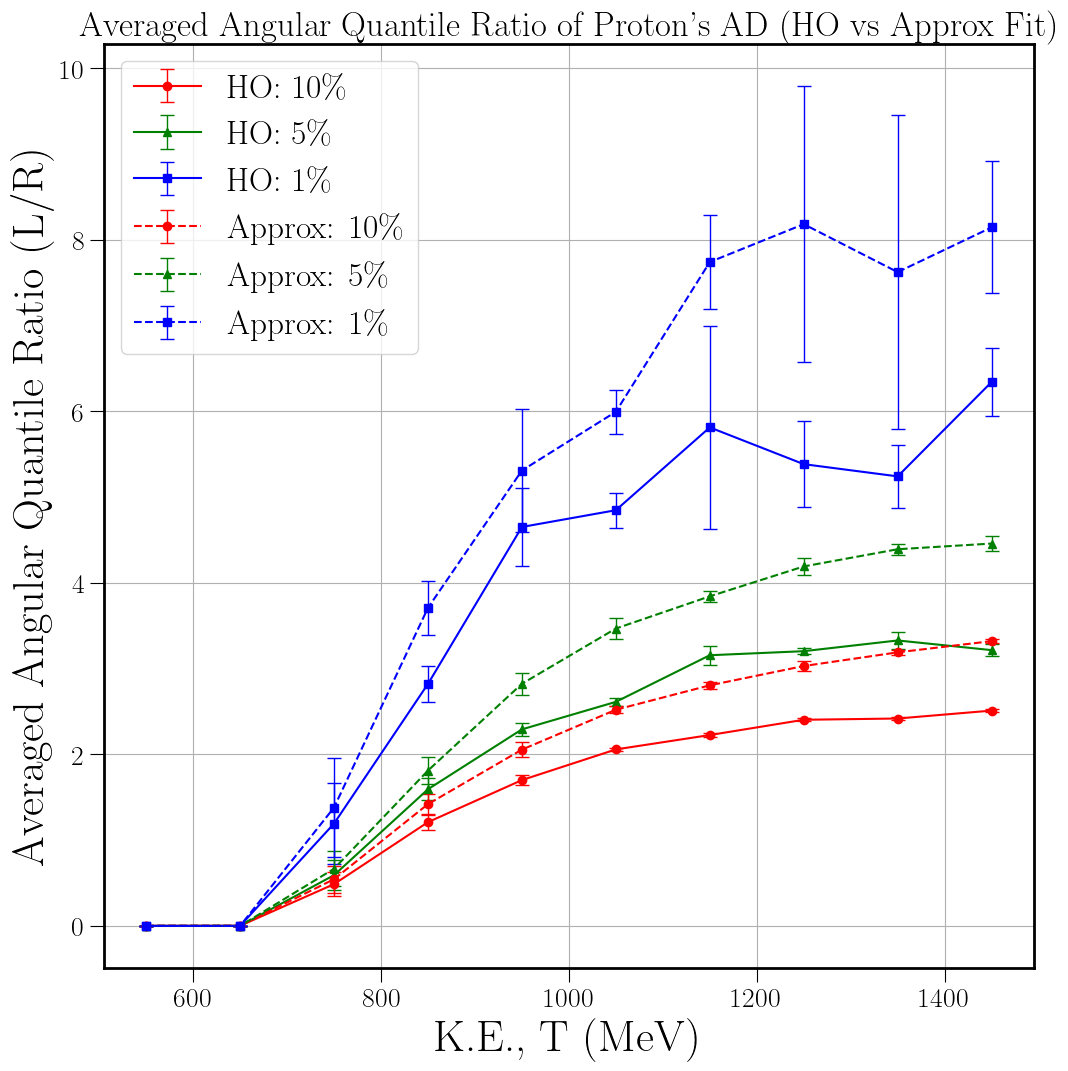}
\includegraphics[width=0.49\textwidth]{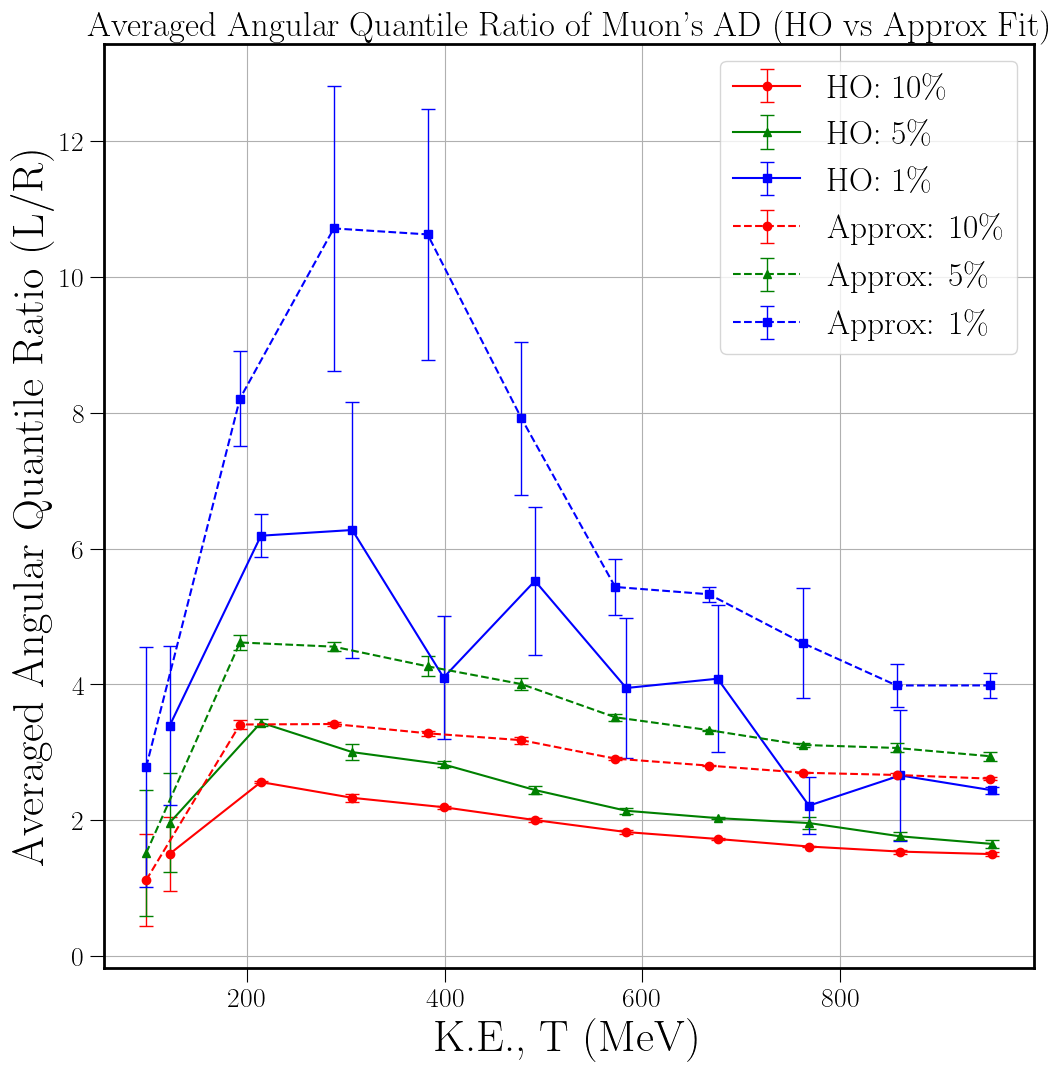}
\caption{Averaged Angular Quantile Ratio (AQR) of Proton (Left Plot) and Muon (Right Plot) from their respective integrated ADs.
\label{f:avgangquantratiop&muhovsapprox_integrated}
}
\end{centering}
\end{figure}

\begin{figure}[ht] 
\begin{centering}
\includegraphics[width=0.49\textwidth]{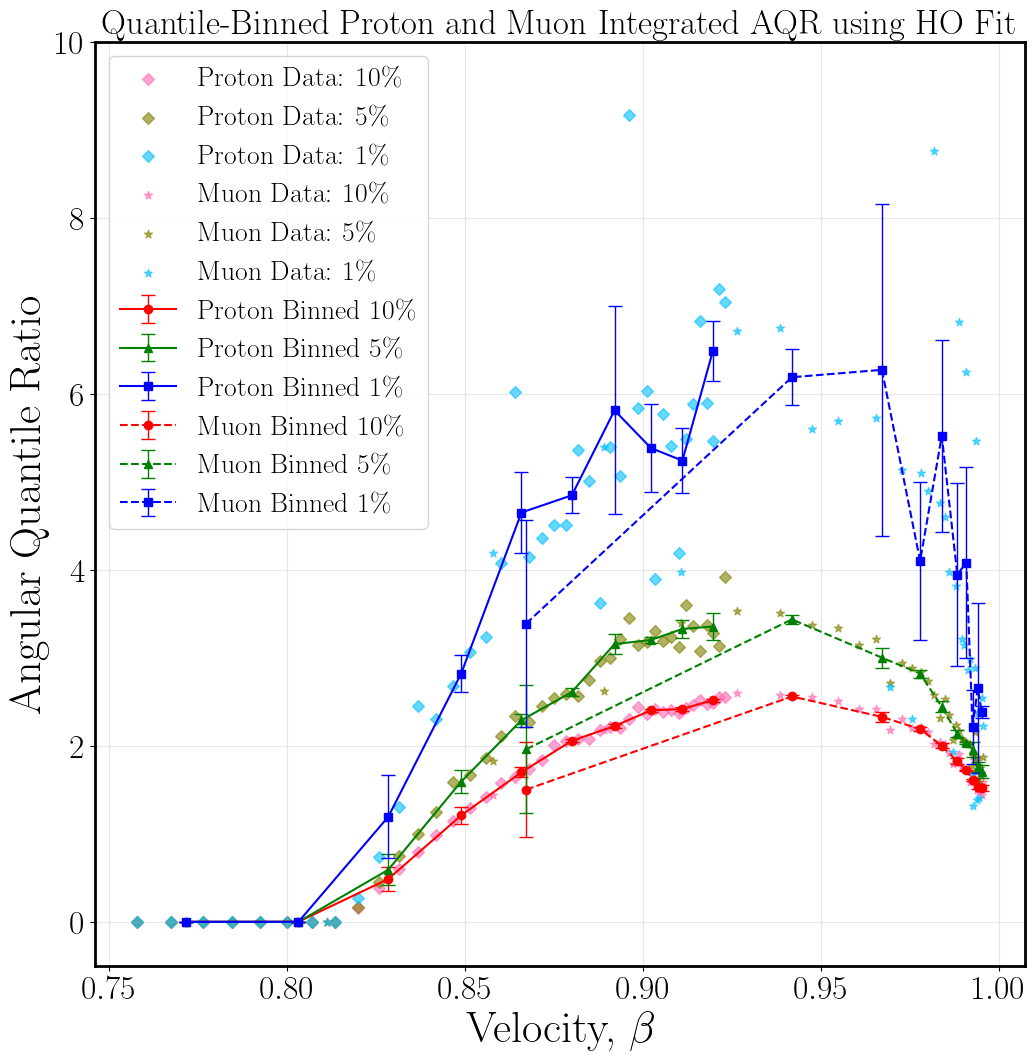}
\includegraphics[width=0.49\textwidth]{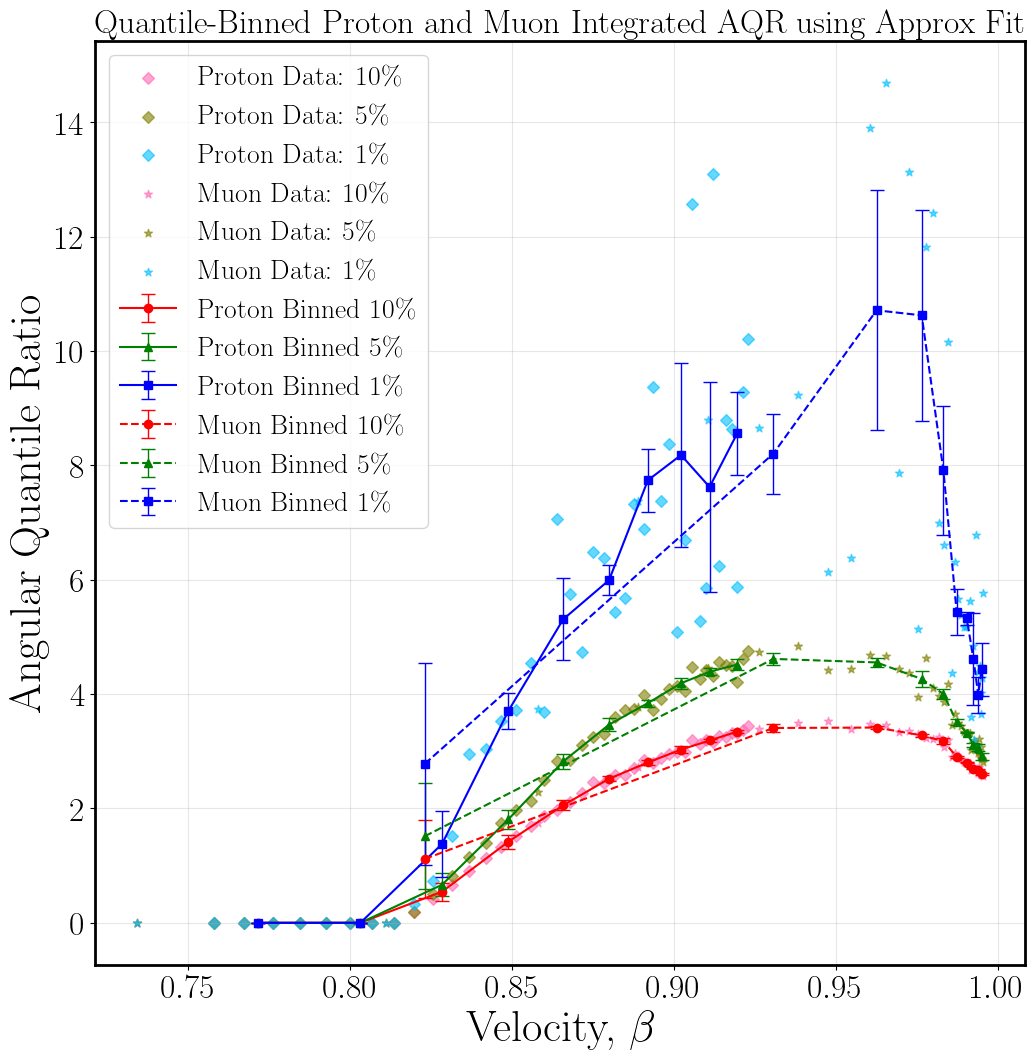}
\caption{Binned AQR of proton and muon with jackknife error as a function of velocity $\beta$ plotted together with the respective raw AQR data for HO fit (left panel) and Approx fit (right panel).
\label{f:AQRbetaplot_integrated}
}
\end{centering}
\end{figure}


In Figs.~\ref{f:avgangquantratiop&muhovsapprox_integrated} and \ref{f:AQRbetaplot_integrated}, we present a
comparison of AQR for protons (left panel) and muons (right panel) for two different models of $n(\lambda)$. Readily we notice that the size of the AQR is an order of magnitude larger than the AQR for instantaneous distributions, reflecting the peak-to-shoulder comparison as opposed to the peak-to-peak comparison. The AQR shows even more model discrimination ability, with the discrimination power increasing with energy and increasing with with tighter cuts. Unlike the instantaneous IAQR, the AQR grows in both magnitude and in model discrimination power as the cut is tightened. The only downside or limitation to tightening the cut further is the limitation of statistics in a given bin.

Although the IAQR is guaranteed to be identical for different particles (obeying the same energy loss mechanism) at the same $\beta$, it need not be so for the AQR. This is because different particles will stop at different rates due to their different masses (inertia).    
    
Interestingly, the difference in AQR seems to be small between protons and muons at the same initial $\beta$ shown in Fig.~\ref{f:AQRbetaplot_integrated}, suggesting that differences due to the deceleration rate $\frac{d\beta}{dx}$ are negligible. This makes it even easier to combine data on the Cherenkov distribution from different particle species.
This is a consequence of our use of the same Bethe-Bloch electromagnetic stopping power formula for both protons and muons. Adding in the distinct nuclear stopping power (or discrete nuclear scattering events, as described by Beacom) will likely cause significant changes in the integrated quantities for the proton vs muon. This would mean that the species dependence of the AQR is sensitive to details about the nuclear interactions of the proton. This strongly suggests investigating the effect of adding nuclear stopping power in the next study. A meaningful / significant future direction of the study can potentially analyze the effects on AQR from continuous energy loss and from discrete stopping events \cite{Beacom:2003zu}.

There may be a small but systematic effect leading to a slight enhancement of the AQR for protons versus muons, but this requires further investigation to confirm. The AQR is the most robust, most sensitive observable we have found for model discrimination in the refractive index $n(\lambda)$.

\begin{figure}[t] 
\begin{centering}
\includegraphics[width=0.48\textwidth]{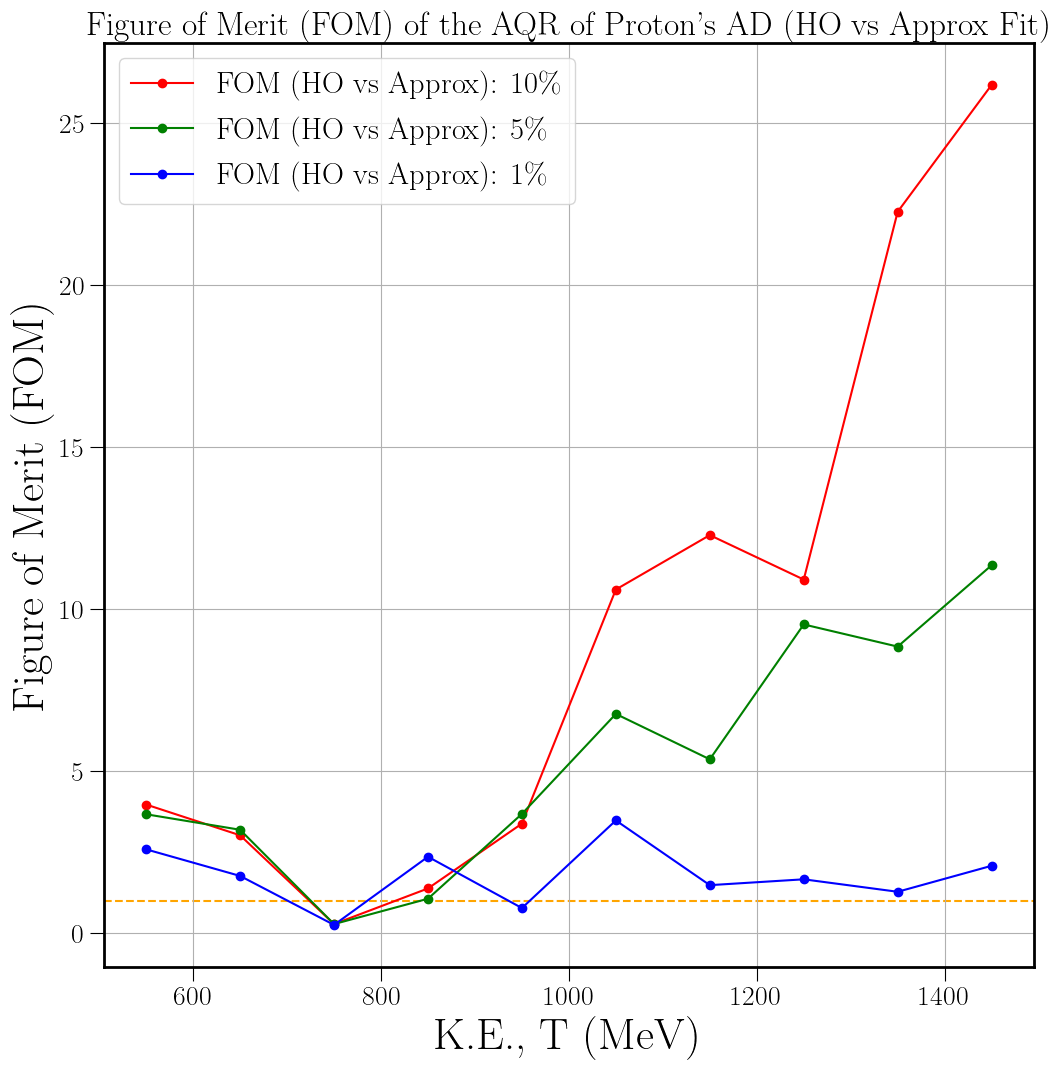}
\includegraphics[width=0.49\textwidth]{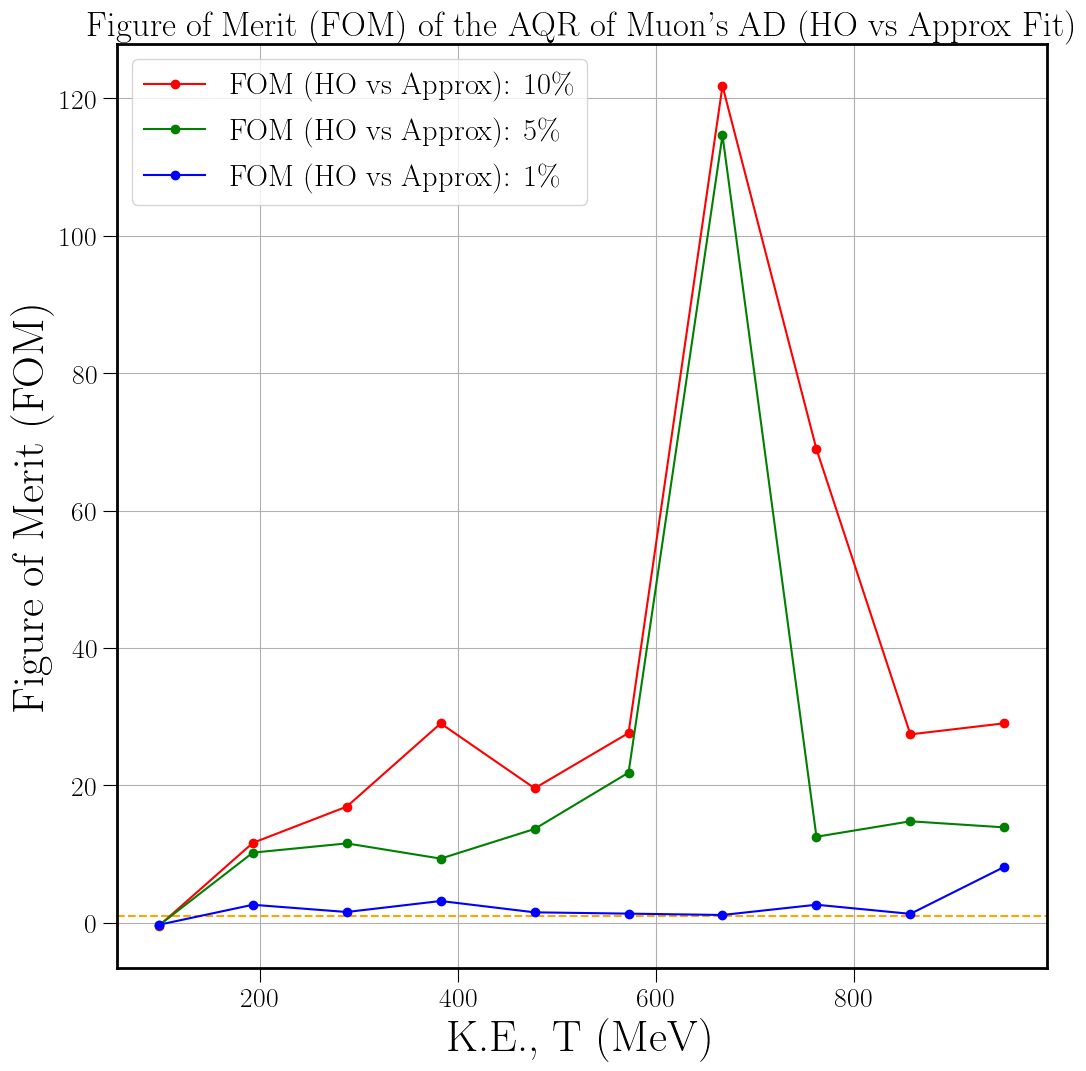}
\caption{Figure of Merit (FoM) of the AQR for protons (left) and muons (right) integrated AD. The dashed orange horizontal line is set at $FoM=1$; points above this line indicate that the difference between the two models is greater than their combined uncertainty.
\label{f:aqr_AD_fom_proton_muon}
}
\end{centering}
\end{figure}

Fig.~\ref{f:aqr_AD_fom_proton_muon} presents an equivalent definition of the FoM for the integrated AQR as shown previously for the instantaneous AQR. FoM is a (mostly) universal function of $\beta$ (a reflection of the use of the Bethe-Bloch formula to compute only the electromagnetic stopping power) and increases with kinetic energy. Unlike the instantaneous AQR, for the integrated AQR, the FoM increases for larger cuts $\Delta\theta$. Also, the separation between curves with different values of the cut $\Delta\theta$ is more significant compared to the FoM of the instantaneous AD shown in Fig.~\ref{f:aqr_iAD_fom_proton}. Some large fluctuations due to bins with unusually small error bars are seen. This feature can be better analyzed with a Monte Carlo simulation in the future. 

The overall size of the FoM is of the same order $(\leq 30)$ for both integrated and instantaneous AQR. But the signal size of the AQR itself is an order of magnitude larger for the integrated case (peak vs. shoulder) compared to instantaneous (peak vs. peak). The FoM suggests that the best model discrimination power in the AQR can be found at higher energies and for larger $\Delta\theta$.

%
\subsection{Conclusions and Outlook}
\label{sec:concl}
%


In this Chapter, we have presented two new absorptive fits to the refractive index of LAr: one (HO) which follows the specific form of the first-principles Lorentz harmonic oscillator model, and the other (Approx) with a qualitatively similar shape but with different peak location and height. Using these fits, we have calculated the instantaneous and integrated Cherenkov yields and angular distributions (AD) for protons and muons.  We have studied the properties of these distributions and illustrated their consequences on an angle-energy heatmap as used for particle identification (PID) purposes. 

By implementing a proper physical treatment of the absorption near a resonance, we were able to quantify the contribution to the Cherenkov yield and AD from the anomalous dispersion component in LAr for the first time. The anomalous dispersion component contributes in a different angular region from the normal dispersion component, which we exploit to characterize the model dependence on the refractive index.  To quantify this, we introduced a new observable, the angular quantile ratio (AQR), which characterizes the changes in shape of the Cherenkov spectrum for different models of the refractive index. We compared the predictions for the AQR and other observables between the two reasonable HO and Approx models for $n(\lambda)$, and we additionally defined a Figure of Merit (FoM) to assess the model discrimination power of the AQR.

Our most important results are (1) constructing the HO and Approx absorptive fits to the LAr refractive index, (2) calculating the corresponding yields and AD's for the two models, (3) constructing the AQR as a new observable to quantify the model-dependent shape of the Cherenkov radiation, and (4) quantifying its model discrimination power with a Figure of Merit (FoM).  Our theoretical analysis also allows a transparent physical interpretation of the features of the Cherenkov spectrum as reflecting specific features in the UV or IR regions of the refractive index.  

This work advances the construction of increasingly powerful observables for model discrimination. We had previously observed the total Cherenkov yields to be weakly model discriminating \cite{rahman2024}. Introducing the AQR adds significant discrimination power to our analysis, as quantified by the FoM. Both the instantaneous and integrated AQR are powerful observables for discrimination between models of the refractive index. The FoM is comparable for both instantaneous and integrated AQR, while the AQR signal itself is an order of magnitude larger for the integrated AQR. Hence, the integrated AQR at high energies seems to be the best possible place to look for extracting information about the refractive index of liquid argon.  These results are significant because they reverse the usual paradigm of PID, where a well-known $n(\lambda)$ is used to discriminate between the identities of unknown particles. Instead, we use the observed Cherenkov spectrum of different particles to set overlapping constraints on a poorly known $n(\lambda)$. We further provide a pathway to measure features of the Cherenkov spectrum and interpret their constraints on the refractive index.

The benefits of this analysis are twofold - first one can use measured Cherenkov distributions in LAr to set more precise constraints on the refractive index, and once the refractive index of LAr is well known, it can be used to provide more accurate predictions for Cherenkov PID in LAr. This theory input can help to mitigate the scarcity of experimental data on the refractive index near the UV resonance at 106.6 nm.  This analysis can also be generalized to consider effects like the molecular composition of LAr or contaminants on the form of the refractive index.  Next steps to continue this analysis in the future include the integration of these distributions into a Monte Carlo simulation, which would allow us to study the effect of statistical fluctuations.  Another important direction will be to generalize the treatment of energy loss beyond the Bethe-Bloch equation to include the effects of hadronic energy loss for protons, including the effects of discrete scattering events \cite{Beacom:2003zu}.

%% file: chp3_Jetdrift.tex
\section{JET DRIFT IN QCD} \label{jetdrift}
\hspace{\parindent}

%
\subsection{Introduction}
\label{sec:jetdrift}
%

%
\subsubsection{Standard Model of Heavy-Ion Collisions}
%

Ultra-relativistic collisions of heavy ions, primarily performed at the Large Hadron Collider (LHC) and Relativistic Heavy-Ion Collider (RHIC), serve as the principal method for studying the Quark-Gluon Plasma (QGP). While the LHC provides significantly higher center-of-mass energies compared to RHIC ($\sqrt{s}_{NN} = 5.02$ TeV or above, versus $\sqrt{s}_{NN} = 200$ GeV), the accelerator design of RHIC permits extensive versatility in its beam species. 

The standard theoretical description of the formation, evolution, and detection of the QGP medium (the ``soft sector'') is as follows. Initially, the nuclei experience extreme relativistic length contraction which compresses them into two-dimensional sheets of energy density.  In the earliest moments of the collision, the system undergoes extreme non-equilibrium dynamics; this phase remains the subject of active research \cite{Iancu:2003xm, Gelis:2010nm, Lappi:2006fp, Berges:2020yzp, Heller:2015dha}. By approximately $\mathcal{O}(1 \, \mathrm{fm}/c)$, the medium reaches a near-equilibrium thermalized state that serves as the initial condition for subsequent evolution via relativistic viscous hydrodynamics \cite{Luzum:2013yya, Heinz:2024jwu}. This fluid-like behavior persists through the confinement phase transition from a QGP into a hadron resonance gas (HRG), a gas of hadrons and hadronic resonance states.  As the HRG expands and cools, the density eventually drops too low to be described using hydrodynamics; at this ``freezeout'' stage, the system expands as a collection of distinct particles. These particles may undergo final-state rescattering at late times, and at sufficiently low densities, they follow simple ballistic trajectories to the detectors.

Simultaneously, high-$p_T$ partons are produced through rare hard processes where partons (quarks and gluons) from the colliding nuclei scatter at short distances. These serve as hard probes of the QGP, propagating through the medium as it forms and evolves.  The medium affects these partons through effects such as radiative energy loss, collisional energy loss, and transverse momentum diffusion. Following these interactions, the hard partons undergo vacuum-like fragmentation into hadrons, described by:
\begin{align} \label{e:sigma_diff}
\frac{d\sigma^{h}}{dydp_{T}^{h}} & = \int dz \frac{1}{z} \, D^{h}_{p}(z) \, \frac{d\sigma^{p}}{dydp_{T}^{p}} \, ,
\end{align}
where $D^{h}_{p}(z)$ is the fragmentation function for a parton $p$ creating a hadron $h$ with momentum fraction $z = p_T^h / p_T^p$.

%
\subsubsection{Theory and Observables}
%

Gluon emission within the medium is governed by the Landau-Pomeranchuk-Migdal (LPM) effect \cite{Landau:1953um, Migdal:1956tc}, which accounts for the suppression of radiation due to coherence effects in multiple scattering.  Various frameworks for calculating the radiative energy loss exist in QCD; we use the perturbative Gyulassy-Levai-Vitev (GLV) opacity expansion formalism \cite{Gyulassy:2000gk}, implemented specifically as the stopping power
\begin{align}   \label{e:ELossRate}
    \frac{dE}{d\ell} &= 
    - \frac{d}{dL} \Bigg(\frac{2 C_R \alpha_s}{\pi} \frac{L}{\lambda} E \int_0^1 dx \,
 \int_{{k}_{min}}^{{k}_{max}} \frac{dk}{k}  \; \int_0^{q_{max}} dq \, q \int_0^{2\pi} d\phi  \notag \\
    & \hspace{1cm} \times 
    \frac{\mu^2}{\pi (q^2 + \mu^2)^2} \frac{2 \mathbf{k}\cdot\mathbf{q} \, (\mathbf{k} - \mathbf{q})^2 L^2}{16 x^2 E^2 + (\mathbf{k} - \mathbf{q})^4 L^2} \Bigg)_{L=\ell} \, . 
\end{align}
%

Here,  $C_R$ is the quadratic Casimir of the representation $R$ of SU(3) ($C_R = 3$ for gluons, $C_R = 4/3$ for quarks). We take $\tau_{0} = 0.37$ fm for the thermalization time and do not treat the short-pathlength corrections as derived in Ref.~\cite{Kolbe:2015rvk}, which may be significant even for large collision systems. Thus, for massless partons, $L = \tau - \tau_0$, with $\tau$ the time relative to the collision in the plasma center of mass rest frame.
This formalism treats the medium as an effective static medium of length $L$, a simplification derived from an assumed density form that allows the interference integrals to be performed analytically. The calculation is performed at the first order of the opacity expansion (assuming a dilute medium) and assumes soft gluon emission ($x \ll 1$). It incorporates finite kinematic bounds, where $k_{min} = \mu$, $k_{max} = \min[ 2 E x \, , \, 2 E \sqrt{x(1-x)}]$, and $q_{max} = \sqrt{6ET}$ \cite{Gyulassy:2000er}, and it utilizes the Gyulassy-Wang potential \cite{Gyulassy:2000gk} for the scattering centers in the medium.

Additionally, collisional energy loss is implemented following the Braaten and Thoma framework \cite{Braaten:1991we}:
\begin{align}   \label{e:EColl}
    \frac{dE}{d\ell} &= -  {2\pi C_R \,\alpha_s^2} \, \mu^2\ln\left( 2^{\frac{N_f}{2(6+N_f)}} 0.920 \frac{\sqrt{3ET}}{\mu} \right).
\end{align}
%
%
A perturbative calculation of parton drift and energy loss requires input for the mean free path $\lambda$ and Debye screening mass $\mu$. The latter reads
$   \mu \approx gT \sqrt{1 + {N_f}/{6}}$ where we use $N_f = 2$,
and by definition
${\lambda}^{-1} = \sigma_{q} \rho_{q} + \sigma_{g} \rho_{g}$,
with $\sigma_{q}$($\sigma_{g}$) the total cross section for the hard parton to interact with a quark (gluon) in the plasma and $\rho_{q}$($\rho_{g}$) the density of quarks (gluons) in the medium. Again, we use the Gyulassy-Wang cross section modeled after static scattering centers in the medium~\cite{Gyulassy:1993hr}, and the cross sections and partial quark and gluon densities are taken from~\cite{Sievert:2019cwq}, which approximates the medium as an ideal gas of gluons and quarks.

As derived in Ref.~\cite{Sadofyev:2021ohn}, jet drift manifests as a net transverse momentum deflection $\langle \vec{q}_{drift} \rangle$ in the direction of the flow vector $\vec{u}_\bot$ transverse to the parton momentum:
\begin{align}\label{e:q_drift_moment_2}
\left\langle \vec{q}_{drift} \right\rangle &= \hat{e}_\perp \int d\ell \, \frac{3}{E} \, \frac{\mu^{2} }{\lambda} \, \ln\frac{E}{\mu} \: 
\frac{u_\perp }{1-u_\parallel } \: .
\end{align}

Here, $\lambda$ is the mean free path, $\mu$ is again the Debye screening  mass, and $\vec{u}$ is the collective flow velocity, with components $\parallel$ and $\perp$ to the parton momentum vector, and $\hat{e}_\perp$ is the corresponding unit vector.

This derivation relies on standard small-$x$ and broad-source approximations and is leading order in the opacity expansion. It accounts for velocity corrections to the first sub-eikonal order ($1/E$) but does not include gradient effects.  The energy suppression $\left\langle \vec{q}_{drift} \right\rangle \propto 1 / E$ is a distinctive feature of sub-eikonal interactions like jet drift.

\begin{figure} [t]
    \centering
    \includegraphics[width=0.4\textwidth,
    trim={0cm 2cm 0cm 0cm}, clip]{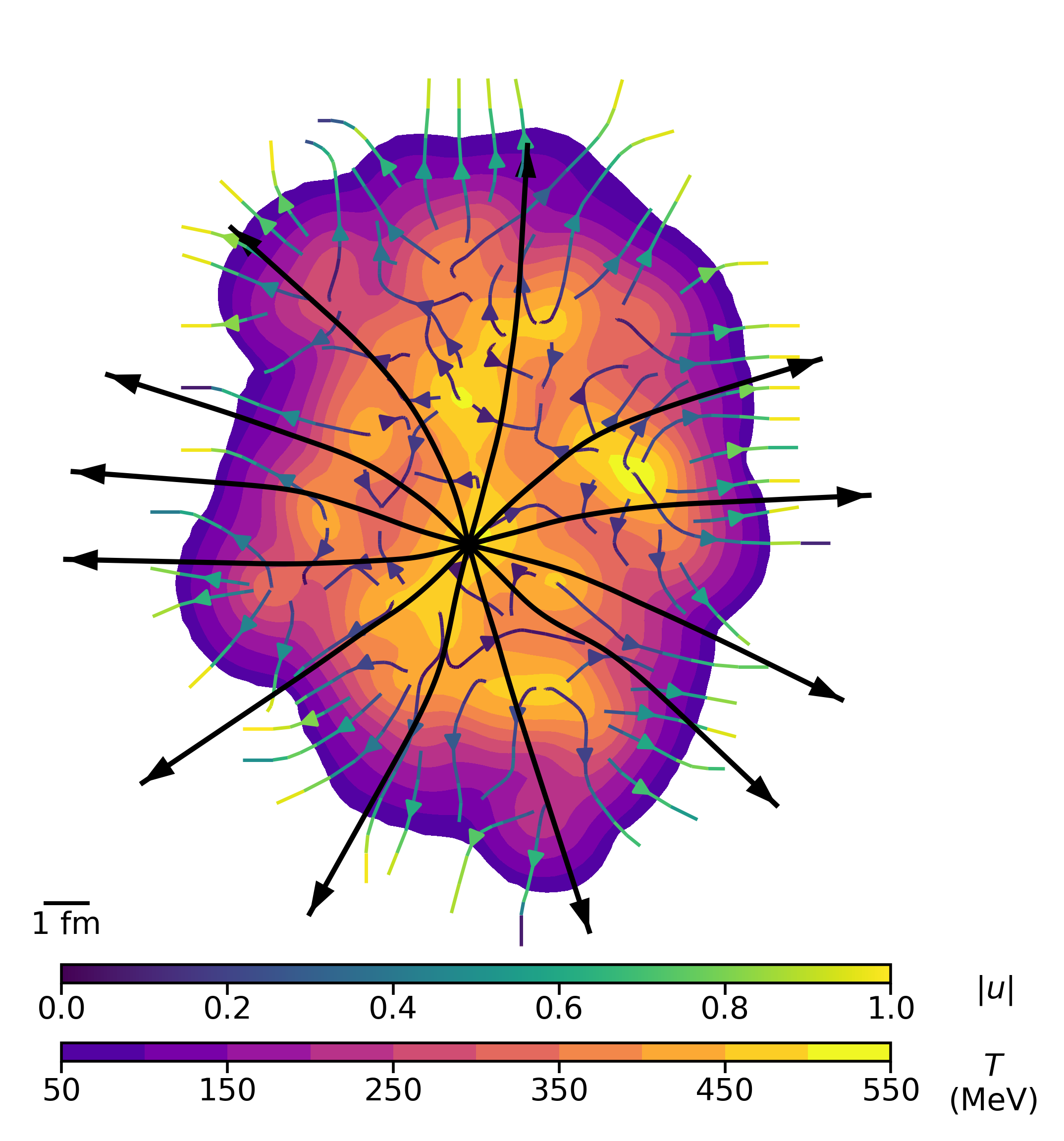}
    \includegraphics[width=0.35\textwidth]{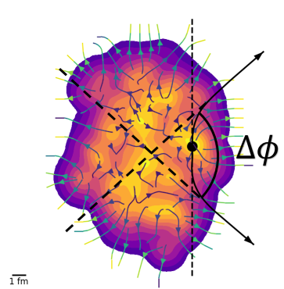}
    \caption{  
    Coupling to the elliptic flow of the medium, attracting partons to the event plane (left panel) and increasing the acoplanarity of initially back-to-back partons (right panel) in 5.02 TeV PbPb collisions. Parameters selected for illustrative purposes. Figures reproduced from Ref.~\cite{Bahder:2024jpa}; see it for further details. \label{fig:sample_drift}
    }
\end{figure}

For this work, we focus on two main hard probe observables sensitive to jet drift: acoplanarity and elliptic flow (see Fig.~\ref{fig:sample_drift}). The acoplanarity measures the deviation of two particles from a back-to-back configuration, as naively expected for an initial hard scattering:
\begin{align}
    \Delta \phi = \pi - \left|\mathrm{mod}\big(\phi_1 - \phi_2 \: , \: \pi\big) \right|.
\end{align}
The single-event anisotropic flow vector $\vec{v}_n$ defines the magnitude and direction of the $n^\mathrm{th}$ Fourier harmonic of the inclusive particle distribution:
\begin{align}
    \vec{v}_n (p_T) 
    \equiv
    v_n (p_T) \, e^{i n \psi_n (p_T)}
    =
    \frac{1}{\frac{dN}{dp_T}}
    \int d\phi \, e^{i n \phi} \, \frac{dN}{dp_T \, d\phi} \, .
\end{align}
This defines a complex vector $\vec{v}_n$ with magnitude $v_n = | \vec{v}_n|$ and phase $n \psi_n = \mathrm{Arg}( \vec{v}_n )$. Since $\vec{v}_n$ is generally not observed directly due to large event-by-event directional fluctuations, it is commonly measured using two- or four-particle cumulants. Experimentally, one can separately calculate $v_n \{2\}$ for different $p_T$ bins or identified particles; theoretically, we can distinguish between flow vectors for particles from the soft sector simulation, $\vec{v}_n^{soft}$, and those from hard particle production, $\vec{v}_n^{hard}$. Two-point correlations of the same particle species with identical kinematics yield the second-order cumulant (root mean square):
\begin{align}
    v_n \{2\} = \sqrt{ \langle v_n^2 \rangle } \, .
\end{align}
In contrast, two-point correlations of different particles or kinematics provide a pure correlation measure. A common experimental case \cite{Giacalone:2020dln, ALICE:2022mno} involves correlating one hard particle with momentum $p_T$ and one soft particle integrated over momentum, given by:
\begin{align} \label{e:vndefn}
    {v}_n^{exp}(p_T) & =  \frac{\mathrm{Re}\left\langle \vec{v}_n^{hard \, *} (p_T) \cdot \vec{v}_n^{soft} \right\rangle}{\sqrt{\left\langle \left|\vec{v}_n^{soft}\right|^2\right\rangle}}.
\end{align}
This quantity measures a genuine correlation between the hard and soft anisotropic flow coefficients, which need not be positive.

Because of the low viscosity of the QGP, the measured final-state anisotropic flow coefficients are highly correlated with the spatial anisotropy of the initial state (see e.g. Ref.~\cite{Rao:2019vgy}).  Theoretically, the geometry of the initial state is characterized by the complex eccentricity vector
\begin{align}
    \vec{\varepsilon}_n = - \frac{\int r dr d\phi \, r^n e^{i n \phi} \, \rho(r,\phi)}{\int r dr d\phi \, r^n \, \rho(r,\phi)}
\end{align}
analogous to $\vec{v}_n$ in the final state.  The eccentricity can be calculated with respect to the distribution of any thermodynamic quantity; here we use the entropy density.  To a remarkably good accuracy, the complex evolution of hydrodynamics simply reflects quasi-linear response $\vec{v}_n \propto \vec{\varepsilon}_n$ to the initial state geometry.  While theoretically, the event geometry is primarily controlled by the impact parameter $b$, the impact parameter is not directly observable in experiment.  Instead, a useful proxy is ``centrality classes,'' which order events into quantiles based on the total multiplicity of produced soft particles.  Because of the strong correlation between impact parameter and multiplicity (direct head-on collisions produce the most particles), a cut which selects on events with the top $0 - 10\%$ number of produced particles also selects on the smallest impact parameters $b$. Dividing the data into centrality classes provides a powerful experimental lever to select on different event geometries.

%
\subsubsection{Anisotropic Parton Evolution (APE) Monte Carlo Simulation}
%

The numerical results presented in this work are generated using the Anisotropic Parton Evolution (APE) framework\footnote{available on \href{https://github.com/Jopacabra/ape}{GitHub}}, an open-source Monte Carlo parton trajectory simulator designed to study the anisotropic jet drift of hard probes to the soft bulk medium~\cite{Bahder:2024jpa}. The APE workflow proceeds in a modular event-by-event fashion in four stages: medium evolution, hard parton production, medium modification, and hadronization.

The medium evolution stage is adapted from the open-source Duke QCD heavy ion collision event generator \cite{Bernhard:2019bmu}. The simulation begins with the generation of the initial condition for the primary hydrodynamic evolution.  We utilize the TRENTO parameterization for the initial entropy density profile, including the relevant nuclear deformation parameters for the case of nonspherical nuclei collisions, followed by a short free streaming phase.  This initial-state profile then evolves according to relativistic viscous hydrodynamics, providing the time-dependent background temperature $T(x)$ and flow velocity field $u^\mu(x)$ through which the hard partons propagate.

Hard scattering reactions at the level are sampled from the PYTHIA Monte Carlo event generator for proton-proton collisions \cite{Sjostrand:2006za, Sjostrand:2014zea} and initialized at $\tau_0$ at transverse positions sampled from the binary collision density profile, $N_{coll}(x,y)$.  In our current setup, the initial partons are generated with Leading Order (LO) kinematics, ensuring they start back-to-back with zero intrinsic acoplanarity. Consequently, any acoplanarity observed in the final state is generated exclusively by jet drift. While this does not account for Next-to-Leading Order (NLO) effects that introduce initial acoplanarity, it provides a clean estimate of the specific contribution arising from medium interactions.

As the partons traverse the medium, APE solves the transport equations for the parton energy and transverse momentum over discrete timesteps $d\tau$, incorporating GLV radiative energy loss \eqref{e:ELossRate},  Braaten-Thoma collisional energy loss \eqref{e:EColl}, and jet drift \eqref{e:q_drift_moment_2}.  The simulation tracks the partons until they exit the medium or their energy drops below a thermal threshold. Finally, the partons are hadronized via independent fragmentation in vacuum.

We note one caveat about the calculation of anisotropic flow coefficients presented here.  As noted above, $v_n^{exp}$ specifically measures the correlation between the soft event plane vector $\vec{v}_n^{soft}$ and the hard particle flow vector $\vec{v}_n^{hard}(p_T)$.  In our calculations, the soft reference flow $\vec{v}_n^{soft}$ is calculated from thermal hadrons in the range $0.2 \, \mathrm{GeV} \leq p_T \leq 5 \, \mathrm{GeV}$. For the hard sector, while experimental measurements of $\vec{v}_n^{hard}(p_T)$ typically include all particles in a given $p_T$ bin, our simulation counts only those arising explicitly from hard processes. Based on the mean $p_T$ analysis in Ref.~\cite{Moreland:2018gsh}, this distinction is only significant for $p_T \leq 2$~GeV, where the soft sector makes a nontrivial contribution.

\begin{figure}[t]
\begin{centering}
\includegraphics[width=0.52\textwidth]{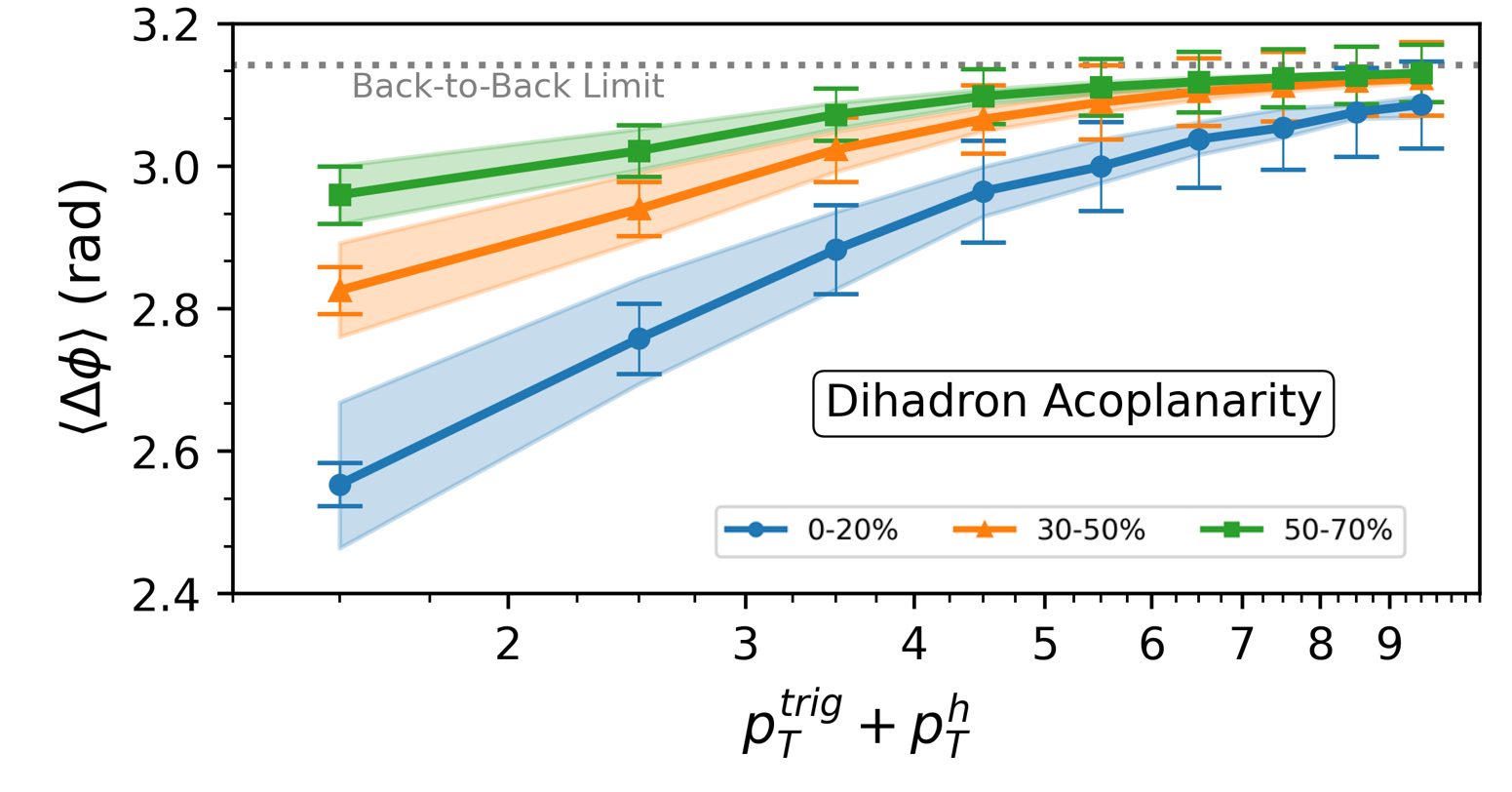}
\includegraphics[width=0.44\textwidth]{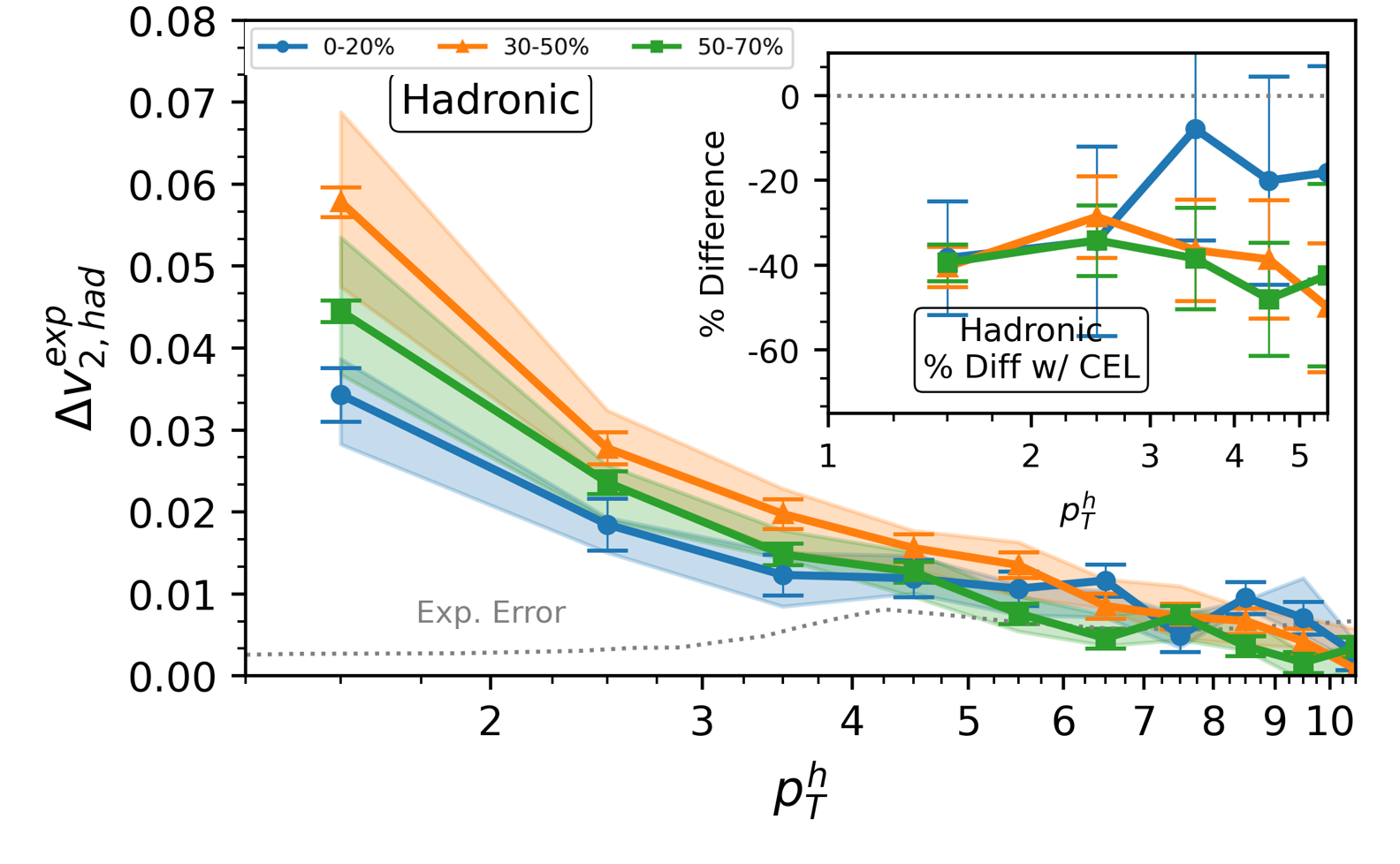}
\caption{
Left:  Mean dihadron acoplanarity $\Delta \phi$ (deviation from $\pi$) produced by jet drift.  Right: Elliptic flow enhancement 
$\Delta v_2^{exp}$ due to jet drift.  Figures reproduced from Ref.~\cite{Bahder:2024jpa}; see that paper for all details.
\label{fig:acoandv2}
}
\end{centering}
\end{figure}

Our previous paper \cite{Bahder:2024jpa} pointed out some striking differences between the centrality ordering of the two complementary observables, acoplanarity of dihadrons and elliptic flow $v_2^{exp}$. As shown in Fig.~\ref{fig:sample_drift}, the acoplanarity observable measures the magnitude of transverse momentum broadening in the medium rather than its directionality. Fig.~\ref{fig:acoandv2} shows that this enhancement is smallest in peripheral collisions and largest in central collisions, reflecting a straightforward increase in momentum broadening due to the greater path length and temperature (or equivalently, density) in central events. In contrast, the elliptic flow $v_2$ reflects not only the magnitude of deflection but also its correlation with the event plane. Consequently, $v_2$ depends on both the temperature and path length, which drive stronger deflection, and the event geometry, which determines the efficiency of correlating that deflection to the event plane. 

This leads to a non-monotonic centrality ordering for $v_2$: the enhancement is smallest in peripheral collisions due to minimal deflection, largest in semi-peripheral collisions where there is a balance of moderate deflection and moderate ellipticity, and decreases again in central collisions where ellipticity is reduced. While the competition between temperature (or path length) and geometry drives $v_2$, acoplanarity provides a means to disambiguate the two effects. This motivates a deeper analysis to understand how jet drift responds to these underlying variables—temperature and path length versus the geometric eccentricity $\varepsilon_2$.

%
\subsection{Results:  PbPb vs. AuAu}
%

Let us begin with a comparison between PbPb collisions at $\sqrt{s} = 5.02 \, \mathrm{TeV}$ at the LHC and AuAu collisions at $\sqrt{s} = 200 \, \mathrm{GeV}$ at RHIC.  As we will show, these systems have nearly identical geometries, but create plasmas with temperatures that differ by about $20\%$.  This makes it possible to study the temperature dependence of jet drift in a systematic way.

%
\subsubsection{System Comparison}
%

\begin{figure}[t]
\begin{centering}
\includegraphics[width=0.48\textwidth]{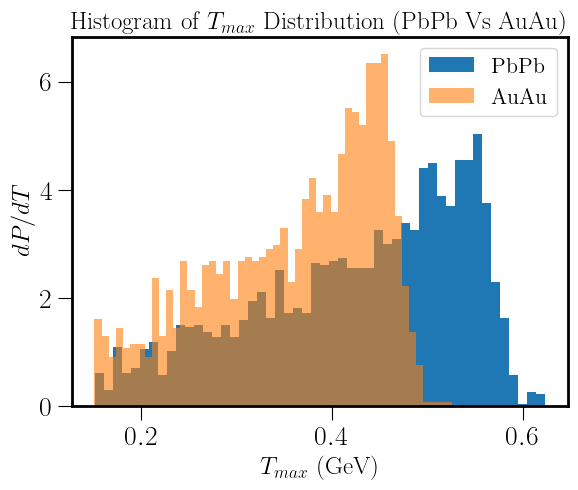}
\includegraphics[width=0.49\textwidth] {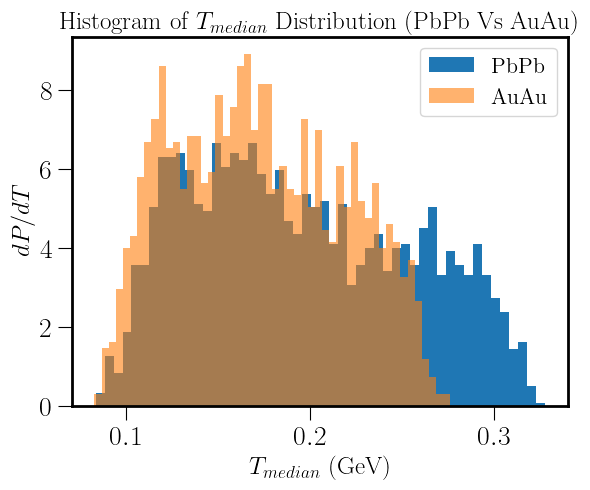}
\caption{Histograms of the maximum temperature $T_{max}$ (left) and median temperature $T_{med}$ (right) of the initial conditions of PbPb and AuAu collisions.  This figure represents approximately 2500 events per data set.
\label{f:pbpbauauhistT}
}
\end{centering}
\end{figure}

In Fig.~\ref{f:pbpbauauhistT} we compare the distribution of the maximum event temperature $T_{max}$ (left panel) and median temperature $T_{median}$ (right panel) in the initial states of 5.02 TeV PbPb collisions at the LHC and 200 GeV AuAu collisions at RHIC as simulated in APE.  Here the median is computed after dropping all cells with temperatures below 10 MeV so as to only average over regions containing the medium.  The maximum $T_{max}$ reached in the initial condition (the hottest hotspot) in AuAu was about 500 MeV, whereas in PbPb it was about 600 MeV. On the other hand, the maximum value of $T_{med}$ reached in the initial condition of AuAu was about 260 MeV, whereas in PbPb it was 320 MeV.  Despite the huge difference in the center-of-mass energy from $\sqrt{s} = 200 \, \mathrm{GeV}$ at RHIC to  $5.02 \, \mathrm{TeV}$ at the LHC, there is only a $\sim 20\%$ increase in $T_{max}$ and a $\sim 23\%$ increase in $T_{med}$ from AuAu to PbPb.  The ratio of $T_{max}$ to $T_{med}$ is a measure of how how the hottest hot spot is in such events compared to the rest of the event.  For AuAu, the ratio of the highest $T_{max}$ to the highest $T_{med}$ was about 1.92, while for PbPb it was about $1.875$, which is comparable between the two systems. The distributions of both $T_{max}$ and $T_{med}$ generally increase by a factor of about $20\%$, but are otherwise comparable.




%
\begin{figure}[t]
\begin{centering}
\includegraphics[width=0.5\textwidth]{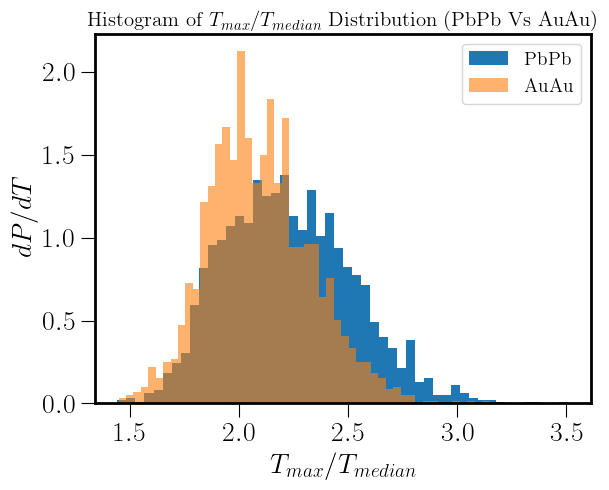}
\caption{Histograms of the ratio of maximum temperature $T_{max}$ and median temperature $T_{med}$ of the initial conditions of PbPb and AuAu collisions. This figure represents approximately 2500 events per data set.
\label{f:pbpbauauhistTmaxbyTmedormean}
}
\end{centering}
\end{figure}

Fig.~\ref{f:pbpbauauhistTmaxbyTmedormean} emphasizes on the qualitative feature from the ratio of maximum temperature $T_{max}$ and median temperature $T_{med}$ of the initial conditions of PbPb and AuAu collisions. This plot represents approximately 2500 events for both PbPb and AuAu datasets (shown in blue and orange, respectively), amplifying the relative temperature measures between the two collisional systems.


\begin{figure}[t] 
    \centering
    \begin{subfigure}{.55\textwidth}
    \includegraphics[width=1\textwidth]{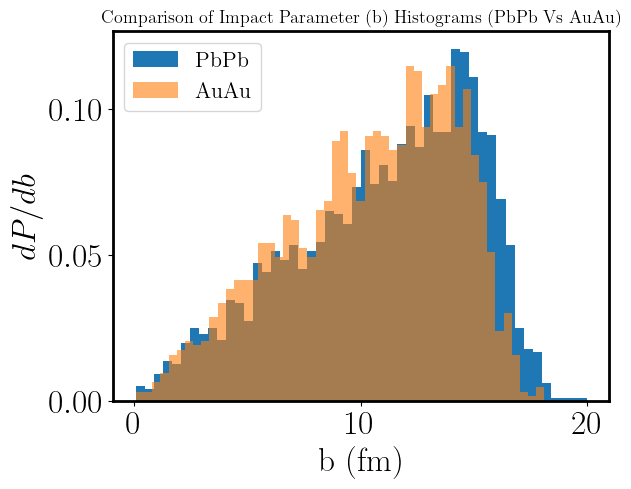}
    \end{subfigure}
    \begin{subfigure}{.43\textwidth}
    \centering
    \includegraphics[width=1\textwidth]       {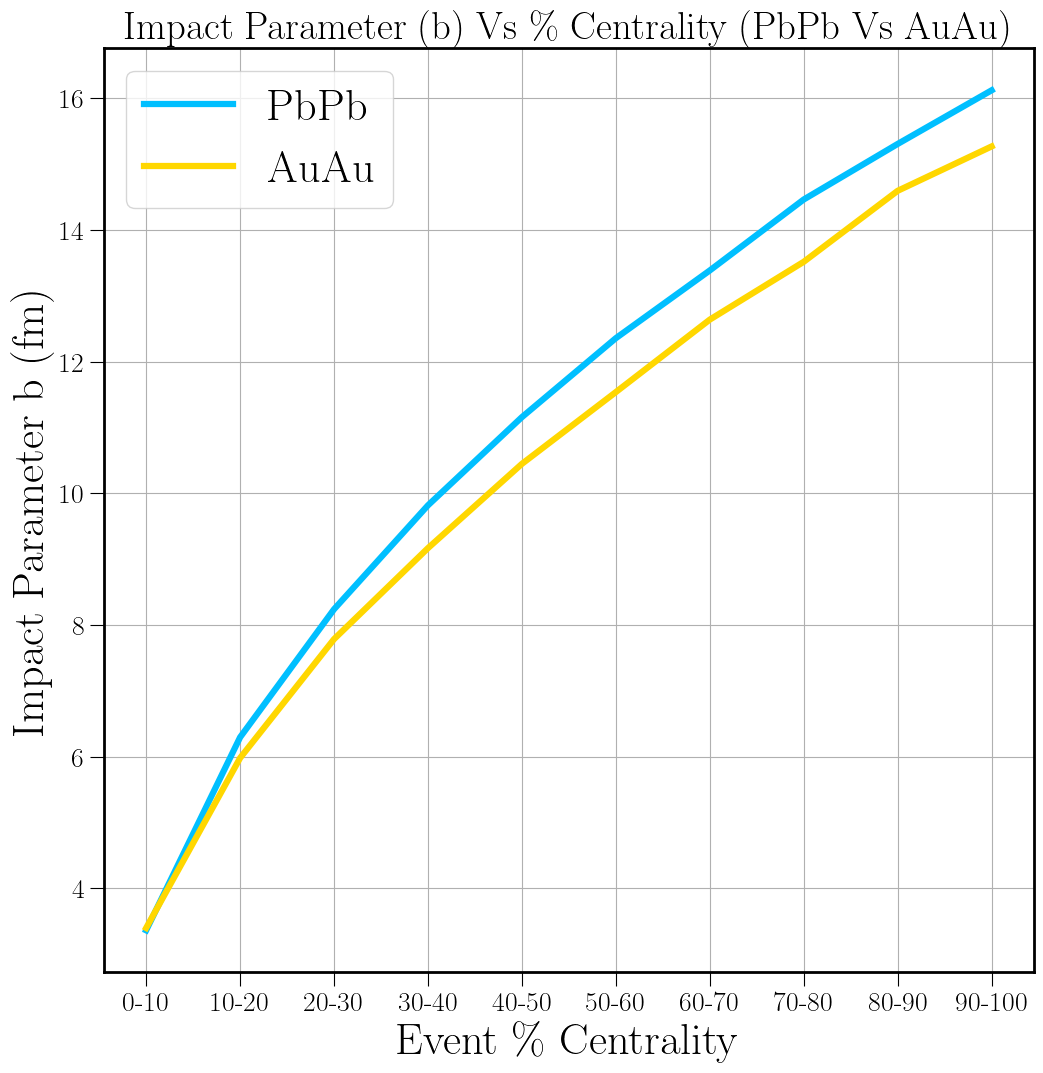}
    \end{subfigure}
    \caption{Histogram [$\sim 2500$ events] of impact parameter $b$  (left) and mean impact parameter as a function of centrality (right) for 5.02 TeV PbPb collisions and 200 GeV AuAu collisions.
    \label{f:pbpbauauhistsbne2}
    }
\end{figure}

In Fig.~\ref{f:pbpbauauhistsbne2}, we compare the geometric profiles of PbPb and AuAu collisions. Au and Pb are both spherically symmetric nuclei, whose densities follow a Woods-Saxon distribution
\begin{align}
    \rho(r, \theta, \phi) = \rho_0 \left[ 1 + \exp \left( \frac{r - R(\theta)}{a} \right) \right]^{-1} \: ,
\end{align}

with radii $R = 6.38$ fm, $R = 6.62$ fm and diffuseness $a = 0.535$ fm, $a = 0.546$ fm, respectively \cite{PHENIX:2015tbb, Acharya_2024}.  For deformed nuclei, the radius can be a function of the polar angle $\theta$, which can be expanded in spherical harmonics (or Legendre polynomials):
\begin{align}
    R(\theta) = R \Big( 1 + \beta_2 \, Y_{20} (\theta) + \beta_4 \, Y_{40} (\theta) + \cdots \Big)
\end{align}
with deformation parameters $\beta_2, \beta_4$, but for Au and Pb we have $\beta_2 = \beta_4 = 0$.
The impact parameter distributions shown in the left panel of Fig.~\ref{f:pbpbauauhistsbne2} are nearly identical, up to a $\sim 4\%$ increase in the radius of Pb. The characteristic shape of $\frac{dP}{db}$ agrees with the naive expectation of uniform impact parameter distribution in two-dimensional space, 
\begin{align}
    \frac{dP}{d^2 b_\bot} = \frac{1}{2 \pi b_\bot} \frac{dP}{db_\bot} = \mathrm{const} \: ,
\end{align}
and therefore, a distribution in impact parameter magnitude that grows linearly, 
\begin{align}
    \frac{dP}{db_\bot} \propto b_\bot \: .
\end{align}
This linear behavior continues up to the peak around $b \approx 2R \approx 13 \, \mathrm{fM}$ where the nuclei begin to miss. The slightly larger radius of Pb also leads to a slight increase in the typical impact parameter as a function of centrality, shown in the right panel of Fig.~\ref{f:pbpbauauhistsbne2}.

\begin{figure}[t] 
    \centering
    \includegraphics[width=.5\textwidth]
    {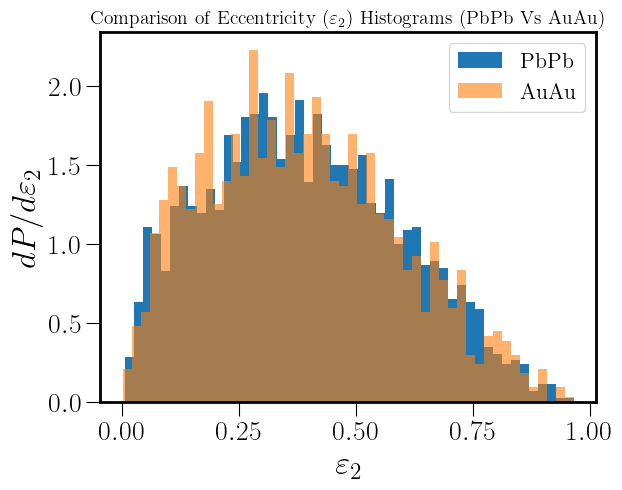}
    \caption{
    Eccentricity $\varepsilon_2$ distribution of 5.02 TeV PbPb collision at the LHC compared to 200 GeV AuAu collision at RHIC.  
    \label{f:pbpbvsauaubnacoenhfncofcent}
    }
\end{figure}

The distributions of ellipticity $\varepsilon_2$ shown in Fig.~\ref{f:pbpbvsauaubnacoenhfncofcent} are also almost identical, with a typical minimum bias ellipticity of $\sim 0.3$.  For these isotropic nuclei, the ellipticity $\varepsilon_2$ is derived entirely from the effect of the impact parameter $b$, with no other geometric effects such as deformation.


\begin{figure}[t]
\begin{centering}
\includegraphics[width=0.48\textwidth]{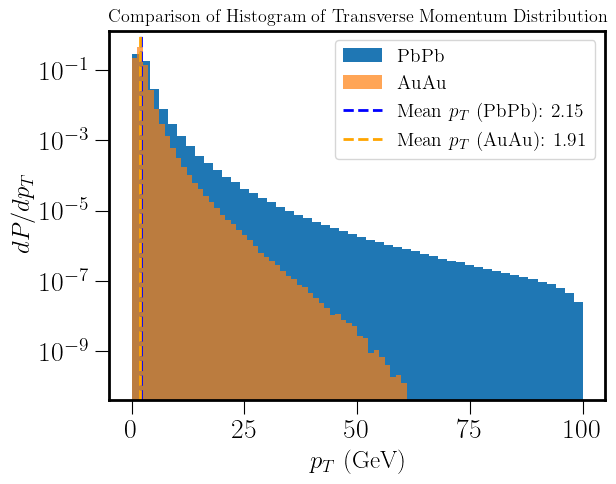}
\includegraphics[width=0.49\textwidth] {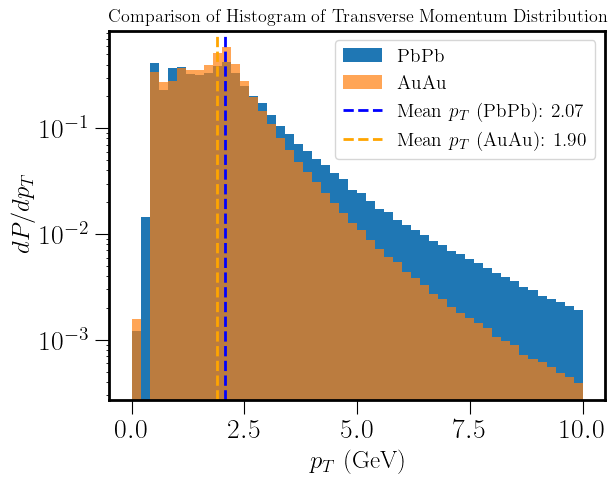}
\caption{Histograms [$\sim 2500$ events] of $p_T$ for PbPb and AuAu for minimum bias collisions (left) and with a cut $p_T \leq 10 \, \mathrm{GeV}$ (right).
\label{f:pbpbauauhistpT}
}
\end{centering}
\end{figure}

The left panel of Fig.~\ref{f:pbpbauauhistpT} clearly shows that, due to the much higher $\sqrt{s}$ in PbPb collisions, the spectrum of partons is significantly different and extends to much higher $p_T$. In minimum bias, the $\langle p_T \rangle$ of partons extends much higher in PbPb than in AuAu, although the mean value $\langle p_T \rangle_\mathrm{PbPb} = 2.15$ GeV in PbPb collisions is only 12.5\% larger than the mean value $\langle p_T \rangle_\mathrm{AuAu} = 1.91$ GeV in AuAu collisions.
With a cut on semi-hard partons $p_T \leq 10 \, \mathrm{GeV}$ (right panel of Fig.~\ref{f:pbpbauauhistpT}), the higher $p_T$ range of PbPb compared to AuAu is reduced, but an excess still remains.  After the cut, the mean values are $\langle p_T \rangle_\mathrm{PbPb} = 2.07$ GeV and $\langle p_T \rangle_\mathrm{AuAu} = 1.90$, with only a 9\% excess for PbPb collisions.


%
\subsubsection{Jet Observables: Acoplanarity}
%

\begin{figure}[t!] 
\centering
    \begin{subfigure}{.49\textwidth}
    \includegraphics[width=1\textwidth]{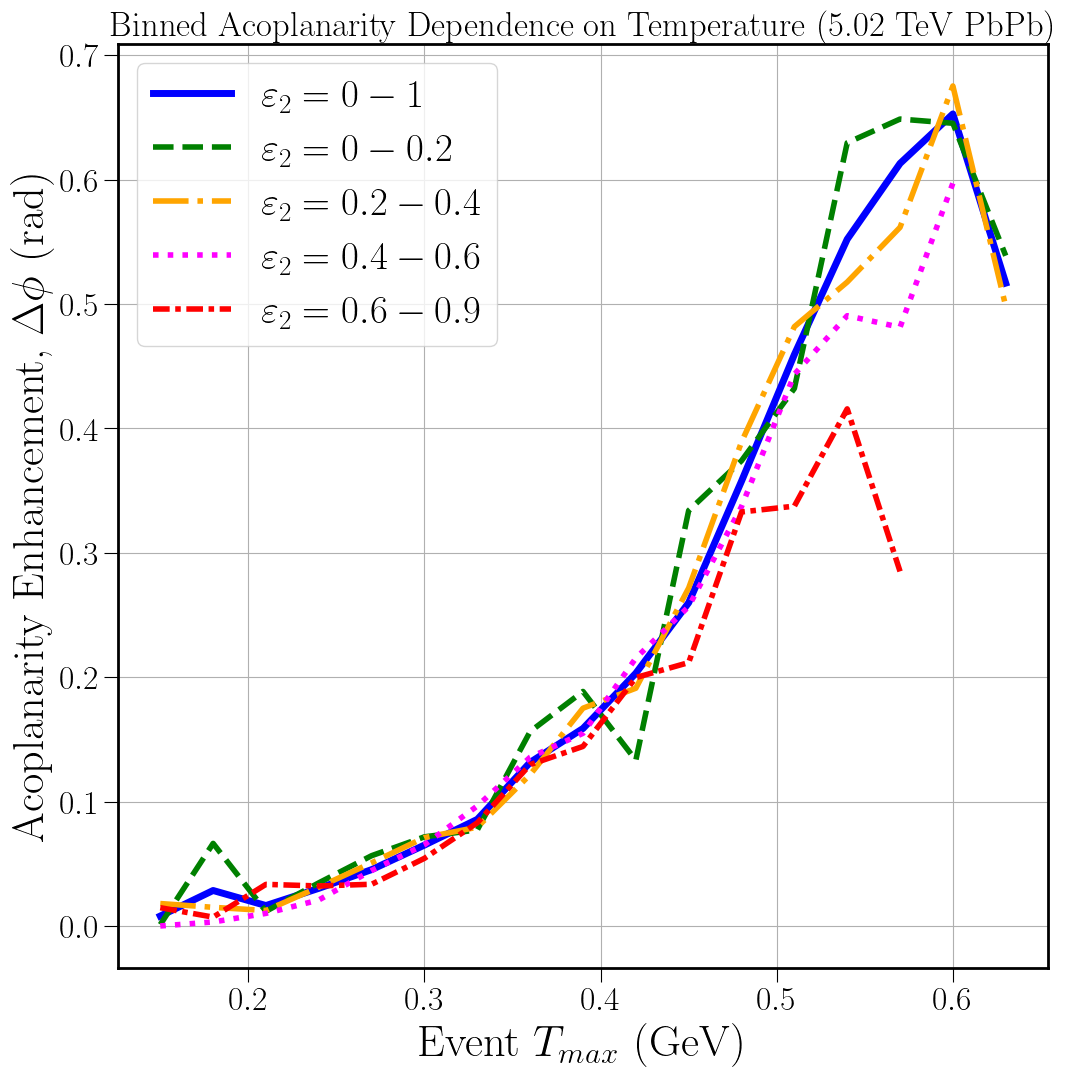}
    \end{subfigure}
    \begin{subfigure}{.49\textwidth}
    \centering
    \includegraphics[width=1\textwidth]        {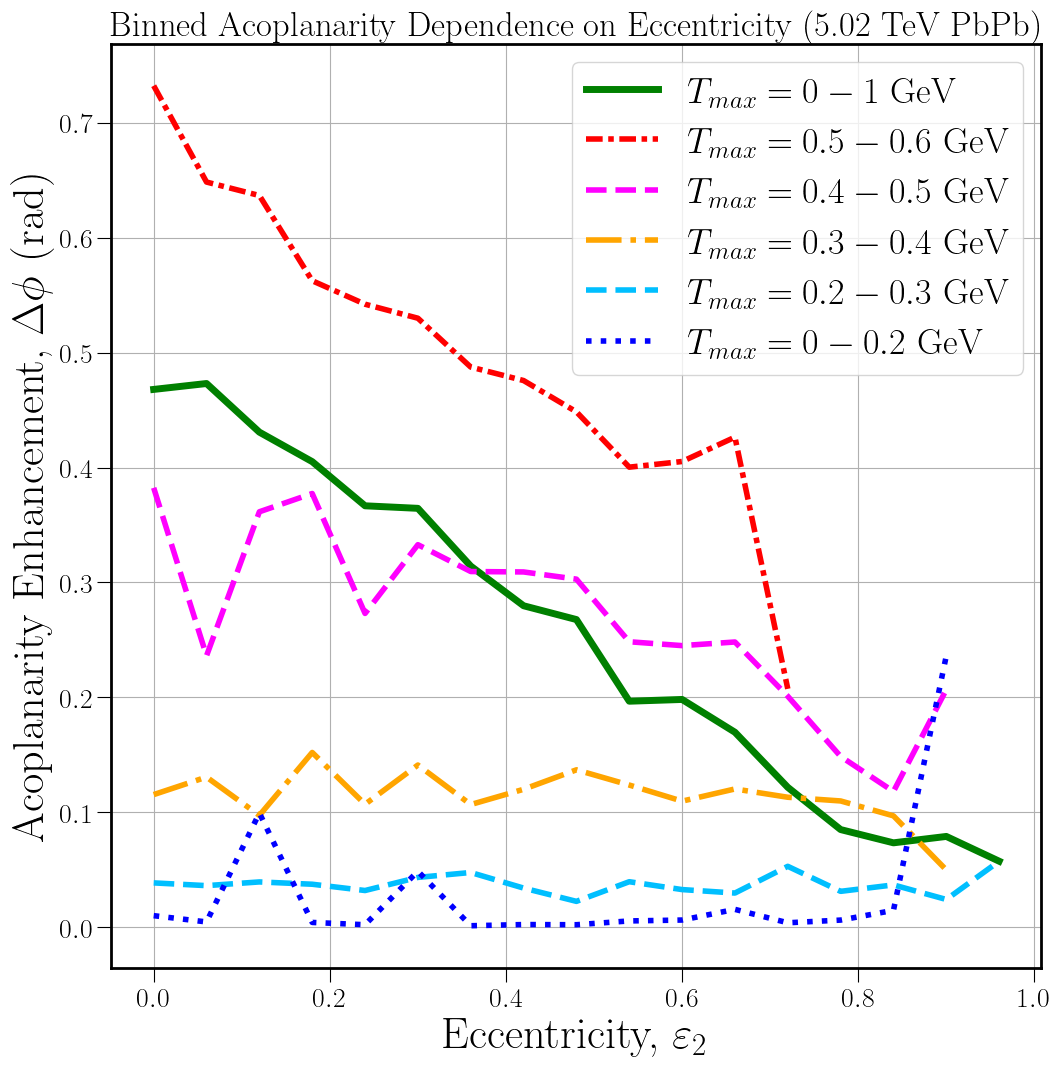}
    \end{subfigure}
    \vspace{1em}
    \begin{subfigure}{.49\textwidth}
    \includegraphics[width=1\textwidth]{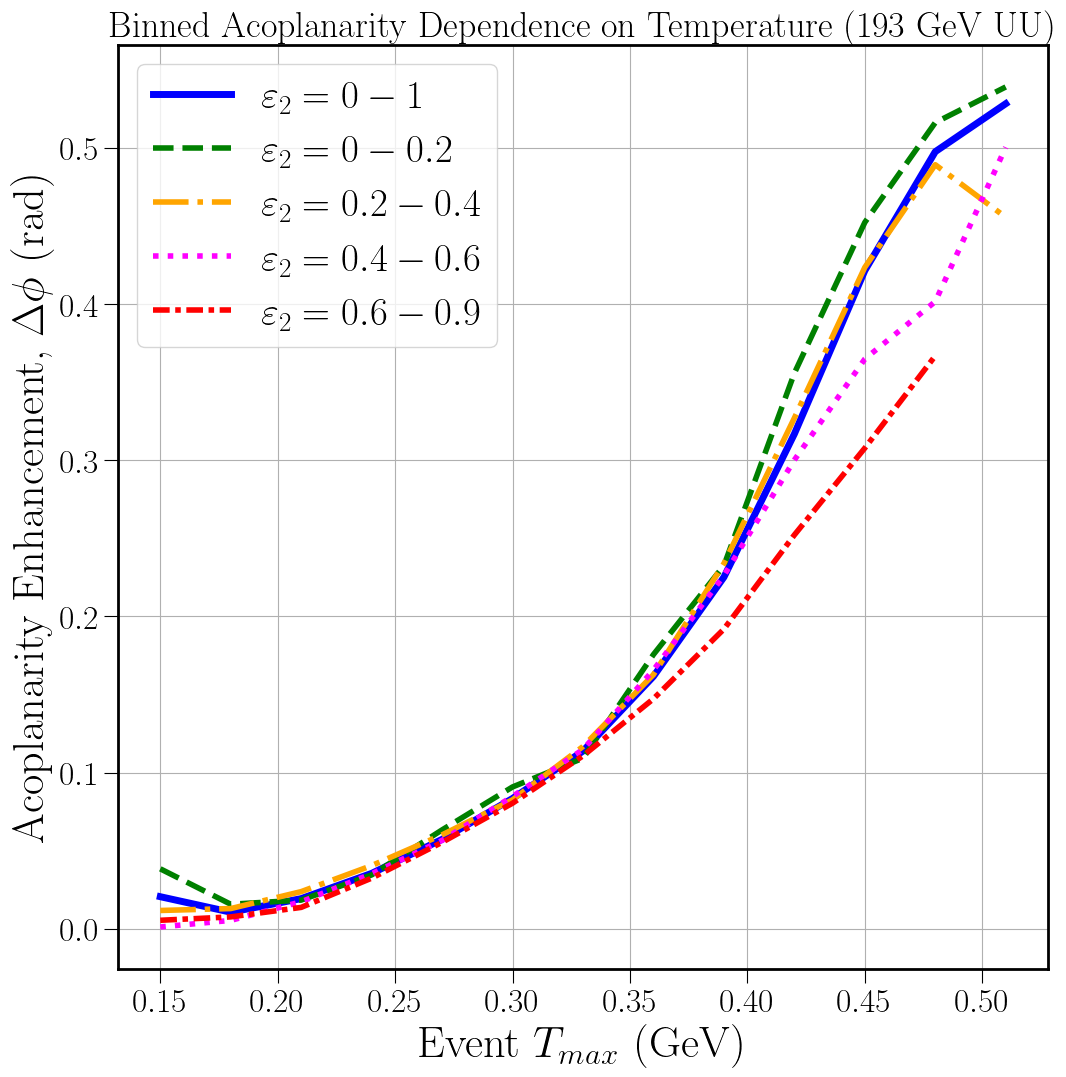}
    \end{subfigure}
    \begin{subfigure}{.49\textwidth}
    \centering
    \includegraphics[width=1\textwidth]        {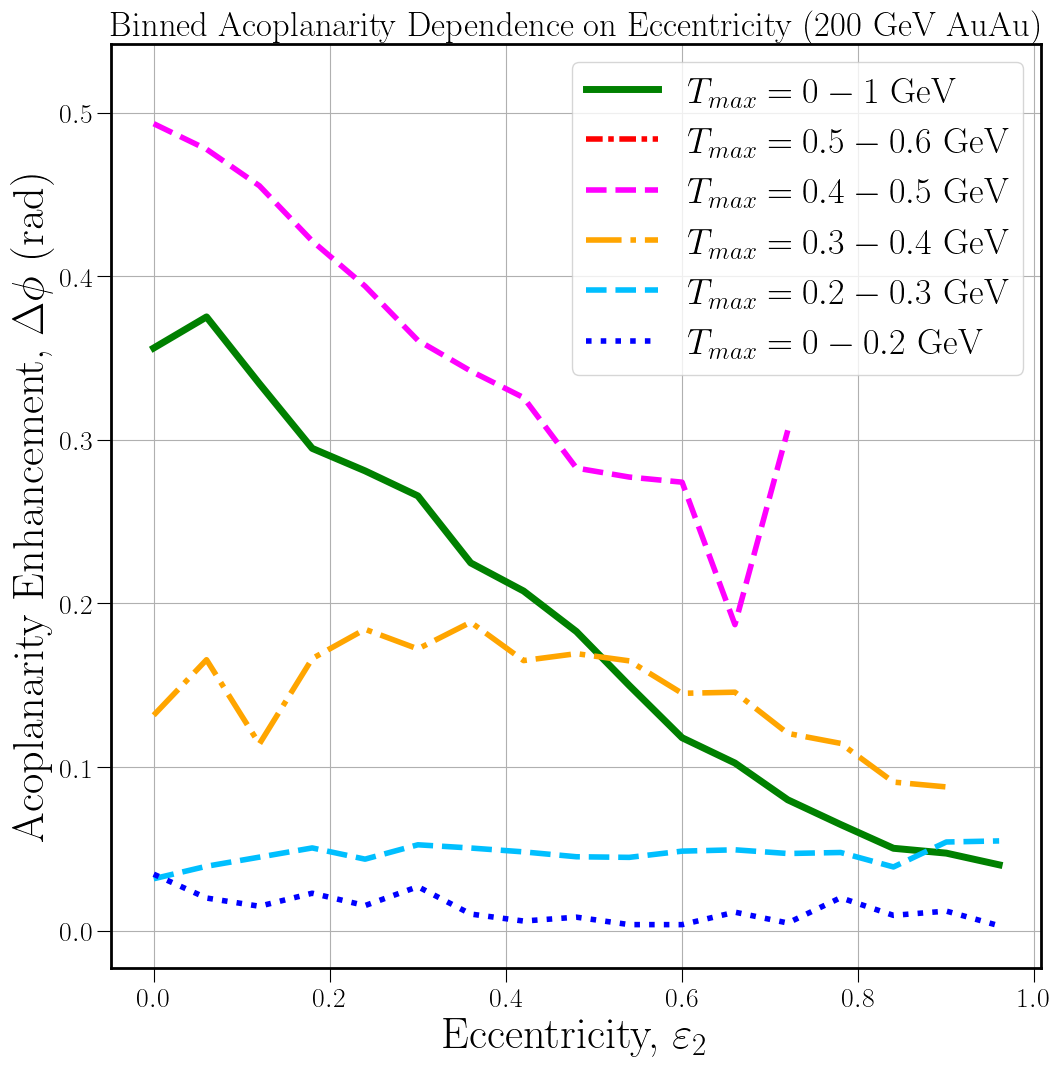}
    \end{subfigure}
\caption{Acoplanarity histograms for 5.02 TeV PbPb collision at the LHC plotted as a function of $T_{max}$ in bins of $\varepsilon_2$ (top left) and plotted as a function of $\varepsilon_2$ in bins of $T_{max}$ (top right). Similar plots for 200 GeV AuAu collisions at RHIC are shown as a function of $T_{max}$ in bins of $\varepsilon_2$ (bottom left) and plotted as a function of $\varepsilon_2$ in bins of $T_{max}$ (bottom right).}
\label{pbpbnauaubinnedhist}
\end{figure}

Next, we study the acoplanarity of dihadrons in PbPb versus AuAu collisions, using a strategy of multi-differential binning in the maximum temperature $T_{max}$ and ellipticity $\varepsilon_2$ to compare one independent variable at a time.  Recall that, as implemented in APE, the initial partons are back to back, so that all acoplanarity in the simulation is the result of jet drift (the acoplanarity \textit{enhancement} due to drift).

In Fig.~\ref{pbpbnauaubinnedhist} (top left panel), we notice that, for 5.02 TeV PbPb collisions, the acoplanarity enhancement is an increasing function of $T_{max}$ at fixed $\varepsilon_2$, reflecting a positive correlation of acoplanarity with the temperature.  In contrast, in Fig.~\ref{pbpbnauaubinnedhist} (top right panel), it is a decreasing function of $\varepsilon_2$ at fixed $T_{max} \geq 300$ MeV, indicating a negative correlation or anticorrelation. Moreover, the anticorrelation with $\varepsilon_2$ increases in strength (slope) with increasing $T_{max}$. A similar trend for acoplanarity enhancement has also been observed for the 200 GeV AuAu collisions, where it is an increasing function of $T_{max}$ at fixed $\varepsilon_2$, reflecting a positive correlation of acoplanarity with the temperature shown in Fig.~\ref{pbpbnauaubinnedhist} (bottom left panel) compared to being a decreasing function of $\varepsilon_2$ at fixed $T_{max} \geq 300$ MeV shown in Fig.~\ref{pbpbnauaubinnedhist} (bottom right panel), indicating a negative correlation or anticorrelation. An important thing to note here is that we have used $p_T < 10$ GeV cut on the spectra for both systems to generate these histograms. 

The positive correlation of acoplanarity with $T$ was expected because the scattering cross section of the jet with the medium increases with temperature. But the \textit{negative} correlation with $\varepsilon_2$ is surprising; at first glance, one would expect \textit{no correlation} of the acoplanarity with the ellipticity whatsoever, since the acoplanarity has no directional preference.  One can speculate that the anticorrelation of acoplanarity with $\varepsilon_2$ may instead be an indirect correlation with the path length, which is expected to be positively correlated with the acoplanarity due to drift. Furthermore, we noticed in the top left plot in Fig.~\ref{pbpbnauaubinnedhist} that the acoplanarity enhancement for $T_{max} \leq 450 \, \mathrm{MeV}$ is independent of $\varepsilon_2$ (uncorrelated to the geometry), while the anticorrelation with $\varepsilon_2$ sets in for $T_{max} \geq 450 \, \mathrm{MeV}$. One can definitely ask why the transition happens at this specific temperature for PbPb collision, but it will be an interesting avenue to further explore.  Note also that with increasing $T_{max}$, the maximum $\varepsilon_2$ decreases.  This is because a cut which selects on large $T_{max}$ is effectively selecting on central collisions with $b \rightarrow 0$, leading to smaller $\varepsilon_2$ for isotropic nuclei like Pb or Au.  

We also see from the binning in the bottom panel plots of Fig.~\ref{pbpbnauaubinnedhist} that the acoplanarity enhancement again exhibits a universal increase with $T$ up until about $375$ MeV, above which the anticorrelation to the eccentricity (positive correlation to the path length) sets in.

\begin{figure}[t]
    \centering
    \begin{subfigure}{.49\textwidth}
    \centering
    \includegraphics[width=1\textwidth]{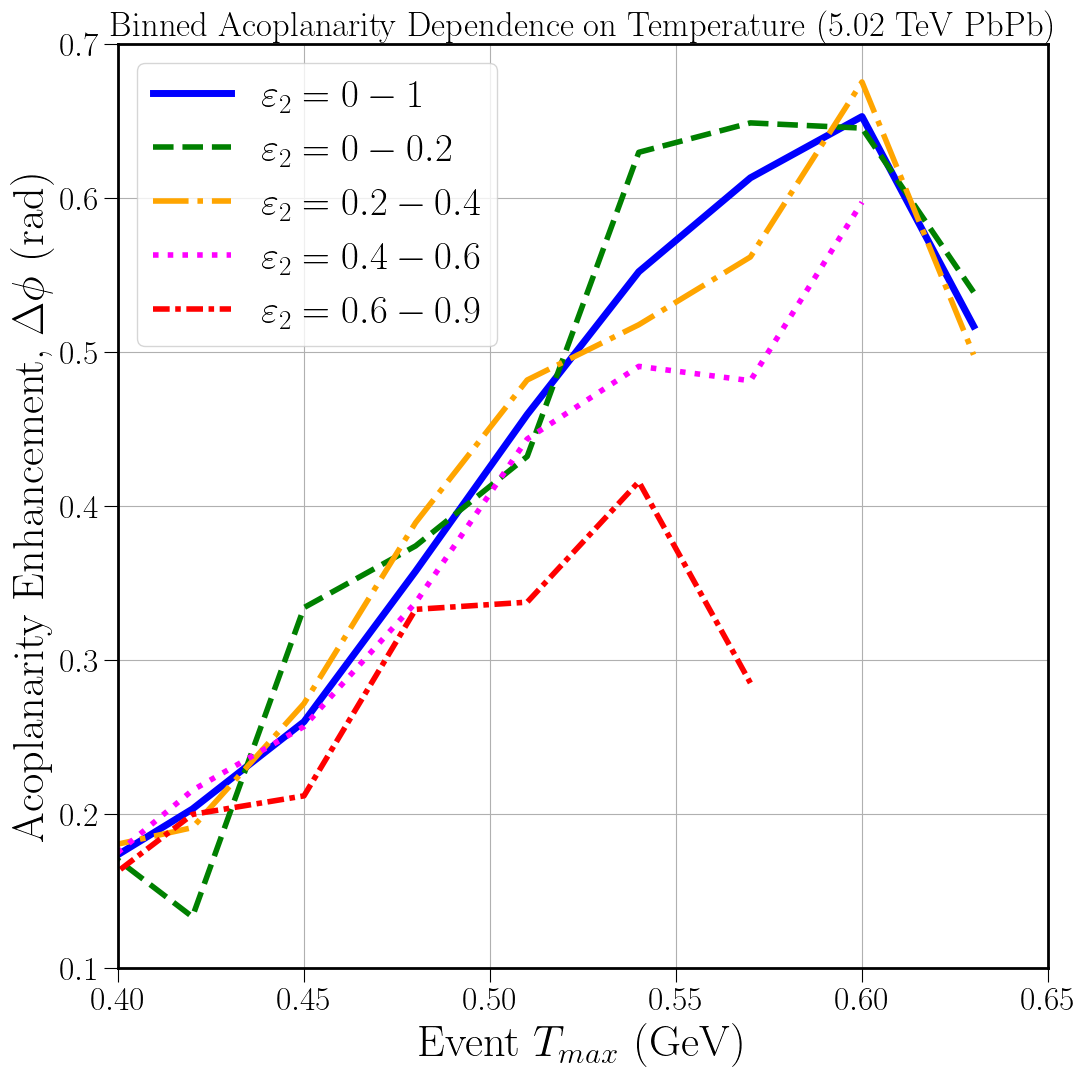}
    \end{subfigure}
    \begin{subfigure}{.49\textwidth}
    \centering
    \includegraphics[width=1\textwidth]        {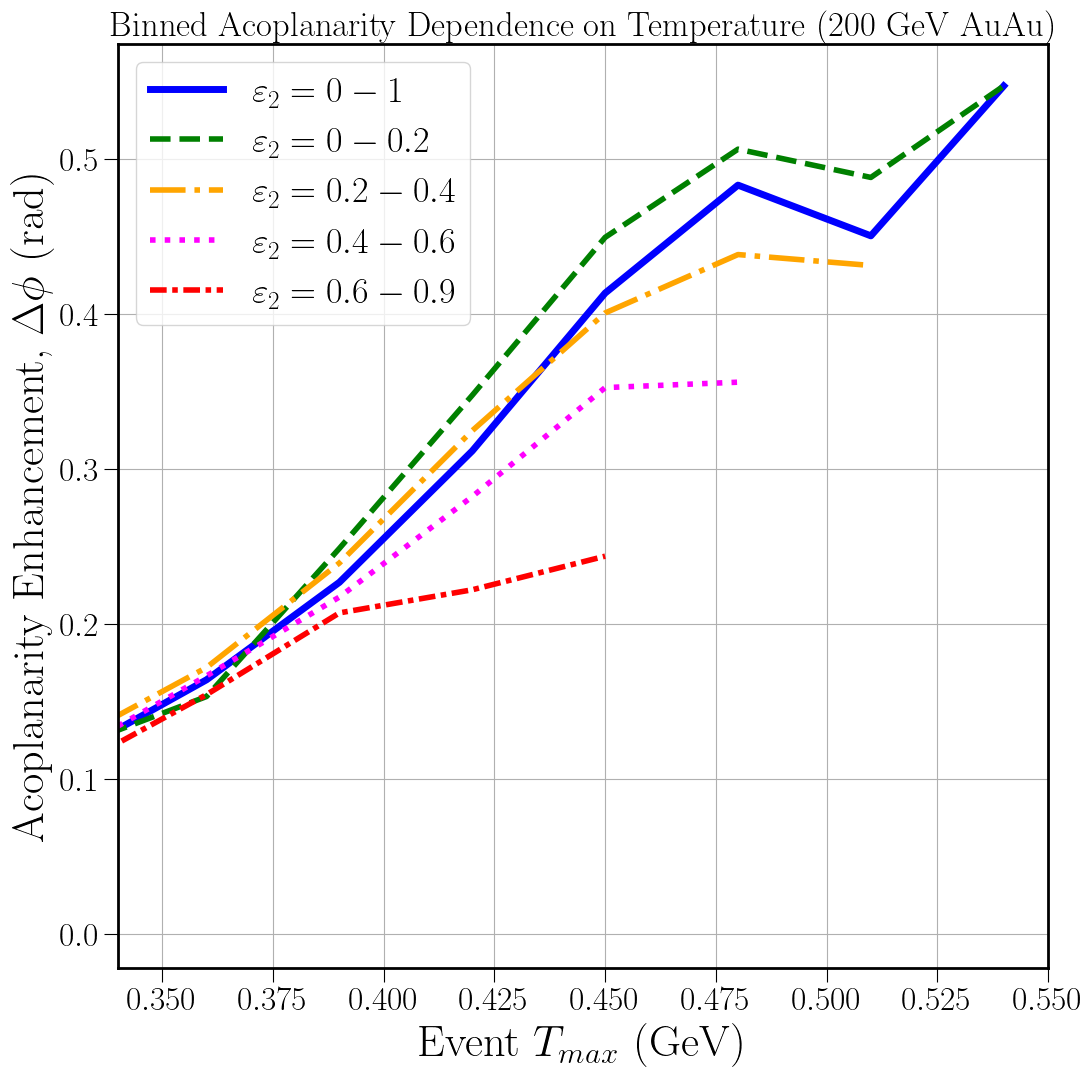}
    \end{subfigure}
    \caption{Acoplanarity enhancement as a function of $T_{max}$ in bins of $\varepsilon_2$ for 5.02 TeV PbPb collisions (left plot) versus 200 GeV AuAu collisions (right plot), focusing especially on the high-temperature region.}
\label{pbpbvsauaubinnedhistTmaxzoom}
\end{figure}

In Fig.~\ref{pbpbvsauaubinnedhistTmaxzoom}, we compare the acoplanarity enhancements ($\Delta\phi$) in PbPb versus AuAu collisions, focusing specially on the high-temperature region. On top of having a difference in the maximum $\Delta\phi$ scale between the PbPb and AuAu systems, we also notice a difference in the transition temperature scale for the two systems. The transition temperature at which the ordering of the acoplanarity enhancements for different $\varepsilon_2$-bins emerges, occurs at a much lower temperature ($T_{max} \sim 375$ MeV) for AuAu compared to the previously discussed PbPb system where $T_{max} \sim 450$ MeV. We also notice that this ordering is more differential between all $\varepsilon_2$-bins in AuAu collisions than PbPb.

\begin{figure}[t] 
    \centering
    \begin{subfigure}{.49\textwidth}
    \includegraphics[width=1\textwidth]{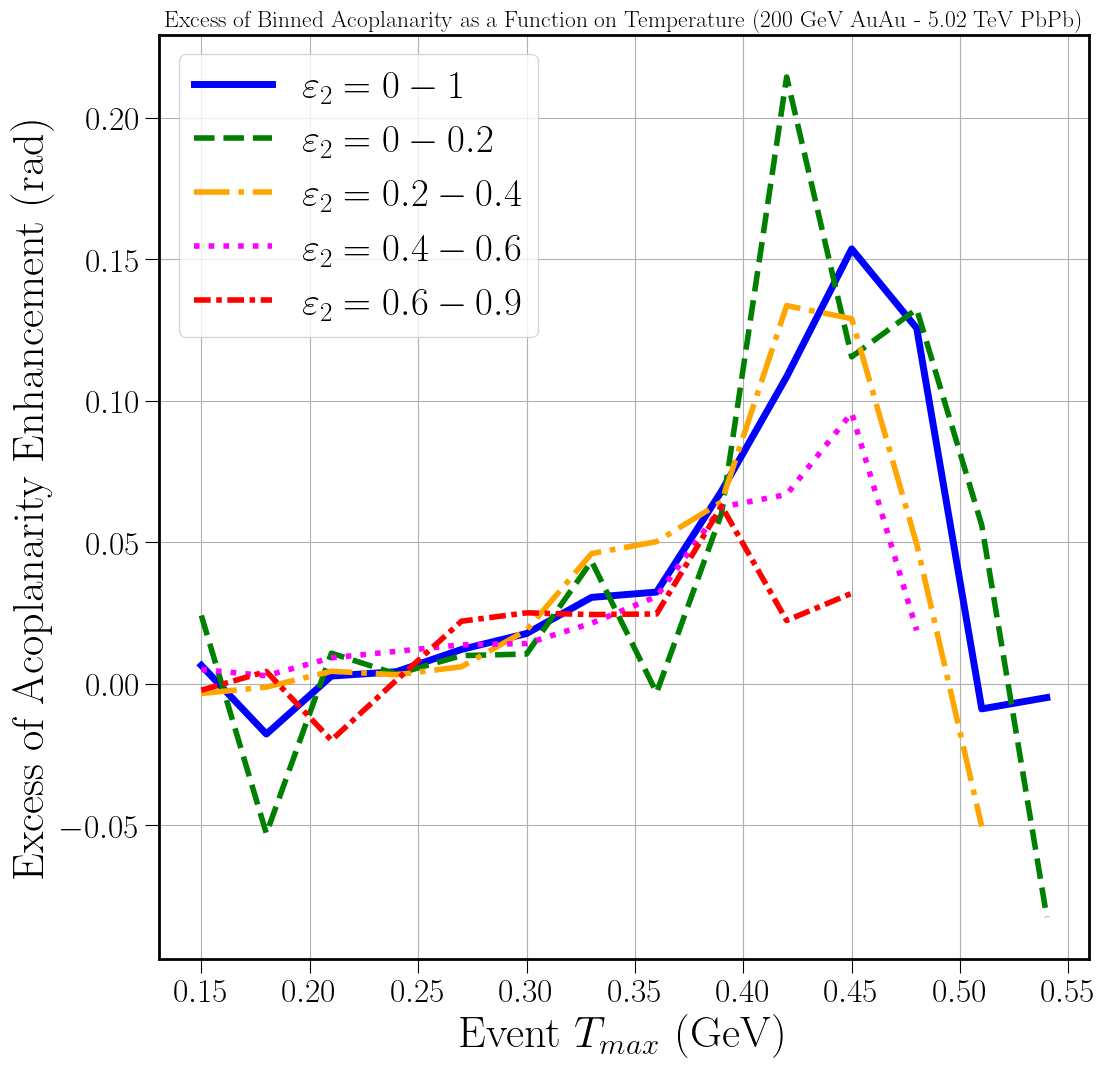}
    \end{subfigure}
    \caption{Excess of acoplanarity enhancement $\Delta\phi$ between 200 GeV AuAu collisions at RHIC and 5.02 TeV PbPb collisions at the LHC.
    \label{f:excessacohistTe2sauauvspbpb}
    }
\end{figure}

Fig.~\ref{f:excessacohistTe2sauauvspbpb} further illustrates the quantitative differences on $\Delta\phi$ as a function of $T_{max}$ between AuAu and PbPb collisions in different $\varepsilon_2$-bins. We see that, in the intermediate temperature regions ($250 MeV \leq T_{max} \leq 500 MeV$), AuAu collisions produce an excess of $\Delta\phi$ compared to PbPb collisions ($\sim 0.15$ radian, peaked at 450 MeV for minimum bias in $\varepsilon_2$ shown by the solid blue curve). For different bins of $\varepsilon_2$, the excess of $\Delta\phi$ is shown by the dashed and dotted lines of different colors. We readily see that more differentiability can be achieved for the excess of acoplanarity enhancement between these two systems when we look into different eccentricity bins.

\begin{figure}[t]
\begin{centering}
    \includegraphics[width=0.48\textwidth] {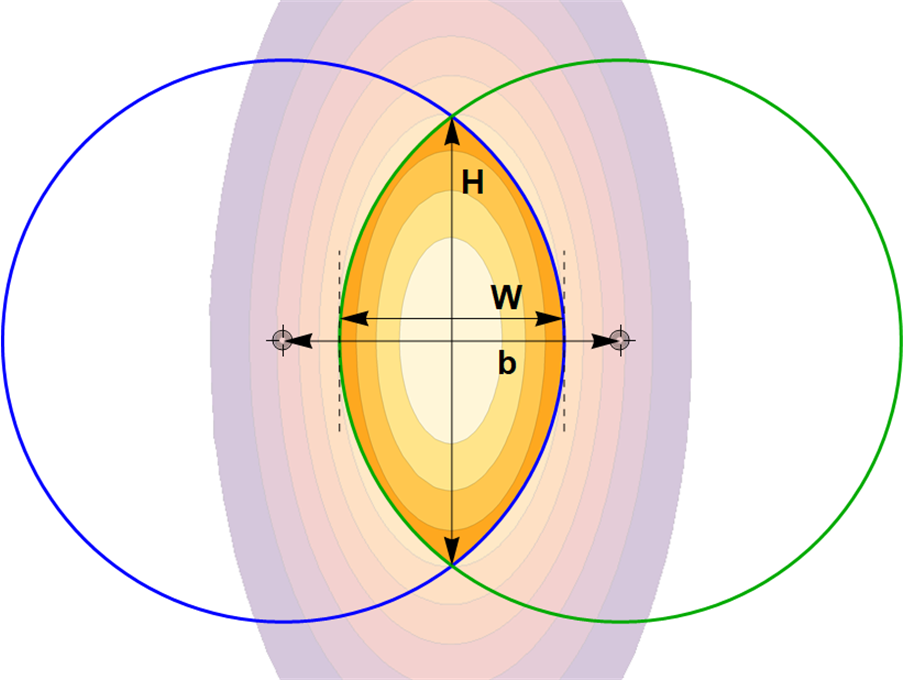}
    \includegraphics[width=0.48\textwidth]     {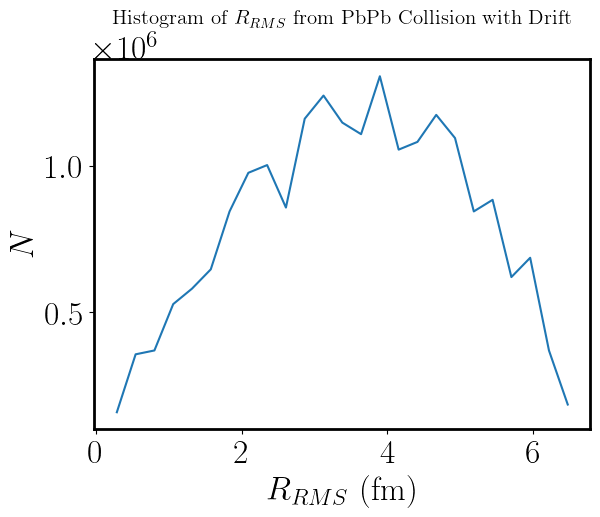}
    \caption{Left: Impact parameter relationship with event geometry. Note profile geometry spills over impact region. Right: Histogram of $R_{RMS}$ of 5.02 TeV PbPb.}
    \label{f:HWB}
\end{centering}
\end{figure}

While we leave a detailed analysis of how the acoplanarity and other observables depend on the time spent in the plasma phase $\tau_{QGP}$, hadron resonance gas phase $\tau_{HRG}$, and low-density ``un-hydrodynamic'' phase $\tau_{unhydro}$ for future work, we can perform an indirect selection on path length using the impact parameter $b$ as follows.  In a smooth optical Glauber model of the collision geometry shown in the left panel of Fig.~\ref{f:HWB}, the height is $H - \sqrt{4 R^2 - b^2}$, the width is $W = 2R - b$, and if we replace this with an ellipse with $a = W/2, b = H/2$, the root-mean-square radius over the ellipse is $R_{RMS} = \sqrt{(a^2 + b^2)/2}$.  Therefore, if we treat the typical initial state at impact parameter $b$ as an ellipse with height $H$ and width $W$ we obtain
\begin{align}   \label{e:R_RMS}
    R_{RMS} (b) = \sqrt{\frac{(2R-b)^2 + (4R^2 - b^2)}{8}} = \sqrt{\frac{R (2R - b)}{2}} \: .
\end{align}
%

This nonlinear transformation on the impact parameter serves as a simple proxy for selecting on path length. From the right panel of Fig.~\ref{f:HWB}, we see for PbPb collisions, the typical expected path length in minimum bias collisions is around 4 fm, with a roughly Gaussian distribution.

\begin{figure}[t!] 
\centering
    \begin{subfigure}{.49\textwidth}
    \includegraphics[width=1\textwidth]{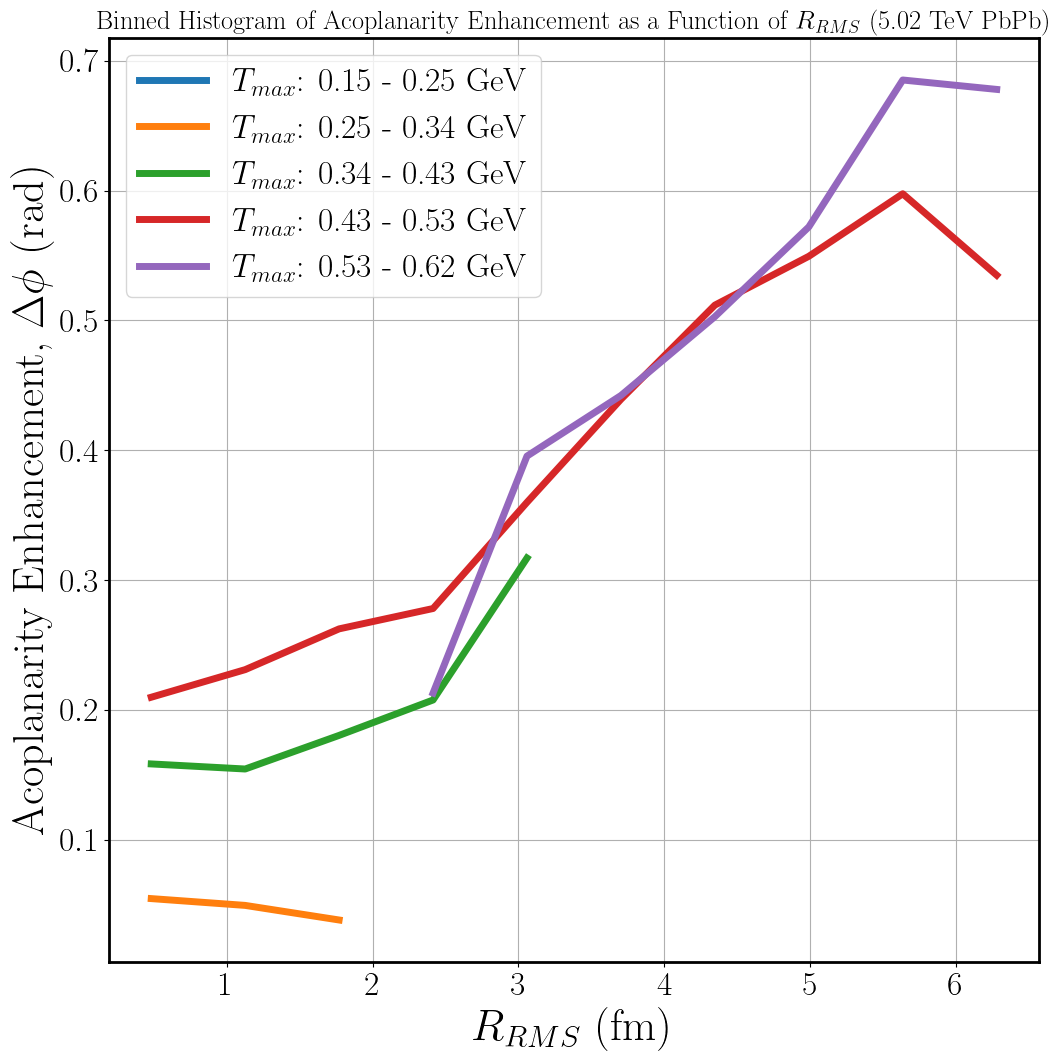}
    \end{subfigure}
    \begin{subfigure}{.49\textwidth}
    \centering
    \includegraphics[width=1\textwidth]        {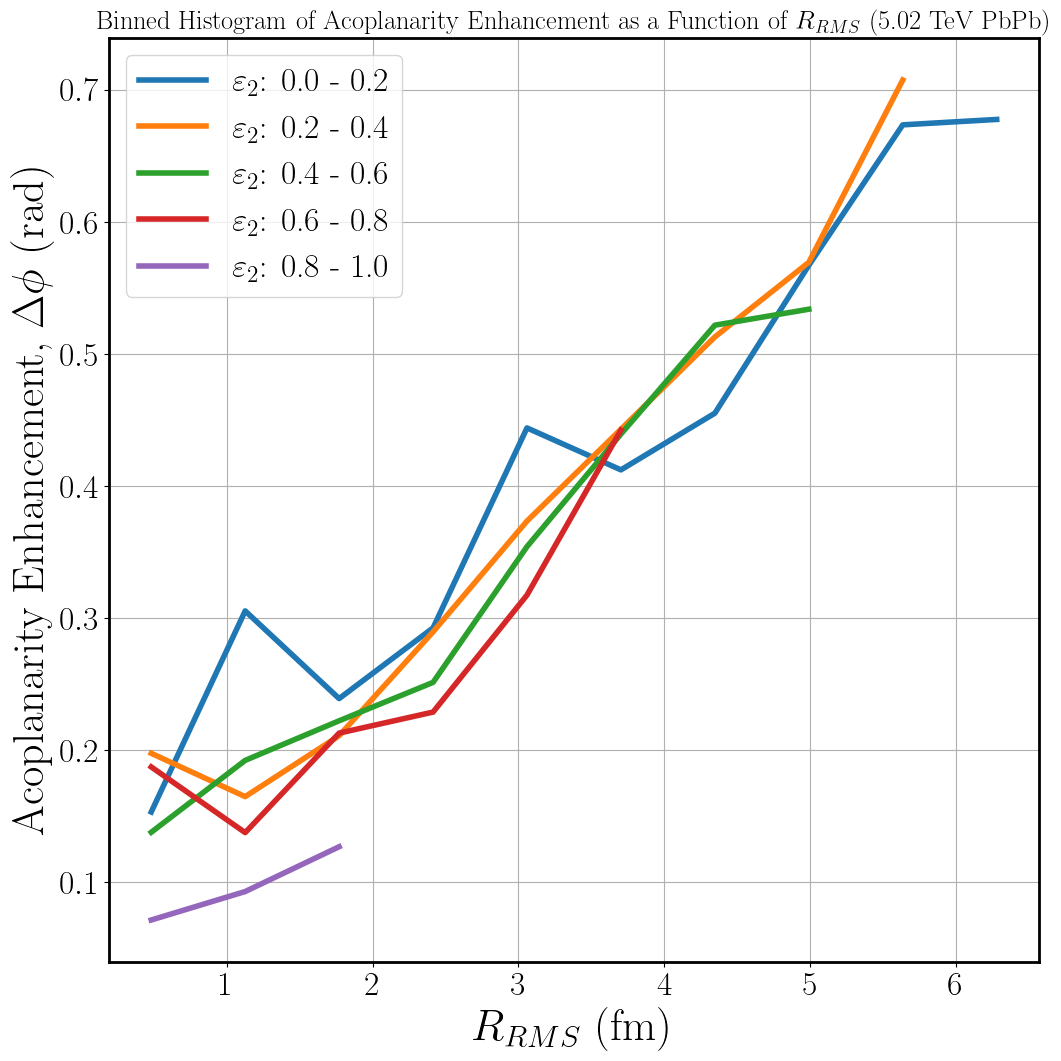}
    \end{subfigure}
    \vspace{1em}
    \begin{subfigure}{.49\textwidth}
    \includegraphics[width=1\textwidth]{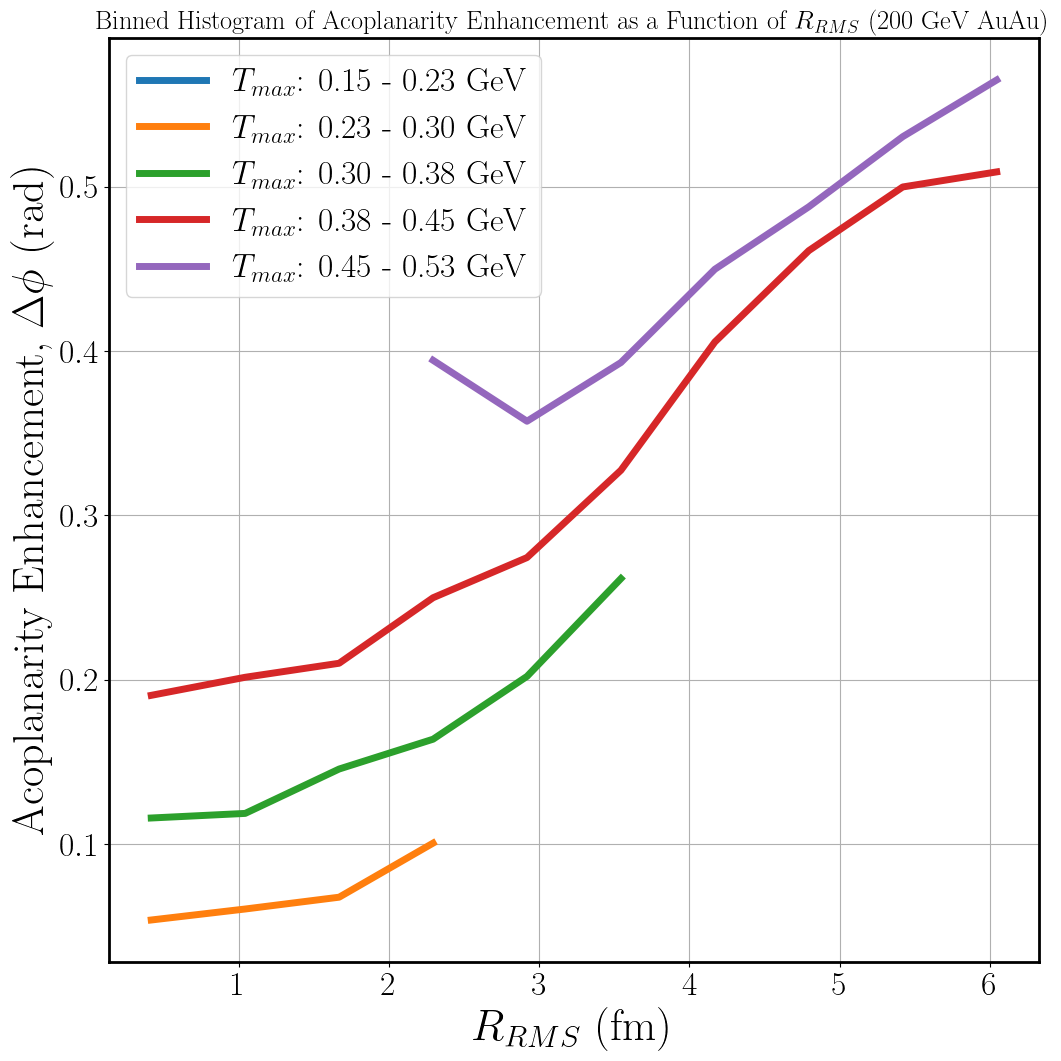}
    \end{subfigure}
    \begin{subfigure}{.49\textwidth}
    \centering
    \includegraphics[width=1\textwidth]        {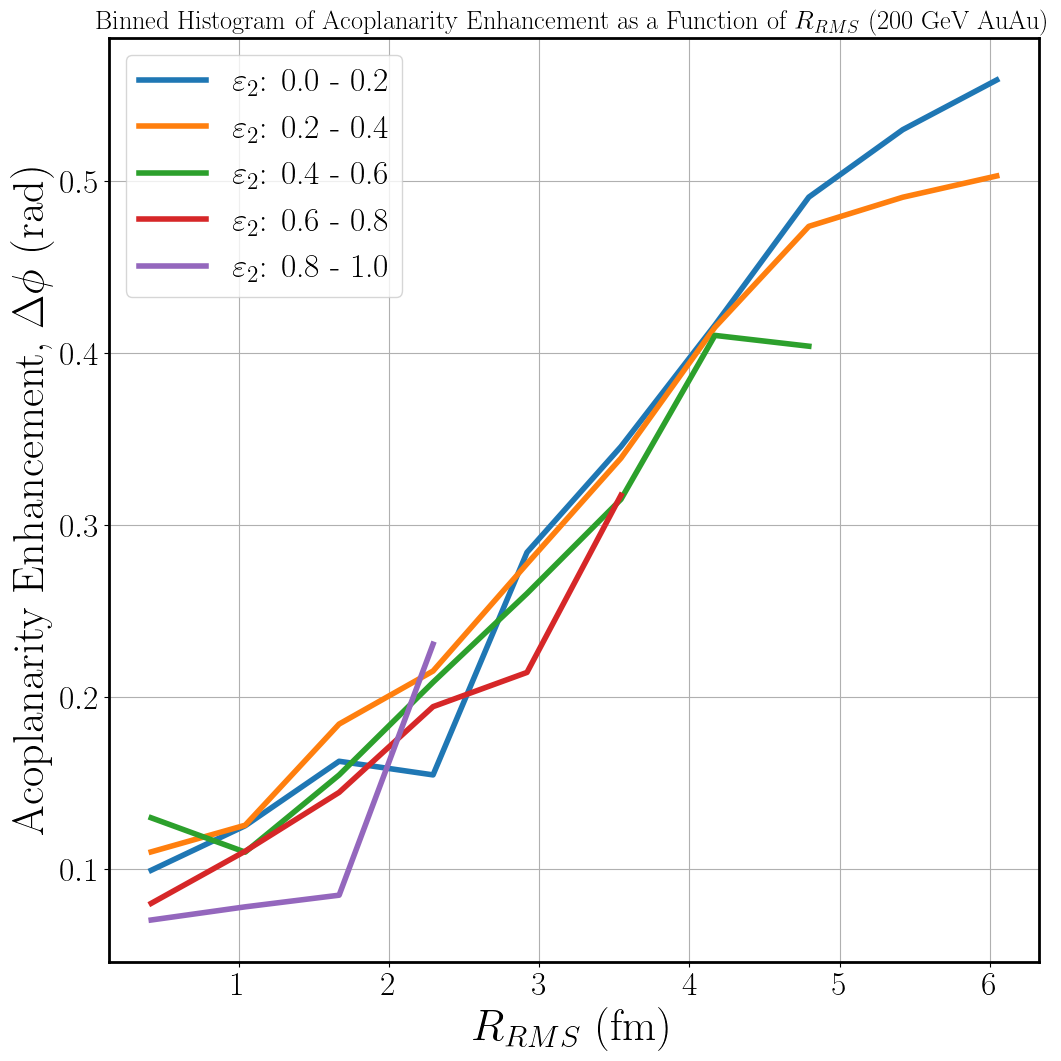}
    \end{subfigure}
\caption{Acoplanarity histograms for 5.02 TeV PbPb collision at the LHC plotted as a function of $R_{RMS}$ in bins of $T_{max}$ (top left) and in bins of $\varepsilon_2$ (top right). Similar plots for 200 GeV AuAu collisions at RHIC are shown in bins of $T_{max}$ (bottom left) and in bins of $\varepsilon_2$ (bottom right) [$\sim 2500$ events]}
\label{pbpbvsauaubinnedhistrrms_1}
\end{figure}

As described in the nonlinear mapping shown in Eq.~\eqref{e:R_RMS}, $R_{RMS}$ is a decreasing function of $b$, so that the two are anticorrelated. As noted in the right panels of Fig.~\ref{pbpbnauaubinnedhist}, the acoplanarity is also anticorrelated with the ellipticity $\varepsilon_2$ at fixed $T_{max}$. Consequently, the acoplanarity is an increasing function of $R_{RMS}$ (positively correlated), as seen in the left panels of Fig.~\ref{pbpbvsauaubinnedhistrrms_1} for PbPb (top left) and AuAu (bottom left). On the other hand, at fixed $R_{RMS}$, the acoplanarity has almost no dependence on $\varepsilon_2$ shown in the right panels of Fig.~\ref{pbpbvsauaubinnedhistrrms_1} for PbPb (top right) and AuAu (bottom right). This confirms the interpretation of the anticorrelation seen in Fig.~\ref{pbpbnauaubinnedhist} (right panel) as actually an indirect effect due to the positive correlation of the acoplanarity with path length. That is, the acoplanarity due to drift is \textit{not} directly sensitive to $\varepsilon_2$; it is only indirectly sensitive through the path length. 


The overall trends for AuAu shown in the bottom panel of Fig.~\ref{pbpbvsauaubinnedhistrrms_1} are qualitatively similar to those observed for PbPb collisions in the top panel of Fig.~\ref{pbpbvsauaubinnedhistrrms_1}. For AuAu collisions, the acoplanarity is also an increasing function of $T$ and an increasing function of $R_{RMS}$, and therefore a decreasing function of $\varepsilon_2$. Moreover, from the bottom left panel of Fig.~\ref{pbpbvsauaubinnedhistrrms_1}, AuAu shows a greater splitting of different temperature bins than previously seen in PbPb in the top left panel of Fig.~\ref{pbpbvsauaubinnedhistrrms_1}. The hierarchy of these temperature bins reflects the fact that the acoplanarity is an increasing function of temperature, suggesting that the coupling of acoplanarity to event temperature may be stronger at the lower collisional energy of RHIC than at the LHC.


\begin{figure}[t] 
    \centering
    \begin{subfigure}{.49\textwidth}
    \includegraphics[width=1\textwidth]{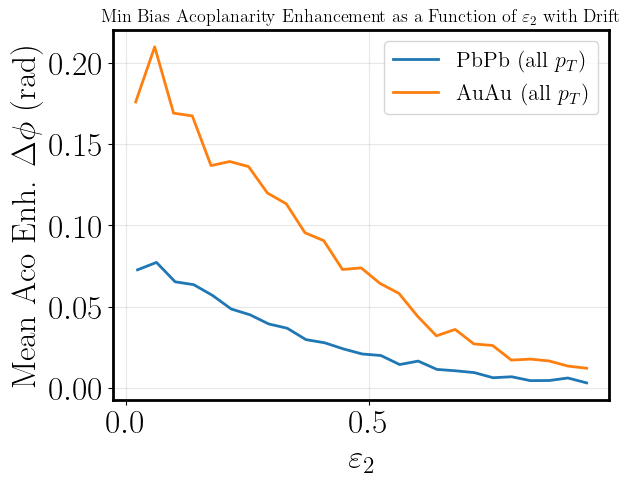}
    \end{subfigure}
    \begin{subfigure}{.49\textwidth}
    \centering
    \includegraphics[width=1\textwidth]        {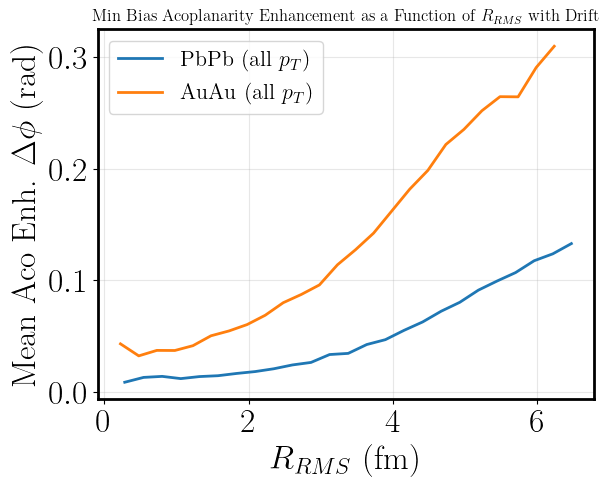}
    \end{subfigure}
    \\
    \begin{subfigure}{.49\textwidth}
    \includegraphics[width=1\textwidth]{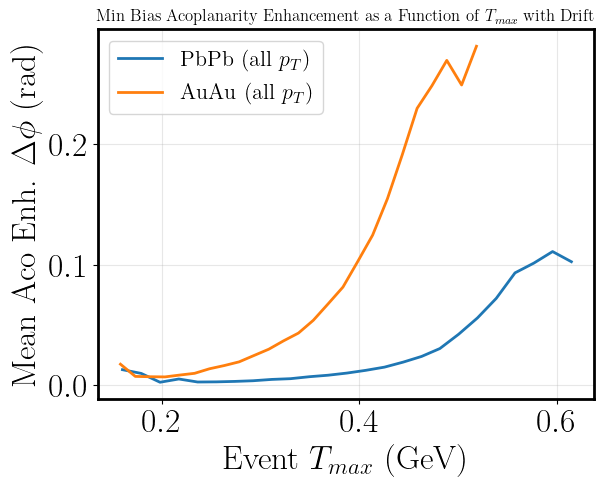}
    \end{subfigure}
    \caption{Min Bias binned histograms of $\varepsilon_2$ (top left), $R_{RMS}$ (top right), and $T_{max}$ (bottom) generated from the full initial $p_T$ spectrum.} 
    \label{pbpbvsauaubinnedacohiste2rrmstmax}
\end{figure}
%

\begin{figure}[t] 
\centering
    \begin{subfigure}{.49\textwidth}
    \includegraphics[width=1\textwidth]{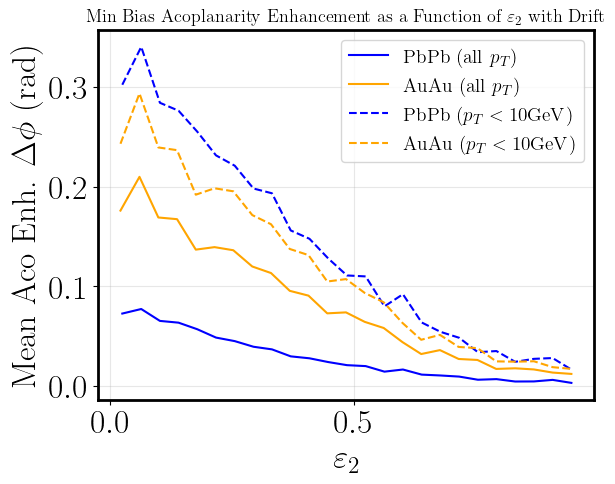}
    \end{subfigure}
    \begin{subfigure}{.49\textwidth}
    \centering
    \includegraphics[width=1\textwidth]        {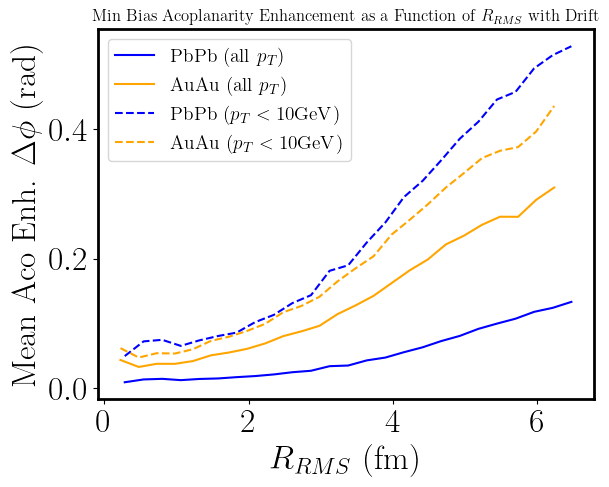}
    \end{subfigure}
    \\
    \begin{subfigure}{.49\textwidth}
    \includegraphics[width=1\textwidth]{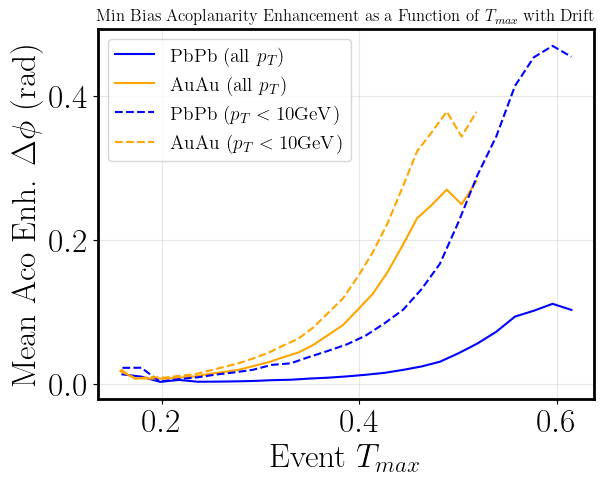}
    \end{subfigure}
    \caption{Min Bias binned histograms of $\varepsilon_2$ (top left), $R_{RMS}$ (top right), and $T_{max}$ (bottom) generated from the full initial $p_T$ spectrum for full $p_T$ (solid) versus $p_T < 10 \mathrm{GeV}$ (dashed).} 
    \label{pbpbvsauaubinnedacohiste2vsrrmspt10}
\end{figure}

In Fig.~\ref{pbpbvsauaubinnedacohiste2rrmstmax} we quantitatively compare the acoplanarity enhancement in AuAu and PbPb collisions as a function of ellipticity $\varepsilon_2$, typical path length $R_{RMS}$, and temperature $T_{max}$.  While the acoplanarity in AuAu (gold curves) has a qualitatively similar dependence on all underlying variables as in PbPb (blue curves), curiously the effect is \textit{larger} for AuAu than for PbPb by a significant margin: a factor of $\sim 3$ when including the full $p_T$ spectrum.  One may be surprised to find that the acoplanarity is larger in AuAu, even after controlling for all medium properties, but this is actually a characteristic signature of jet drift. As derived in Ref.~\cite{Sadofyev:2021ohn} and simulated extensively in Ref.~\cite{Bahder:2024jpa}, jet drift is the \textit{sub-eikonal} correction correlated to the medium flow, which is more prominent for jets with \textit{lower} energy. That is, the significantly larger $p_T$ of jets produced in PbPb collisions at the LHC seen in Fig.~\ref{f:pbpbauauhistpT} makes them less responsive to jet drift, even under identical medium conditions.

The right-hand panel of Fig.~\ref{f:pbpbauauhistpT} suggests an interesting follow-up comparison of acoplanarity in the $p_T \leq 10$ GeV regime, where the excess $p_T$ at the LHC is much less pronounced.  When we impose this cut in Fig.~\ref{pbpbvsauaubinnedacohiste2vsrrmspt10} (dashed lines), we see a striking reversal of the hierarchy. In this lower $p_T$ bin, the acoplanarity in both systems increases due to the sub-eikonal scaling $q_{drift} \propto 1 / p_T$ of jet drift, and the higher temperatures achieved at the LHC dominate. At fixed $\varepsilon_2$ or fixed $R_{RMS}$, the acoplanarity in PbPb collisions is largest due to the higher temperatures. But at fixed $T_{max}$, the acoplanarity of AuAu collisions is nearly the same as in PbPb collisions, with an excess for PbPb in the most central collisions where one is sensitive to the slightly larger path length in PbPb.  The change in AuAu / PbPb ordering under the imposition of the $p_T \leq 10$ GeV cut demonstrates that we have control over the last systematic dependence of jet drift (the parton energy) and  identifies a unique systematic feature due to the sub-eikonal scaling of jet drift.

%
\subsubsection{Jet Observables:  Elliptic Flow}
%

For another measure of the effects of jet drift in heavy ion collisions, we study the elliptic flow $v_2$ of hadrons in PbPb versus AuAu collisions.  We again use binning in $T_{max}$ and ellipticity $\varepsilon_2$ to study the independent contributions of the underlying variables.  While the absolute magnitudes of $v_2$ may be subject to normalization uncertainties that only a large-scale fit to experimental data could constrain, the difference $\Delta v_2$ with and without the effect of jet drift is a measure of the drift-mediated contribution alone.

\begin{figure}[t]
    \centering
    \includegraphics[width=0.5\linewidth]{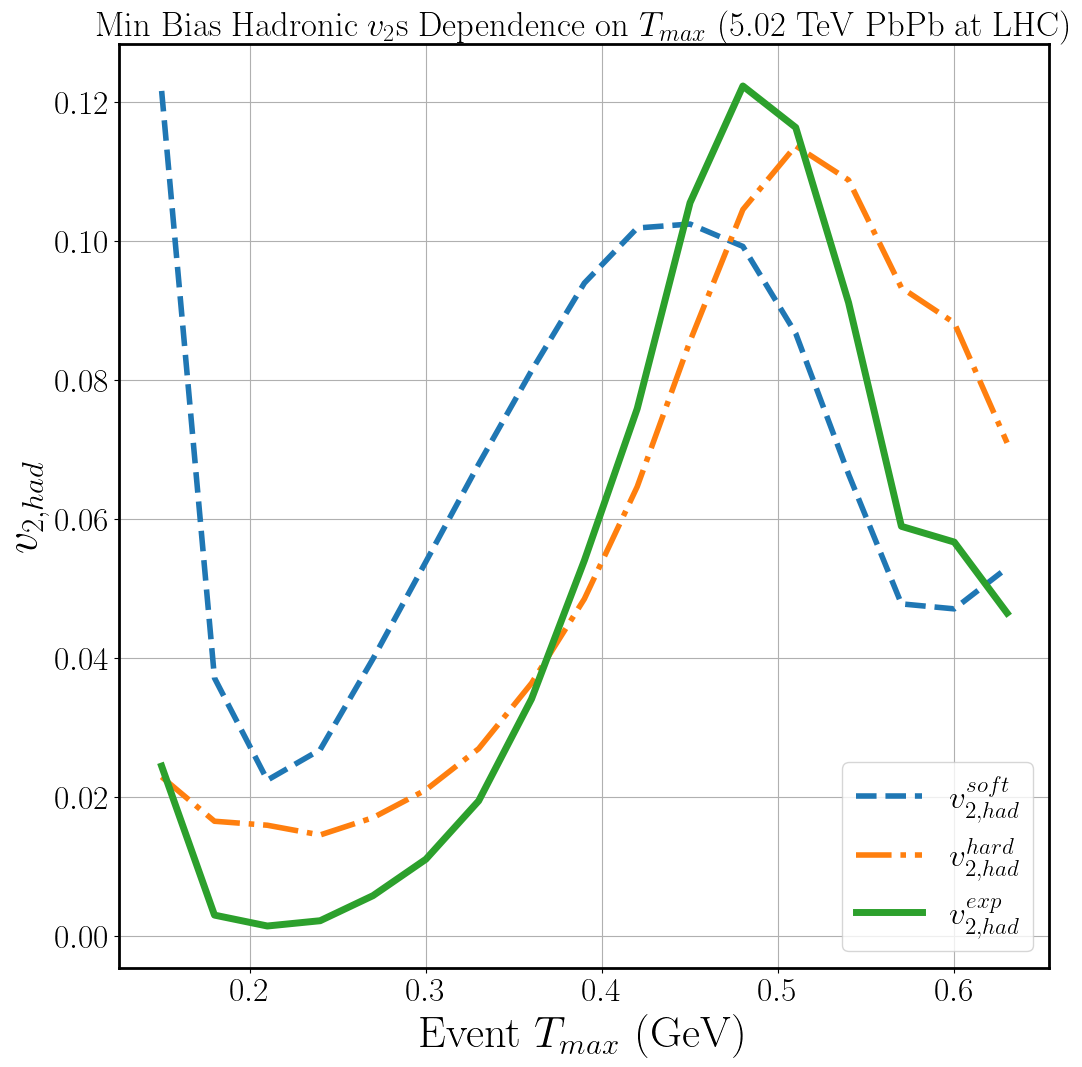}
    \caption{Hadronic $v_2$s of 5.02 TeV PbPb Collision at RHIC}
    \label{fig:v2spbpb}
\end{figure}

Fig.~\ref{fig:v2spbpb} compares three different types of elliptic flow $v_2$, which carry different information about the medium produced in  5.02 TeV PbPb collisions.  The two-particle cumulant of hard particles, 
\begin{align}
    v_2^{hard} \{2\} = \langle | v_2^{hard} |^2 \rangle^{1/2}
\end{align}
is a positive-definite measure of elliptic flow of hard particles alone, and is shown in the orange dot-dashed curve.  Recall that in our present analysis, we include only particles generated from the hard scattering events in $v_2^{hard}$.  We see that $v_2^{hard} \{2\}$ is an increasing function of $T_{max}$ which peaks around 500 MeV for PbPb collisions.  Likewise, the two-particle cumulant of soft particles,
\begin{align}
    v_2^{soft} \{2\} = \langle | v_2^{soft} |^2 \rangle^{1/2}
\end{align}
is a positive-definite measure of elliptic flow of soft particles alone shown in the blue dashed curve.  In our calculation of $v_2^{soft}$ we include all final-state particles produced from the hydrodynamic evolution with $0.2 \, \mathrm{GeV} \leq p_T \leq 5 \, \mathrm{GeV}$.  The soft elliptic flow $v_2^{soft} \{2\}$ is an increasing function of $T_{max}$ as well and peaks around 400 MeV, but also increases again below 200 MeV.  Simulated events with $T_{max} \leq 200$ MeV are unlikely to be well described by the assumed physics of the QGP, so the soft elliptic flow generated in these events should be interpreted skeptically. Unlike the other two cumulants, $v_2^{exp}$ is a \textit{correlation} \eqref{e:vndefn} between $\vec{v}_2^{\: hard}$ and $\vec{v}_2^{\: soft}$, which can be positive or negative depending on the alignment between the hard and soft. The experimentally measured $v_2^{exp}$ (shown in the green solid curve) blends the features of $v_2^{hard}$ and $v_2^{exp}$, with a sharp peak around 475 MeV and an upturn below 200 MeV.

\begin{figure}[t] 
\centering
    \begin{subfigure}{.50\textwidth}
    \includegraphics[width=1\textwidth]        {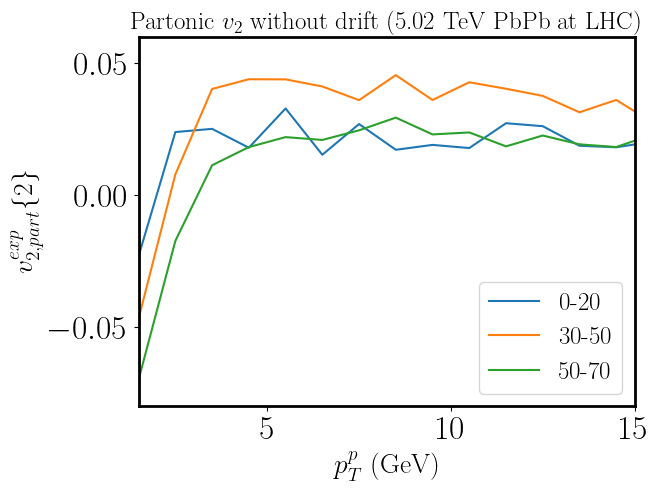}
    \end{subfigure}
    \begin{subfigure}{.49\textwidth}
    \centering
    \includegraphics[width=1\textwidth]{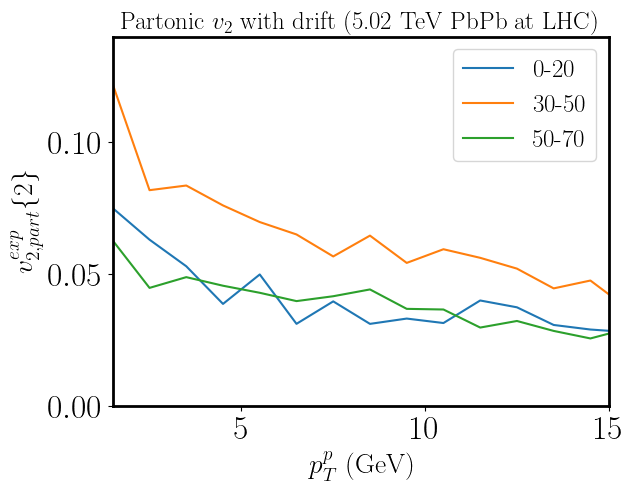}
    \end{subfigure}
    \\
    \begin{subfigure}{.49\textwidth}
    \includegraphics[width=1\textwidth]{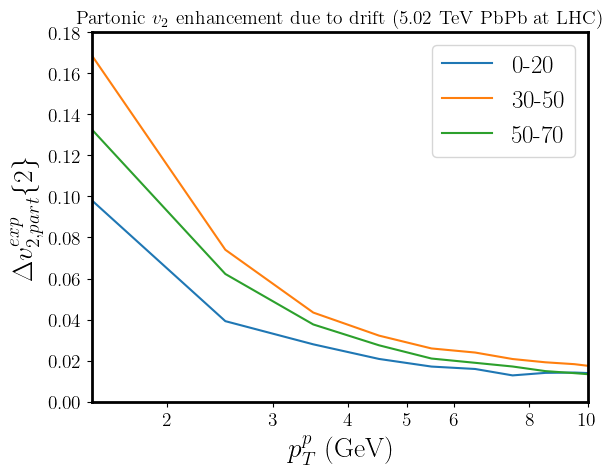}
    \end{subfigure}
    \begin{subfigure}{.49\textwidth}
    \centering
    \end{subfigure}
    \caption{Partonic $v_2^{exp}$ of 5.02 TeV PbPb with drift off (top left), on (top right), along with the net change in $v_2^{exp}$ (bottom) resulting from drift}
    \label{pbpbdeltav2partnpsi2_1}
\end{figure}

Fig.~\ref{pbpbdeltav2partnpsi2_1} compares $v_2^{exp}$ for partons produced in 5.02 TeV PbPb collisions, with and without the effect of jet drift.  From the top left panel of Fig.~\ref{pbpbdeltav2partnpsi2_1}, we see that without drift, $v_2^{exp}$ goes negative at low $p_T$ for all centrality classes.  This can only occur with a genuine correlation like $v_2^{exp}$, if 
\begin{align}
    \cos 2(\psi_2^{hard} - \psi_2^{soft}) < 0
    \qquad, \: \mathrm{or} \qquad
    \Delta\psi_2 > \pi /4 \:. 
\end{align}
Negativity of $v_2^{exp}$ means that the \textit{directions} of $\vec{v}_2^{hard}$ and $\vec{v}_2^{soft}$ have become misaligned by greater than $\pi / 4$, corresponding to event plane decorrelation between the hard and soft sectors.  Event plane decorrelation is seen at the partonic level, at low $p_T$, in several centrality bins 0-20\%, 30-50\%, and 50-70\%.

In contrast, as shown in the top right panel of Fig.~\ref{pbpbdeltav2partnpsi2_1}, with drift turned \textit{on}, the event plane decorrelation disappears completely, and $v_2^{exp}$ becomes positive for all centralities.  Drift effects are strong enough, especially at low $p_T$, to completely overturn the event plane decorrelation and restore the directional correlation of $\vec{v}_2^{hard}$ to the direction of the event plane defined by $\vec{v}_2^{soft}$. When subtracting $v_2^{exp}$ with and without drift, the restoration of event plane correlations and resultant sign change in $v_2^{exp}$ leads to a large enhancement $\Delta v_2^{exp}$ due to drift, as shown in the bottom panel of Fig.~\ref{pbpbdeltav2partnpsi2_1}. One can also note that, $\Delta v_2^{exp}$ shows a non-monotonic dependence on centrality, with the largest enhancement for mid central (30-50$\%$) events.  This non-monotonic centrality ordering reflects the competition between high temperatures in central collisions and high ellipticities in peripheral collisions, as noted in our previous paper Ref.~\cite{Bahder:2024jpa}. 
%


\begin{figure}[t!] 
    \centering
    \begin{subfigure}{.49\textwidth}
    \includegraphics[width=1\textwidth]
    {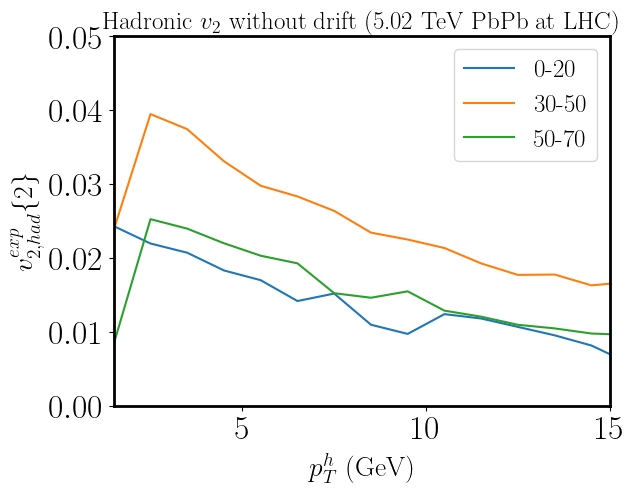}
    \end{subfigure}
    \begin{subfigure}{.49\textwidth}
    \centering
    \includegraphics[width=1\textwidth] 
    {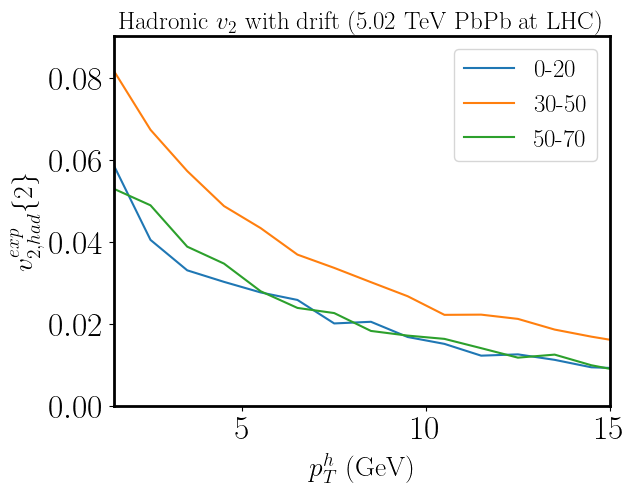}
    \end{subfigure}
    \\
    \begin{subfigure}{.49\textwidth}
    \includegraphics[width=1\textwidth]{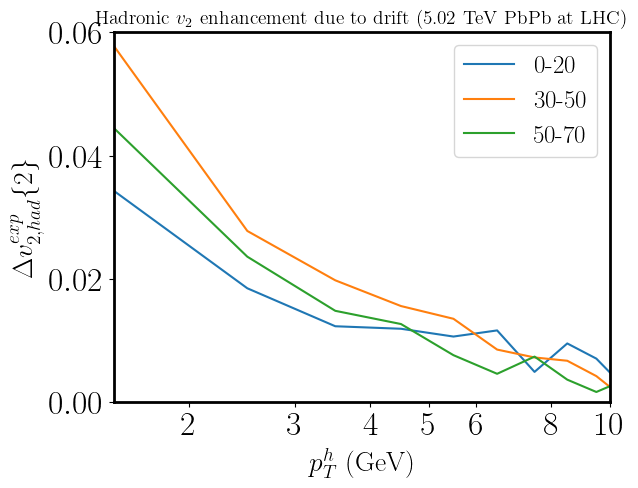}
    \end{subfigure}
    \begin{subfigure}{.49\textwidth}
    \centering
    \includegraphics[width=1\textwidth]{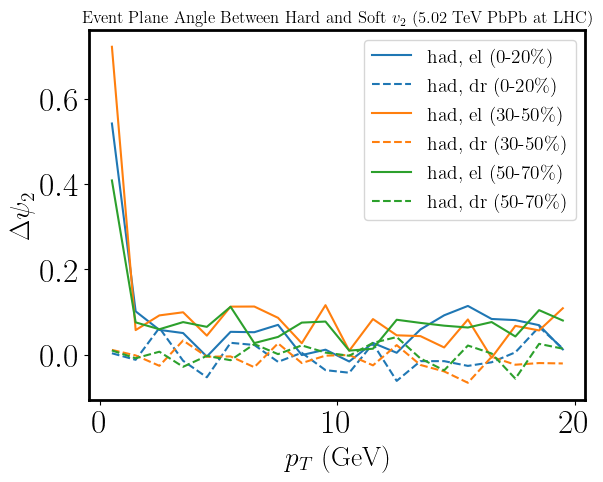}  
    \end{subfigure}
    \caption{Hadronic $v_2^{exp}$ for 5.02 TeV PbPb collisions with drift off (top left), on (top right), along with the net change in $v_2^{exp}$ (bottom left) and event plane angle $\psi_2$ due to drift (bottom right).}
    \label{pbpbv2hadworwodrift}
\end{figure}

The picture of event plane decorrelation and drift-based restoration is clearest at the partonic level shown in Fig.~\ref{pbpbdeltav2partnpsi2_1}.  But the qualitative features still persist in at the level of final-state hadrons, as shown in Fig.~\ref{pbpbv2hadworwodrift}.  Without drift (top left panel), we see a turnaround and decrease of $v_2^{exp}$ at low $p_T$ in for $>30\%$ centrality, a sign that event plane decorrelation is starting to set in, but not enough to cause $v_2^{exp}$ to fully become negative, as in the partonic case. With drift turned on (top right panel), that downturn in $v_2^{exp}$ disappears, reflecting a restored correlation to the soft event plane.  The result is a significant enhancement of $v_2^{exp}$ at low $p_T$, even for hadrons after fragmentation. Now we understand that this enhancement reflects the fact that jet drift can reverse the beginnings of event plane decorrelation at the hadronic level.  We can confirm this picture of event plane decorrelation and restoration due to jet drift by examining the difference $\Delta\psi_2$ between hard and soft event plane angles.  As shown in the bottom-right panel of Fig.~\ref{pbpbdeltav2partnpsi2_1}, in the absence of drift (solid curves), all three centrality classes show large $\Delta \psi_2$ at small $p_T$ where the downturn in $v_2^{exp}$ occurs. With drift turned on (dashed curves) $\Delta \psi_2$ falls to zero, directly revealing the restored correlation between $\psi_2^{hard}$ and $\psi_2^{soft}$.

\begin{figure}[t] 
    \centering
    \begin{subfigure}{.48\textwidth}
    \includegraphics[width=1\textwidth]{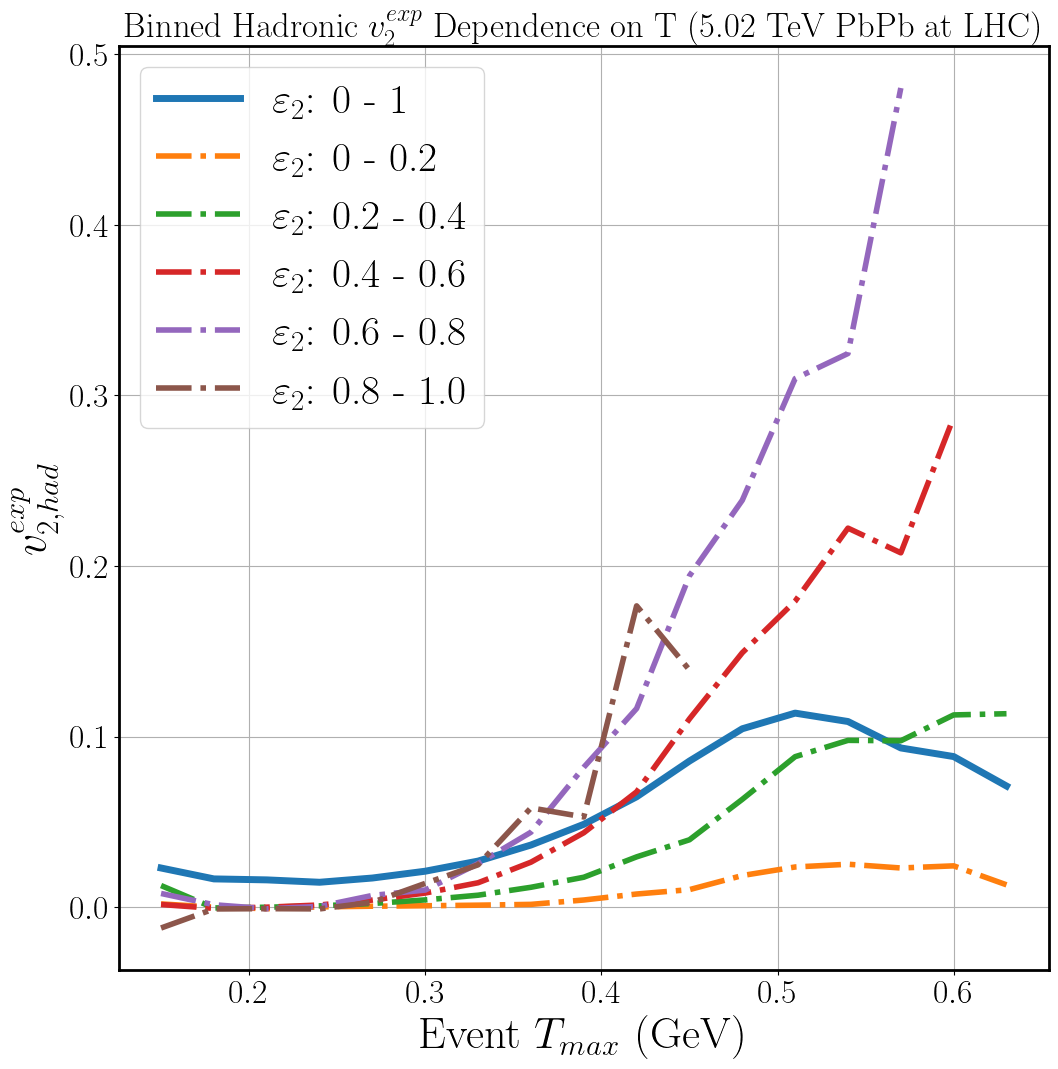}
    \caption{5.02 TeV PbPb
    \label{f:pbpbbinnedv2histTfore2s}
    }
    \end{subfigure}
    \begin{subfigure}{.49\textwidth}
    \centering
    \includegraphics[width=1\textwidth]        {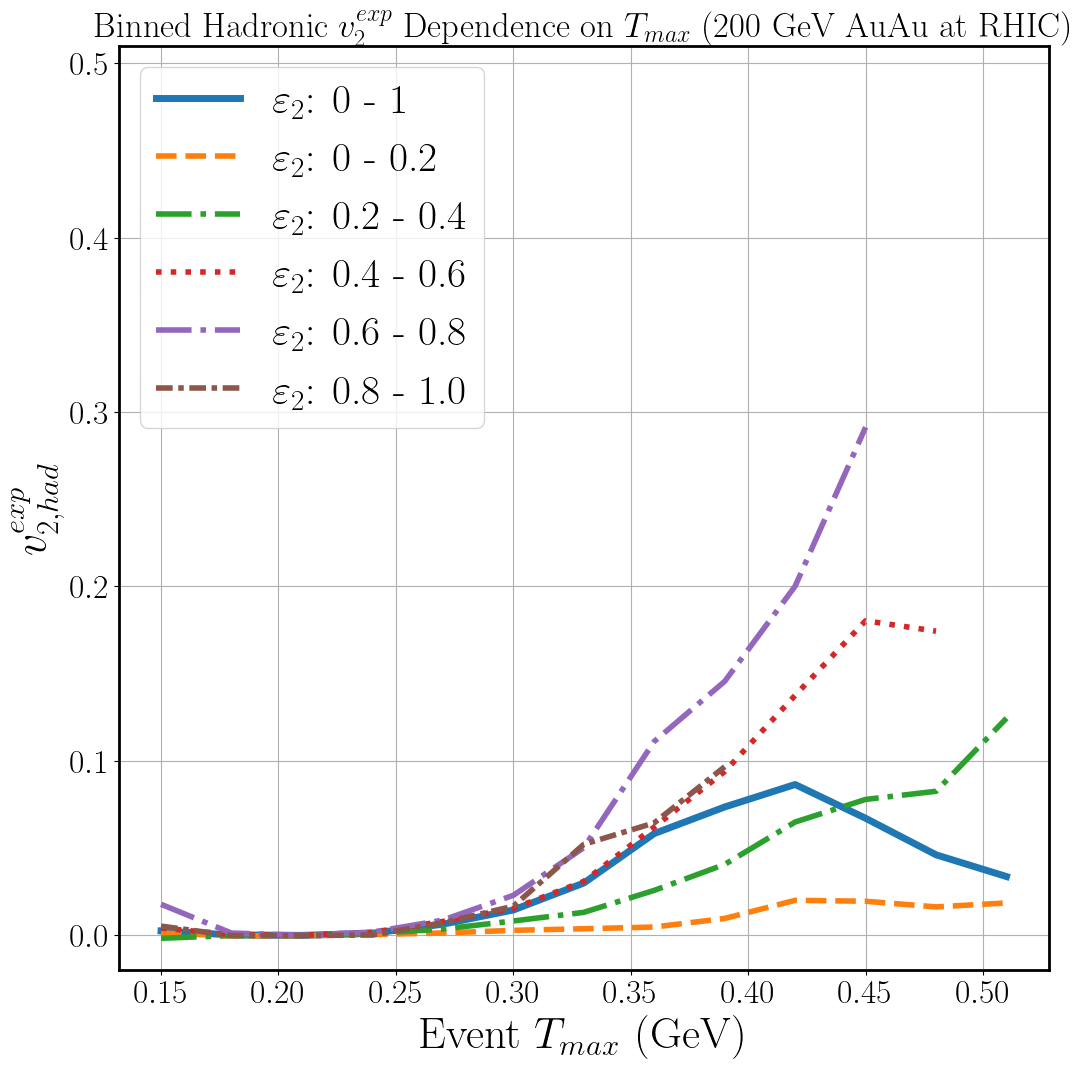}
    \caption{200 GeV AuAu 
    \label{f:auaubinnedv2histTfore2s}
    }
    \end{subfigure}
    \caption{Binned histograms of $v_2$ for 5.02 TeV PbPb (left) and 200 GeV AuAu (right). This figure represents approximately 5000 total events per data set.
    \label{pbpbvsauaubinnedv2hist}
    }
\end{figure}

Next, we move to the multidifferential binning of $v_2^{exp}$ for hadrons as a function of $T_{max}$ and $\varepsilon_2$.  Fig.~\ref{pbpbvsauaubinnedv2hist} compares the binned $v_2^{exp}$ histograms as a function of $T_{max}$ in bins of $\varepsilon_2$ for PbPb collisions (left panel) versus AuAu collisions (right panel).  The qualitative dependence of $\varepsilon_2$ on these variables is the same for both systems.  We see that the $v_{2, had}^{exp}$ is an increasing function of temperature (although not necessarily monotonic), as well as an increasing function of $\varepsilon_2$.  The increase of elliptic flow with ellipticity holds at least up until $\varepsilon_2 \leq 0.8$.  Moreover, the largest absolute $v_{2, had}^{exp}$ comes from events both with large ellipticity $0.6 \leq \varepsilon_2 \leq 0.8$ and the highest possible temperatures $T_{max}$.  At the same $T_{max}$ and the same $\varepsilon_2$, the absolute scale of $v_{2,had}^{exp}$ is \textit{larger} in AuAu collisions than in PbPb collisions. Again, we anticipate that this is due to the effect of the different $p_T$ spectra beween RHIC and the LHC as seen for the acoplanarity in Fig.~\ref{pbpbvsauaubinnedacohiste2vsrrmspt10}.

\begin{figure}[t] 
    \centering
    \begin{subfigure}{.49\textwidth}
    \includegraphics[width=1\textwidth]{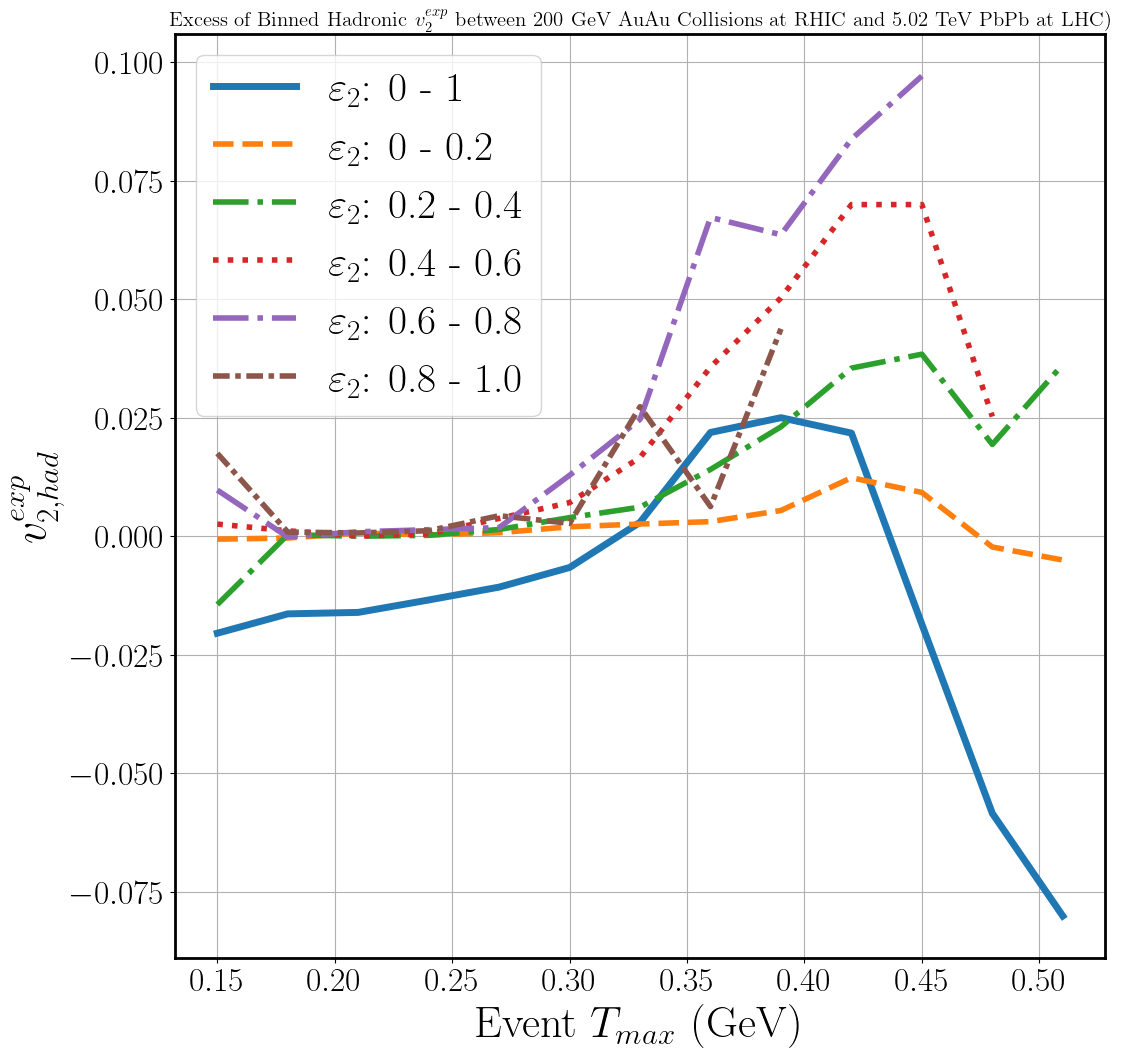}
    \end{subfigure}
    \caption{Excess of $v_2^{exp}$ between 200 GeV AuAu at RHIC and 5.02 TeV PbPb at the LHC for minimum bias (solid blue curve) and in bins of $\varepsilon_2$ (colored dashed and dotted lines).
    \label{f:excessv2histTe2sauauvspbpb}
    }
\end{figure}

Finally, Fig.~\ref{f:excessv2histTe2sauauvspbpb} shows this excess of $v_2$ in AuAu compared to PbPb for minimum bias (solid blue curve) and in bins of $\varepsilon_2$ (colored dashed and dotted lines). Similar to the acoplanarity excess between AuAu and PbPb previously shown in Fig.~\ref{f:excessacohistTe2sauauvspbpb}, we see an excess of $v_2$ in AuAu collisions compared to PbPb collisions for minimum bias only in the intermediate temperature region ($325 \mathrm{MeV} \leq T_{max} \leq 435 \mathrm{MeV}$) for min bias in $\varepsilon_2$. Again, greater differentiability can be achieved by examining the excess $v_2$ across different $\varepsilon_2$-bins. For all $\varepsilon_2$-bins, we observe no excess of $v_2$ in the low temperature region (universality at $T_{max} < 250$ MeV). For $T_{max} \geq 250$ MeV, the ordering or hierarchy in $\varepsilon_2$-bins emerges, where we see higher $v_2$ excess for higher eccentricity bins.




%
\subsection{Results: UU vs. AuAu}
%

Next, we move to comparing UU collisions with AuAu collisions at RHIC. In this case, both systems have nearly the same energy ($\sqrt{s} = 200$ GeV versus 193 GeV), but uranium is a deformed nucleus, ($\beta_2 = 0.280, \beta_4 = 0.093$).  (See for instance Ref.~\cite{Wertepny:2019yye} and references therein.)  The comparable energies but meaningfully different geometries between UU and AuAu collisions make for a complementary comparison to the analysis of PbPb versus AuAu collisions.

%
\subsubsection{System Comparison}
%

\begin{figure}[ht]
\begin{centering}
\includegraphics[width=0.49\textwidth]{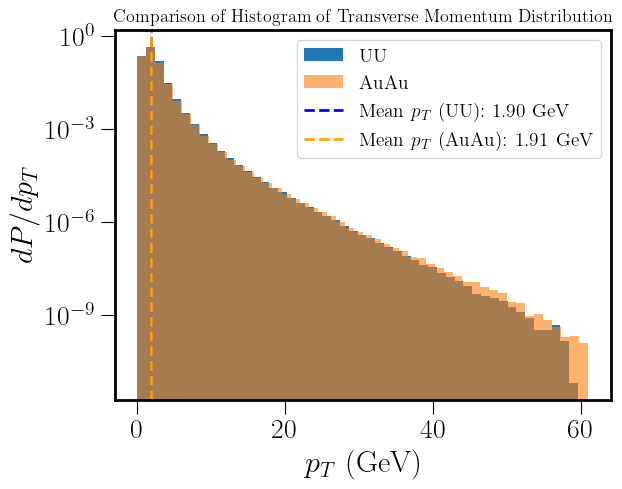}
\includegraphics[width=0.49\textwidth] {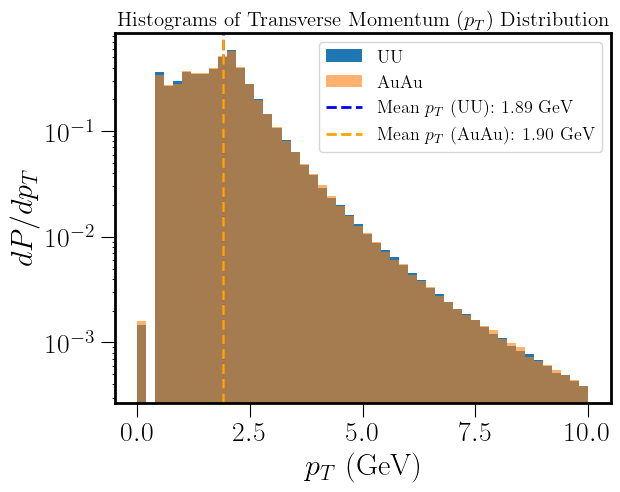}
\caption{Comparison of the histograms of the transverse momentum $p_T$ distribution of AuAu and UU collisions for $0 \leq p_T \leq 100 \mathrm{GeV}$ (left) and $0 \leq p_T \leq 10 \mathrm{GeV}$ (right).  
\label{f:uuauauhistpT}
}
\end{centering}
\end{figure}

\begin{figure}[ht] 
\centering
    \begin{subfigure}{.49\textwidth}
    \includegraphics[width=1\textwidth]{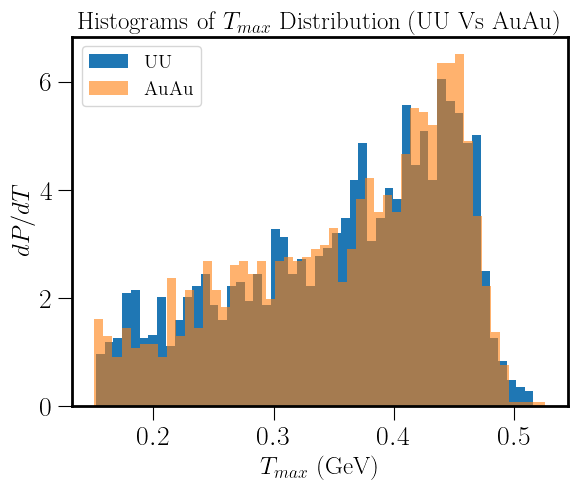}
    \end{subfigure}
    \begin{subfigure}{.50\textwidth}
    \centering
    \includegraphics[width=1\textwidth]        {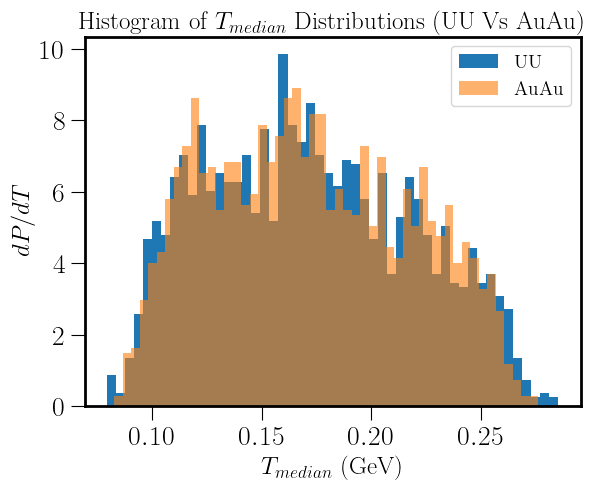}
    \end{subfigure}
\caption{Histograms of the maximum temperature $T_{max}$ (left) and median temperature $T_{med}$ (right) of the initial conditions of UU and AuAu collisions. This figure represents approximately 2500 events per data set.
\label{f:auaunuuhistsT}
}
\end{figure}

\begin{figure}[ht] 
    \centering
    \begin{subfigure}{.55\textwidth}
    \includegraphics[width=1\textwidth]{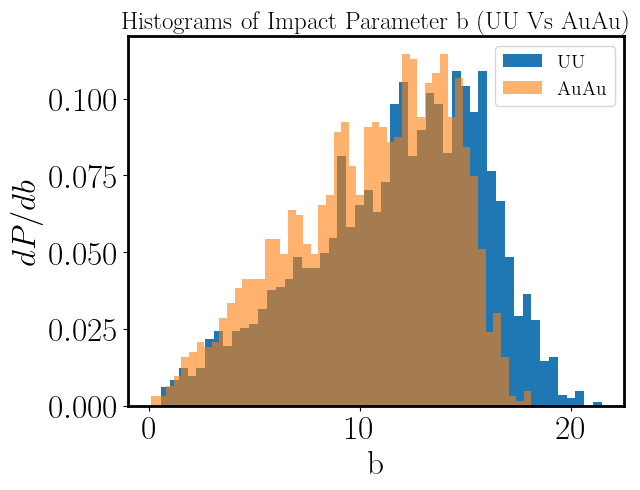}
    \end{subfigure}
    \begin{subfigure}{.44\textwidth}
    \centering
    \includegraphics[width=1\textwidth]            {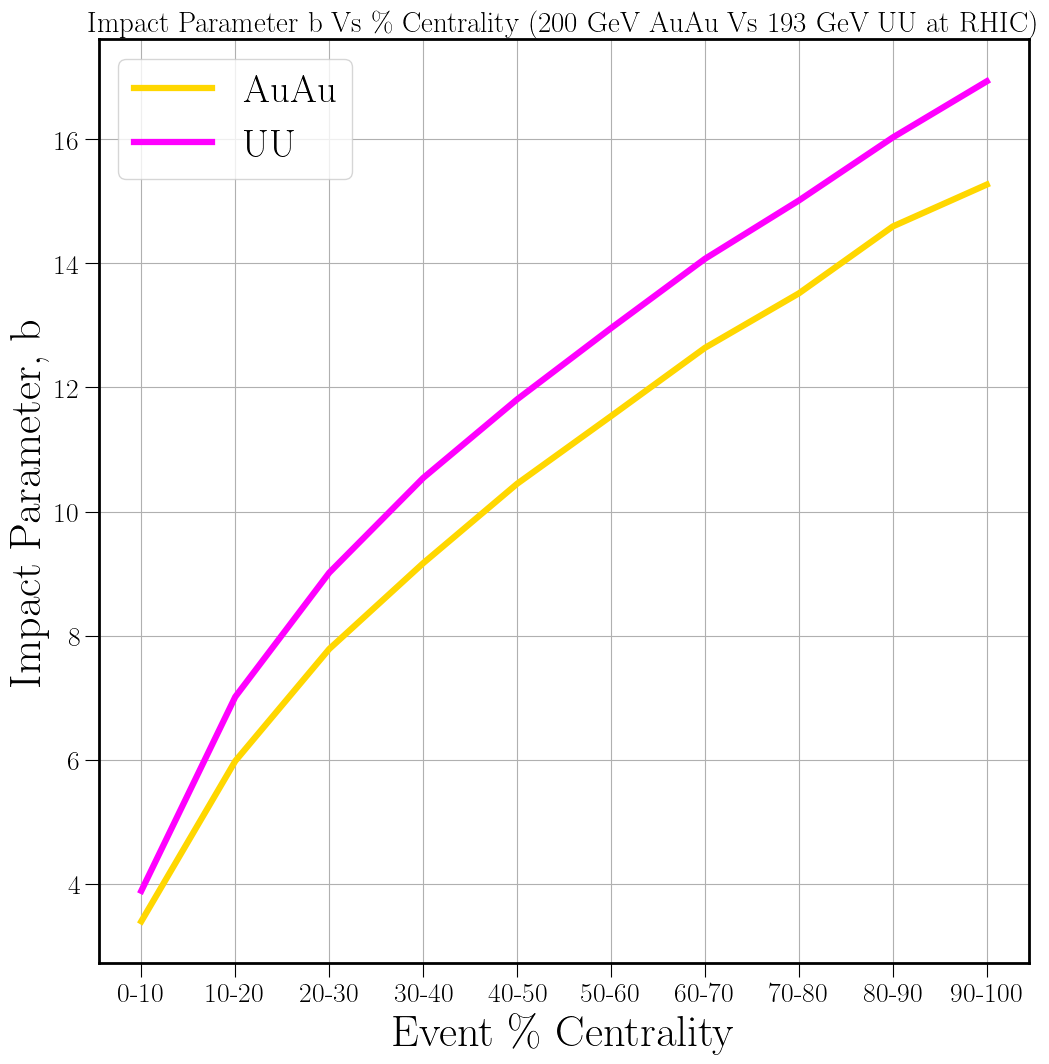}
    \end{subfigure}
    \caption{Impact parameter $b$ histograms (left) and as a function of centrality (right) for 193 GeV UU collisions and 200 GeV AuAu collisions at RHIC.  
    \label{f:uuhistsbne2}
    }
\end{figure}
%



Fig.~\ref{f:uuauauhistpT} confirms that UU and AuAu have similar $p_T$ spectra, with a tiny excess in the $p_T$ spectrum of AuAu at the highest $p_T$ due to its slightly higher energy. The mean $p_T$ values, $\langle p_T \rangle_\mathrm{AuAu} = 1.911$ GeV versus $\langle p_T \rangle_\mathrm{UU} = 1.903$ GeV, are equal to within $0.4\%$ for minimum bias.  If we impose a momentum cut $p_T \leq 10$ GeV, the new averages are $\langle p_T \rangle_\mathrm{AuAu} = 1.902$ GeV, and $\langle p_T \rangle_\mathrm{UU} = 1.894$ GeV respectively, an excess of only 3.9\% for AuAu over UU. Likewise, Fig.~\ref{f:auaunuuhistsT} confirms our prediction that UU and AuAu have similar temperature profiles. There are no significant changes in the distributions of either $T_{max}$ or $T_{med}$ histograms due to the small change in $\sqrt{s}$, or to the distribution of the ratio $T_{max} / T_{med}$ shown in Fig.~\ref{f:uuhistsbne2}, which measures the relative importance of a hot spot compared to the background.

At least in minimum bias collisions, UU and AuAu also have similar geometry profiles, as shown in Fig.~\ref{f:uuhistsbne2}.  We see clearly the effect of the slightly larger radius in UU than in AuAu \cite{PHENIX:2015tbb}. From AuAu to UU, the parameters are $R = 6.81$ fm versus $R = 6.38$ fm (6.7\% larger) for the radius and $a = 0.535$ fm versus $a = 0.600$ fm. The slightly larger radius of uranium also leads to a slight increase in the typical impact parameter as a function of centrality, shown in the right panel of Fig.~\ref{f:uuhistsbne2}.

\begin{figure}[t!] 
    \centering
    \begin{subfigure}{.5\textwidth}
    \includegraphics[width=1\textwidth]
    {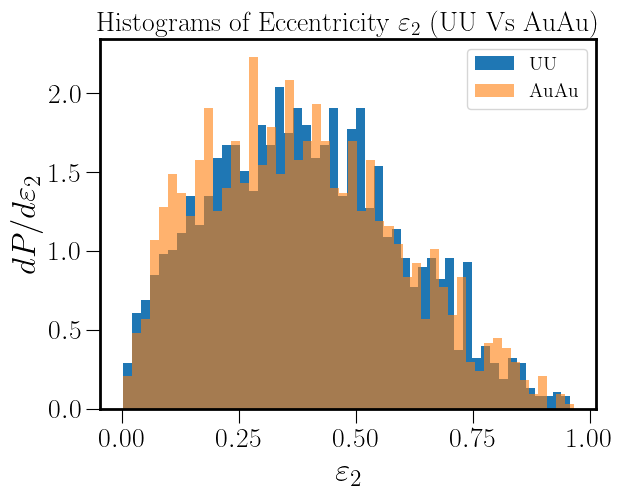}
    \end{subfigure}
    \\
    \begin{subfigure}{.50\textwidth}
    \centering
    \includegraphics[width=1\textwidth]        {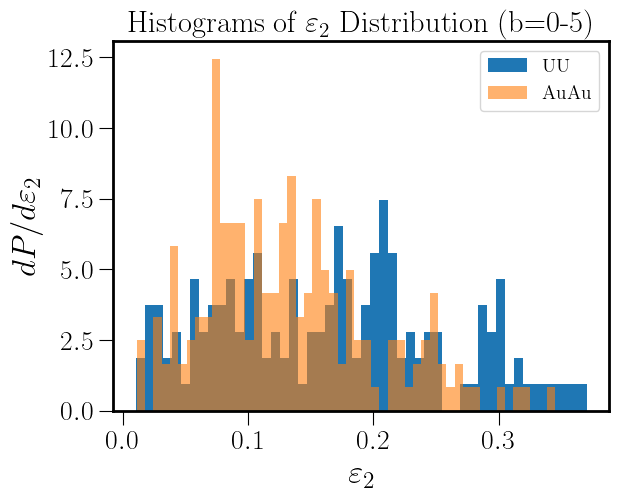}
    \end{subfigure}
    \begin{subfigure}{.48\textwidth}
    \centering
    \includegraphics[width=1\textwidth]
    {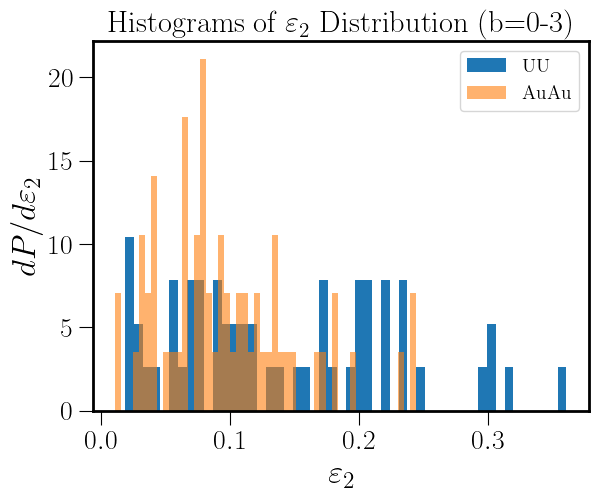}
    \end{subfigure}
    \caption{Histograms of the eccentricity $\varepsilon_2$ of 193 GeV UU collisions and 200 GeV AuAu collisions at RHIC with all impact parameter b (top), $b = 0-5$ (bottom left), and $b = 0-3$ (bottom right).  
    \label{f:uuvsauaubnacoenhfncofcent}
}
\end{figure}

Although UU is deformed \cite{Wertepny:2019yye}, whereas AuAu is not, we don't see an effect of the deformation in a minimum bias dataset.  At minimum bias, the large impact parameter dominates the geometry and the deformation plays little role, as shown in the left panel of Fig.~\ref{f:uuvsauaubnacoenhfncofcent}. But if we zoom in to central and ultracentral collisions $b \rightarrow 0$, then the impact parameter does not play a role, and one can clearly see the effect of the deformation: UU achieves larger $\varepsilon_2$ than AuAu in these bins, as shown in the bottom panels of Fig.~\ref{f:uuvsauaubnacoenhfncofcent}. This lets us use central UU vs AuAu collisions to test the effect of different $\varepsilon_2$ with the same other medium properties: the more tightly we cut on $b \rightarrow 0$, the more we emphasize the role of the deformation in UU collisions shown for $b = 0-5$ (bottom left), and $b = 0-3$ (bottom right).

%
\subsubsection{Jet Observables: Acoplanarity}
%



\begin{figure}[h!] 
\centering
    \begin{subfigure}{.49\textwidth}
        \centering
        \includegraphics[width=\textwidth]{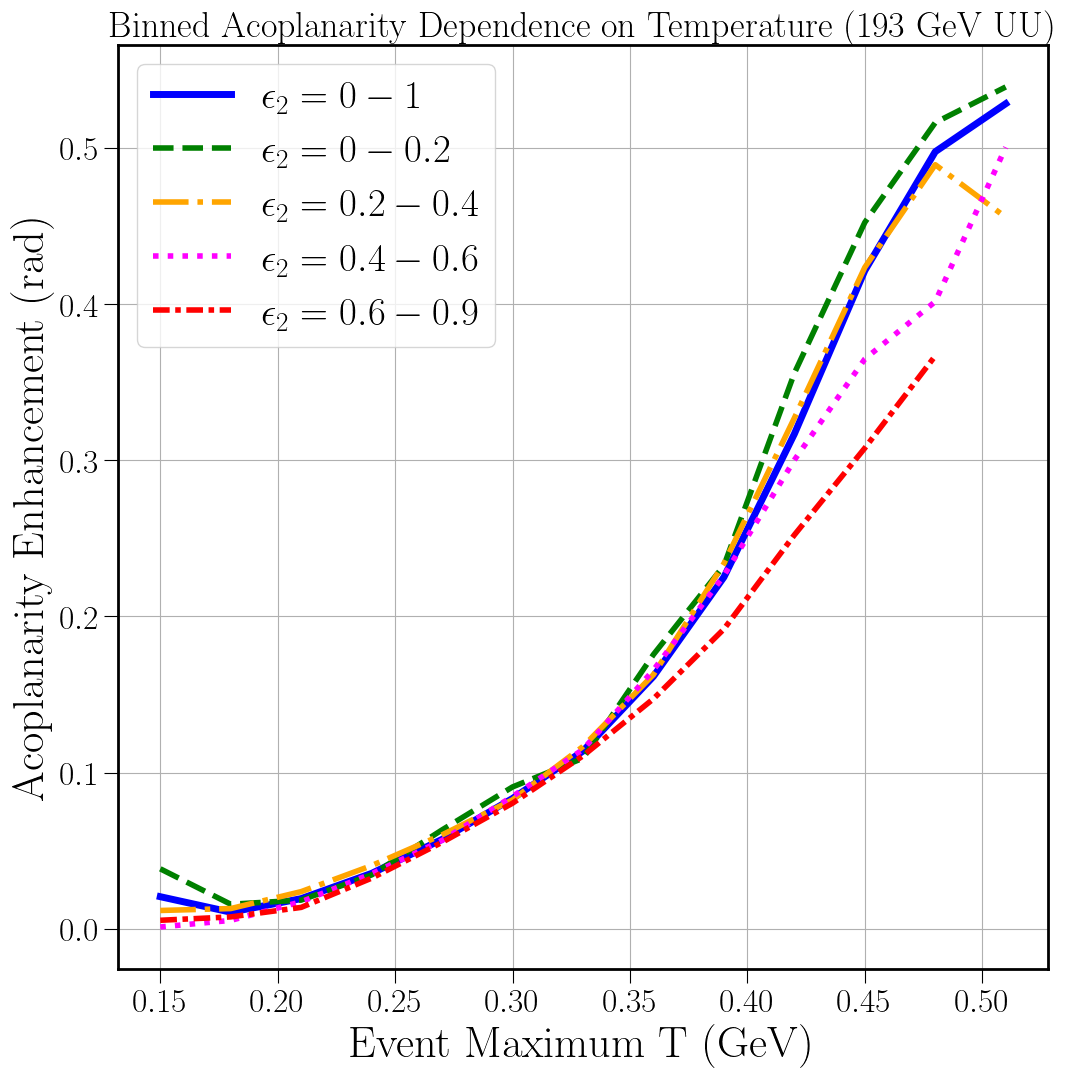}
    \end{subfigure}
    \hfill
    \begin{subfigure}{.49\textwidth}
        \centering
        \includegraphics[width=\textwidth]{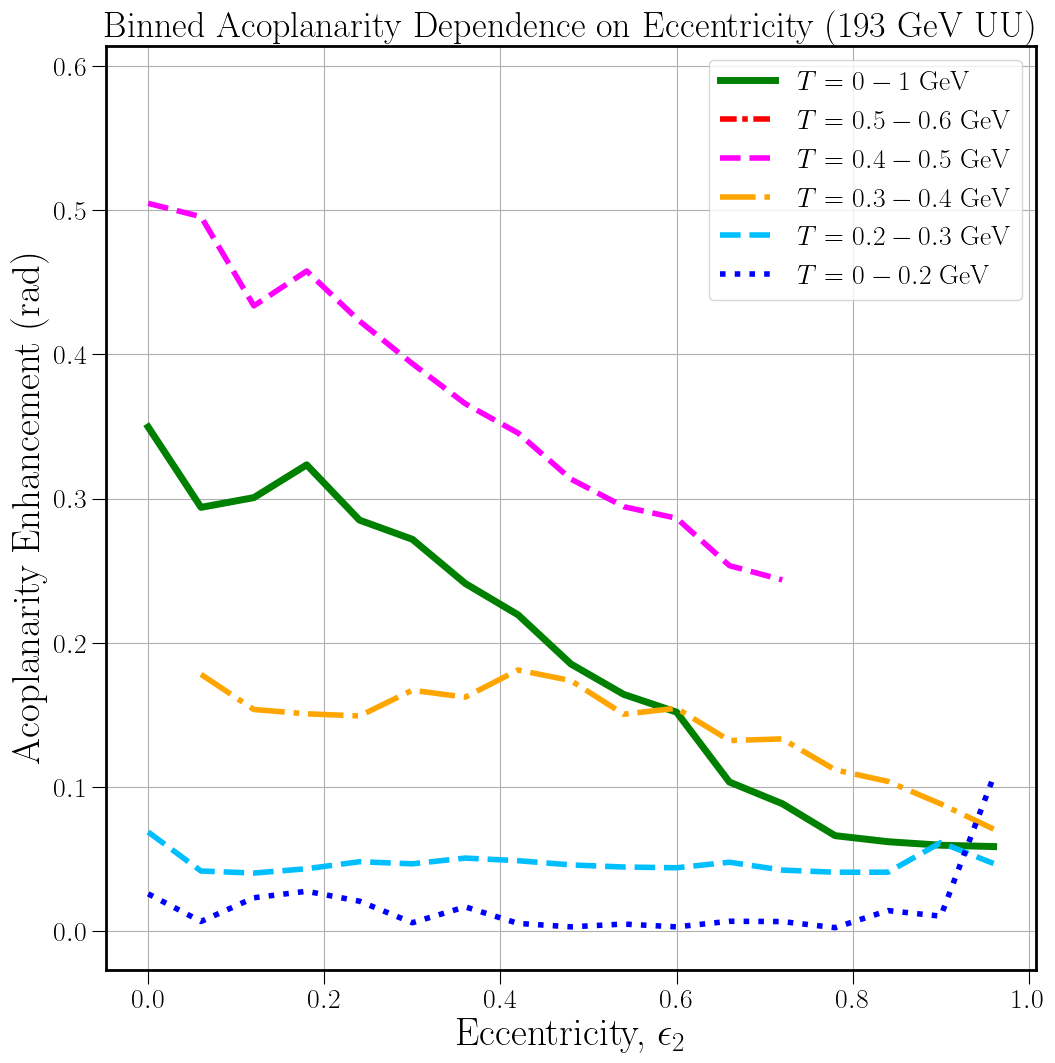}
    \end{subfigure}

    \vspace{1em} 

    \begin{subfigure}{.49\textwidth}
        \centering
        \includegraphics[width=\textwidth]{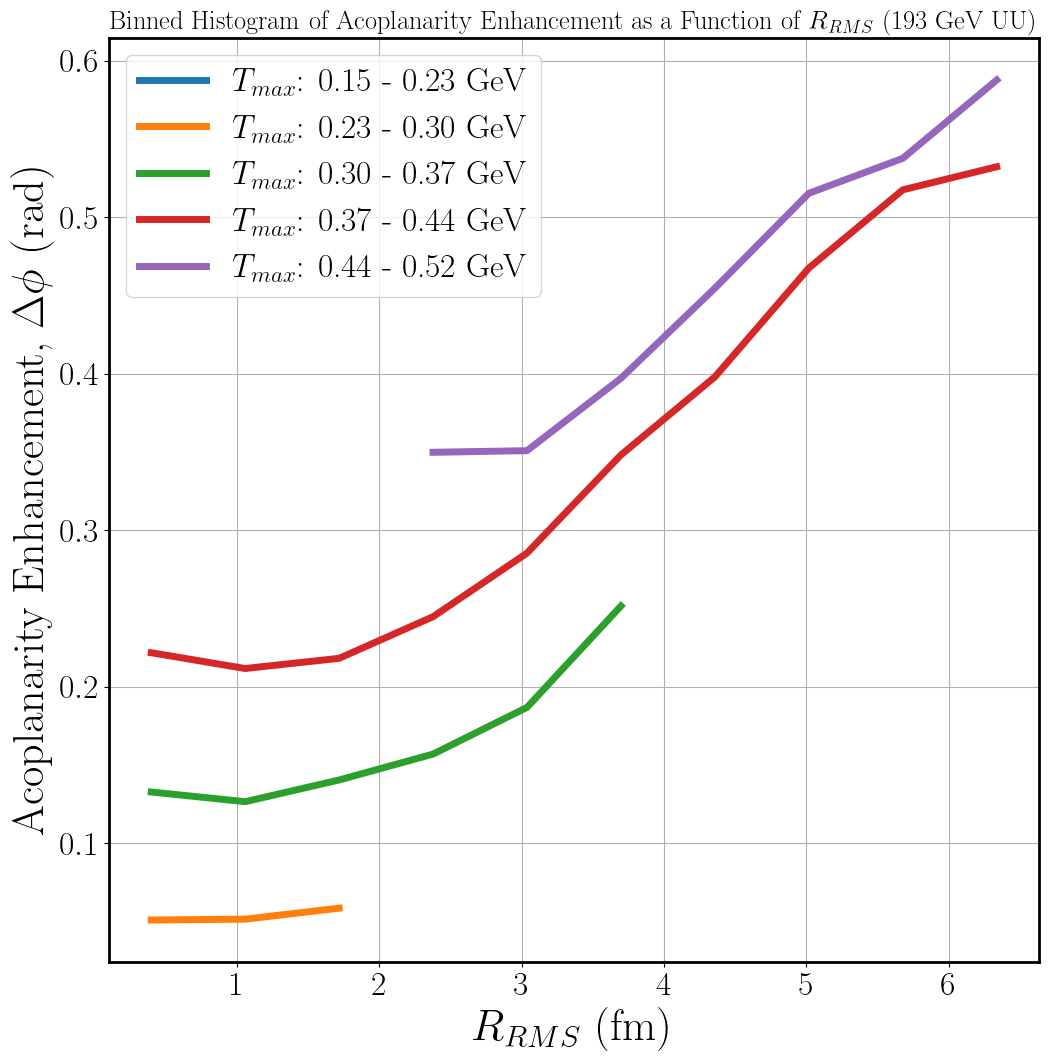}
    \end{subfigure}
    \hfill
    \begin{subfigure}{.49\textwidth}
        \centering
        \includegraphics[width=\textwidth]{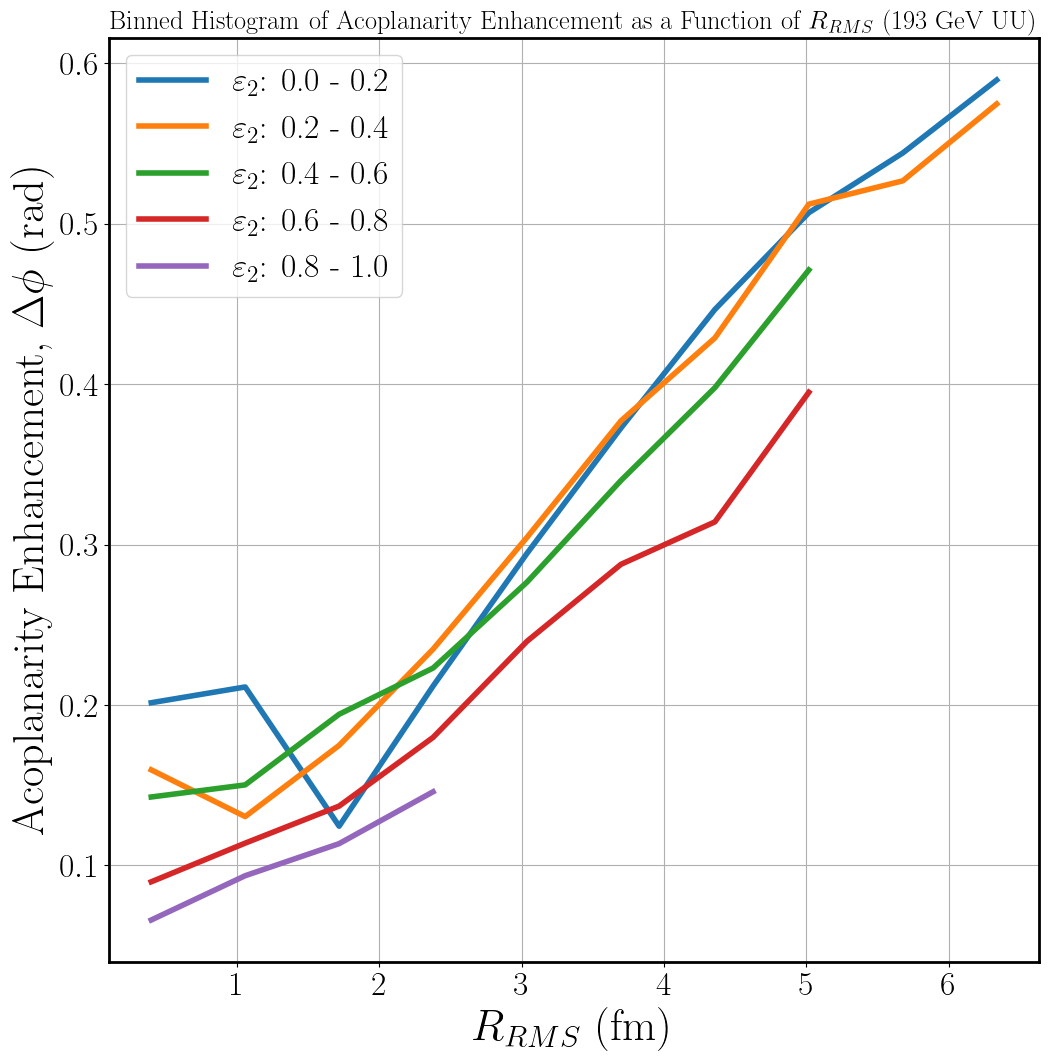}
    \end{subfigure}
    
    \caption{Comparison of binned histograms of acoplanarity enhancement of 193 GeV UU as a function of $T_{max}$ (top left) and $\epsilon_2$ (top right), as a function of $R_{RMS}$ in bins of $T_{max}$ (bottom left) and $\varepsilon_2$ (bottom right). This figure represents approximately 2500 events per data set.} 
    \label{f:uubinnedacohistTne2}
\end{figure}

Fig.~\ref{f:uubinnedacohistTne2} shows that, as with PbPb and AuAu, the acoplanarity in UU collisions is an increasing function of $T_{max}$ (top left) and a decreasing function of $\varepsilon_2$ (top right). We also plotted the dependence on the estimated path length $R_{RMS}$ in Fig.~\ref{f:uubinnedacohistTne2} in bins of $T_{max}$ (bottom left) and in bins of $\varepsilon_2$ (bottom right), but we do not expect this estimate to be as accurate here (as in symmetric systems like PbPb and AuAu) because of the deformation of uranium.  Nevertheless, APE in principle tracks certain path length spent in the QGP phase, the HRG phase, and in the late time ``un-hydrodynamic'' phase. It will be interesting to study the dependence on these detailed quantities in the future, rather than the simple estimate $R_{RMS}$. We see that there is a similar magnitude of acoplanarity enhancement in both UU and AuAu collisions at a given $T_{max}$ and $\varepsilon_2$, but UU can achieve a simultaneous combination of both higher $T_{max}$ and higher $\varepsilon_2$ than AuAu because of the deformation.

\begin{figure}[t] 
    \centering
    \begin{subfigure}{.49\textwidth}
    \includegraphics[width=1\textwidth]{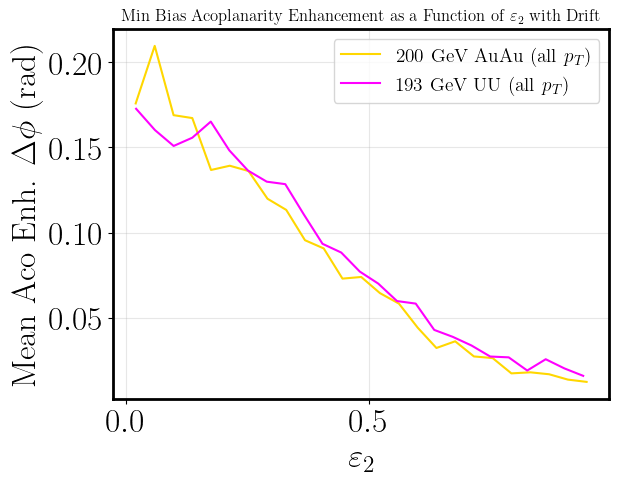}
    \end{subfigure}
    \begin{subfigure}{.49\textwidth}
    \centering
    \includegraphics[width=1\textwidth]        {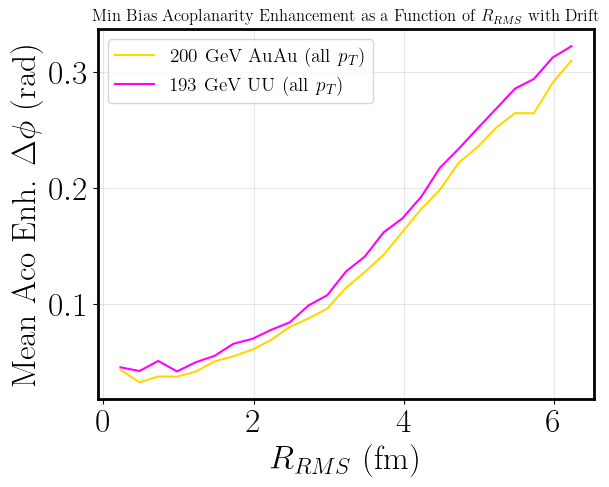}
    \end{subfigure}
    \\
    \begin{subfigure}{.49\textwidth}
    \includegraphics[width=1\textwidth]{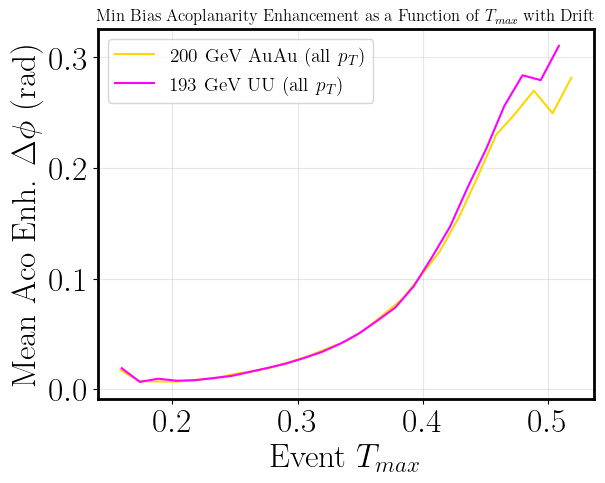}
    \end{subfigure}
    \caption{Mean acoplanarity as a function of $\varepsilon_2$ (top left), estimated $R_{RMS}$ (top right), and event $T_{max}$ (bottom) for 193 GeV UU collisions and 200 GeV AuAu collisions at RHIC.}
    \label{uuvsauaubinnedhiste2vsrrms}
\end{figure}

As a function of all three variables $\varepsilon_2$, estimated $R_{RMS}$, and $T_{max}$ shown in the top left, top right, and bottom panels of Fig.~\ref{uuvsauaubinnedhiste2vsrrms}, respectively, the acoplanarity in UU collisions is nearly the same as in AuAu collisions.  At fixed $\varepsilon_2$, UU collisions have a slightly larger path length and a slightly lower $p_T$ than AuAu collisions; both contribute to a small excess of acoplanarity in UU compared to AuAu.  Meanwhile their temperatures are about the same.  At fixed $R_{RMS}$, AuAu collisions have slightly higher $p_T$, leading to a slight reduction in acoplanarity compared to UU. However, at fixed $T_{max}$, the agreement is even more precise, despite the fact that UU still has a slightly larger radius and slightly lower $p_T$.

\begin{figure}[t] 
    \centering
    \begin{subfigure}{.49\textwidth}
    \includegraphics[width=1\textwidth]{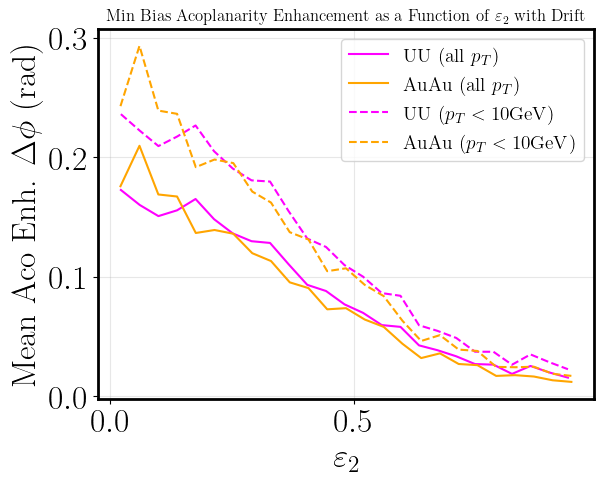}
    \end{subfigure}
    \begin{subfigure}{.49\textwidth}
    \centering
    \includegraphics[width=1\textwidth]        {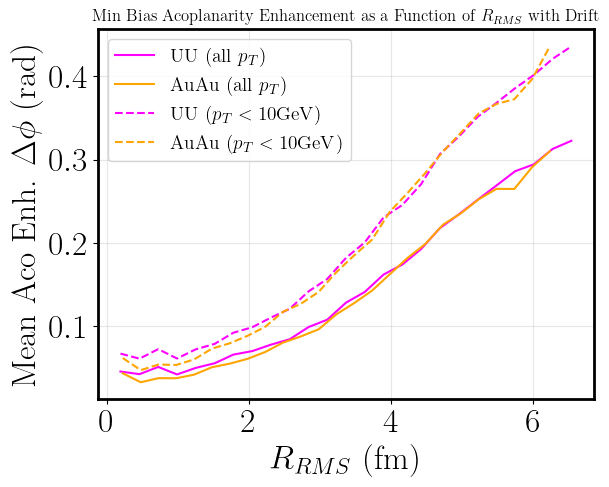}
    \end{subfigure}
    \\
    \begin{subfigure}{.49\textwidth}
    \includegraphics[width=1\textwidth]{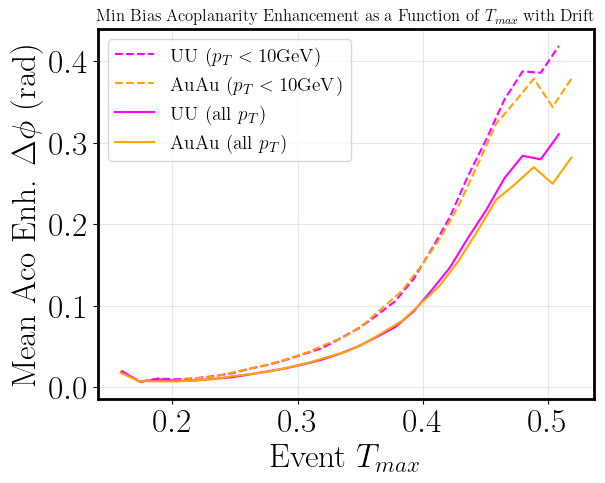}
    \end{subfigure}
    \caption{Mean acoplanarity as a function of $\varepsilon_2$ (top left), estimated $R_{RMS}$ (top right), and event $T_{max}$ for all $p_T$ (solid curves) vs $p_T < 10 \mathrm{GeV}$ (dashed curves) for 193 GeV UU collisions and 200 GeV AuAu collisions at RHIC.  
    \label{uuvsauauacohiste2vsrrmspt10}
    }
\end{figure}

In Fig.~\ref{uuvsauauacohiste2vsrrmspt10} we compare the magnitudes of the acoplanarity enhancement in AuAu and UU collisions, with and without a momentum cut $p_T \leq 10$ GeV.  After imposing the $p_T$ cut, the mean acoplanarity for both systems goes up, in accordance with sub-eikonal scaling of drift. Cutting on $p_T$ increases the overall magnitude of the effect but does not change the agreement or margin of separation between the curves. This would suggest that the differences in the $p_T$ spectra are not significant and that the small excess of UU is due to other factors which will be interesting to explore further.

\begin{figure}[htbp] 
    \centering
    \begin{subfigure}{.49\textwidth}
    \includegraphics[width=1\textwidth]{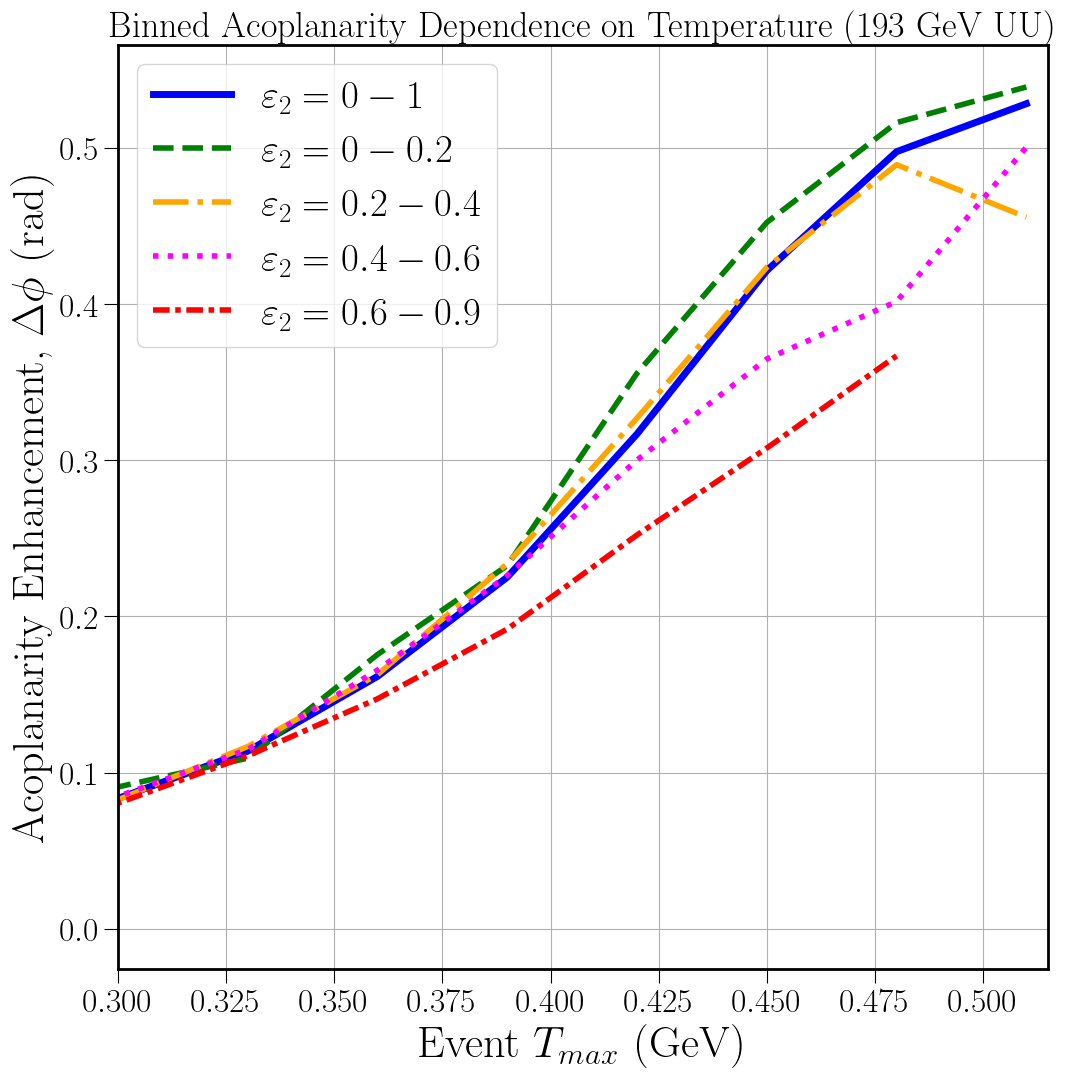}
    \end{subfigure}
    \begin{subfigure}{.50\textwidth}
    \centering
    \includegraphics[width=1\textwidth]        {images/aco_enh_weighted_funcT_pT_10GeV_ecc2_all_200GeV_AuAu_v1.3.1_g_rad=2.2_g_col=2.0_zoom.png}
    \end{subfigure}
    \caption{Mean acoplanarity enhancement for 193 GeV UU (left) and 200 GeV AuAu (right) collisions as a function of $T_{max}$, in bins of ellipticity $\varepsilon_2$. This figure represents approximately 2500 total events per data set.}
    \label{uuvsauaubinnedhistzoomin}
\end{figure}
\begin{figure}[h!]
    \centering
    \begin{subfigure}{.49\textwidth}
    \centering
    \includegraphics[width=1\textwidth]
    {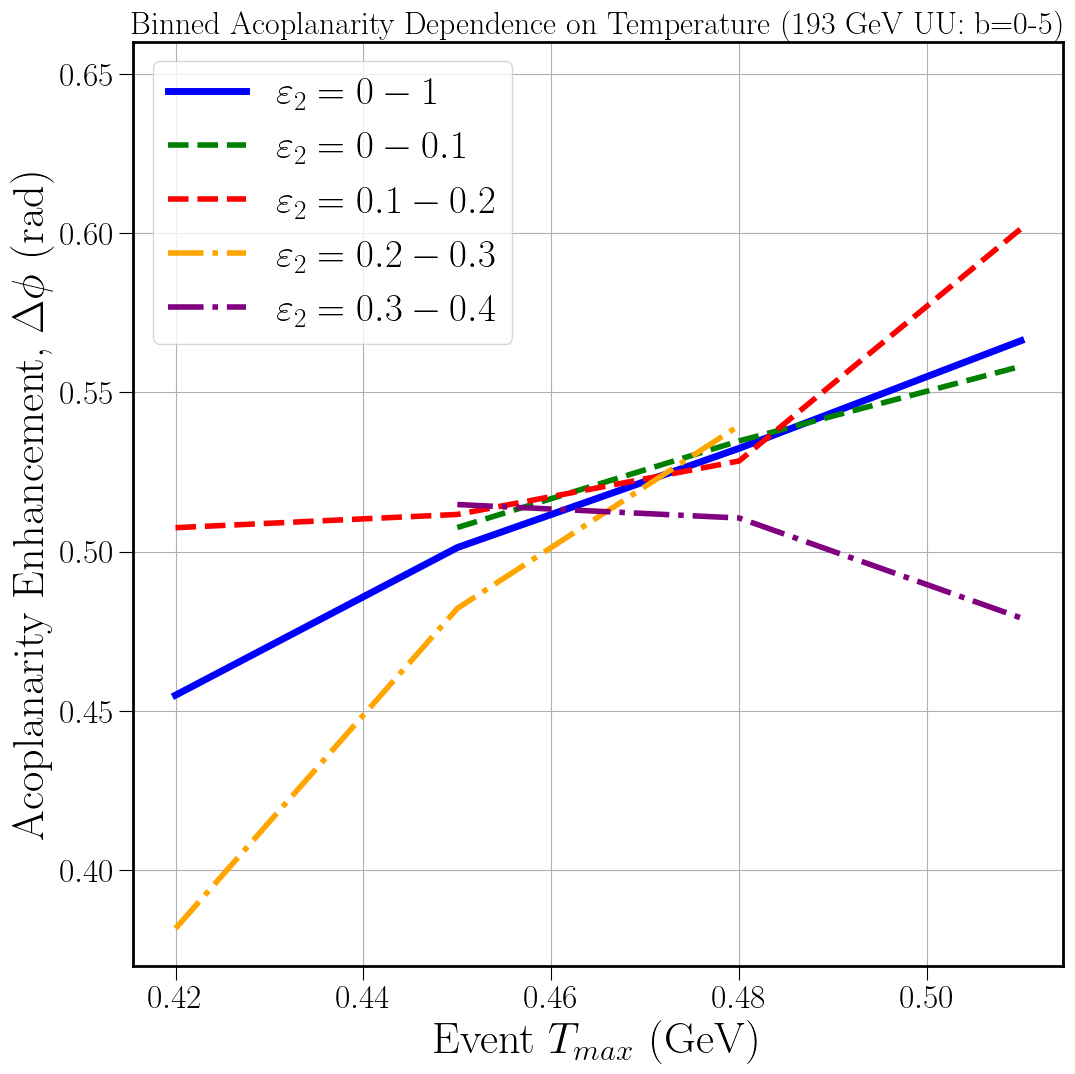}
    \end{subfigure}
    \begin{subfigure}{.49\textwidth}
    \centering
    \includegraphics[width=1\textwidth]        {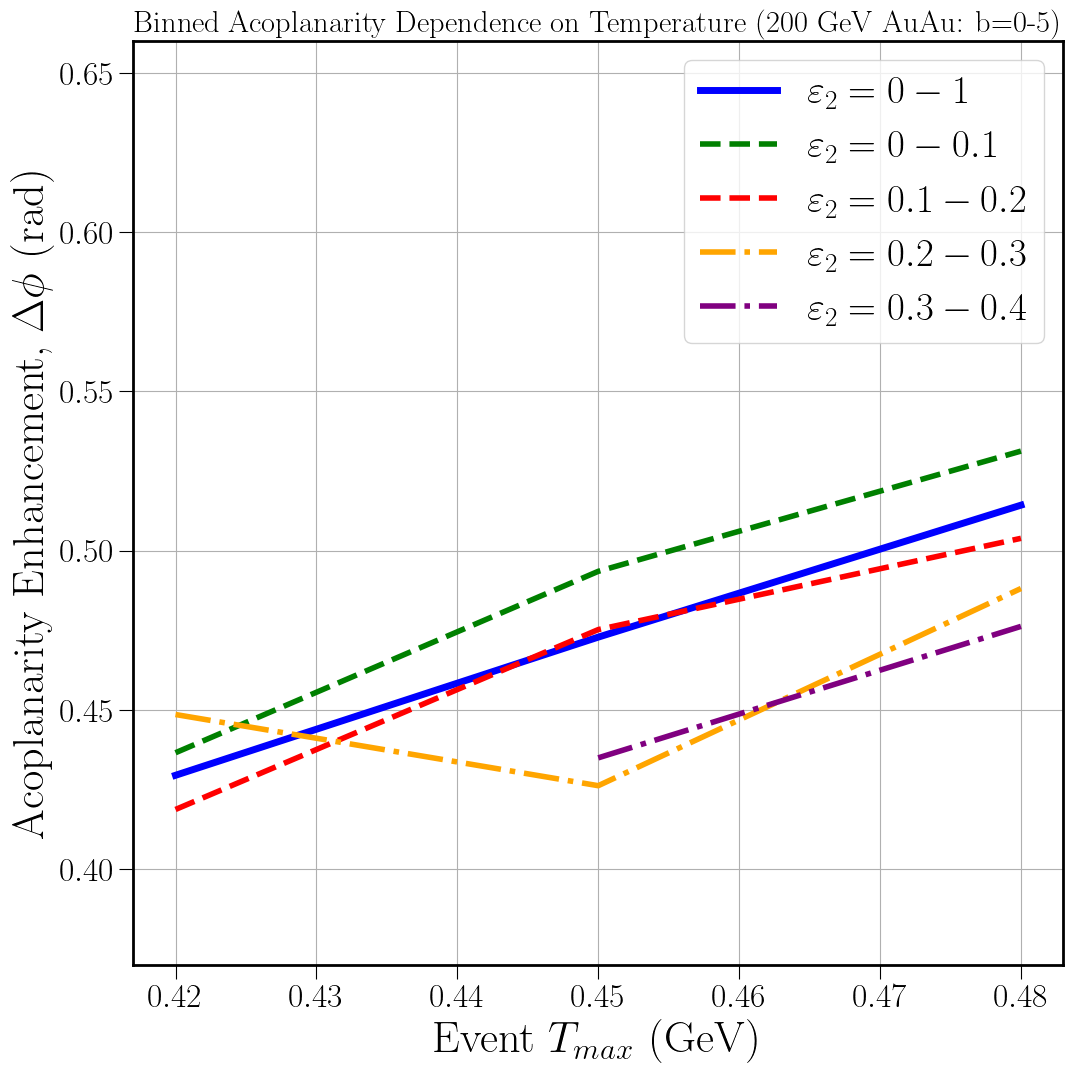}
    \end{subfigure}
    \\
    \begin{subfigure}{.49\textwidth}
    \centering
    \includegraphics[width=1\textwidth]
    {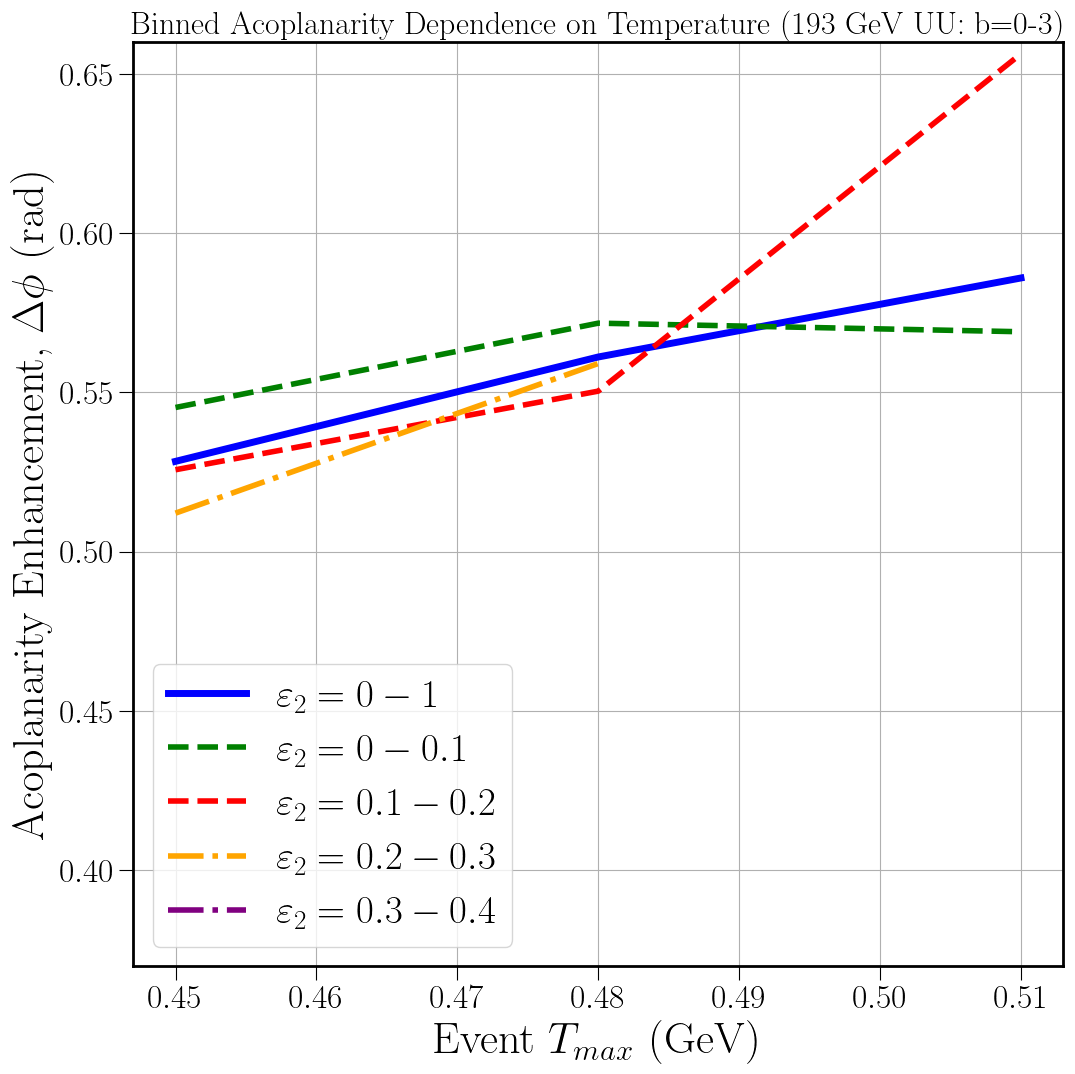}
    \end{subfigure}
    \begin{subfigure}{.49\textwidth}
    \centering
    \includegraphics[width=1\textwidth]        {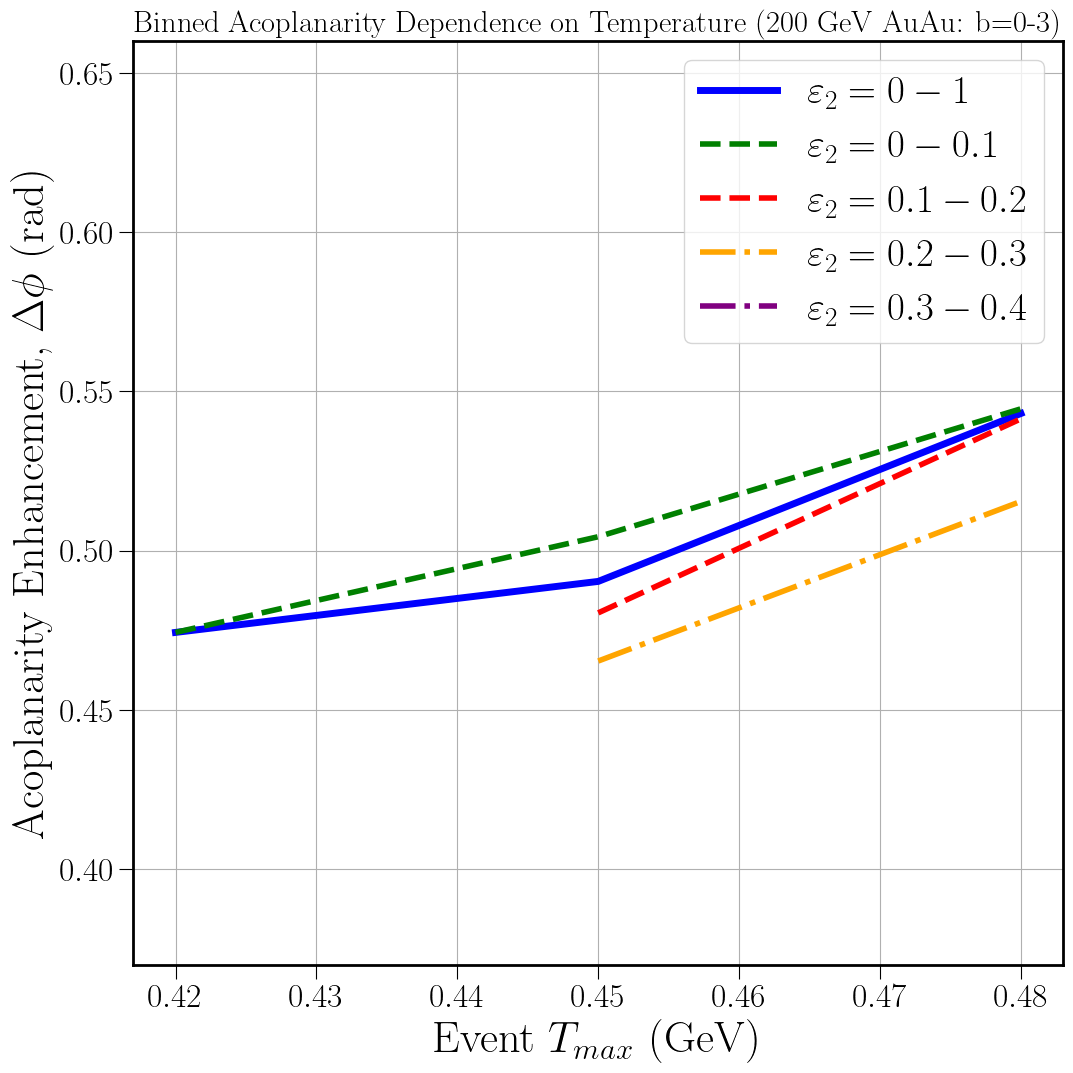}
    \end{subfigure}
    \caption{
    Histogram of acoplanarity enhancement of 193 GeV UU (left) and 200 GeV AuAu (right) at RHIC as a function of $T_{max}$ for different bins of eccentricities with impact parameter cuts  $b \leq 5$ fm (top) and $b \leq 3$ fm (bottom).  
    \label{fig:binnedacoUUvsAuAuhistTb0to5n3}
    }
\end{figure}
As shown in Fig.~\ref{uuvsauaubinnedhistzoomin}, zooming in on the high-$T$ region reveals a significant path length correlation ($\varepsilon_2$ anticorrelation), with $T_{max} \geq 300$ MeV for UU and $T_{max} \geq 350$ MeV for AuAu. Due to nuclear deformation, bins of higher $\varepsilon_2$ extend to larger $T_{max}$ in UU than in AuAu, demonstrating that UU events can simultaneously exhibit large $\varepsilon_2$ and large $T_{max}$. Specifically, the most eccentric events ($0.6 \leq \varepsilon_2 \leq 0.9$) persist up to 480~MeV in UU compared to 450~MeV in AuAu, increasing the maximum acoplanarity from $\sim 0.25$ to $\sim 0.37$. Similarly, the $0.4 \leq \varepsilon_2 \leq 0.6$ bin expands its reach from 480~MeV in AuAu to 520~MeV in UU. Interestingly, while the $0.2 \leq \varepsilon_2 \leq 0.4$ bin remains unchanged at 520~MeV in both systems, the least eccentric events ($0 \leq \varepsilon_2 \leq 0.2$) actually see a decrease in $T_{max}$ reach from 540~MeV in AuAu to 520~MeV in UU, which could be the result of the slightly lower $\sqrt{s}$ for UU collisions.

Furthermore, there is a systematic change in the slopes of the acoplanarity versus $T_{max}$. At lower temperatures ($T_{max} \leq 375$ MeV), the acoplanarity increases roughly linearly with $T_{max}$ in both systems, with a slope of order $\sim 10^{-3}$ rad / MeV across most centralities. However, in AuAu, the highest temperatures in each $\varepsilon_2$ bin trigger a drastic decrease in this slope; for example, in the most eccentric bin, the acoplanarity slope decreases by an order of magnitude to $\sim 10^{-4}$ rad / MeV above 380~MeV, with similar downturns occurring in the other $\varepsilon_2$ bins above 450~MeV. In contrast, the acoplanarity in UU maintains its roughly linear trend with the same original slope of order $\sim 10^{-3}$ rad / MeV over a much higher extent in $T_{max}$. This suggests that the flattening of the acoplanarity slope in AuAu is a consequence of the strong mean field elliptical geometry breaking down in the most central AuAu collisions where $b \rightarrow 0$. Because of the deformation in UU collisions, the ellipticity $\varepsilon_2$ remains substantial even in central collisions as $b \rightarrow 0$, a difference that can be systematically tested by imposing cuts on the impact parameter $b$.

At a given $T_{max}$ and $\varepsilon_2$, the acoplanarity in UU is significantly larger than in AuAu as shown in Fig.~\ref{fig:binnedacoUUvsAuAuhistTb0to5n3}. As the $b$ cut is tightened, the acoplanarities generally increase across all systems. The minimum bias slopes in UU collisions are of order $\sim 5 \times 10^{-4}$~rad / MeV for both choices of $b_{cut}$, while the slopes in AuAu collisions are of order $10^{-3}$~rad / MeV for both cuts. Notably, the $0.1 \leq \varepsilon_2 \leq 0.2$ bin in UU collisions shows a distinctive upturn in the acoplanarity at the highest temperatures, while other bins show the downturn characteristic of central AuAu collisions. The magnitude of these effects shows the greatest sensitivity to the $b_{cut}$ parameter and highlights the enhancement in UU collisions due to the deformation. While a definitive statement about the correlation with the slopes cannot be made with this dataset alone, a dedicated high-statistics dataset from central collisions could clarify the dependence of the slope on event parameters in the future.

%
\subsubsection{Jet Observables: Elliptic Flow}
%


\begin{figure}[t] 
    \centering
    \begin{subfigure}{.48\textwidth}
    \includegraphics[width=1\textwidth]{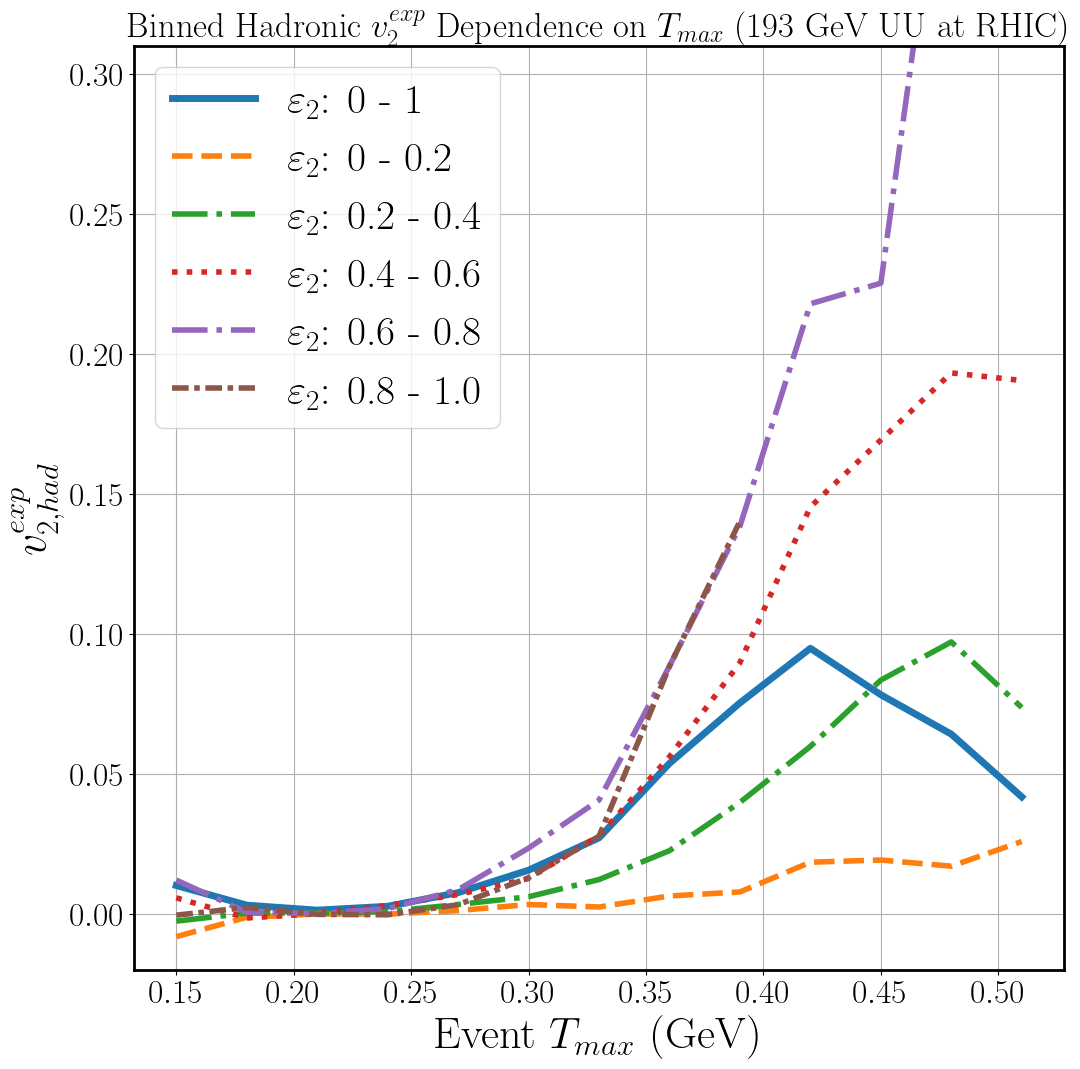}
    \caption{193 GeV UU
    \label{f:uubinnedv2histTfore2s}
    }
    \end{subfigure}
    \begin{subfigure}{.49\textwidth}
    \centering
    \includegraphics[width=1\textwidth]        {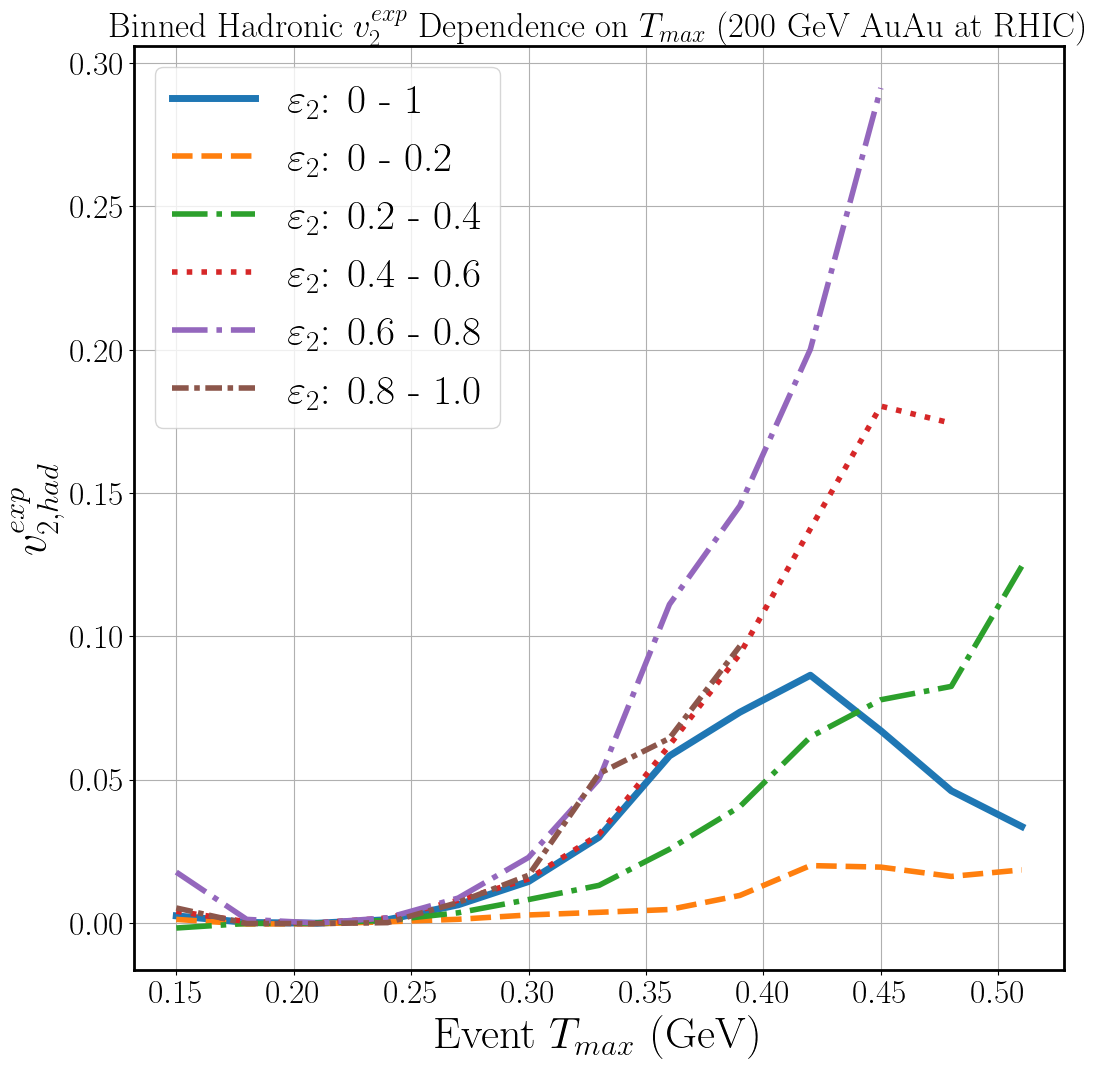}
    \caption{200 GeV AuAu 
    \label{f:auaubinnedv2histTfore2s2}
    }
    \end{subfigure}
    \caption{Binned histograms of $v_2$ for 193 GeV UU (left) and 200 GeV AuAu (right). This figure represents approximately 2500 events per data set. 
    }
    \label{uuvsauaubinnedv2hist}
\end{figure}

As illustrated in Fig.~\ref{uuvsauaubinnedv2hist}, the qualitative features of $v_2^{exp}$ as a function of $T_{max}$ in bins of $\varepsilon_2$ are similar to those in AuAu collisions; specifically, $v_2^{exp}$ is an increasing function of both $T_{max}$ and $\varepsilon_2$, with large $\varepsilon_2$ determining the slope of the correlation with $T_{max}$. As previously noted, the deformation causes bins of $\varepsilon_2$ to extend to larger $T_{max}$ in UU than in AuAu; for instance, the $0.4 \leq \varepsilon_2 \leq 0.6$ bin reaches $T_{max} = 480$~MeV in AuAu but extends up to $T_{max} = 510$~MeV in UU. At identical $T_{max}$ and $\varepsilon_2$ values, UU and AuAu yield comparable effects on $v_2^{exp}$.  One could further examine whether the deformation has a more significant impact in central collisions by again imposing cuts on small $b$.

\newpage

%
\subsection{Conclusions and Outlook}
%

In this chapter, we have performed a comprehensive analysis of the initial geometry and local temperature which drive the correlations observed in acoplanarity and elliptic flow $v_2$ for 5.02 TeV PbPb collisions, 200 GeV AuAu collisions, and 193 GeV UU collisions.  We have used multi-differential binning to pinpoint the dependencies of these observables on the underlying variables and the systematics which reveal the underlying mechanisms.  First, we explored the role of the large energy difference between the LHC and RHIC for two symmetric systems: PbPb and AuAu.  While these two systems have similar shapes, PbPb collisions have a somewhat larger path length and significantly higher reach in $p_T$  compared to AuAu.  Counterintuitively, AuAu collisions produced acoplanarities and elliptic flow enhancement that were \textit{greater} than in PbPb collisions, owing to the sub-eikonal scaling $q_{drift} \propto 1/E$ of jet drift.  This enhancement from lower $p_T$ in AuAu also leads to a more pronounced ordering from the path length dependence compared to PbPb.  We confirm that these effects are governed by jet drift by implementing a $p_T<10$ GeV cut on the spectra; then the acoplanarities increased for both systems differently, leading to a role reversal in which the acoplanarity is larger for PbPb than in AuAu \textit{at the same $p_T$}.

While the transition from RHIC to LHC energies significantly alters the medium's lifetime and its initial temperature profile, both Au and Pb are round, isotropic nuclei.  To study the role of geometric deformation, we performed a similar comparison between 193 GeV UU collisions versus 200 GeV AuAu at RHIC.  A key finding in this work is the enhancement of drift-mediated acoplanarity in central UU collisions due to the prolate deformation of the $^{238}$U nucleus.  We see that even with central cuts on the impact parameter, UU collisions cover a much wider range of ellipticities than AuAu collisions due to the  transition between ``tip-on-tip'' and ``body-on-body'' geometries, even if $b = 0$.  Because of the ability to attain both high temperatures and high ellipticities in central events, UU collisions are able to attain higher acoplanarities due to jet drift.

The insights gained from this study will facilitate a larger-scale system scan and provide the necessary guidance to apply ``event engineering'' to the dataset. By identifying the specific regions of phase space where the jet drift effect is maximal, this work offers a roadmap for isolating signatures of jet-medium interactions. Furthermore, this analysis provides experimentalists with concrete signatures of jet drift to search for in current and future heavy-ion collision data, serving as a foundation for future global analyses and more granular jet substructure investigations.

Despite these advancements, several questions remain to be addressed in subsequent research. A primary next step is to more study the dependence of jet propagation on medium properties over a range of symmetric systems like AuAu and deformed systems like UU, of similar size and radius.  Other systems of interest include XeXe and OO collisions at the LHC.  Comparing these systems will further clarify the role of nuclear deformation in jet quenching and jet drift.  Additionally, the system scan should also be extended to smaller collision systems where the mechanism of jet-medium interactions is less understood.  OO collisions are of particular current interest, and smaller anisotropic systems like proton-gold (pAu), deuteron-gold (dAu), and helium-gold (${}^3$HeAu) collisions have revealed striking signals of a collective flow response to the initial-state geometry \cite{PHENIX:2018lia}. Fluctuations of energy loss and drift, already important in large systems, will become even more dominant in small systems.  

%% file: chp4_Conclusion.tex
\section{CONCLUSION \& OUTLOOK} \label{conclusion}
\hspace{\parindent}

%
\subsection{Overview: Hard Probes of a Background Medium}
%

The overarching theme of this dissertation is the demonstration that \textit{hard probes}, whether high-momentum partons in a quark-gluon plasma or relativistic particles in liquid argon, serve as precision diagnostics for extracting the properties of an underlying medium. This approach unifies physics across vastly different energy scales and interaction types through the universal kinematic principle of the \textit{separation of scales}. While the specific theoretical details vary, the core methodology remains consistent: the medium's characteristics are imprinted upon the probe's energy loss, transverse momentum diffusion, and stimulated radiation.

%
\subsection{Induced Cherenkov Radiation as a Probe of Liquid Argon}
%

In Ch.~\ref{cherenkov}, we established a robust theoretical framework for the high-precision characterization of the liquid argon (LAr) refractive index using the angular distribution of induced Cherenkov radiation. Treating Cherenkov radiation as a \textit{hard probe} of the LAr dielectric properties, we have shown that the stimulated Cherenkov radiation can be used to characterize the refractive index, even in regimes where direct measurements are scarce. We then leveraged this general principle by developing custom observables to quantify changes in the angular distribution. This approach is emblematic of a universal path forward for medium characterization, in which the well-known kinematics of a hard probe are leveraged to resolve the non-perturbative or poorly known features of the surrounding matter.

The work presented in Ch.~\ref{cherenkov} provides a critical theoretical foundation for the experimental programs at DUNE and CCM, where the precise characterization of the optical properties of liquid argon remains a primary challenge. By constructing HO and Approx absorptive fits to the LAr refractive index, we have moved beyond traditional resonant models to a framework that remains physically consistent near the 106.6 nm UV resonance.  We have successfully characterized the optical properties of LAr by mapping the features of the ultraviolet resonances to the resulting Cherenkov yields and angular distributions.

The centerpiece of this analysis is the introduction of the Angular Quantile Ratio (AQR), a novel observable designed to quantify model-dependent spectral shapes. Our findings demonstrate that while total Cherenkov yields offer limited sensitivity, the AQR provides significant discrimination power, particularly in its integrated form at high energies. This is quantitatively validated by a Figure of Merit (FoM) that identifies the integrated AQR as a primary candidate for experimental extraction of the refractive index.

Significantly, this research inverts the traditional Particle Identification (PID) paradigm: rather than using a known refractive index to identify particles, we demonstrate how the observed Cherenkov spectra of known particles can be used to constrain a poorly known refractive index. This approach provides a critical pathway to bridge the current gap in experimental data near the 106.6 nm UV resonance. Furthermore, by laying the groundwork for future Monte Carlo integration and the inclusion of hadronic energy loss effects, this study advances the development of more accurate, model-independent PID strategies for the next generation of LAr-based detectors for neutrino and particle physics. The initial results from this study, as published in Ref.~\cite{rahman2024}, have already been incorporated by the CCM Collaboration into 
their measurement of the liquid argon scintillation pulse shape\cite{CCM:2025dbq}. Even before the addition of novel measures like the AQR and associated FoM, major experimental collaborations recognized the implications for the experimental neutrino program of characterizing the optical properties of LAr.

%
\subsection{Jet Drift as a Probe of QGP Temperature and Geometry}
%

In Ch.~\ref{jetdrift}, we demonstrated that high-momentum partons or jets function as dynamic probes of the Quark-Gluon Plasma (QGP), capable of discriminating between the effects of initial-state geometry versus temperature and singling out novel sub-eikonal interactions with the medium.  By performing a comprehensive system scan across PbPb, AuAu, and deformed UU collisions, we showed that the acoplanarity and elliptic flow $v_2$ are not merely signatures of energy loss, but are sensitive to the local velocity field, which has distinct systematic dependences on the local temperature, path length, and ellipticity. 

A key finding of this work is the identification of experimental signatures of the \textit{sub-eikonal scaling} ($q_{drift} \propto 1/E$) of jet drift. We observed the counterintuitive result that, despite the much larger collision energy $\sqrt{s}$ in PbPb collisions at the LHC, AuAu collisions at RHIC energies exhibit greater acoplanarity.  Moreover, we showed that this excess arising from the larger $p_T$ accessible at the LHC can be removed by imposing cuts which select on lower momenta. This unusual scaling of jet drift provides a precise tool for distinguishing between traditional energy loss effects and the genuine sub-eikonal interactions responsible for jet drift.

Furthermore, by utilizing the prolate deformation of the $^{238}$U nucleus, we demonstrated that ``event engineering'' can isolate specific geometric configurations, such as tip-on-tip versus body-on-body, to maximize the visibility of drift-mediated effects. Deformed nuclei appear to be especially sensitive to jet drift due to their ability to simultaneously achieve both high temperatures and large ellipticities in central collisions, and may be a useful testbed for singling out the effects of jet drift.  

Ultimately, these results provide a roadmap for future experimental searches in deformed and small systems like OO or pAu collisions. By identifying characteristic signatures of hard probes which add a new sensitivity to the medium velocity, this work contributes to advancing the toolkit for characterizing the QGP in novel ways.

%
\subsection{Takeaway: A Universal Paradigm}
%

In this dissertation, we have argued that the theory of hard probes possesses a remarkable universality, providing a powerful toolkit capable of addressing a wide range of fundamental physics questions. By recognizing common features among superficially different problems, from the subatomic densities of the quark-gluon plasma to the dielectric environment of liquid argon, we have demonstrated that the core methodology of medium characterization remains consistent. This perspective allows us to move beyond narrow specialization, fostering \textit{cross-fertilization} \cite{Nambu:2009zza} across fields and identifying novel opportunities for innovation.

The work presented herein achieves this goal on two distinct fronts. First, we have successfully characterized the optical properties of a LAr medium using Cherenkov radiation as a hard probe. The AQR provides a roadmap to neutrino experimentalists and others for extracting the LAr refractive index in regimes where direct measurements are limited. Second, we have successfully characterized the underlying properties of the QGP and their sensitivity to jet drift. Our analysis pinpointed specific signatures of drift-mediated acoplanarity and their systematics as a function of path length, temperature, and initial-state ellipticity across multiple collision systems: PbPb, AuAu, and UU. In both of these radically different physical systems, the hard probe acted as a microscopic diagnostic, resolving the complex transport properties of the surrounding matter.

Far from being a collection of independent projects, this research represents the application of a synergistic, interdisciplinary approach to theoretical physics based on the separation of scales. The unique value of the experience accumulated in the course of this work is the ability to tackle physics problems from underlying first principles rather than relying on domain-specific approximations. The tools, insights, and strategies presented herein offer the ability to flexibly navigate the boundaries between nuclear physics, particle physics, astrophysics, and cosmology.

This broad theoretical framework can be applied to a vast landscape of future problems at the forefront of physics research. This framework could be applied to study neutrino transport in the extreme environments of core-collapse supernovae \cite{Most:2022yhe}, or to ultra-high-energy cosmic rays, or many challenging interdisciplinary physics problems. Furthermore, as both neutrinos and the QGP are precursors to the very early universe, continuing this work will deepen our understanding of phase transitions in the early cosmos and the building blocks of matter at the same time.

Whether conducting precision investigations of the Standard Model or searching for Beyond the Standard Model (BSM) signatures, a generalist's perspective on disparate problems offers new opportunities to turn these theoretical predictions into observable discoveries.  This is indeed the mission of physics itself: to contribute to our collective knowledge of the matter that constitutes our universe, from the ``Little Bangs'' created in particle accelerators to the grand evolution of the universe itself.

%% file: chp5_Appendix.tex
\section*{APPENDIX} \label{appendix}
\hspace{\parindent}

\section{Harmonic Oscillator Model: The Frequency Dependence of Refractive Index} \label{sec:HarmonicOscillator}

\hspace{\parindent}

\begin{figure}[h!] 
\begin{centering}
\includegraphics[width=0.5\textwidth]{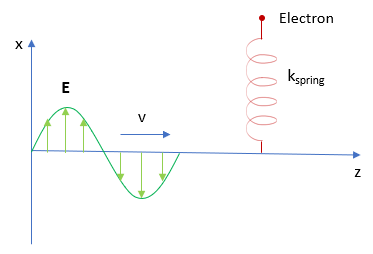}
\caption{Simplified Model of Electrons in Dielectrics  
\label{f:electronsindielectrics}
}
\end{centering}
\end{figure}

In this Appendix, we summarize the derivation of the particular (absorptive) form of the refractive index used in this work, following the canonical textbook treatment \cite{Jackson:1998nia, griffiths_2017}.

Three parameters govern the propagation of electromagnetic waves through matter: the permittivity $\epsilon$, the permeability $\mu$, and the conductivity $\sigma$.  In general, all three quantities can be frequency dependent: $\epsilon (\omega), \mu(\omega), \sigma(\omega)$.  The frequency dependence of the refractive index $n(\omega)$ inherited from these three parameters leads to the phenomenon of dispersion, such as the different angles taken by different colors of light in a prism.



Consider an elementary, classical model of in a dielectric material.  For small displacements from equilibrium, the forces which bind electrons to their atomic nuclei can be approximated as a simple harmonic oscillator (HO) (~Fig \ref{f:electronsindielectrics}): 
\begin{align}   \label{e:fbinding}
    F_{binding} = - k_{spring} \, x = - m\omega_{0}^2 \, x,
\end{align}
where $x$ is the displacement from equilibrium, $m$ is the electron's mass, and $\omega_{0} = \sqrt{k_{spring}/m}$ is the natural oscillation frequency.  (Geometrically, the statement is that any function can be approximated by a suitable parabola near a local minimum.)  In the presence of an electromagnetic wave of frequency $\omega$ propagating through the medium (Fig. \ref{f:electronsindielectrics}), the electron is also subject to a driving force due to the oscillating electric field, 
\begin{align}   \label{e:fdriving}
    F_{driving} = qE = qE_0 \cos(\omega t) \: ,
\end{align}
where q is the charge of the electron and $E_0$ is the amplitude of the electric field.  Meanwhile, dissipative effects can be approximated through a linear damping term of the form
\begin{align}   \label{e:fdamping}
    F_{damping} = -m\gamma \frac{dx}{dt} \: ,
\end{align}
with $\gamma$ the damping constant.  Altogether, this gives for the equation of motion
\begin{align}   \label{e:ftot}
    \qquad m \frac{d^2x}{dt^2} + m\gamma \frac{dx}{dt} +  m\omega_{0}^2 x = qE_0 \cos(\omega t)  \: .
\end{align}
We emphasize that the ``harmonic oscillator model'' (HO) encapsulated in the equation of motion \eqref{e:ftot} is derived from first principles as approximations to the exact physics; if desired, Eq.~\eqref{e:ftot} can be systematically improved to higher accuracy.

Eq.~\eqref{e:ftot} may be regarded as the real part of a complex equation $x(t) = \mathrm{Re} \: \tilde{x} (t)$ given by
\begin{align}   \label{e:ftotcomplex}
    \frac{d^2\tilde{x}}{dt^2} + \gamma \frac{d\tilde{x}}{dt} +  \omega_{0}^2 \tilde{x} = \frac{q}{m} E_0 \: e^{-i\omega t} \: .
\end{align}
In the steady state, the system will oscillate at the driving frequency $\omega$ with some amplitude $\tilde{x}_0$:
\begin{align}   \label{e:xcomplex}
    \tilde{x} (t) = \tilde{x}_0 \: e^{-i\omega t}, 
\end{align}
which gives the solution for the amplitude as
\begin{align}   \label{e:x0complex}
    \tilde{x}_0 = \frac{q/m}{\omega_{0}^2 - \omega^2 - i\gamma\omega} E_{0}. 
\end{align}
Then the induced (complex) dipole moment $\tilde{p} (t)$ is just proportional to the displacement:
\begin{align}   \label{e:dipolemoment}
    \tilde{p} (t) = q \: \tilde{x} (t) = \frac{q^2/m}{\omega_{0}^2 - \omega^2 - i\gamma\omega} 
    \: E_{0} \: e^{-i\omega t}. 
\end{align}
We note that the imaginary term in the denominator produces a phase shift of the induced dipole moment with respect to the external field.  The electron oscillations being out of phase with the driving force is associated with the dissipation of energy as the external field now does some amount of negative work on the oscillating electron.

To generalize Eq.~\eqref{e:dipolemoment} to a material, we consider $N$ molecules per unit volume with $f_j$ of electrons per molecule\footnote{We use the uppercase $N$ for the number of molecules per unit volume because we reserve the lowercase $n$ to refer to the refractive index.}.  Rather than assuming that each electron can oscillate about a single local minimum, we consider oscillations about a range $j$ of possible minima.  Describing each of these possible oscillation modes by a   frequency $\omega_j$ and damping factor $\gamma_j$, the total polarization density $\textbf{P}$ generated by the electromagnetic wave is
\begin{align}   \label{e:dipolemomentN}
    \tilde{\textbf{P}} = \frac{Nq^2}{m} {\left( \sum_{j}\frac{f_j}{\omega_{j}^2 - \omega^2 - i\gamma_j \omega}\right)} \tilde{\textbf{E}}. 
\end{align}
Eq.~\eqref{e:dipolemomentN} treats the many electrons as independent; this is valid for the case of a dilute gas and may receive corrections at high density.

The complex susceptibility $\tilde{\chi}_e$ is defined as the constant of proportionality between the electric field and polarization:
\begin{align}   \label{e:dipolemomentcomplexN}
    \tilde{\textbf{P}} = \epsilon_0 \tilde{\chi_e} \tilde{\textbf{E}},
\end{align}
so we may read off from Eq.~\eqref{e:dipolemomentN}
\begin{align}   \label{e:complxsusc}
    \tilde{\chi}_e = \frac{Nq^2}{m \epsilon_0} {\left( \sum_{j}\frac{f_j}{\omega_{j}^2 - \omega^2 - i\gamma_j \omega}\right)} \: . 
\end{align}
Then the susceptibility enters the permittivity through
\begin{align}   \label{e:idielectricconst}
    \frac{\tilde{\epsilon}}{\epsilon_0} = 
    1 + \tilde{\chi}_e =
    1 + \frac{Nq^2}{m\epsilon_0} {\left( \sum_{j}\frac{f_j}{\omega_{j}^2 - \omega^2 - i\gamma_j \omega}\right)}. 
\end{align}
Ordinarily, the imaginary term is negligible; however, when $\omega$ is very close to one of the resonant frequencies ($\omega_j$) it describes absorption, as we shall see. In a dispersive medium, the plane wave solution of the wave equation ($\nabla^2\tilde{\textbf{E}} = \tilde{\epsilon} \mu_0 \frac{\partial^2\tilde{\textbf{E}}}{\partial t^2}$) for a given frequency reads, 



%
\begin{align}   \label{e:planewavesoln}
    \tilde{\textbf{E}} (z, t) = \tilde{\textbf{E}}_0 \: e^{i(\tilde{k} z - \omega t)}, 
\end{align}

with the complex wave number given by, 

\begin{align}   \label{e:iwaveno}
    \tilde{k} = k + i \kappa \equiv \sqrt{\tilde{\epsilon} \mu_0} \: \omega. 
\end{align}
%


Thus, Eq.~\eqref{e:planewavesoln} becomes, 
\begin{align}   \label{e:planewavesolnfinal}
    \tilde{\textbf{E}} (z, t) = \tilde{\textbf{E}}_0 \: e^{-\kappa z} \: e^{i(\tilde{k} z - \omega t)}, 
\end{align}

The wave is attenuated due to the absorption of energy from damping. Because the intensity is proportional to $E^2$ (and hence to $e^{-2\kappa z})$, the quantity $\alpha \equiv 2 \kappa$, is called the \textit{absorption coefficient}.  

Moreover, the wave velocity is $\omega/k$, and the refractive index, $n = \frac{ck}{\omega}$; where k and $\kappa$ are determined by the parameters of our damped harmonic oscillator. The second term in Eq.~\eqref{e:idielectricconst} is small for gases, and we can approximate the square root (Eq.~\eqref{e:iwaveno}) by the first term in the binomial expansion. $ \sqrt{1+\epsilon} \cong 1 + \frac{1}{2}\epsilon$. Then,


%
\begin{align}   \label{e:iwavenofinal}
    \tilde{k} = \frac{\omega}{c} \sqrt{\tilde{\epsilon}_r} \cong \frac{\omega}{c} {\left[1 + \frac{Nq^2}{2m\epsilon_0} \sum_{j}\frac{f_j}{\omega_{j}^2 - \omega^2 - i\gamma_j \omega} \right]}, 
\end{align}

Finally, we derive general expressions for $n$ and $\alpha$ given by,   

\begin{align}   \label{e:nfinal}
     n = \frac{ck}{\omega} \cong 1 + \frac{Nq^2}{2m\epsilon_0} \sum_{j}\frac{f_j (\omega_{j}^2 - \omega^2)}{(\omega_{j}^2 - \omega^2)^2 + \gamma_j^2 \omega^2}, 
\end{align}

and 

\begin{align}   \label{e:alphafinal}
    \alpha \equiv 2 \kappa \cong \frac{Nq^2 \omega^2}{m \epsilon_0 c} \sum_{j}\frac{f_j \gamma_j}{(\omega_{j}^2 - \omega^2)^2 + \gamma_j^2 \omega^2}. 
\end{align}

If we stay far from any resonances and ignore damping, the formula for refractive index further simplifies to, 

\begin{align}   \label{e:nfinalsimplified}
     n \cong 1 + \frac{Nq^2}{2m\epsilon_0} \sum_{j}\frac{f_j }{\omega_{j}^2 - \omega^2}.
\end{align}

For most substances, the natural frequencies $\omega_j$ are scattered all over the spectrum in a somewhat chaotic fashion. But the nearest significant resonances typically lie in the ultraviolet for transparent materials, so that $\omega < \omega_j$. 


Now we will derive the wavelength equivalent expression of \eqref{e:nfinal} and \eqref{e:alphafinal} by simply noting that the angular frequency ($\omega$) is related to the wavelength ($\lambda$) as, $\omega = 2\pi c/\lambda$
and $\omega_j = 2\pi c/\lambda_j$ so by substituting this in \eqref{e:nfinal} and \eqref{e:alphafinal} we have, 

\begin{align}   \label{e:nlambda}
     n = 1 + \frac{Nq^2}{2m\epsilon_0} \sum_{j}\frac{f_j \lambda^2 \lambda_{j}^2 (\lambda^2 - \lambda_{j}^2)}{(2 \pi c)^2 (\lambda^2 - \lambda_{j}^2)^2 + \gamma_j^2 \lambda^2 \lambda_{j}^2}, 
\end{align}

and 

\begin{align}   \label{e:alphalambda}
    \alpha \cong \frac{Nq^2}{m \epsilon_0 c} \sum_{j}\frac{f_j \gamma_j \lambda^2 \lambda_{j}^4}{(2 \pi c)^2 (\lambda^2 - \lambda_{j}^2)^2 + \gamma_j^2 \lambda^2 \lambda_{j}^4}. 
\end{align}

Now, if we consider only UV and IR resonances, we have the corresponding UV and IR terms from the summation from \eqref{e:nfinal} and \eqref{e:alphafinal} as follows,

\begin{align}   \label{e:nlambdauvir}
     n = 1 + \frac{Nq^2}{2m\epsilon_0} \left( \frac{f_{UV} \lambda^2 \lambda_{UV}^2 (\lambda^2 - \lambda_{UV}^2)}{(2 \pi c)^2 (\lambda^2 - \lambda_{UV}^2)^2 + \gamma_{UV}^2 \lambda^2 \lambda_{UV}^2} + \frac{f_{IR} \lambda^2 \lambda_{IR}^2 (\lambda^2 - \lambda_{IR}^2)}{(2 \pi c)^2 (\lambda^2 - \lambda_{IR}^2)^2 + \gamma_{IR}^2 \lambda^2 \lambda_{IR}^2} \right), 
\end{align}

and 

\begin{align}   \label{e:alphalambdauvir}
    \alpha \cong \frac{Nq^2}{m \epsilon_0 c} \left( \frac{f_{UV} \gamma_{UV} \lambda^2 \lambda_{UV}^4}{(2 \pi c)^2 (\lambda^2 - \lambda_{UV}^2)^2 + \gamma_{UV}^2 \lambda^2 \lambda_{UV}^4} + \frac{f_{IR} \gamma_{IR} \lambda^2 \lambda_{IR}^4}{(2 \pi c)^2 (\lambda^2 - \lambda_{IR}^2)^2 + \gamma_{IR}^2 \lambda^2 \lambda_{IR}^4} \right). 
\end{align}

\begin{figure}[h!] 
\begin{centering}
\includegraphics[width=0.5\textwidth]{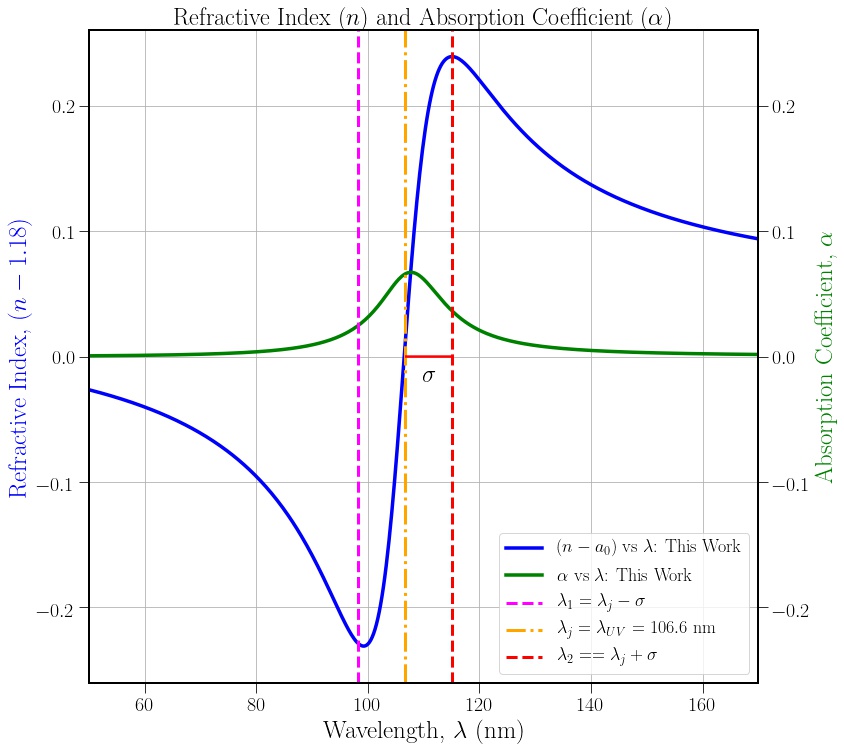}
\caption{Refractive Index and Absorption Coefficients in the vicinity of UV resonance ($\lambda_{UV} = $ 106.6 nm)   
\label{f:absorption_coeff_theoryplot2}
}
\end{centering}
\end{figure}

The index of refraction and the absorption coefficient given by Eq.~\eqref{e:nlambdauvir} and \eqref{e:alphalambdauvir} in the vicinity of one of the resonances ($\lambda_{UV} = $ 106.6 nm) is plotted here in Fig. \ref{f:absorption_coeff_theoryplot2}. 


%

Generally, the index of refraction rises gradually with increasing frequency or decreasing wavelength, however, near a resonance the index of refraction drops sharply. This is called \textit{anomalous dispersion} because of this unusual behavior. Notice that the region of anomalous dispersion ($\lambda_1 < \lambda_{UV} < \lambda_2$, in the figure) coincides with the region of maximum absorption. In this wavelength range, the material can be nearly impenetrable. Because we are now driving the electrons at their ``preferred'' frequency/wavelength, their oscillations have a comparatively significant amplitude, which causes the damping process to dissipate a substantial amount of energy.


\vspace{1cm}

\noindent

\newpage

%% file: chp6_glossary.tex
\section{Glossary of Muon Cherenkov PID} 

\subsection{Muon Yield} \label{glossary:muonyield}

\begin{figure}[!ht] 
    \centering
    \includegraphics[width=0.45\textwidth]{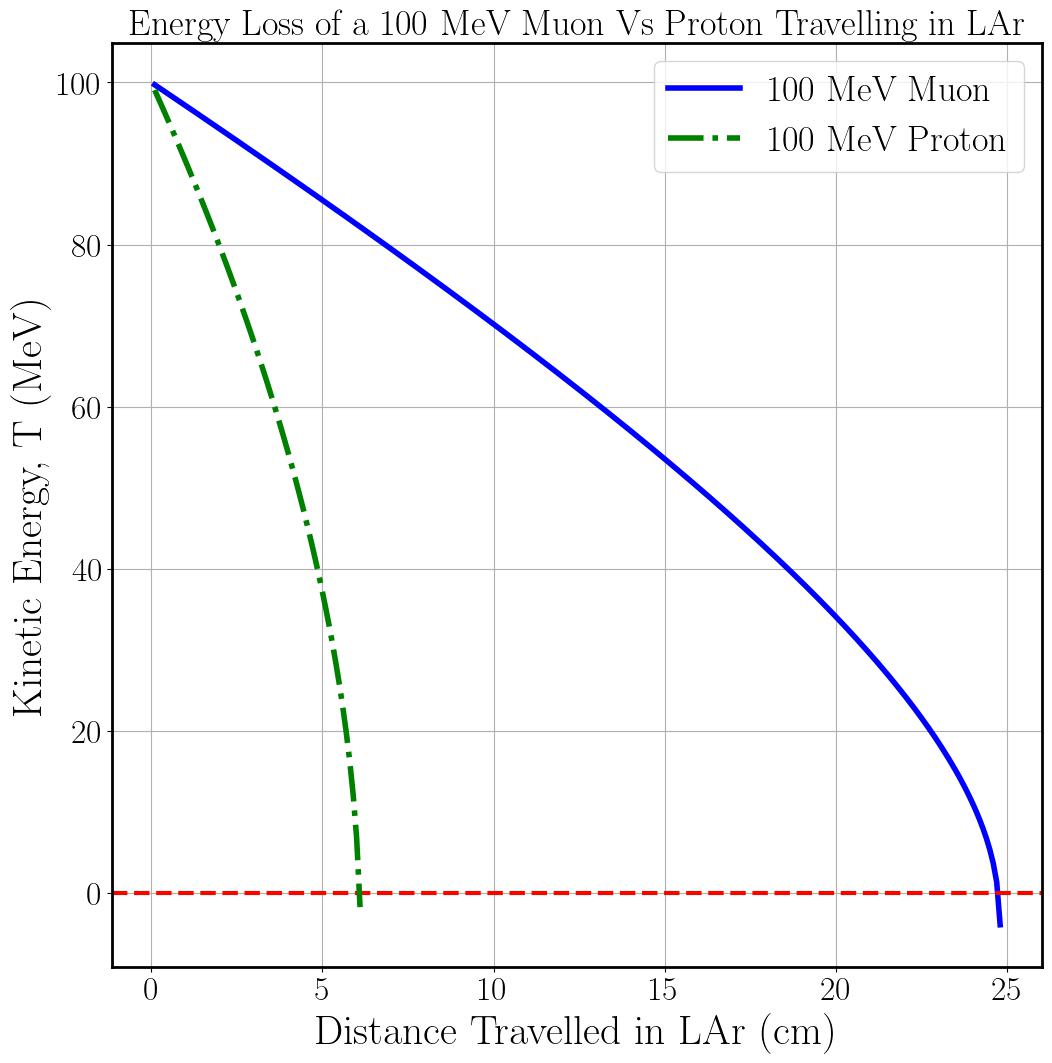}
    \hspace{0.5cm}
    \includegraphics[width=0.45\textwidth]{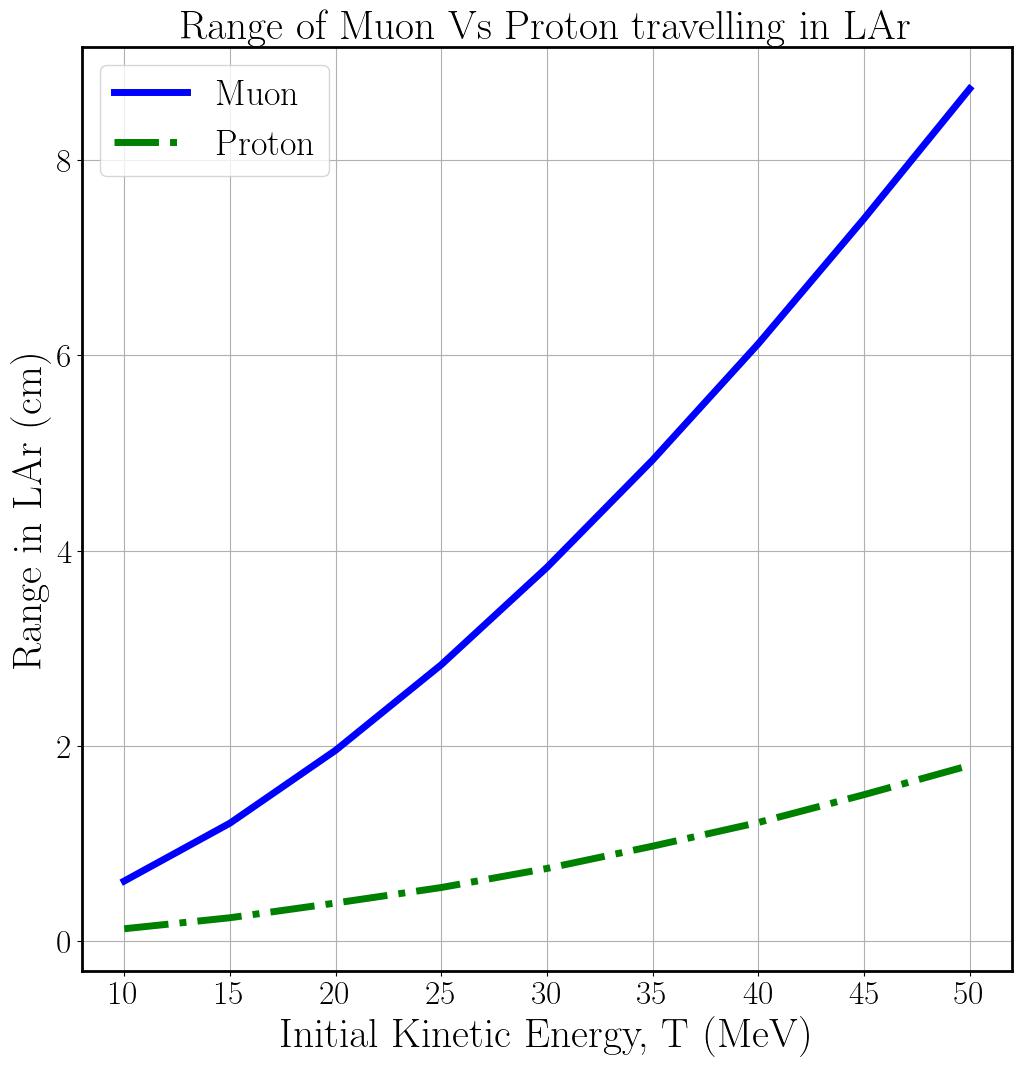}
    \caption{Comparison of Energy loss (Left) and Range (Right) of a 100 MeV Muon Vs 100 MeV Proton in LAr.}
    \label{f:KElossmuonvsprotonrel}
\end{figure}


\begin{figure}[!ht]
    \centering
    \begin{subfigure}{.48\textwidth}
        \includegraphics[width=\textwidth]{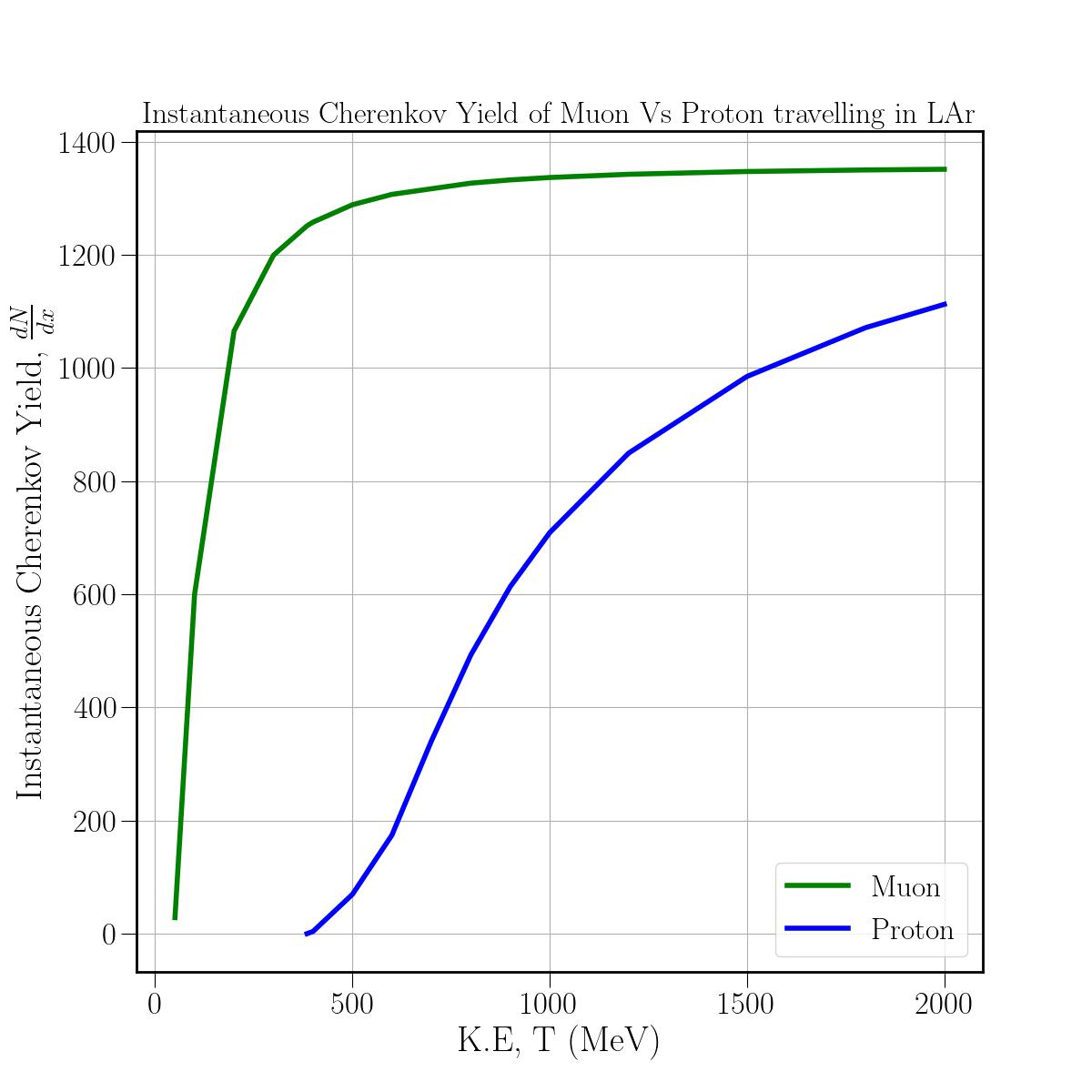}
        \caption{Instantaneous Yield}
        \label{f:instyieldmuonvsprotonaprxfit}
    \end{subfigure} 
    \hfill
    \begin{subfigure}{.48\textwidth}
        \includegraphics[width=\textwidth]{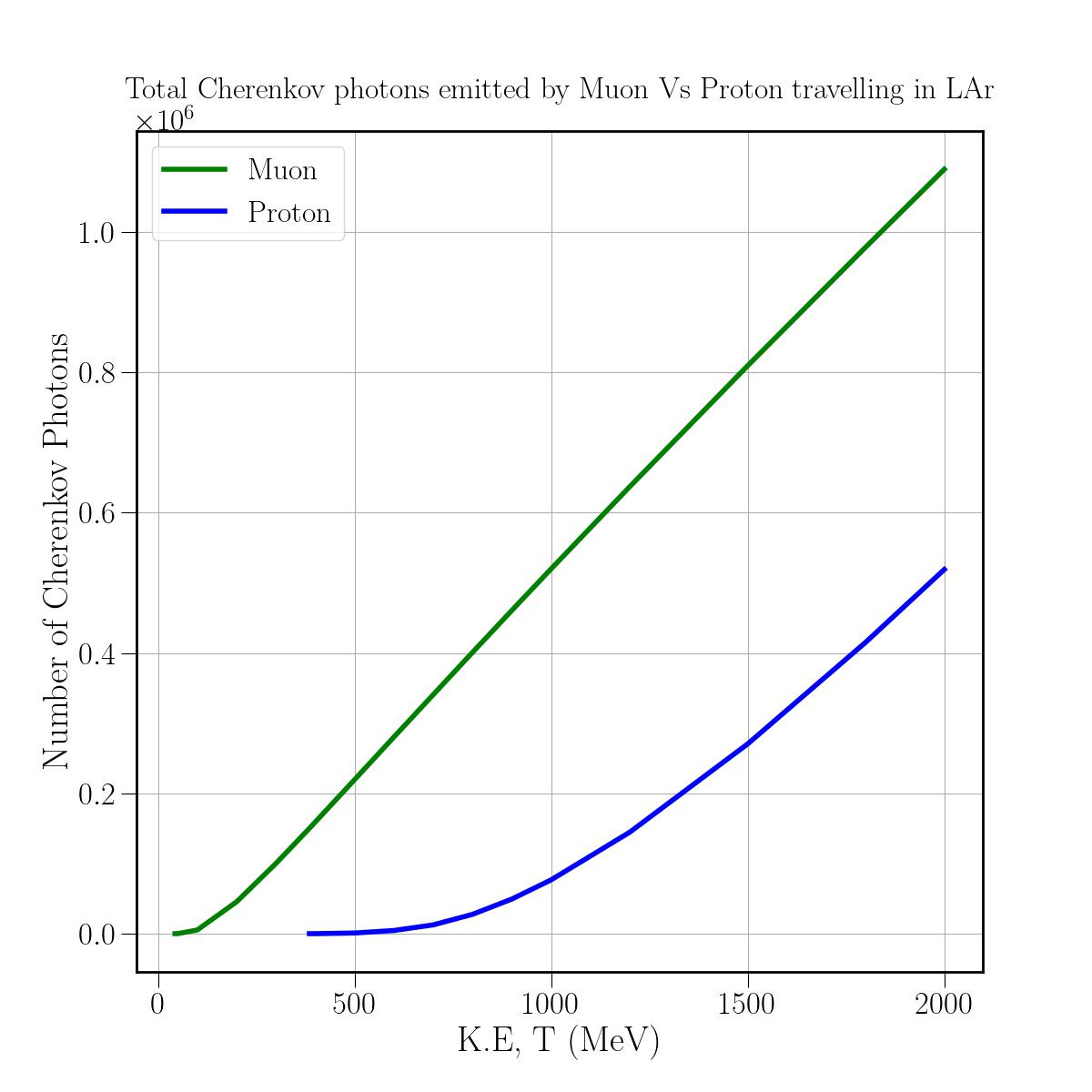}
        \caption{Integrated Yield}
        \label{f:totalyieldmuonvsprotonaprxfit}
    \end{subfigure}
    \caption{Comparison of Total Cherenkov Yield: Muons vs Protons (Different K.E.s).}
\end{figure}

\newpage


\subsection{PID from Instantaneous AD}
\label{sec:muoniadpid}

\begin{figure}[!ht] 
    \centering
    \begin{subfigure}{.47\textwidth}
        \centering
        \includegraphics[width=\textwidth]{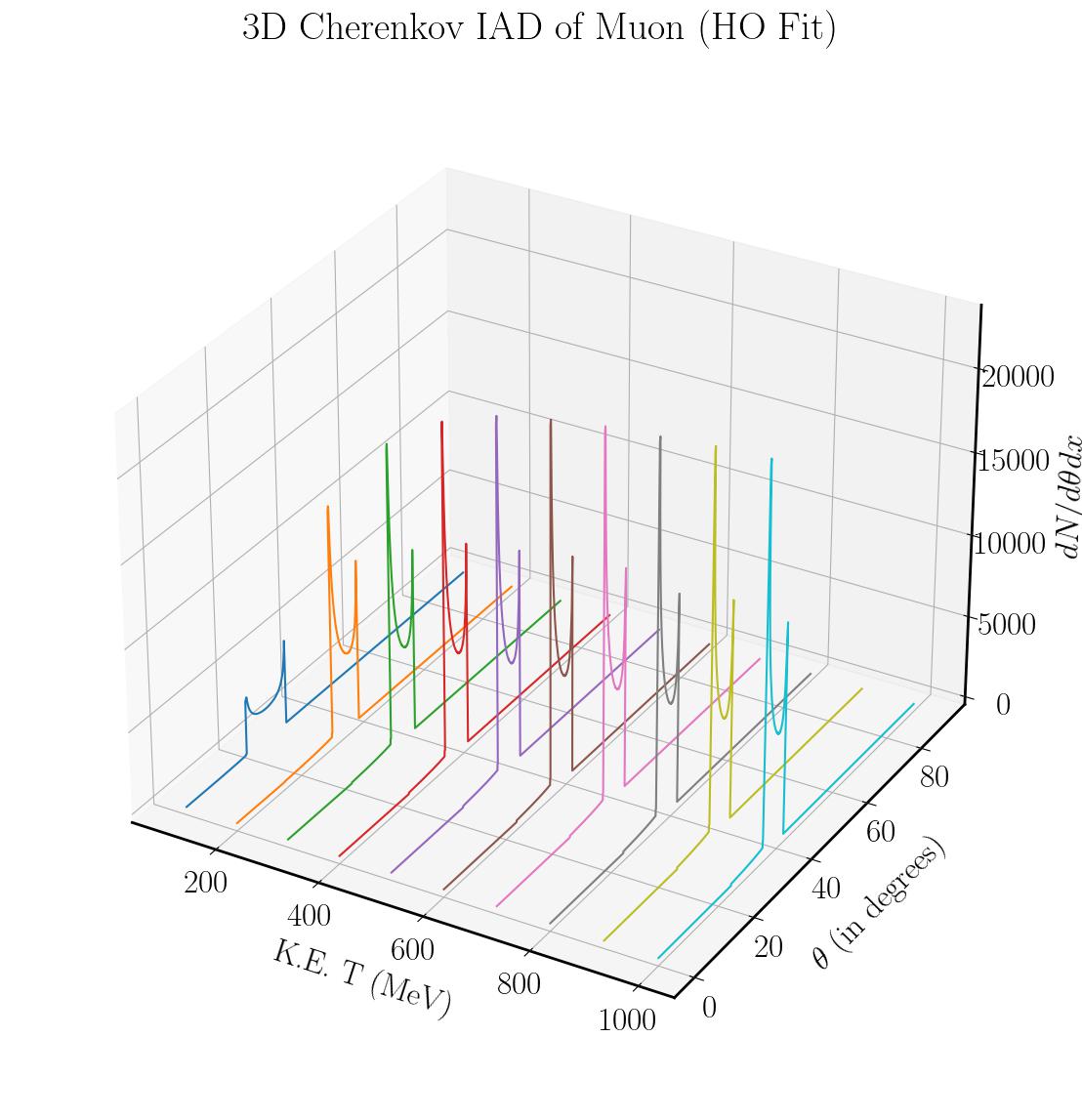}
    \end{subfigure}       
    \hfill
    \begin{subfigure}{.47\textwidth}
        \centering
        \includegraphics[width=\textwidth]{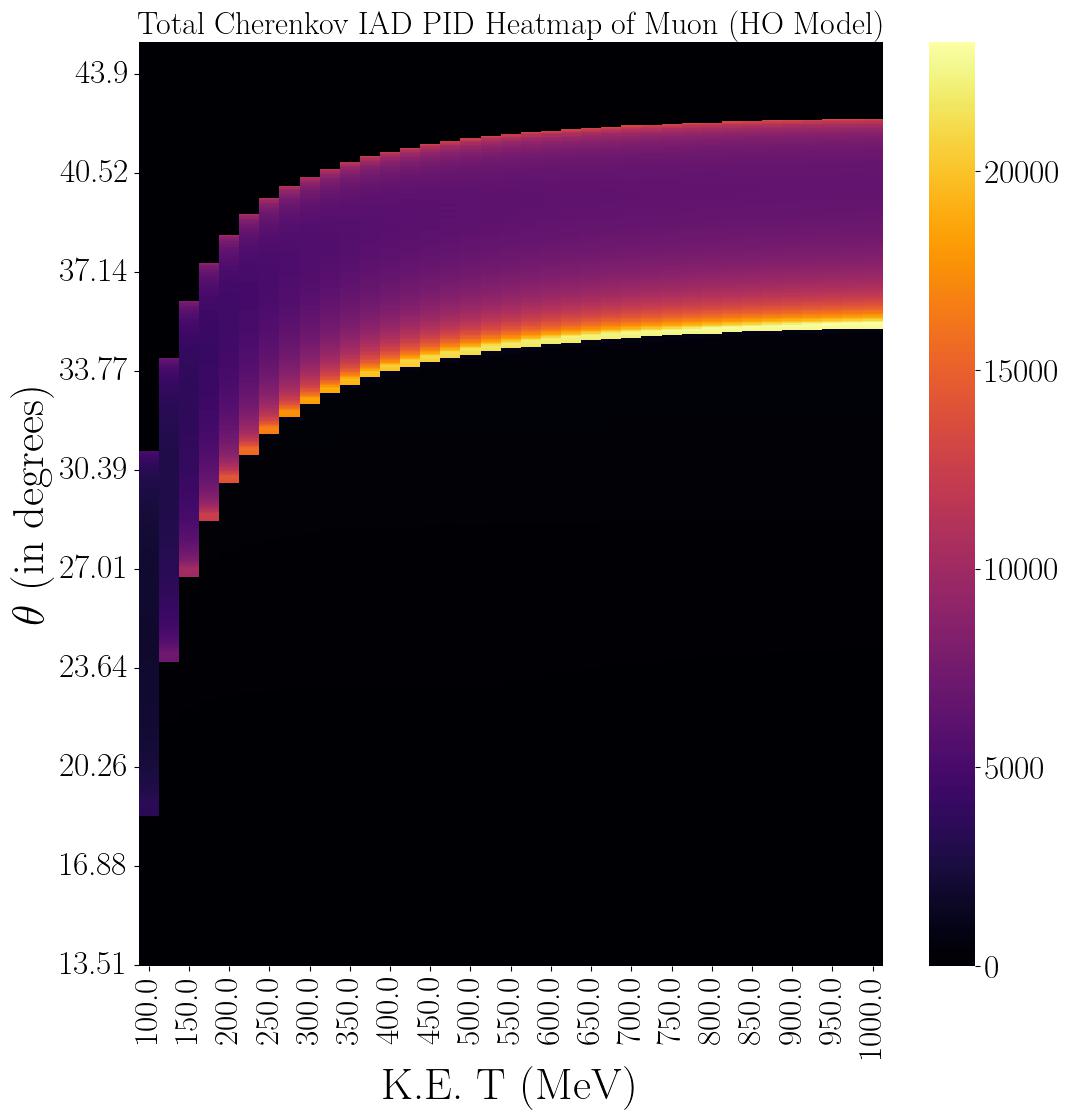}
    \end{subfigure}
    \caption{Muon's instantaneous AD for different K.E.s using HO model fit (left) and the associated PID heatmap (right).}
    \label{f:muonpidiadho}
\end{figure}

\begin{figure}[!ht] 
    \centering
    \begin{subfigure}{.47\textwidth}
        \centering
        \includegraphics[width=\textwidth]{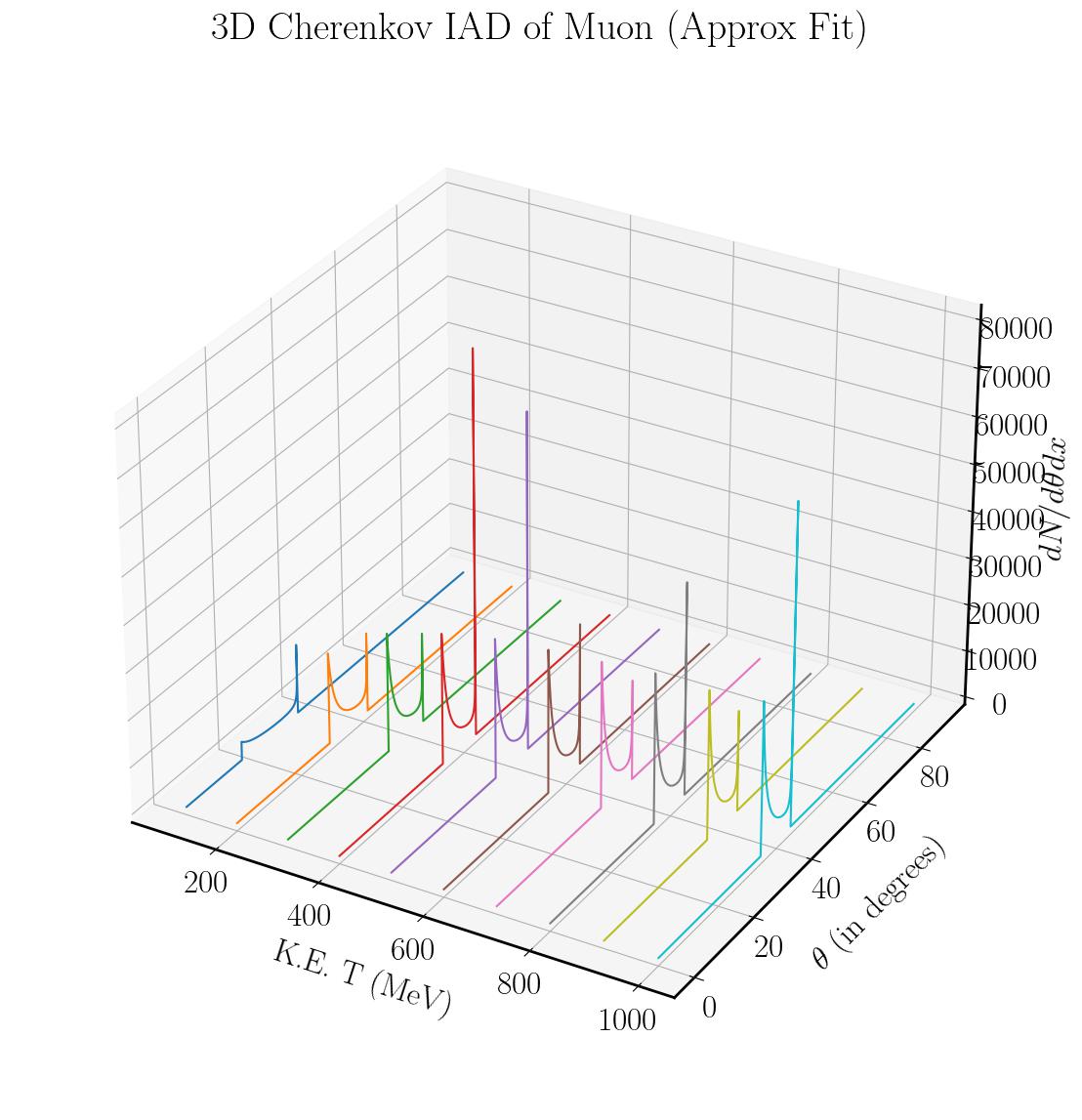}
    \end{subfigure}       
    \hfill
    \begin{subfigure}{.47\textwidth}
        \centering
        \includegraphics[width=\textwidth]{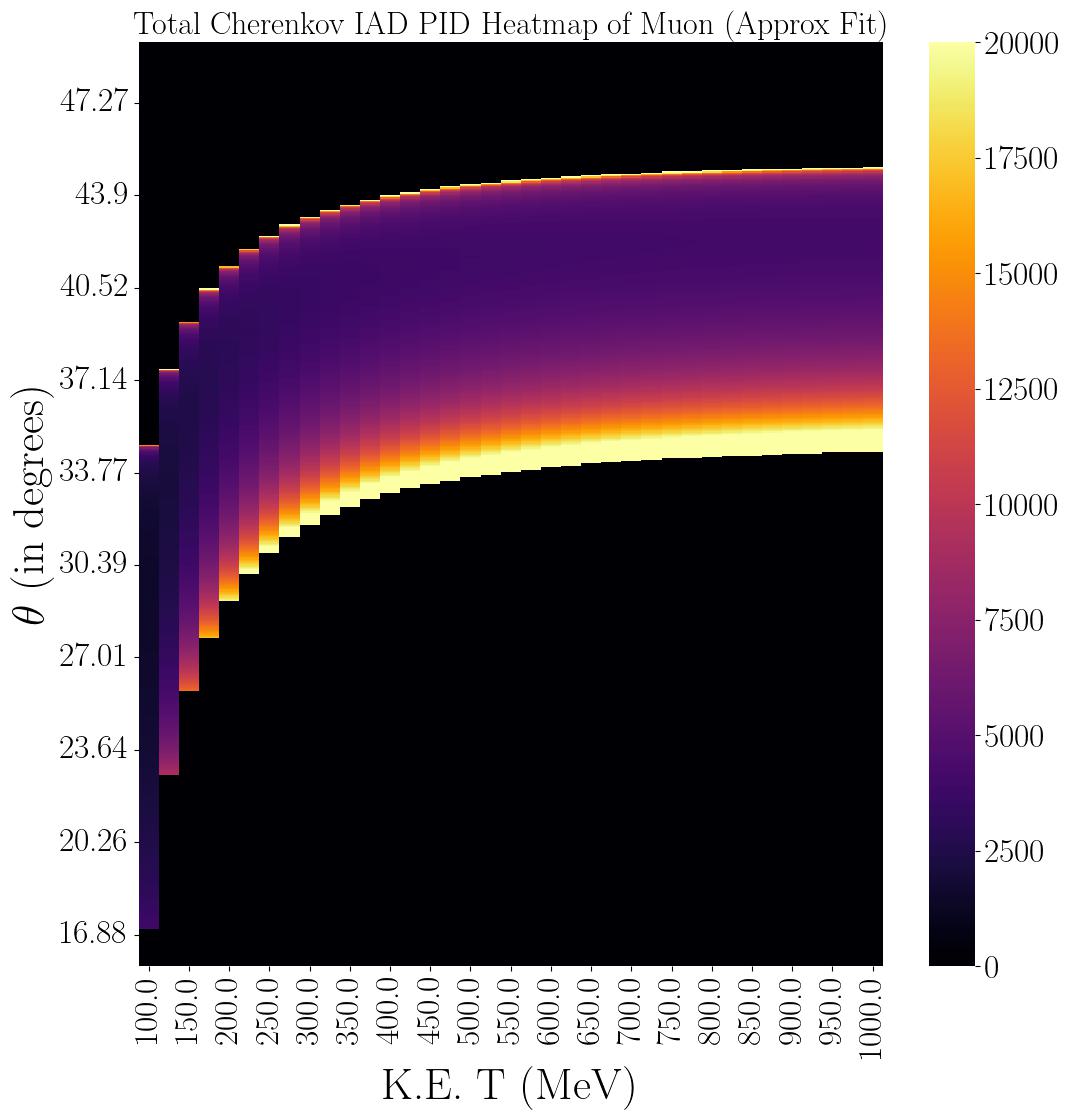}
    \end{subfigure}
    \caption{Muon's instantaneous AD for different K.E.s using the approx. fit (left) and the associated PID heatmap (right).}
    \label{f:muonpidiadapprox}
\end{figure}

\newpage

\subsection{PID from Integrated AD}
\label{sec:muonadpid}

\begin{figure}[!ht] 
    \centering
    \begin{subfigure}{.49\textwidth}
        \centering
        \includegraphics[width=\textwidth]{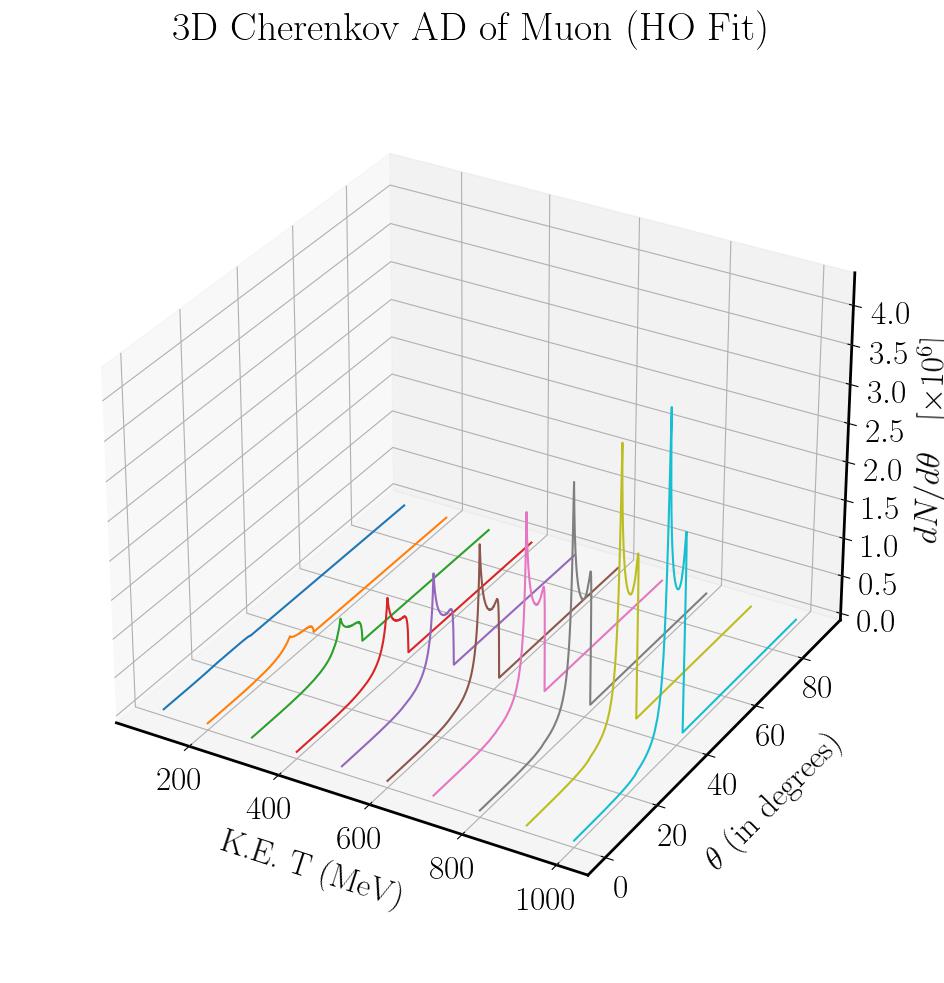}
    \end{subfigure}   
    \hfill
    \begin{subfigure}{.49\textwidth}
        \centering
        \includegraphics[width=\textwidth]{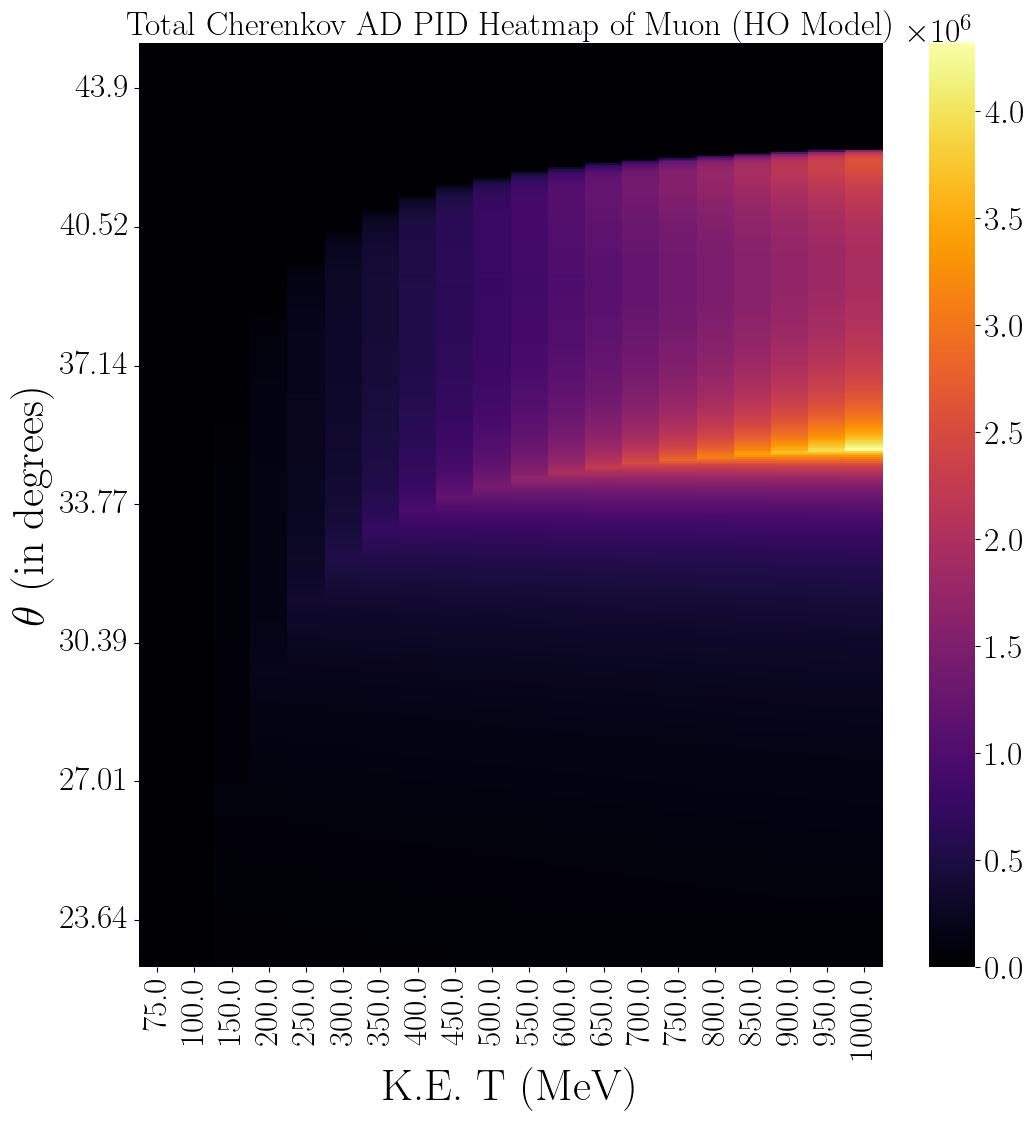}
    \end{subfigure}
    \caption{Muon's total integrated AD for different K.E.s using HO fit (left) and the associated PID heatmap (right).}
    \label{f:muonadpidho}
\end{figure}

\begin{figure}[!ht] 
    \centering
    \begin{subfigure}{.49\textwidth}
        \centering
        \includegraphics[width=\textwidth]{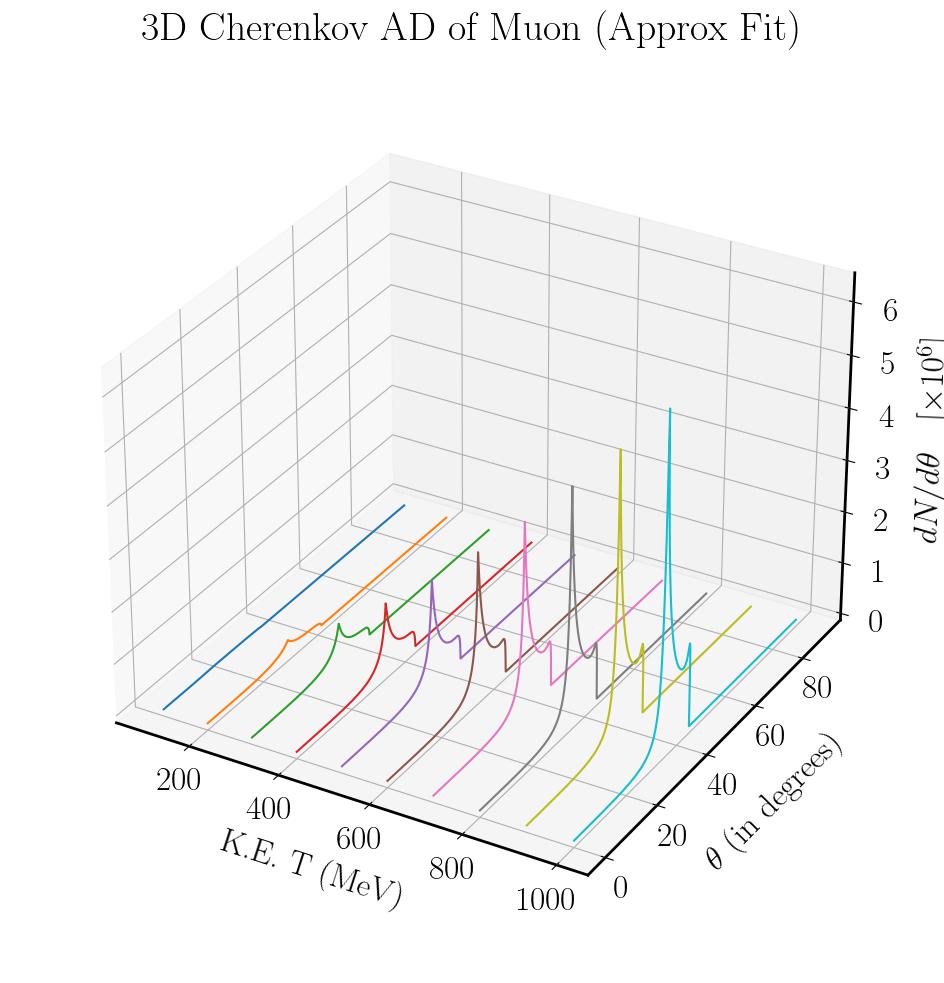}
    \end{subfigure}   
    \hfill
    \begin{subfigure}{.49\textwidth}
        \centering
        \includegraphics[width=\textwidth]{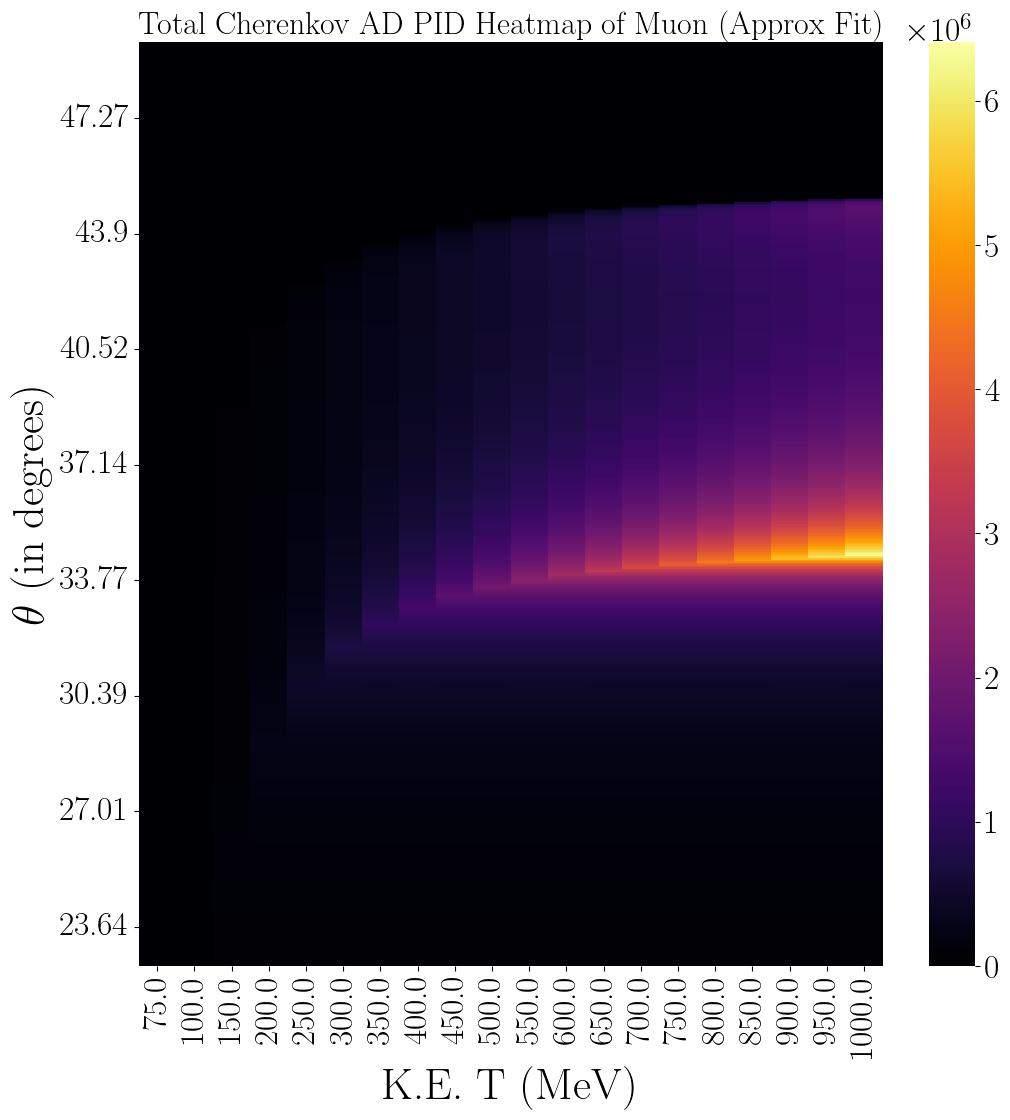}
    \end{subfigure}
    \caption{Muon's total integrated AD for different K.E.s using the approx. fit (left) and the associated PID heatmap (right).}
    \label{f:muonadpidapprox}
\end{figure}


%% file: Refs.bib
@article{Rutherford1911,
  author = {Rutherford, Ernest},
  title = {The scattering of $\alpha$ and $\beta$ particles by matter and the structure of the atom},
  journal = {The London, Edinburgh, and Dublin Philosophical Magazine and Journal of Science},
  volume = {21},
  number = {125},
  pages = {669--688},
  year = {1911},
  doi = {10.1080/14786440508637080}
}

@article{Feynman:1969ej,
    author = "Feynman, Richard P.",
    title = "{Very High-Energy Collisions of Hadrons}",
    journal = "Phys. Rev. Lett.",
    volume = "23",
    pages = "1415--1417",
    year = "1969",
    doi = "10.1103/PhysRevLett.23.1415"
}

@article{Petersen:2008kb,
      title={UrQMD-2.3 - Changes and Comparisons}, 
      author={Hannah Petersen and Marcus Bleicher and Steffen A. Bass and Horst Stöcker},
      year={2008},
      eprint={0805.0567},
      archivePrefix={arXiv},
      primaryClass={hep-ph},
      url={https://arxiv.org/abs/0805.0567}, 
}

@article{Bass:1998ca,
    author = "Bass, S. A. and others",
    title = "{Microscopic models for ultrarelativistic heavy ion collisions}",
    eprint = "nucl-th/9803035",
    archivePrefix = "arXiv",
    doi = "10.1016/S0146-6410(98)00058-1",
    journal = "Prog. Part. Nucl. Phys.",
    volume = "41",
    pages = "255--369",
    year = "1998"
}

@article{Bahder:2024jpa,
    author = "Bahder, Joseph and Rahman, Hasan and Sievert, Matthew D. and Vitev, Ivan",
    title = "{Signatures of jet drift in quark-gluon plasma hard-probe observables}",
    eprint = "2412.05474",
    archivePrefix = "arXiv",
    primaryClass = "nucl-th",
    doi = "10.1103/8yzj-2l5c",
    journal = "Phys. Rev. Res.",
    volume = "8",
    number = "1",
    pages = "L012016",
    year = "2026"
}

@incollection{Iancu:2003xm,
    author        = "Iancu, Edmond and Venugopalan, Raju",
    title         = "{The Color Glass Condensate and High Energy Scattering in {QCD}}",
    booktitle     = "{Quark-Gluon Plasma 3}",
    editor        = "Hwa, Rudolph C. and Wang, Xin-Nian",
    publisher     = "World Scientific",
    address       = "Singapore",
    year          = "2004",
    pages         = "249-336",
    eprint        = "hep-ph/0303204",
    archivePrefix = "arXiv",
    doi           = "10.1142/9789812702708_0005"
}

@Article{Lappi:2006fp,

     author    = "Lappi, T. and McLerran, L.",

     title     = "{Some features of the glasma}",

     journal   = "Nucl. Phys.",

     volume    = "A772",

     year      = "2006",

     pages     = "200-212",

     eprint    = "hep-ph/0602189",

     archivePrefix = "arXiv",

     doi       = "10.1016/j.nuclphysa.2006.04.001",

     SLACcitation  = "%%CITATION = HEP-PH/0602189;%%"

}

@Article{Gyulassy:1993hr,
     author    = "Gyulassy, Miklos and Wang, Xin-nian",
     title     = "{Multiple collisions and induced gluon Bremsstrahlung in
                  QCD}",
     journal   = "Nucl. Phys.",
     volume    = "B420",
     year      = "1994",
     pages     = "583-614",
     eprint    = "nucl-th/9306003",
     archivePrefix = "arXiv",
     doi       = "10.1016/0550-3213(94)90079-5",
     SLACcitation  = "%%CITATION = NUCL-TH/9306003;%%"
}

@Article{Vitev:2002pf,
     author    = "Vitev, Ivan and Gyulassy, Miklos",
     title     = "{High $p_{T}$ tomography of $d$ + Au and Au+Au at SPS,
                  RHIC, and LHC}",
     journal   = "Phys. Rev. Lett.",
     volume    = "89",
     year      = "2002",
     pages     = "252301",
     eprint    = "hep-ph/0209161",
     archivePrefix = "arXiv",
     doi       = "10.1103/PhysRevLett.89.252301",
     SLACcitation  = "%%CITATION = HEP-PH/0209161;%%"
}

@article{Gyulassy:2000er,
      author         = "Gyulassy, M. and Levai, P. and Vitev, I.",
      title          = "{Reaction operator approach to nonAbelian energy loss}",
      journal        = "Nucl.Phys.",
      volume         = "B594",
      pages          = "371-419",
      doi            = "10.1016/S0550-3213(00)00652-0",
      year           = "2001",
      eprint         = "nucl-th/0006010",
      archivePrefix  = "arXiv",
      primaryClass   = "nucl-th",
      reportNumber   = "CU-TP-979",
      SLACcitation   = "%%CITATION = NUCL-TH/0006010;%%",
}

@article{Gelis:2010nm,
      author         = "Gelis, Francois and Iancu, Edmond and Jalilian-Marian,
                        Jamal and Venugopalan, Raju",
      title          = "{The Color Glass Condensate}",
      journal        = "Ann.Rev.Nucl.Part.Sci.",
      volume         = "60",
      pages          = "463-489",
      doi            = "10.1146/annurev.nucl.010909.083629",
      year           = "2010",
      eprint         = "1002.0333",
      archivePrefix  = "arXiv",
      primaryClass   = "hep-ph",
      SLACcitation   = "%%CITATION = ARXIV:1002.0333;%%",
}

@article{Rao:2019vgy,
      author         = "Rao, S. and Sievert, M. and Noronha-Hostler, J.",
      title          = "{Baseline predictions of elliptic flow and fluctuations
                        at the RHIC Beam Energy Scan using response coefficients}",
      year           = "2019",
      eprint         = "1910.03677",
      archivePrefix  = "arXiv",
      primaryClass   = "nucl-th",
      SLACcitation   = "%%CITATION = ARXIV:1910.03677;%%"
}

@Article{	  bernhard:2016tnd,
  author	= "Bernhard, Jonah E. and Moreland, J. Scott and Bass,
		  Steffen A. and Liu, Jia and Heinz, Ulrich",
  title		= "{Applying Bayesian parameter estimation to relativistic
		  heavy-ion collisions: simultaneous characterization of the
		  initial state and quark-gluon plasma medium}",
  journal	= "Phys. Rev.",
  volume	= "C94",
  year		= "2016",
  number	= "2",
  pages		= "024907",
  doi		= "10.1103/PhysRevC.94.024907",
  eprint	= "1605.03954",
  archiveprefix	= "arXiv",
  primaryclass	= "nucl-th",
  slaccitation	= "%%CITATION = ARXIV:1605.03954;%%"
}

@article{Bernhard:2019bmu,
      author         = "Bernhard, Jonah E. and Moreland, J. Scott and Bass,
                        Steffen A.",
      title          = "{Bayesian estimation of the specific shear and bulk
                        viscosity of quark–gluon plasma}",
      journal        = "Nature Phys.",
      volume         = "15",
      year           = "2019",
      number         = "11",
      pages          = "1113-1117",
      doi            = "10.1038/s41567-019-0611-8",
      SLACcitation   = "%%CITATION = NPAHA,15,1113;%%"
}

@article{Luzum:2013yya,
      author         = "Luzum, Matthew and Petersen, Hannah",
      title          = "{Initial State Fluctuations and Final State Correlations
                        in Relativistic Heavy-Ion Collisions}",
      journal        = "J. Phys.",
      volume         = "G41",
      year           = "2014",
      pages          = "063102",
      doi            = "10.1088/0954-3899/41/6/063102",
      eprint         = "1312.5503",
      archivePrefix  = "arXiv",
      primaryClass   = "nucl-th",
      SLACcitation   = "%%CITATION = ARXIV:1312.5503;%%"
}

@article{Sadofyev:2021ohn,
    author = "Sadofyev, Andrey V. and Sievert, Matthew D. and Vitev, Ivan",
    title = "{Ab Initio Coupling of Jets to Collective Flow in the Opacity Expansion Approach}",
    eprint = "2104.09513",
    archivePrefix = "arXiv",
    primaryClass = "hep-ph",
    reportNumber = "LA-UR-21-21420",
    month = "4",
    year = "2021"
}

@article{Landau:1953um,
    author = "Landau, L. D. and Pomeranchuk, I.",
    title = "{Limits of applicability of the theory of bremsstrahlung electrons and pair production at high-energies}",
    journal = "Dokl. Akad. Nauk Ser. Fiz.",
    volume = "92",
    pages = "535--536",
    year = "1953"
}

@article{Migdal:1956tc,
    author = "Migdal, A. B.",
    title = "{Bremsstrahlung and pair production in condensed media at high-energies}",
    doi = "10.1103/PhysRev.103.1811",
    journal = "Phys. Rev.",
    volume = "103",
    pages = "1811--1820",
    year = "1956"
}

@article{JET:2013cls,
    author = "Burke, Karen M. and others",
    collaboration = "JET",
    title = "{Extracting the jet transport coefficient from jet quenching in high-energy heavy-ion collisions}",
    eprint = "1312.5003",
    archivePrefix = "arXiv",
    primaryClass = "nucl-th",
    reportNumber = "NT-LBNL-13-011",
    doi = "10.1103/PhysRevC.90.014909",
    journal = "Phys. Rev. C",
    volume = "90",
    number = "1",
    pages = "014909",
    year = "2014"
}

@article{Cao:2020wlm,
    author = "Cao, Shanshan and Wang, Xin-Nian",
    title = "{Jet quenching and medium response in high-energy heavy-ion collisions: a review}",
    eprint = "2002.04028",
    archivePrefix = "arXiv",
    primaryClass = "hep-ph",
    doi = "10.1088/1361-6633/abc22b",
    journal = "Rept. Prog. Phys.",
    volume = "84",
    number = "2",
    pages = "024301",
    year = "2021"
}

@article{CMS:2013lhm,
    author = "Chatrchyan, Serguei and others",
    collaboration = "CMS",
    title = "{Modification of Jet Shapes in PbPb Collisions at $\sqrt {s_{NN}} = 2.76$ TeV}",
    eprint = "1310.0878",
    archivePrefix = "arXiv",
    primaryClass = "nucl-ex",
    reportNumber = "CERN-PH-EP-2013-189, CMS-HIN-12-002",
    doi = "10.1016/j.physletb.2014.01.042",
    journal = "Phys. Lett. B",
    volume = "730",
    pages = "243--263",
    year = "2014"
}

@article{CMS:2019btm,
    author = "The CMS Collaboration",
    collaboration = "CMS",
    title = "{Measurement of Jet Nuclear Modification Factor in PbPb Collisions at $\sqrt{s_{NN}}$ = 5.02 TeV with CMS}",
    reportNumber = "CMS-PAS-HIN-18-014",
    year = "2019"
}

@article{Chien:2016led,
    author = "Chien, Yang-Ting and Vitev, Ivan",
    title = "{Probing the Hardest Branching within Jets in Heavy-Ion Collisions}",
    eprint = "1608.07283",
    archivePrefix = "arXiv",
    primaryClass = "hep-ph",
    doi = "10.1103/PhysRevLett.119.112301",
    journal = "Phys. Rev. Lett.",
    volume = "119",
    number = "11",
    pages = "112301",
    year = "2017"
}

@article{Larkoski:2014wba,
    author = "Larkoski, Andrew J. and Marzani, Simone and Soyez, Gregory and Thaler, Jesse",
    title = "{Soft Drop}",
    eprint = "1402.2657",
    archivePrefix = "arXiv",
    primaryClass = "hep-ph",
    reportNumber = "MIT-CTP-4531, DCPT-14-24, IPPP-14-12",
    doi = "10.1007/JHEP05(2014)146",
    journal = "JHEP",
    volume = "05",
    pages = "146",
    year = "2014"
}

@article{Moreland:2018gsh,
    author = "Moreland, J. Scott and Bernhard, Jonah E. and Bass, Steffen A.",
    title = "{Bayesian calibration of a hybrid nuclear collision model using p-Pb and Pb-Pb data at energies available at the CERN Large Hadron Collider}",
    eprint = "1808.02106",
    archivePrefix = "arXiv",
    primaryClass = "nucl-th",
    doi = "10.1103/PhysRevC.101.024911",
    journal = "Phys. Rev. C",
    volume = "101",
    number = "2",
    pages = "024911",
    year = "2020"
}

@article{Antiporda:2021hpk,
    author = "Antiporda, Logan and Bahder, Joseph and Rahman, Hasan and Sievert, Matthew D.",
    title = "{Jet Drift and Collective Flow in Heavy-Ion Collisions}",
    eprint = "2110.03590",
    archivePrefix = "arXiv",
    primaryClass = "hep-ph",
    month = "10",
    year = "2021"
}

@article{Sievert:2019cwq,
    author = "Sievert, Matthew D. and Vitev, Ivan and Yoon, Boram",
    title = "{A complete set of in-medium splitting functions to any order in opacity}",
    eprint = "1903.06170",
    archivePrefix = "arXiv",
    primaryClass = "hep-ph",
    doi = "10.1016/j.physletb.2019.06.019",
    journal = "Phys. Lett. B",
    volume = "795",
    pages = "502--510",
    year = "2019"
}

@article{Noronha-Hostler:2016eow,
    author = "Noronha-Hostler, Jacquelyn and Betz, Barbara and Noronha, Jorge and Gyulassy, Miklos",
    title = "{Event-by-event hydrodynamics $+$ jet energy loss: A solution to the $R_{AA} \otimes v_2$ puzzle}",
    eprint = "1602.03788",
    archivePrefix = "arXiv",
    primaryClass = "nucl-th",
    doi = "10.1103/PhysRevLett.116.252301",
    journal = "Phys. Rev. Lett.",
    volume = "116",
    number = "25",
    pages = "252301",
    year = "2016"
}

@article{Bierlich:2022pfr,
    author = "Bierlich, Christian and others",
    title = "{A comprehensive guide to the physics and usage of PYTHIA 8.3}",
    eprint = "2203.11601",
    archivePrefix = "arXiv",
    primaryClass = "hep-ph",
    reportNumber = "LU-TP 22-16, MCNET-22-04, FERMILAB-PUB-22-227-SCD",
    month = "3",
    year = "2022"
}

@article{Gyulassy:2000gk,
    author = "Gyulassy, M. and Vitev, I. and Wang, X. N.",
    title = "{High p(T) azimuthal asymmetry in noncentral A+A at RHIC}",
    eprint = "nucl-th/0012092",
    archivePrefix = "arXiv",
    doi = "10.1103/PhysRevLett.86.2537",
    journal = "Phys. Rev. Lett.",
    volume = "86",
    pages = "2537--2540",
    year = "2001"
}

@article{Braaten:1991we,
    author = "Braaten, Eric and Thoma, Markus H.",
    title = "{Energy loss of a heavy quark in the quark - gluon plasma}",
    reportNumber = "LBL-30998, NUHEP-TH-91-14",
    doi = "10.1103/PhysRevD.44.R2625",
    journal = "Phys. Rev. D",
    volume = "44",
    number = "9",
    pages = "R2625",
    year = "1991"
}

@article{Molnar:2013eqa,
    author = "Molnar, Denes and Sun, Deke",
    title = "{High-pT suppression and elliptic flow from radiative energy loss with realistic bulk medium expansion}",
    eprint = "1305.1046",
    archivePrefix = "arXiv",
    primaryClass = "nucl-th",
    month = "5",
    year = "2013"
}

@article{Zhao:2021vmu,
    author = "Zhao, Wenbin and Ke, Weiyao and Chen, Wei and Luo, Tan and Wang, Xin-Nian",
    title = "{From Hydrodynamics to Jet Quenching, Coalescence, and Hadron Cascade: A Coupled Approach to Solving the RAA\ensuremath{\otimes}v2 Puzzle}",
    eprint = "2103.14657",
    archivePrefix = "arXiv",
    primaryClass = "hep-ph",
    doi = "10.1103/PhysRevLett.128.022302",
    journal = "Phys. Rev. Lett.",
    volume = "128",
    number = "2",
    pages = "022302",
    year = "2022"
}

@article{Zhang:2013oca,
    author = "Zhang, Xilin and Liao, Jinfeng",
    title = "{Jet Quenching and Its Azimuthal Anisotropy in AA and possibly High Multiplicity pA and dA Collisions}",
    eprint = "1311.5463",
    archivePrefix = "arXiv",
    primaryClass = "nucl-th",
    month = "11",
    year = "2013"
}

@article{Kopeliovich:2012sc,
    author = "Kopeliovich, B. Z. and Nemchik, J. and Potashnikova, I. K. and Schmidt, Ivan",
    title = "{Quenching of high-pT hadrons: Energy Loss vs Color Transparency}",
    eprint = "1208.4951",
    archivePrefix = "arXiv",
    primaryClass = "hep-ph",
    reportNumber = "USM-TH-302",
    doi = "10.1103/PhysRevC.86.054904",
    journal = "Phys. Rev. C",
    volume = "86",
    pages = "054904",
    year = "2012"
}

@article{Andres:2019eus,
    author = "Andres, Carlota and Armesto, N\'estor and Niemi, Harri and Paatelainen, Risto and Salgado, Carlos A.",
    title = "{Jet quenching as a probe of the initial stages in heavy-ion collisions}",
    eprint = "1902.03231",
    archivePrefix = "arXiv",
    primaryClass = "hep-ph",
    reportNumber = "JLAB-THY-19-2888, CERN-TH-2019-012",
    doi = "10.1016/j.physletb.2020.135318",
    journal = "Phys. Lett. B",
    volume = "803",
    pages = "135318",
    year = "2020"
}

@inproceedings{osg07,
  title  = {The open science grid},
  author = {
    Pordes, Ruth
    and Petravick, Don
    and Kramer, Bill
    and Olson, Doug
    and Livny, Miron
    and Roy, Alain
    and Avery, Paul
    and Blackburn, Kent
    and Wenaus, Torre
    and W{\"u}rthwein, Frank
    and Foster, Ian
    and Gardner, Rob
    and Wilde, Mike
    and Blatecky, Alan
    and McGee, John
    and Quick, Rob
  },
  doi       = {10.1088/1742-6596/78/1/012057},
  booktitle = {J. Phys. Conf. Ser.},
  volume    = {78},
  series    = {78},
  pages     = {012057},
  year      = {2007},
}

@inproceedings{osg09,
  title        = {The pilot way to grid resources using glideinWMS},
  author       = {
    Sfiligoi, Igor
    and Bradley, Daniel C
    and Holzman, Burt
    and Mhashilkar, Parag
    and Padhi, Sanjay
    and Wurthwein, Frank
  },
  doi          = {10.1109/CSIE.2009.950},
  booktitle    = {2009 WRI World Congress on Computer Science and Information Engineering},
  volume       = {2},
  series       = {2},
  pages        = {428--432},
  year         = {2009},
}

@misc{https://doi.org/10.21231/906p-4d78,
  doi = {10.21231/906P-4D78},
  author = {{OSG}},
  title = {\href{https://osg-htc.org/services/open\_science\_pool.html}{OSPool}},
  publisher = {OSG},
  year = {2006}
}

@misc{https://doi.org/10.21231/0kvz-ve57,
  doi = {10.21231/0KVZ-VE57},
  url = {https://osdf.osg-htc.org/},
  author = {{OSG}},
  title = {Open Science Data Federation},
  publisher = {OSG},
  year = {2015}
}

@article{Connors:2017ptx,
    author = "Connors, Megan and Nattrass, Christine and Reed, Rosi and Salur, Sevil",
    title = "{Jet measurements in heavy ion physics}",
    eprint = "1705.01974",
    archivePrefix = "arXiv",
    primaryClass = "nucl-ex",
    doi = "10.1103/RevModPhys.90.025005",
    journal = "Rev. Mod. Phys.",
    volume = "90",
    pages = "025005",
    year = "2018"
}

@article{Kolbe:2015rvk,
    author = "Kolbe, Isobel and Horowitz, W. A.",
    title = "{Short path length corrections to Djordjevic-Gyulassy-Levai-Vitev energy loss}",
    eprint = "1511.09313",
    archivePrefix = "arXiv",
    primaryClass = "hep-ph",
    doi = "10.1103/PhysRevC.100.024913",
    journal = "Phys. Rev. C",
    volume = "100",
    number = "2",
    pages = "024913",
    year = "2019"
}

@article{CMS:2024krd,
    author = "Hayrapetyan, Aram and others",
    collaboration = "CMS",
    title = "{Overview of high-density QCD studies with the CMS experiment at the LHC}",
    eprint = "2405.10785",
    archivePrefix = "arXiv",
    primaryClass = "nucl-ex",
    reportNumber = "CMS-HIN-23-011, CERN-EP-2024-057",
    month = "5",
    year = "2024"
}

@article{Valle:2013,
author = {Valle, Jose and Boer, Jan},
year = {2013},
month = {07},
pages = {},
title = {Grand challenges in High Energy and Astroparticle Physics},
volume = {1},
journal = {Frontiers in Physics},
doi = {10.3389/fphy.2013.00005}
}

@article{Bazavov2012,
  author = {Bazavov, A. and others},
  collaboration = {HotQCD Collaboration},
  title = {The chiral and deconfinement aspects of the QCD transition},
  journal = {Physical Review D},
  volume = {85},
  number = {5},
  pages = {054503},
  year = {2012},
  doi = {10.1103/PhysRevD.85.054503}
}

@article{LESGOURGUES2006307,
  title = {Massive neutrinos and cosmology},
  journal = {Physics Reports},
  volume = {429},
  number = {6},
  pages = {307--379},
  year = {2006},
  issn = {0370-1573},
  doi = {10.1016/j.physrep.2006.04.001},
  url = {https://www.sciencedirect.com/science/article/pii/S0370157306001359},
  author = {Julien Lesgourgues and Sergio Pastor}
}

@article{FUKUGITA198645,
  title = {Baryogenesis without grand unification},
  journal = {Physics Letters B},
  volume = {174},
  number = {1},
  pages = {45--47},
  year = {1986},
  issn = {0370-2693},
  doi = {10.1016/0370-2693(86)91126-3},
  url = {https://www.sciencedirect.com/science/article/pii/0370269386911263},
  author = {M. Fukugita and T. Yanagida}
}

@article{KATRIN:2024cdt,
    author = "Aker, Max and others",
    collaboration = "KATRIN",
    title = "{Direct neutrino-mass measurement based on 259 days of KATRIN data}",
    eprint = "2406.13516",
    archivePrefix = "arXiv",
    primaryClass = "nucl-ex",
    doi = "10.1126/science.adq9592",
    journal = "Science",
    volume = "388",
    number = "6743",
    pages = "adq9592",
    year = "2025"
}

@article{LandauPomeranchuk1953,
  author = {Landau, L. D. and Pomeranchuk, I. J.},
  title = {The limits of applicability of the theory of bremsstrahlung by electrons and creation of pairs at high energies},
  journal = {Dokl. Akad. Nauk Ser. Fiz.},
  volume = {92},
  pages = {535--536},
  year = {1953}
}

@article{Migdal1956,
  author = {Migdal, A. B.},
  title = {Bremsstrahlung and Pair Production in Condensed Media at High Energies},
  journal = {Physical Review},
  volume = {103},
  number = {6},
  pages = {1811--1820},
  year = {1956},
  doi = {10.1103/PhysRev.103.1811}
}

@misc{Heinz:2024jwu,
    author = {Heinz, Ulrich and Schenke, Bj{\"o}rn},
    title = "{Hydrodynamic Description of the Quark-Gluon Plasma}",
    eprint = "2412.19393",
    archivePrefix = "arXiv",
    primaryClass = "nucl-th",
    month = "12",
    year = "2024"
}

@article{PHENIX:2015tbb,
    author = "Adare, A. and others",
    collaboration = "PHENIX",
    title = "{Transverse energy production and charged-particle multiplicity at midrapidity in various systems from $\sqrt{s_{NN}}=7.7$ to 200 GeV}",
    eprint = "1509.06727",
    archivePrefix = "arXiv",
    primaryClass = "nucl-ex",
    doi = "10.1103/PhysRevC.93.024901",
    journal = "Phys. Rev. C",
    volume = "93",
    number = "2",
    pages = "024901",
    year = "2016"
}

@article{Wertepny:2019yye,
    author = "Wertepny, Douglas and Noronha-Hostler, Jacquelyn and Sievert, Matthew and Rao, Skandaprasad and Paladino, Noah",
    editor = "Liu, Feng and Wang, Enke and Wang, Xin-Nian and Xu, Nu and Zhang, Ben-Wei",
    title = "{Ultracentral Collisions of Small and Deformed Systems at RHIC}",
    eprint = "1905.13323",
    archivePrefix = "arXiv",
    primaryClass = "hep-ph",
    doi = "10.1016/j.nuclphysa.2020.121839",
    journal = "Nucl. Phys. A",
    volume = "1005",
    pages = "121839",
    year = "2021"
}

@article{Acharya_2024,
   title={Measurement of the radius dependence of charged-particle jet suppression in Pb–Pb collisions at $\sqrt{s_{NN}} = 5.02$ TeV},
   volume={849},
   ISSN={0370-2693},
   url={http://dx.doi.org/10.1016/j.physletb.2023.138412},
   DOI={10.1016/j.physletb.2023.138412},
   journal={Physics Letters B},
   publisher={Elsevier BV},
   author={Acharya, S. and others},
   collaboration = "ALICE",
   year={2024},
   month={feb}, 
   pages={138412} 
}

@article{Berges:2020yzp, 
    author = "Berges, J. and Heller, M. P. and Mazeliauskas, A. and Venugopalan, R.",
    title = "{QCD thermalization: Ab initio approaches and interdisciplinary connections}",
    journal = "Rev. Mod. Phys.",
    volume = "93",
    number = "3",
    pages = "035003",
    year = "2021",
    eprint = "2005.12299",
    archivePrefix = "arXiv",
    primaryClass = "hep-ph",
    doi = "10.1103/RevModPhys.93.035003"
}

@article{Heller:2015dha,
    author = "Heller, Michal P. and Spalinski, Michal",
    title = "{Hydrodynamics Beyond the Hydrodynamic Gradient Expansion}",
    journal = "Phys. Rev. Lett.",
    volume = "115",
    number = "7",
    pages = "072501",
    year = "2015",
    eprint = "1503.07514",
    archivePrefix = "arXiv",
    primaryClass = "hep-th",
    doi = "10.1103/PhysRevLett.115.072501"
}

@article{Giacalone:2020dln, 
    author = "Giacalone, Giuliano and Schenke, Bj{\"{o}}rn and Teaney, Derek",
    title = "{Correlation between mean transverse momentum and anisotropic flow in heavy-ion collisions}",
    eprint = "2006.15721",
    archivePrefix = "arXiv",
    primaryClass = "nucl-th",
    doi = "10.1103/PhysRevC.103.024910",
    journal = "Phys. Rev. C",
    volume = "103",
    number = "2",
    pages = "024910",
    year = "2021"
}

@article{ALICE:2022mno,
    author = "{ALICE Collaboration}",
    title = "{Characterizing the initial conditions of heavy-ion collisions at the LHC with mean transverse momentum and anisotropic flow correlations}",
    journal = "Phys. Lett. B",
    volume = "834",
    pages = "137393",
    year = "2022",
    doi = "10.1016/j.physletb.2022.137393"
}

@article{Sjostrand:2006za,
    author        = "Sj{\"o}strand, Torbj{\"o}rn and Mrenna, Stephen and Skands, Peter Z.",
    title         = "{PYTHIA 6.4 Physics and Manual}",
    journal       = "JHEP",
    volume        = "05",
    pages         = "026",
    year          = "2006",
    doi           = "10.1088/1126-6708/2006/05/026",
    eprint        = "hep-ph/0603175",
    archivePrefix = "arXiv"
}

@article{Sjostrand:2014zea,
    author        = "Sj{\"o}strand, Torbj{\"o}rn and Ask, Stefan and Christiansen, Jesper R. and Corke, Richard and Desai, Nishita and Ilten, Philip and Mrenna, Stephen and Prestel, Stefan and Rasmussen, Christine O. and Skands, Peter Z.",
    title         = "{An introduction to PYTHIA 8.2}",
    journal       = "Comput. Phys. Commun.",
    volume        = "191",
    pages         = "159--177",
    year          = "2015",
    doi           = "10.1016/j.cpc.2015.01.024",
    eprint        = "1410.3012",
    archivePrefix = "arXiv",
    primaryClass  = "hep-ph"
}

@article{PHENIX:2018lia,
    author         = "Aidala, Christine and others",
    collaboration  = "PHENIX",
    title          = "{Creation of quark-gluon plasma droplets with three distinct geometries}",
    journal        = "Nature Phys.",
    volume         = "15",
    pages          = "214--220",
    year           = "2019",
    doi            = "10.1038/s41567-018-0360-0",
    eprint         = "1805.02973",
    archivePrefix  = "arXiv",
    primaryClass   = "nucl-ex"
}

@article{Nambu:2009zza,
    author = "Nambu, Yoichiro",
    title = "{Nobel Lecture: Spontaneous symmetry breaking in particle physics: A case of cross fertilization}",
    doi = "10.1103/RevModPhys.81.1015",
    journal = "Rev. Mod. Phys.",
    volume = "81",
    pages = "1015--1018",
    year = "2009"
}

@article{Most:2022yhe,
    author = "Most, Elias R. and others",
    title = "{Fast Neutrino Flavor Conversions Can Prevent and Defer Supernova Explosions}",
    eprint = "2207.00569",
    archivePrefix = "arXiv",
    primaryClass = "astro-ph.HE",
    doi = "10.1103/PhysRevLett.129.251102",
    journal = "Phys. Rev. Lett.",
    volume = "129",
    number = "25",
    pages = "251102",
    year = "2022"
}

@article{CCM:2025dbq,
    author = "Aguilar-Arevalo, A. A. and others",
    collaboration = "CCM",
    title = "{Measurement of the liquid argon scintillation pulse shape using differentiable simulation in the coherent CAPTAIN-Mills experiment}",
    eprint = "2507.08887",
    archivePrefix = "arXiv",
    primaryClass = "physics.ins-det",
    reportNumber = "LA-UR-25-26474, FERMILAB-PUB-25-0463-PPD",
    doi = "10.1103/34mf-z6l9",
    journal = "Phys. Rev. D",
    volume = "112",
    number = "7",
    pages = "072010",
    year = "2025"
}


%% file: references.bib
@article{Bahcall_2005,
   author = {Bahcall, John N. and Serenelli, Aldo M. and Basu, Sarbani},
   title = {New Solar Opacities, Abundances, Helioseismology, and Neutrino Fluxes},
   journal = {The Astrophysical Journal},
   volume = {621},
   number = {1},
   pages = {L85--L88},
   year = {2005},
   month = {jan},
   doi = {10.1086/428929}
}

@article{1934ZPhy...88..161F,
    author = {Fermi, E.},
    title = "{Versuch einer Theorie der {\ensuremath{\beta}}-Strahlen. I}",
    journal = {Zeitschrift f{\"u}r Physik},
    volume = {88},
    pages = {161--177},
    year = {1934},
    month = {mar},
    doi = {10.1007/BF01351864}
}

@ARTICLE{1978PhT....31i..23B,
       author = {{Brown}, Laurie M.},
        title = "{The idea of the neutrino}",
      journal = {Physics Today},
         year = 1978,
        month = {sep},
       volume = {31},
       number = {9},
        pages = {23-28},
          doi = {10.1063/1.2995181},
       adsurl = {https://ui.adsabs.harvard.edu/abs/1978PhT....31i..23B}
}

@book{FClose:2012,
    author = "Close, Frank",
    title = "{Neutrino}",
    isbn = "978-0-199-69599-7",
    publisher = "Oxford University Press",
    year = "2012"
}

@book{RJayawardhana:2015,
    author = "Jayawardhana, Ray",
    title = "{The Neutrino Hunters: The chase for the ghost particle and the secrets of the universe}",
    isbn = "978-1-780-74647-0",
    publisher = "Oneworld Publications",
    year = "2015"
}

@article{Formaggio_2012,
   title={From eV to EeV: Neutrino cross sections across energy scales},
   volume={84},
   ISSN={1539-0756},
   url={http://dx.doi.org/10.1103/RevModPhys.84.1307},
   DOI={10.1103/revmodphys.84.1307},
   number={3},
   journal={Reviews of Modern Physics},
   publisher={American Physical Society (APS)},
   author={Formaggio, J. A. and Zeller, G. P.},
   year={2012},
   month={sep}, pages={1307–1341} }

@article{Lesgourgues:2006nd, 
    author = "Lesgourgues, Julian and Pastor, Sergio",
    title = "{Massive neutrinos and cosmology}",
    journal = "Phys. Rept.",
    volume = "429",
    pages = "307--379",
    year = "2006",
    doi = "10.1016/j.physrep.2006.04.001",
    eprint = "astro-ph/0603494",
    archivePrefix = "arXiv"
}

@book{Weinberg:2008zzc,
    author = "Weinberg, Steven",
    title = "{Cosmology}",
    publisher = "Oxford University Press",
    year = "2008",
    isbn = "978-0-19-852682-7"
}

@book{Weinberg:1995mt, author = "Weinberg, Steven", title = "{The Quantum theory of fields. Vol. 1: Foundations}", doi = "10.1017/CBO9781139644167", isbn = "978-0-521-67053-1, 978-0-511-25204-4", publisher = "Cambridge University Press", month = "6", year = "2005" }

@article{Katori:2016yel,
    author = "Katori, Teppei and Martini, Marco",
    title = "{Neutrino\textendash{}nucleus cross sections for oscillation experiments}",
    eprint = "1611.07770",
    archivePrefix = "arXiv",
    primaryClass = "hep-ph",
    doi = "10.1088/1361-6471/aa8bf7",
    journal = "J. Phys. G",
    volume = "45",
    number = "1",
    pages = "013001",
    year = "2018"
}

@article{PhysRevD.9.1389,
  title = {Coherent effects of a weak neutral current},
  author = {Freedman, Daniel Z.},
  journal = {Phys. Rev. D},
  volume = {9},
  issue = {5},
  pages = {1389--1392},
  numpages = {0},
  year = {1974},
  month = {Mar},
  publisher = {American Physical Society},
  doi = {10.1103/PhysRevD.9.1389},
  url = {https://link.aps.org/doi/10.1103/PhysRevD.9.1389}
}

@article{coherentcollaboration2016coherent,
  title   = "{The {COHERENT} Experiment at the Spallation Neutron Source}",
  author  = "{COHERENT Collaboration} and Akimov, D. and others",
  journal = "arXiv",
  volume  = "1509.08702",
  pages   = "v1",
  year    = "2016",
  note    = "physics.ins-det"
}

@article{Strumia_2003,
   title={Precise quasielastic neutrino/nucleon cross-section},
   volume={564},
   ISSN={0370-2693},
   url={http://dx.doi.org/10.1016/S0370-2693(03)00616-6},
   DOI={10.1016/s0370-2693(03)00616-6},
   number={1–2},
   journal={Physics Letters B},
   publisher={Elsevier BV},
   author={Strumia, Alessandro and Vissani, Francesco},
   year={2003},
   month={jul}, pages={42–54} }

@inproceedings{Sobczyk_2011,
   title={Quasi-elastic Neutrino Scattering—an Overview},
   ISSN={0094-243X},
   url={http://dx.doi.org/10.1063/1.3661558},
   DOI={10.1063/1.3661558},
   booktitle={AIP Conference Proceedings},
   publisher={AIP},
   author={Sobczyk, Jan T. and Singh, S. K. and Morfin, J. G. and Sakuda, Makoto and Purohit, K. D.},
   year={2011} }

@inproceedings{LBNE:2013dhi,
    author = "Adams, Corey and others",
    collaboration = "LBNE",
    title = "{The Long-Baseline Neutrino Experiment: Exploring Fundamental Symmetries of the Universe}",
    booktitle = "{Snowmass 2013}: {Workshop on Energy Frontier}",
    eprint = "1307.7335",
    archivePrefix = "arXiv",
    primaryClass = "hep-ex",
    reportNumber = "BNL-101354-2013-JA, BNL-101354-2014-JA, FERMILAB-PUB-14-022, LA-UR-14-20881, BNL-101354-2014-JA, FERMILAB-PUB-14-022, LA-UR-14-20881",
    month = "7",
    year = "2013"
}

@article{snowmass,
    author  = {Arg{\"u}elles, C. A. and others},
    title   = "{Snowmass White Paper: Beyond the Standard Model Effects on Neutrino Flavor}",
    journal = "Eur. Phys. J. C",
    volume  = "83",
    pages   = "15",
    year    = "2023",
    doi     = "10.1140/epjc/s10052-022-11049-7",
    note    = "arXiv:2203.10811 [hep-ph]"
}

@article{shoemaker2021sailing,
    author  = {Shoemaker, Ian M. and Welch, Eli},
    title   = "{Sailing the {CE$\nu$NS} Seas of Non-Standard Neutrino Interactions with the {COHERENT} {CAPTAIN-Mills} Experiment}",
    journal = "arXiv",
    volume  = "2103.08401",
    pages   = "v1",
    year    = "2021",
    note    = "hep-ph"
}

@article{Dunton:2022dez,
    author  = "Dunton, Edward C.",
    title   = "{A Search for Axion-like Particles at the {COHERENT} {CAPTAIN-Mills} Experiment}",
    journal = "PhD Thesis, Columbia University",
    volume  = "2022",
    pages   = "10.7916/x9x1-ka48",
    year    = "2022",
    doi     = "10.7916/x9x1-ka48"
}

@article{PhysRev.52.378,
  title = {Visible Radiation Produced by Electrons Moving in a Medium with Velocities Exceeding that of Light},
  author = {\ifmmode \check{C}\else \v{C}\fi{}erenkov, P. A.},
  journal = {Phys. Rev.},
  volume = {52},
  issue = {4},
  pages = {378--379},
  numpages = {0},
  year = {1937},
  month = {Aug},
  publisher = {American Physical Society},
  doi = {10.1103/PhysRev.52.378},
  url = {https://link.aps.org/doi/10.1103/PhysRev.52.378}
}

@article{Babicz:2020den,
    author = "Babicz, M. and others",
    title = "{A measurement of the group velocity of scintillation light in liquid argon}",
    eprint = "2002.09346",
    archivePrefix = "arXiv",
    primaryClass = "physics.ins-det",
    reportNumber = "FERMILAB-PUB-20-397-ND",
    doi = "10.1088/1748-0221/15/09/P09009",
    journal = "JINST",
    volume = "15",
    number = "09",
    pages = "P09009",
    year = "2020"
}

@article{Grace:2015yta,
    author = "Grace, Emily and Nikkel, James A.",
    title = "{Index of refraction, Rayleigh scattering length, and Sellmeier coefficients in solid and liquid argon and xenon}",
    eprint = "1502.04213",
    archivePrefix = "arXiv",
    primaryClass = "physics.ins-det",
    doi = "10.1016/j.nima.2017.06.031",
    journal = "Nucl. Instrum. Meth. A",
    volume = "867",
    pages = "204--208",
    year = "2017"
}

@article{Sinnock:1969zz,
    author = "Sinnock, A. C. and Smith, B. L.",
    title = "{Refractive Indices of the Condensed Inert Gases}",
    doi = "10.1103/PhysRev.181.1297",
    journal = "Phys. Rev.",
    volume = "181",
    pages = "1297--1307",
    year = "1969"
}

@book{Jackson:1998nia,
    author = "Jackson, John David",
    title = "{Classical Electrodynamics}",
    isbn = "978-0-471-30932-1",
    publisher = "Wiley",
    year = "1998"
}

@book{griffiths_2017, place={Cambridge}, edition={4}, title={Introduction to Electrodynamics}, DOI={10.1017/9781108333511}, publisher={Cambridge University Press}, author={Griffiths, David J.}, year={2017}}

@article{BNL:2023,
    author  = "{Brookhaven National Laboratory}",
    title   = "{Liquid Argon Properties (Tables and Calculators)}",
    journal = "BNL Technical Website",
    volume  = "Online",
    pages   = "Accessed 03-02-2023",
    year    = "2023",
    note    = "\url{https://lar.bnl.gov/properties/}"
}

@article{Workman:2022ynf,
    author = "Workman, R. L. and others",
    collaboration = "Particle Data Group",
    title = "{Review of Particle Physics}",
    doi = "10.1093/ptep/ptac097",
    journal = "PTEP",
    volume = "2022",
    pages = "083C01",
    year = "2022"
}

@article{Lippmann_2012,
  title = {Particle identification},
  author = {Christian Lippmann},
  journal = {Nuclear Instruments and Methods in Physics Research Section A: Accelerators, Spectrometers, Detectors and Associated Equipment},
  volume = {666},
  pages = {148--172},
  year = {2012},
  month = {feb},
  publisher = {Elsevier {BV}},
  doi = {10.1016/j.nima.2011.03.009},
  url = {https://doi.org/10.1016/j.nima.2011.03.009}
}

@article{Segreto_2021,
  title = {Properties of liquid argon scintillation light emission},
  author = {Segreto, Ettore},
  journal = {Physical Review D},
  volume = {103},
  number = {4},
  pages = {043001},
  year = {2021},
  month = {feb},
  publisher = {American Physical Society ({APS})},
  doi = {10.1103/physrevd.103.043001},
  url = {https://doi.org/10.1103/PhysRevD.103.043001}
}

@article{athar2022neutrinosinteractionsmatter,
  author = {Sajjad Athar, M. and Fatima, A. and Singh, S. K.},
  title = {Neutrinos and their interactions with matter},
  journal = {arXiv preprint},
  volume = {2206.13792},
  pages = {1--212},
  year = {2022},
  doi = {10.48550/arXiv.2206.13792},
  url = {https://arxiv.org/abs/2206.13792},
  note = {hep-ph}
}

@article{aguilararevalo2025measurementliquidargonscintillation,
  author = {Aguilar-Arevalo, A. A. and Biedron, S. and Boissevain, J. and others},
  title = {Measurement of the {Liquid Argon} Scintillation Pulse Shape Using Differentiable Simulation in the {Coherent CAPTAIN-Mills} Experiment},
  journal = {arXiv preprint},
  volume = {2507.08887},
  pages = {1--15},
  year = {2025},
  doi = {10.48550/arXiv.2507.08887},
  url = {https://arxiv.org/abs/2507.08887},
  note = {physics.ins-det}
}

@article{rahman2024,
    author        = {Rahman, Hasan R.},
    title         = "{Calculating the Total Cherenkov Radiation Emitted by Low Energy Protons in Liquid Argon and Comparing with Argon Scintillation Light at 128 nm}",
    journal       = "arXiv preprint",
    volume        = "2408.00817",
    pages         = "v1",
    year          = "2024",
    eprint        = "2408.00817",
    archivePrefix = "arXiv",
    primaryClass  = "hep-ph",
    url           = {https://arxiv.org/abs/2408.00817}
}

@article{Beacom:2003zu,
    author = "Beacom, John F. and Palomares-Ruiz, Sergio",
    title = "{Neutral Current Atmospheric Neutrino Flux Measurement Using Neutrino Proton Elastic Scattering in Super-Kamiokande}",
    eprint = "hep-ph/0301060",
    archivePrefix = "arXiv",
    reportNumber = "FERMILAB-PUB-03-003-A, FTUV-03-0109, IFIC-03-01",
    doi = "10.1103/PhysRevD.67.093001",
    journal = "Phys. Rev. D",
    volume = "67",
    pages = "093001",
    year = "2003"
}
